# A flexible approach to sequential prediction under intervention

Matthew Sperrin[1], Bowen Jiang[1], Joyce Huang[1,2], Pedro Henrique Da Costa Avelar[1], Niels Peek[3], Alexander Pate[1].

1: Division of Informatics, Imaging and Data Science, Faculty of Biology, Medicine and Health, University of Manchester, UK

2: Centre for Pharmacoepidemiology and Drug Safety, Division of Pharmacy and Optometry, University of Manchester

3: THIS Institute, Department of Public Health and Primary Care, University of Cambridge, UK.

Correspondence to: Matthew Sperrin, Vaughan House, University of Manchester, Manchester, M13 9PL, matthew.sperrin@manchester.ac.uk,

## Abstract

**Background**: Prognostic clinical prediction models (CPMs) inform decisions about interventions targeting modifiable risk factors, but do not respect causality. Proposed solutions, such as fixing coefficients to causal values, lead to issues making multiple predictions over time, and lack flexibility for multiple interventions. We aimed to develop a prediction under intervention approach that allows for sequential predictions, with flexibility for multiple interventions, while retaining good predictive performance.

**Methods**: We propose a causal-predictive framework for estimating risk under preventative interventions. The Unexposed Mediator Model maintains mediators that are also predictors at their unexposed level, removing double-counting of intervention effects at follow-up visits. The Modifiable Risk Factor Model handles multiple interventions flexibly by modelling their effects via mediators that are also predictors, assuming a known causal structure. The Two-Component Model combines a predictive baseline model with an intervention model to improve predictive performance. We illustrate the framework in primary prevention of cardiovascular disease.

**Results**: Models were applied to 4 million individuals from the UK Clinical Practice Research Datalink. For anti-hypertensive intervention, the Treatment Offset model produced inconsistent risk estimates over time; the Unexposed Mediator Model resolved this while maintaining discrimination. The Modifiable Risk Factor Model accommodated multiple interventions but calibration was degraded compared with earlier models (integrated calibration index 0.005); this was resolved by the Two-Component Model (integrated calibration index 0.001).

**Conclusion**: The proposed models allow arbitrary interventions to be evaluated within a prediction-under-intervention framework, with causally consistent risk estimates across repeated visits. Limitations include reliance on predictor values from an arbitrary first visit, requirements for causal structural knowledge, and a consistency assumption — that interventions with identical effects on predictors have identical effects on outcomes — which warrant further investigation.

## 1. Introduction

Clinical prediction models (CPMs) estimate individual risk based on known characteristics. They are usually developed using supervised learning techniques such as regression, while use of flexible AI approaches such as transformers shows significant potential [1]. A common application of such models, and the motivation for this work, is primary prevention. Here, interventions are proposed, usually targeting modifiable risk factors, to reduce future disease risk. A well-known example is QRISK, which predicts 10-year risk of stroke or heart attack [2], and is used to help decide upon use of statins and promotion of lifestyle interventions.

A limitation of such CPMs is that they do not contain the causal machinery required to make 'prediction under intervention' (PUI). For example, "what is my risk if I take statins" versus "what is my risk if I improve my lifestyle" versus "what is my risk if I make no changes" [3,4]. This is important because these models are often used to support such decisions and therefore availability of such outputs would be of clear benefit to decision support. As such, there has been growing interest in methods for PUI, a field at the intersection of causal inference and prediction, which allows us to make accurate predictions under specific interventions.

Most approaches to PUI place predictor variables in two categories – intervention and non-intervention variables. Intervention variables are fixed to their causal values in the model – the 'offset' approach (so-called because the fixed terms are specified as offsets in statistical software). Implementations of this include allowing for heterogeneity in treatment effects, and estimating effects from the same observational data used to fit the model versus acquiring them elsewhere – as summarised by Lin et al [5].

While these approaches are a step forward, several limitations and challenges remain to be addressed. We will use primary prevention in cardiovascular disease (CVD) as an example. First, people are evaluated repeatedly by the same model. For example, in UK primary care, individuals should undergo cardiovascular risk evaluation every five years. Therefore, models must appropriately estimate causal contrasts across time – *sequential prediction under intervention* [6]. Second, many interventions are possible, including pharmaceutical (statins, anti-hypertensives, etc), and non-pharmaceutical (including an array of diet/exercise related interventions). We therefore require *flexibility for multiple interventions.* Third, we would like the model to retain *predictive accuracy.* As we will go on to show, existing approaches fail to address these problems.

In this paper we aim to develop flexible approaches to PUI that address these concerns: allowing for sequential PUI and flexibility for multiple interventions, while retaining predictive accuracy. The paper is organised as follows. In Section 2, we introduce notation and give precise definitions of the three targeted characteristics. Correspondingly we introduce modelling approaches that possess these characteristics. We have chosen to introduce a series of approaches both for simpler exposition, and because intermediate approaches (that satisfy some but not all the targeted characteristics) are useful under certain circumstances. Section 3 contains a clinical example. The purpose of the example is illustration of the approaches – a companion applied paper [7] describes the development and internal validation of a PUI model for primary prevention of CVD using the techniques described in this paper. Section 4 concludes with a discussion and future directions.

# 2. Methods

## 2.1. Setting and targeted features for a sequential PUI model

We begin by formally establishing the setting and outlining the set of targeted characteristics that a PUI model should possess. We follow notation outlined in Luijken et al [6]. Suppose we are concerned with predicting an outcome $Y$ (e.g., binary or time to event) using predictors $X$ and potential interventions $A$, and we have data comprising a random sample of the target population. We consider evaluating risk at time points $t_0, t_1, t_2, \ldots$ (which may be regularly or irregularly spaced), where $t_0$ is the initial visit time and $t_1, t_2, \ldots$ represent follow-ups (this distinction will be critical later). Define the time horizon in which we evaluate $Y$ relative to the prediction time as $w$, and $h_k = t_k + w$. Therefore, our models are of the form $f_w(X_{t_k}, A_{t_k})$. In some circumstances it is necessary to separate predictors $X$ into modifiable risk factors $M$ and non-modifiable risk factors $Z$, i.e., $X = (M, Z)$. A risk factor is modifiable if at least one intervention in $A$ can change its value. For clarity and simplicity, we will not consider predictor or intervention history in this exposition, but see the Discussion for further remarks on this. We drop the $w$ and $t_k$ where this does not lead to ambiguity. We use potential outcome notation to denote causal estimands, for example $Y(A = a)$ or $Y(a)$ denote the value of $Y$ that would be observed if we (potentially counterfactually) intervened to fix $A$ to $a$. We are primarily interested in evaluating absolute risk under intervention, i.e., causal estimands of the form $E[Y(a)|X]$.

As is well-known, a standard **Non-Causal Model** does not target $E[Y(a)|X]$. One solution is to use a **Treatment Offset Model**, in which the intervention's coefficient is fixed to a causal estimate, $\beta_A^{causal}$, that may be obtained from existing evidence (e.g., a meta-analysis of randomised controlled trials), or the same data used to develop the model [3]. The required causal estimate is a conditional average treatment effect derived from a model with the same conditioning set $X$. Even if there is no treatment effect heterogeneity, using a marginal average treatment effect (or any treatment effect estimated with a different conditioning set) introduces bias, unless the scale is collapsible [8,9]. Transportability between the population in which the causal estimate was estimated and the target population, and no unmeasured confounding, are also required. With this conditions met, the treatment offset model targets [8]

$$f^{TO}(X, A = a) = g(E[Y(a)|X]).$$

The function $g$ is a link function matching the scale of the causal estimate and model (e.g., the logit function for binary outcomes, complementary log-log for proportional hazards models). For survival outcomes, estimands of interest often correspond to treatment remaining constant during follow-up [3,4], so the model must adjust for changes in treatment during follow-up in the development data. This can be achieved by deriving interval censored data for when an individual is considered "on" or "off" treatment, and the cumulative hazard adjusted when the treatment changes using the appropriate causal estimate [8,10]. Further details on these points are in Supplementary file 1, Section 4.2.1.

In developing our sequential PUI model we target the following characteristics:

1. The model should behave in a concordant way across predictions for the same individual at multiple times: *sequential PUI*. Therefore, PUI made at an initial timepoint should agree (in expectation) with the actual risk calculated at a subsequent timepoint, if the individual followed that intervention and all else remained the same.
2. The model should allow *flexibility for multiple interventions* – $A$ should not need to be pre-specified and $\dim(A)$ is allowed to be large.
3. The model should reconcile the tension between predictive and causal optimality, ensuring that intervention variables contribute to robust prediction (maintaining low bias and high accuracy in this sense), while adhering to characteristics 1 and 2.

We will show why the Treatment Offset Model does not achieve these characteristics and develop models that do so.

## 2.2. Sequential Prediction Under Intervention (Characteristic 1)

### 2.2.1. Outline of the issue

If the Treatment Offset Model is required to be used more than once in the same individual, we desire predictions that are concordant over time. Consider predicting 10-year cardiovascular risk, $Y$, with predictors including systolic blood pressure (SBP), $X$, and interventions include use of anti-hypertensives, $A$. Figure 1 gives a simplified directed acyclic graph (DAG). An individual's risk is predicted at $t = t_k$; they are not on anti-hypertensives and their SBP is high at 140mmHg ('SBP before treatment' in Figure 1). Suppose this individual commences anti-hypertensives, then are re-evaluated at $t = t_{k+1}$ and have successfully brought their SBP down to 130mmHg ('SBP after treatment' in Figure 1). Under the Treatment Offset Model, the new prediction will not be concordant with that at $t = t_k$, because the benefit of the anti-hypertensives is being double-counted: the treatment offset is the total effect of anti-hypertensives, and we are re-calculating risk based on the new value of the mediator SBP. The standard model therefore sees both the treatment and the lowered SBP, giving the effect of the risk reduction twice (both the total effect, and the indirect effect via SBP).

*Figure 1: A Causal Directed Acyclic Graph considering the relationship between systolic blood pressure, antihypertensive treatment, and cardiovascular disease. U represents an unmeasured confounder.*

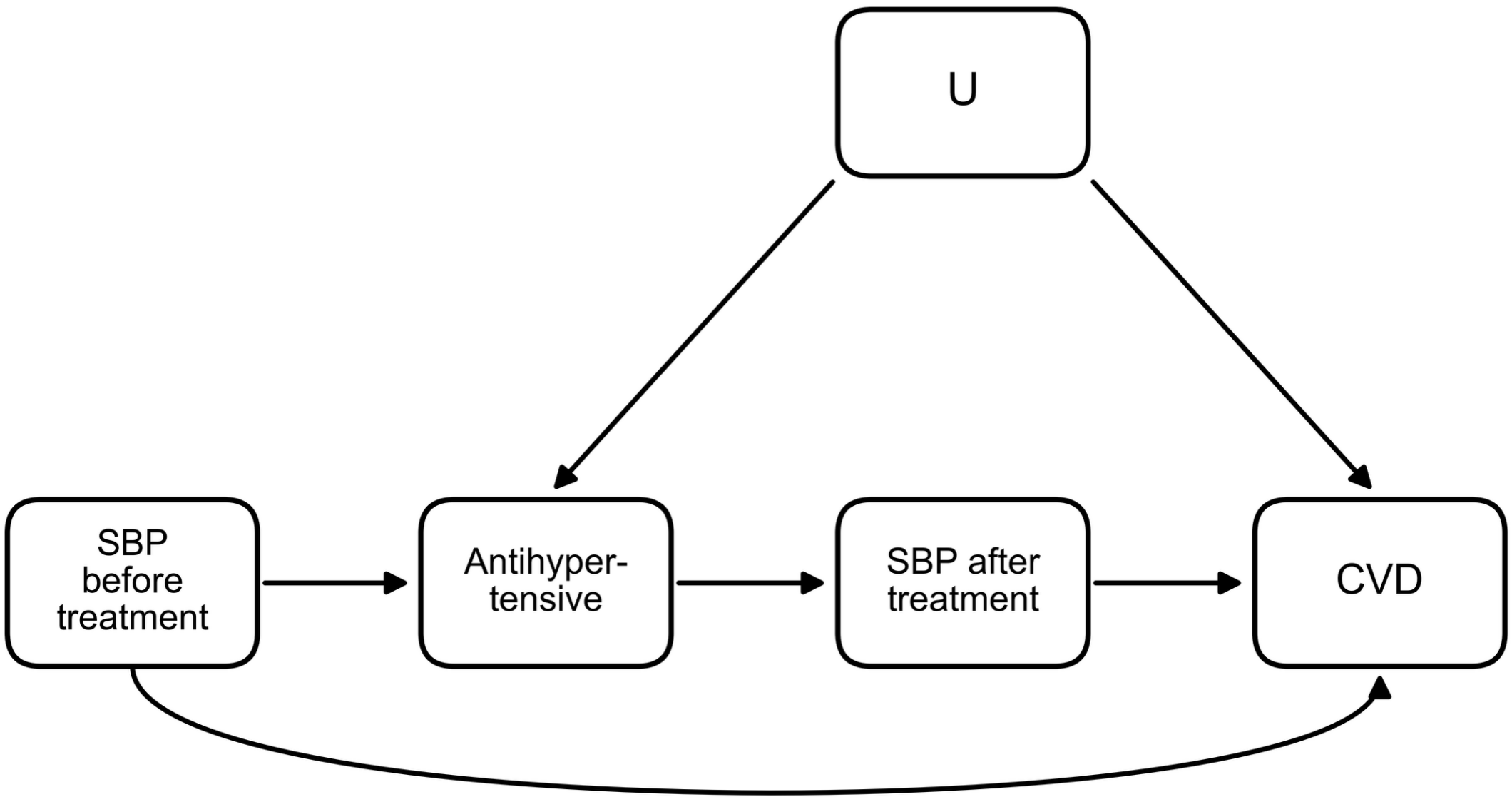

More generally, suppose an intervention is applied at time $t_k$, $A_{t_k} = a$. At time $t_{k+1}$ suppose the modifiable risk factors $M$ have responded to the intervention only, and all other aspects of the individual remain unchanged. Then, in the limit as $t_{k+1} - t_k \rightarrow 0$, the PUI under $A_{t_k} = a$ evaluated at $t_k$ should match the factual risk evaluated at $t_{k+1}$ using the updated risk factors:

$$\lim_{t_{k+1}-t_k \rightarrow 0} \{ f(M_{t_k}, Z_{t_k}, A_{t_k} = a, h_k) - f(M_{t_{k+1}}(A_{t_k} = a), Z_{t_{k+1}}, A_{t_{k+1}} = a, h_{k+1}) \} = 0, \tag{1}$$

where we use potential outcome notation to make explicit that mediators at time $t_{k+1}$ are responding to the intervention at time $t_k$, and recall that outcomes are evaluated later at times $h_k$ and $h_{k+1}$.

### 2.2.2. Proposed solution

This can be addressed by modelling the mediators differently. In the above example, rather than use SBP for prediction, we use 'SBP in absence of exposure to antihypertensives'. At follow-up, a counterfactual value of SBP in absence of antihypertensives is used (e.g., carry forward the value 140), resulting in a consistent estimate of risk on and off antihypertensives. All other aspects of the model remain the same as in the Treatment Offset Model. We call this the **Unexposed Mediator Model**.

Specifically, the model has the form

$$f^{UM}(M, Z, A = a) = f^{TO}(M(A = 0), Z, A = a),$$

i.e., the treatment offset model with the mediator set to its potentially counterfactual treatment-absent value.

It is immediate that this model satisfies Equation (1), and it will still target the causal estimand $g(E[Y(a)|X])$ provided that the original offset model did so.

Various approaches could be used to obtain $M(A = 0)$. We take the pragmatic approach of treating the unexposed mediator as a time-fixed variable, and use the most recent value available prior to commencing the intervention; if this is not available it is treated as missing (and ultimately imputed). This assumes no 'natural' progression in the value of the mediator over time (see Discussion).

This issue could also be dealt with by not including the mediator (in this case SBP) as a predictor in the model (and the issue would not occur if there were no mediators in the model as predictors). However, this could lead to an unacceptable drop in predictive performance (Characteristic 3). Estimating unexposed mediators in this way is helpful in cases where a small number of interventions act through simple pathways, but quickly becomes impractical when several mediators are potentially affected.

## 2.3. Flexibility for Multiple Interventions (Characteristic 2)

### 2.3.1. Outline of the issue

Specifying the model as in the Treatment Offset or Unexposed Mediator Models limits flexibility to add additional interventions. We could iteratively add them in to the model, but this would become cumbersome. Moreover, data is required on whether an individual is exposed or unexposed to each intervention. For medication (such as statins and antihypertensives) this is realistic, but for lifestyle interventions (such as changes to diet and exercise), such data is unavailable in most routinely collected data sources. Further, robust effect estimates for these interventions are not generally available.

### 2.3.2. Proposed solution

We model interventions through their effects on the modifiable risk factors $M$ (i.e., the modifiable risk factors are mediators for the interventions). An example DAG is in Figure 2, with example interventions. When fitting the model, the effects of the modifiable risk factors on the outcome $Y$ are fixed to their causal values. As these factors act as potential mediators for each other, we use the causal *direct* effects with reference to the DAG. Direct effects of any intervention (not through any of $M$) are also incorporated.

*Figure 2: A Directed Acyclic Graph capturing proposed interventions and the modifiable risk factors on which they act*

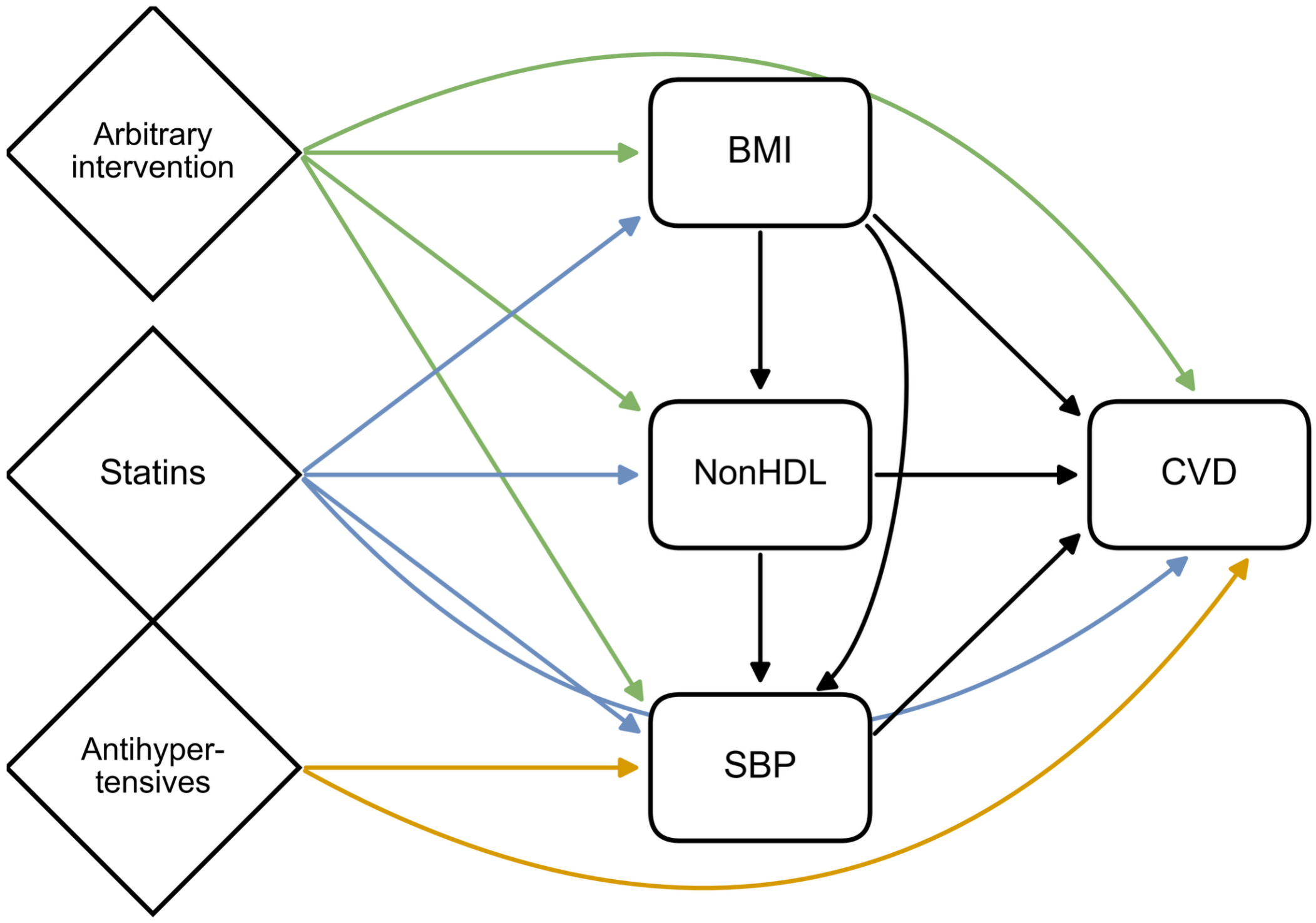

This requires the following assumption.

> **Assumption - Consistency of indirect intervention effects**: Consider any two interventions, $A_1$ and $A_2$, affecting an outcome $Y$ via a modifiable risk factor (mediator) $M$. If $A_1$ and $A_2$ have identical effects on the mediator $M$, then their indirect effect on the outcome $Y$ via $M$ is also the same.

This assumption enables recovery of intervention effects based on incomplete information: if an anti-hypertensive produces a 20% risk reduction acting only via SBP, and reduces SBP by 10mmHg, we can recover the causal effect of the 10mmHg SBP reduction regardless of mechanism. Other effects can be resolved through simultaneous equations constrained by this assumption. The effect of any arbitrary intervention then follows from estimating its effect on each modifiable risk factor. If the intervention is believed to affect the outcome through pathways other than those via the identified mediators, this can be incorporated as a direct effect (which is complementary to the above consistency assumption). Alternatively, the intervention itself can be left unspecified, but rather the effect of changing the risk factors can be estimated (e.g., revised CVD risk if you achieve a healthy weight and healthy blood pressure, regardless of how). Further details are provided in Supplementary file 3 (Sections 3 and 7), including further details of derivations and assumptions for computing the mediator effects on outcome.

For interventions where we can reasonably ascertain changes during follow-up (such as use of statins and anti-hypertensives), we again follow a similar approach to Xu et al [10] to handle changes during follow-up in the development data.

We call this the **Modifiable Risk Factor Model.** It is similar to the treatment offset model except that the appropriate causal effects are now the direct effects specified with respect to the chosen DAG, including the mediators (drawn from a model on the same scale and conditioning set, as before). We denote this $f^{MO}(M(A = a), Z, A = a)$. The model still satisfies the sequential condition given in Equation (1), and to recover the causal estimand $g(E[Y(a)|X])$ the following additional conditions are necessary: The DAG must be correct in the sense that every pathway from an intervention to an outcome passes through an included risk factor or is captured as a direct effect, the effects via each mediator must combine additively on the scale of the model with no interactions between them, and the individual direct effects must be correct (as described above and in Supplementary file 3 Sections 3 and 5). Whether these conditions are sufficient requires verification.

## 2.4. Predictive Accuracy (Characteristic 3)

### 2.4.1. Outline of the issue

For the models described so far, there will be a loss in predictive performance and accuracy (and hence discrimination and calibration) because one or more of the coefficients are being fixed to a causal value, rather than being estimated in a flexible manner. Therefore, the model is not optimised for prediction. For example, it is common for the predictive antihypertensive coefficient to be estimated in the opposite direction to its causal value. The loss in performance will vary on a case-by-case basis and can be assessed during internal validation.

### 2.4.2. Proposed solution

In scenarios where fixing coefficients to their causal values causes an unacceptable drop in performance, we propose an approach comprising two main components. The first component is a *baseline predictive model* that provides a factual prediction the first time an individual’s risk is calculated (i.e., the risk if all pre-initial interventions are maintained at their current level), and provides a reference for all other calculations. The second component is an *intervention model* that handles all future changes in interventions, both potential (looking forward to evaluate risk under different interventions), and actual (accounting for actual changes in interventions since initial visit). We call this the **Two-Component Model.**

*Baseline Component*

The baseline component is a prediction model that we estimate in the usual (non-causal) way [11], except that treatment changes during follow-up are accounted for using the Xu et al [10] approach as described earlier. Any predictors are eligible for inclusion - including non-modifiable risk factors $Z$ and modifiable risk factors $M$. It can include visit 0 values of interventions as predictors (e.g., whether a patient is taking anti-hypertensives the first time their risk is evaluated can act as a proxy for unobserved but pertinent factors).

We denote the baseline predictive model $b(X; \beta_{\text{baseline}})$ where $\beta_{\text{baseline}}$ collects the parameters required.

*Intervention Component*

For the intervention component the DAG is the same as in Section 2.3, with an example DAG in Figure 2. We denote the intervention model $i(A = a, \Delta M;\ \beta_{\text{int}})$, where $\beta_{\text{int}}$ denotes the required parameters, $a$ denotes proposed value(s) of intervention(s) (compared with their actual values $a^*$) and $\Delta M$ denotes the (possibly hypothetical) change in modifiable risk factors compared with baseline because of interventions. We will focus on absolute differences, and the model evaluated at time $t_k$ has the general form

$$i(X, A = a, \Delta M;\ \beta_{\text{int}}) = \phi(X; \beta_{int})\psi\big(m_{t_k} - m_{t_0}, a - a^*\big),$$

for some functions $\phi$ and $\psi$. For example, use of an assumed homogeneous causal effect (on an appropriate scale) for a change in each risk factor independently, with no direct intervention effects, would be written simply as

$$i(X, A, \Delta M;\ \beta_{\text{int}}) = \beta_{int}^{T}\big(m_{t_k} - m_{t_0}\big).$$

Like the Modifiable Risk Factor Model, the appropriate values for the causal estimates $\beta_{int}$ are those corresponding to the direct effects of the mediators and the interventions with respect to the DAG.

*Combination of baseline and intervention model*

At the initial visit (time $t_0$), the baseline model alone estimates factual risk – the risk if no changes are made to current lifestyle or medication. We use $f^{2C^*}$ to denote the two-component model with the $*$ denoting actual intervention status.

$$f^{2C^*}\big(X = x_{t_0}, A = a^*_{t_0}\big) = b\big(X_{t_0}\big).$$

Combining the baseline and intervention model provides PUI. For example, a patient not on anti-hypertensives can calculate their PUI of commencing anti-hypertensives. This is modelled via the modifiable risk factors $M$:

$$f^{2C}\big(X = x_{t_0}, A = a_{t_0}\big) = b\big(X = x_{t_0}\big) + i\big(X = x_{t_0}, A = a_{t_0}, \Delta M = m_{t_0}(a_{t_0}) - m_{t_0}\big),$$

and a series of different potential interventions in $A$ could be evaluated and compared to support decisions. Here the composition is additive as $f$ would be on a log-odds or log-hazard scale for binary or time-to-event outcomes respectively.

At follow-up visits, the factual risk calculation requires use of the intervention model to account for the change in *modifiable* risk factors compared with the initial visit. The baseline model uses the modifiable risk factors from the initial visit, but updated with any revised non-modifiable risk factors (such as age):

$$f^{2C^*}(Y, M = m_{t_k}, X = x_{t_k}, A = a^*_{t_k}) \\ = b(M = m_{t_0}, Z = z_{t_k}) + i(X = x_{t_k}, A = a^*_{t_k}, \Delta M = m_{t_k} - m_{t_0}).$$

PUIs are made at subsequent visits in a similar way:

$$f^{2C}(Y, M = m_{t_k}, X = x_{t_k}, A = a_{t_k}) \\ = b(M = m_{t_0}, Z = z_{t_k}) + i(X = x_{t_k}, A = a_{t_k}, \Delta M = m_{t_k}(a_{t_k}) - m_{t_0}).$$

In follow-up visits, the predicted risk depends on the modifiable risk factors' values from the initial visit. Specifically, for PUI, it is the difference between the modifiable risk factor and its value at the initial visit, rather than the follow-up visit, that is used in the calculation.

This model satisfies the sequential condition given in Equation (1). For correctly targeting the estimand $g(E[Y(a)|X])$, the intervention component inherits the requirements of the modifiable risk factor model. It is also necessary that the baseline model consistently estimates factual risk under continuation of the initial strategy provided, as before, that the causal effects are from models on the same scale and with the same covariates in the conditioning set. Similarly, sufficiency of these conditions needs to be verified.

## 2.5. Model validation and evaluation

Validating PUI models, where the interest is in absolute risk estimates of the form $E[Y(a)|X]$, can target two aspects: first, whether the causal effects are correct; second, whether the interventional risks are accurately predicted conditional on the causal effects being correct. We address the second kind of validation by constructing counterfactual survival times for different interventions, applying the same causal effects that are fixed as offsets in the models, adjusting the individual's cumulative hazard for time to event outcomes or the linear predictor for binary outcomes.

The first kind of validation is more challenging; we suggest two approaches. First, to quantify the effect of a causal effect being incorrect, we propose sensitivity analysis that uses different sizes of causal effects (e.g., 10% or 20% smaller or larger) to generate validation data. Second, where feasible, we suggest triangulating with an approach with complementary assumptions, such as that described by Keogh and van Geloven [12]. This approach produces a pseudo-population for validation by artificially censoring individuals when they deviate from the intervention of interest, and reweighting to represent the entire population. It requires that enough individuals sustain the strategy across the prediction horizon for the weights to be estimable, and that the artificial censoring is conditionally independent of the outcome given the measured covariates.

A complementary literature addresses validation of treatment contrasts including conditional average treatment effects, which we do not consider here [13,14].

# 3. Clinical example

## 3.1. Aims and objectives

Suppose the objective is prediction of 10-year incident cardiovascular risk under several potential interventions. We choose this setting because cardiovascular risk prediction models are used widely around the world [2,15–17], their use is often recommended in clinical guidelines (e.g., National Institute for Health and Care Excellence, 2023), and methodology on PUI has to-date focused on cardiovascular risk prediction. The three models described in Section 2 are developed, alongside the Non-Causal and Treatment Offset Models for reference.

As an illustrative example, we consider a 65-year old woman, Jane, with type 2 diabetes, an SBP of 140 mmHg, a BMI of 30 and non-HDL cholesterol of 5mmol/L. She attends a clinical appointment which we refer to as visit 0, and we consider various interventions (statins, antihypertensives, and lifestyle interventions that operate through the modification of SBP, BMI and non-HDL cholesterol), and follow-up visits for Jane.

## 3.2. Data and Approach

We use data from the Clinical Practice Research Datalink Aurum linked with Hospital Episode Statistics and Office for National Statistics death outcomes [19]. Cohort entry date for each individual was defined as the maximum of: 1 Jan 2005; one year of up to standard registration with a contributing practice; date of attaining age 18. Follow-up ended at the minimum of: 1 March 2020 (to avoid impacts of COVID-19), date of diagnosis of first CVD; death; deregistration with practice; last data upload from practice or HES. The same set of predictors used in QRISK4 [2] were extracted for each individual at the cohort entry date, with the following exclusions: SBP variability, non-high density lipoprotein (HDL), and cholesterol replaced total cholesterol/HDL ratio. SBP unexposed to antihypertensive treatment was additionally extracted. Details of the data extraction process, code lists, and operational definitions of all variables are provided in Supplementary file 1 Section 1.

All analyses were run separately for female and male cohorts, using a random sample of 2,000,000 females and 2,000,000 males, for computational reasons. The data is presented in Table 1. Note that the male/female field is described by the data provider as ‘gender’ rather than sex; hence pre-eclampsia and postnatal depression have non-zero prevalence in the male cohort while impotence has non-zero prevalence in the female cohort. Missing data on SBP, SBP unexposed to antihypertensive treatment (required for the Unexposed Mediator Model), BMI, non-HDL cholesterol, smoking status, ethnicity and index of multiple deprivation vigintiles was imputed using multiple imputation. We used one imputation chain with 20 iterations to obtain one stochastically imputed dataset for both female and male cohorts, which was then analysed as a complete case dataset for clarity of exposition (a model to be implemented should follow best practice for missing data handling [20]). After imputation, the data was split into random halves to create sample data for development and model evaluation. A sensitivity analysis stratified by whether unexposed SBP was observed or imputed is described in Section 3.7.1.

More details on the derivation of all the causal treatment effects is provided in Supplementary file 3. The models do not all target the same estimand, and consequently cannot be evaluated on a common dataset. The Treatment Offset and Unexposed Mediator Models target risk under a specified antihypertensive strategy. The Modifiable Risk Factor and Two-Component Models target baseline risk under continuation of the current intervention strategy. Counterfactual survival times are therefore constructed separately under each strategy.

Calibration of predicted risks at $t = 10$ years was assessed using the proportional hazards approach [21], and discrimination using Harrell's C-index (Supplementary file 1, Section 5.2) [22]. For the causal models, evaluations involve generating counterfactual times: the results presented in Sections 3.3 to 3.6 to assume that the imported causal effect is correct. Section 3.7.2 describes sensitivity analyses where this assumption is relaxed. We present calibration plots for the female models throughout; plots for the male models are in Supplementary file 2.

*Table 1: Visit 0 cohort characterisation*

| **Characteristic** | **Male** N = 2,000,000 | **Female** N = 2,000,000 |
|---|---|---|
| Age | | |
| Median (5% Centile, Q1, Q3, 95% Centile) | 35 (18, 25, 48, 69) | 34 (18, 25, 49, 73) |
| N Missing (% Missing) | 0 (0%) | 0 (0%) |
| Ethnicity | | |
| Bangladeshi | 14,506 (0.7%) | 12,242 (0.6%) |
| Black African | 48,277 (2.4%) | 52,568 (2.6%) |
| Black Caribbean | 20,937 (1.0%) | 24,747 (1.2%) |
| Chinese | 23,290 (1.2%) | 33,083 (1.7%) |
| Indian | 56,195 (2.8%) | 53,243 (2.7%) |
| Other Asian | 44,646 (2.2%) | 43,692 (2.2%) |
| Other ethnic group | 89,503 (4.5%) | 98,187 (4.9%) |
| Pakistani | 33,170 (1.7%) | 28,497 (1.4%) |
| White | 1,248,769 (62%) | 1,417,417 (71%) |
| Missing | 420,707 (21%) | 236,324 (12%) |
| Hypertension | 147,779 (7.4%) | 170,527 (8.5%) |
| Rheumatoid arthritis | 4,901 (0.2%) | 11,942 (0.6%) |
| Atrial Fibrillation | 12,894 (0.6%) | 9,289 (0.5%) |
| Chronic kidney disease | 10,475 (0.5%) | 14,318 (0.7%) |
| Serious mental illness | 27,844 (1.4%) | 30,643 (1.5%) |
| Family history of CVD | 111,733 (5.6%) | 141,336 (7.1%) |
| Migraine | 57,149 (2.9%) | 141,607 (7.1%) |
| Systemic lupus erythematosus | 310 (<0.1%) | 2,408 (0.1%) |
| Diabetes | | |
| Absent | 1,940,429 (97%) | 1,952,605 (98%) |
| Type1 | 8,905 (0.4%) | 6,633 (0.3%) |
| Type2 | 50,666 (2.5%) | 40,762 (2.0%) |
| Impotence | 79,738 (4.0%) | 380 (<0.1%) |
| Oral corticosteroid use | 9,146 (0.5%) | 14,543 (0.7%) |
| Atypical antipsychotic use | 9,950 (0.5%) | 7,711 (0.4%) |
| Smoking | | |
| Non-smoker | 843,901 (42%) | 1,069,024 (53%) |
| Ex-smoker | 365,350 (18%) | 389,589 (19%) |
| Current | 461,452 (23%) | 375,534 (19%) |
| Missing | 329,297 (16%) | 165,853 (8.3%) |
| Statins | 70,850 (3.5%) | 64,600 (3.2%) |
| Antihypertensives | 132,063 (6.6%) | 159,859 (8.0%) |
| Chronic obstructive pulmonary disease | 17,643 (0.9%) | 16,179 (0.8%) |

| **Characteristic** | **Male** N = 2,000,000 | **Female** N = 2,000,000 |
|---|---|---|
| Intellectual disability | 9,387 (0.5%) | 6,186 (0.3%) |
| Downs syndrome | 1,039 (<0.1%) | 1,027 (<0.1%) |
| Oral cancer | 1,248 (<0.1%) | 679 (<0.1%) |
| Brain cancer | 645 (<0.1%) | 517 (<0.1%) |
| Lung cancer | 810 (<0.1%) | 743 (<0.1%) |
| Blood cancer | 4,800 (0.2%) | 3,961 (0.2%) |
| Pre eclampsia | <10 (<0.1%) | 5,597 (0.3%) |
| Postnatal depression | 69 (<0.1%) | 26,441 (1.3%) |
| Body mass index | | |
| Median (5% Centile, Q1, Q3, 95% Centile) | 25.4 (19.8, 22.8, 28.5, 34.5) | 24.2 (19.1, 21.5, 28.3, 36.6) |
| N Missing (% Missing%) | 940,543 (47%) | 723,494 (36%) |
| Systolic blood pressure | | |
| Median (5% Centile, Q1, Q3, 95% Centile) | 130 (105, 120, 139, 155) | 120 (99, 110, 131, 150) |
| N Missing (% Missing%) | 834,775 (42%) | 412,096 (21%) |
| Non-HDL cholesterol | | |
| Median (5% Centile, Q1, Q3, 95% Centile) | 3.79 (2.11, 3.05, 4.50, 5.70) | 3.60 (2.10, 2.90, 4.40, 5.70) |
| N Missing (% Missing) | 1,657,709 (83%) | 1,658,295 (83%) |
| Index multiple deprivation (twentile) | | |
| Median (5% Centile, Q1, Q3, 95% Centile) | 11.0 (2.0, 6.0, 15.0, 19.0) | 11.0 (2.0, 6.0, 15.0, 19.0) |
| N Missing (% Missing) | 2,562 (0.1%) | 2,467 (0.1%) |
| Systolic blood pressure, unexposed to antihypertensives | | |
| Median (5% Centile, Q1, Q3, 95% Centile) | 129 (105, 120, 140, 160) | 120 (99, 110, 130, 155) |
| N Missing (% Missing) | 883,666 (44%) | 479,215 (24%) |

3.3. Non-Causal Model and Treatment Offset Model

We develop Cox proportional hazards models with study entry as time zero, spline terms for continuous variables, and all variables interacted with the linear term for age. The anti-hypertensive coefficient for the Treatment Offset model and follow-up adjustment is derived from Karmali et al [23]. The Non-Causal Model is identical, except that the coefficient for anti-hypertensives is unconstrained. Ten-year absolute risks are estimated through a Breslow estimate of the 10-year hazard.

The calibration of these models (along with other models that will be introduced in sections 3.4 – 3.6) is given in Figure 3. The Integrated Calibration Index (ICI), E50 and E90 [21], and discriminations are in Table 2. The Treatment Offset Model has a minor drop in performance compared with the Non-Causal Model.

*Figure 3: Calibration plots for models.*

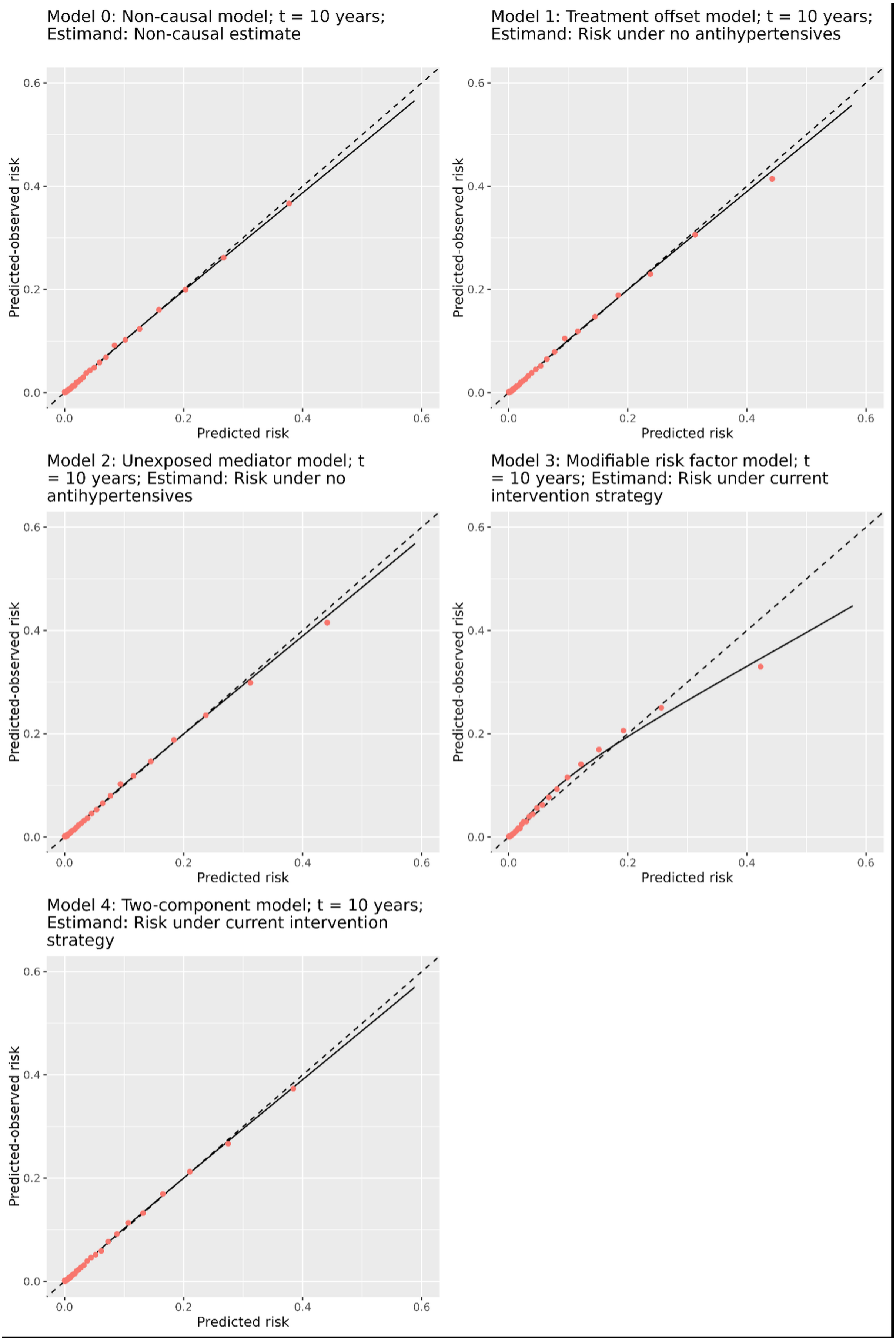

Model 0: Non-causal model; t = 10 years; Estimand: Non-causal estimate
Model 1: Treatment offset model; t = 10 years; Estimand: Risk under no antihypertensives
Model 2: Unexposed mediator model; t = 10 years; Estimand: Risk under no antihypertensives
Model 3: Modifiable risk factor model; t = 10 years; Estimand: Risk under current intervention strategy
Model 4: Two-component model; t = 10 years; Estimand: Risk under current intervention strategy
Predicted-observed risk
Predicted risk
0.0
0.2
0.4
0.6

*Table 2: Calibration and discrimination of the models at t = 10 years.*

| Model | Estimand* | ICI | E50 | E90 | Discrimination |
|---|---|---|---|---|---|
| Non-Causal | Non-causal estimate | 0.001 (0.001, 0.001) | 0.001 (0, 0.001) | 0.001 (0.001, 0.002) | 0.869 (0.869, 0.869) |
| Treatment Offset | No antihypertensive use | 0.001 (0.001, 0.001) | 0.001 (0, 0.001) | 0.002 (0.001, 0.003) | 0.880 (0.880, 0.880) |
| Unexposed Mediator | No antihypertensive use | 0.001 (0.001, 0.001) | 0.001 (0, 0.001) | 0.002 (0.001, 0.003) | 0.880 (0.880, 0.880) |
| Causal MFR | Continue on current interventions | 0.005 (0.005, 0.005) | 0.001 (0.001, 0.001) | 0.014 (0.012, 0.014) | 0.868 (0.868, 0.868) |
| Two Component | Continue on current interventions | 0.001 (0.001, 0.001) | 0.001 (0, 0.001) | 0.002 (0.001, 0.003) | 0.871 (0.871, 0.871) |

**Intervention strategy under which risks are predicted; for treatment offset and unexposed mediator: no antihypertensive use over the prediction period; for causal MFR and two component: continue on current interventions with respect antihypertensives and statins over the prediction period.*

Recall Jane, our 65-year old woman with type 2 diabetes, an SBP of 140 mmHg, a BMI of 30 and non-HDL cholesterol of 5mmol/L. Given her raised SBP, we may want to estimate her risk under antihypertensive treatment, and contrast this with her risk without antihypertensive treatment.

Jane's estimated risk from the non-causal model with antihypertensives "yes" is 17.68%, and for antihypertensives "no" is 15.32%. This reversal in the expected effect of antihypertensives highlights why this cannot be used for PUI.

The risk from the treatment offset model under antihypertensive treatment for 10-years is 12.79%, and the risk under no antihypertensive treatment is 17.25%. As expected, the risks on and off treatment lie on either side of the risk estimated from the non-causal model. However, suppose this individual decides to initiate antihypertensive treatment based on their risk assessment. They return to practice 3 months later (follow-up visit 1) having seen a reduction in their SBP to 120. If we again use the treatment offset model for PUI (with all predictors except SBP remaining the same), we estimate a risk under antihypertensive treatment for 10-years of 11.05%, and 14.96% without. There is a lack of consistency when using this model at multiple visits, highlighting the double counting of the effect of antihypertensives.

### 3.4. Unexposed Mediator Model

We developed the model in the same way as the treatment offset model, except that the SBP variable was replaced with SBP unexposed to antihypertensive treatment. The model is again well calibrated and discriminates well (Figure 3, Table 2), with little drop in performance compared with the non-causal model.

We estimate Jane's risk under antihypertensive treatment for 10-years as 13.01%, and 17.54% without. We then hypothesise Jane initiating antihypertensive treatment and returning at follow-up visit 1 after achieving a reduction in SBP to 120. The inputs when making the prediction remain the same as we use the SBP value prior to initiating antihypertensive treatment, and the risks on and off treatment therefore also remain the same.

### 3.5. Modifiable Risk Factor Model

This model is designed to achieve Characteristic 2: Flexibility for multiple interventions. We assume that the interventions of interest are antihypertensives, statin medication, and changes in lifestyle through diet and exercise, all operating through three modifiable risk factors: SBP, BMI and non-HDL cholesterol. Following the process outlined in section 2.3.2, and based on effect estimates reported in a variety of meta-analyses and systematic reviews, a DAG was developed to estimate the direct effects of changes in each of these three modifiable risk factors (Figure 2, Table 3). See Supplementary file 3 for details of the derivation of these effects. For statins, a judgement is required about pleiotropic effects on cardiovascular risk not mediated by non-HDL cholesterol, SBP or BMI. Our main analysis assumes no direct effect (consistent with our approach described in Supplementary file 3, Section 5), but we report a sensitivity analysis in which a direct effect is introduced such that the total effect of statins is 20% larger on the hazard ratio scale (Supplementary file 7, Sections 2 and 3).

Table 3: Estimated direct effect of modifiable risk factors on CVD, and total expected effect of statins on each modifiable risk factor.

| **Pathway** | **Direct effect (HR)** |
|---|---|
| SBP -> CVD | 1.025 per 1mmHg increase |
| Non-HDL -> CVD | 1.214 per 1mmol/L increase |
| BMI -> CVD | 1.022 per 1 kg/m$^2$ increase |
| **Pathway** | **Total effect** |
| Statins -> SBP | 2.62 mmHg increase |
| Statins -> Non-HDL | 1.3268 mmol/L decrease |
| Statins -> BMI | 0.33 1 kg/m$^2$ increase |

Cox proportional hazards model was developed with a spline on continuous variables, and all variables interacted with the linear term for age. This is except for SBP, BMI and non-HDL cholesterol, for which we created cut-offs at 120 (SBP), 25 (BMI) and 2.6 (non-HDL cholesterol) and did not consider reductions below those values as they would not be routinely considered in primary prevention. The coefficients of these variables were then fixed to their direct causal effects estimated from the DAG. To target the estimand of "risk under current intervention strategy", we adjust for changes in statin and antihypertensive use during follow-up as described in Section 2.1. The estimated risks under intervention can then be obtained by altering the inputs for SBP, BMI and non-HDL cholesterol.

Jane's baseline risk at visit 0 is 18.71%. Under antihypertensive treatment reducing SBP to 120, her estimated risk drops to 11.95% (or 14.99% if SBP only reaches 130). Under statin treatment, her estimated risk drops to 14.05%. We can model other interventions: a diet-induced BMI reduction of 5 (to BMI=25) would also decrease non-HDL cholesterol by 0.522 mmol/L and SBP by 3.5 mmHg, yielding 16.87% risk. Optimistically, combining diet, exercise, and medications to achieve healthy levels across all risk factors could reduce risk to 6.94%. At follow-up visit 1, we can measure actual changes in risk factors and update the model to provide factual rather than counterfactual risk estimates.

While this model is flexible in its use, there is a large drop in performance, particularly calibration (Figure 3, Table 2). We now move onto our final proposed model that addresses this performance drop.

## 3.6. Two-Component model

This model is designed to achieve Characteristic 3: Good predictive accuracy (Section 2.4.2). For the baseline risk estimation component, a Cox proportional hazards model was developed with a spline on all continuous variables (including SBP, BMI and non-HDL cholesterol), and all variables interacted with the linear term for age. To target the estimand of “risk under current intervention strategy”, we adjust for changes in statin and antihypertensive use during follow-up using the approaches detailed in Supplementary file 1.

Using the DAG (Figure 2), the effect of changes in the modifiable risk factors are applied through the intervention component using hazard ratios. Contrary to the modifiable risk factor model, this model does not have a drop in calibration compared with the other models (Figure 3, Table 2).

For the same individual and hypothetical scenarios described in section 3.5, the baseline risk at visit 0 is estimated at 16.88%. Antihypertensive intervention reducing SBP to 120 mmHg would yield an estimated risk of 10.74%, while reduction to 130 mmHg would result in 13.49%. Statins would reduce her risk to 12.64%, whereas a dietary intervention producing a 5-unit BMI reduction, including associated knock-on effects, estimates risk at 15.20%. If combined interventions successfully bring all modifiable risk factors to healthy levels, the estimated risk is 6.22%.

Suppose that Jane initiates a weight loss programme, with planned changes to both diet and exercise regime. They then return for a check-up at follow-up visit 1, three months later. We can estimate a factual risk based on the achieved levels of each of the modifiable risk factors. Suppose Jane achieved the reduction in BMI of 5, and also achieved a reduction in non-HDL cholesterol of 1mmol/L, but their SBP remained the same. Their estimated risk given the achieved level of their modifiable risk factors is 12.80%. Further hypothetical interventions can then be considered, for example, the risk under antihypertensive treatment is 10.18% if a reduction in SBP to 130 mmHg is achieved, and 8.07% if a reduction to 120 mmHg is achieved.

## 3.7. Sensitivity analyses

### 3.7.1. Missingness in unexposed SBP for unexposed mediator model

Unexposed SBP was missing for 22.6% of women and 45.0% of men not receiving antihypertensives at visit 0, and for 37.4% of women and 34.4% of men who were. Validation of the Unexposed Mediator Model stratified by whether unexposed SBP was observed or imputed gave similar calibration curves, with visible differences confined to predicted risks above 20%; discrimination was higher in the group with imputed values, particularly in men, likely reflecting the wider case-mix among those without a recorded untreated SBP (Supplementary file 4, Section 2).

#### 3.7.2. Evaluation under different ways to construct counterfactual outcomes

Allowing the intervention effects used to construct the counterfactual outcomes to differ from those in the fitted models produced negligible changes in calibration and discrimination across the full cohort (Supplementary file 5, Section 1). Within the subgroups receiving each treatment at visit 0, assuming a larger true effect produced greater apparent over-prediction, and a smaller effect less. The magnitude of these differences was small and confined to the upper tail of the predicted risk distribution (Supplementary file 5, Section 2).

Following the Keogh and van Geloven approach [12] the Two-Component Model appeared miscalibrated, consistently in the same direction across estimators and in both cohorts (Supplementary file 6, Section 3). The diagnostics indicate that combined weights were extremely heavy-tailed and the effective sample size after weighting remained a small fraction of the nominal sample size, reflecting positivity issues. Moreover, attempting to recover the treatment effects under the same assumptions led to qualitative reversal of effect sizes (Supplementary file 6, Section 4), suggesting residual confounding that would also affect the counterfactual outcome generation.

#### 3.7.3. Evaluation at post-baseline time points

Given the focus on sequential prediction, we also evaluated performance of the baseline component of the two-component model at later time points (Supplementary file 8). This shows a small calibration drift, and a drop in discrimination as index dates move further away from the start of follow-up.

## 4. Discussion

### 4.1. Main findings

We present a collection of related methods for fitting prediction under intervention (PUI) models that can be used sequentially in a decision support context. The developed approaches deal flexibly with sparse data, target useful causal estimands, and maintain good predictive performance.

The **Unexposed Mediator Model** is useful where risk is repeatedly evaluated in the presence of few interventions whose causal pathways comprise direct effects plus only a small number of mediators that are also used as predictors. It relies on being able to measure or impute the mediators in absence of intervention; this may be plausible for simple applications of PUI in primary prevention, where data is available on individuals before primary prevention commenced, but will become unmanageable for multiple interventions, and rely on increasingly strong assumptions about the (potentially imputed) unexposed mediator values.

Where flexibility for multiple interventions is required, the **Modifiable Risk Factor Model** can be used as this can model interventions abstractly via their effects on the mediators that are also predictors (plus any direct effects).

In each of these three models, fixing model coefficients to their causal value will inevitably lead to lower predictive performance, as the coefficients are no longer estimated to maximise the likelihood function. To avoid this, the **Two-Component Model** can be used, where baseline risk at initial visit is calculated using a purely predictive model, and all subsequent risks, both actual and hypothetical, are calculated in an intervention component. This improves predictive performance, and mitigates the no unmeasured confounding assumption since the baseline risk model is not estimated dependent on the offset terms (other than those handling treatment drop-in). However, it introduces extra considerations around model implementation and model updating, as the values from this first visit must be remembered at subsequent visits. These extra considerations mean that the Two-Component Model should only be preferred to the Modifiable Risk Factor when the improved predictive performance is both substantive and important, and the implementation requirements can be overcome. Table 4 summarises recommendations on circumstances where each approach might be most useful.

*Table 4: Summary of recommendations*

| Number of interventions | Need for sequential re-evaluation | Recommendation |
|---|---|---|
| Small, causal effects known | No | **Treatment Offset Model** |
| Small, causal effects known | Yes | **Unexposed Mediator Model** |
| Large, causal effects not always known | No/Yes | Either **Modifiable Risk Factor Model** or **Two-Component Model**. *Prefer the Two-Component Model if accuracy of absolute risk is important to the application, and substantively superior compared with the Modifiable Risk Factor Model. Prefer the Modifiable Risk Factor Model otherwise.* |

It should be noted that there is a drop in discrimination for the non-causal model, the causal MFR model, and the two-component model, compared with the treatment offset and unexposed mediator models. This is because of the different estimands they are targeting. The latter two models estimate the risk under the treatment strategy of ‘no antihypertensive use over the prediction period’. Antihypertensive use among higher risk individuals reduces the spread of predicted risks in the population, and therefore in this counterfactual setting on no antihypertensive use, there is a greater spread of predicted risks resulting in an increase in discrimination. This is therefore an artifact of the different estimands, rather than a comparative improvement in statistical performance.

### 4.2. Related Literature

Sequential PUI was described by Luijken et al [6], who propose estimands for scenarios where there are multiple timepoints where an intervention can be considered. This work builds on that by providing explicit strategies for estimation and allowing for arbitrarily many interventions.

The most similar existing approach is that proposed in the Million Hearts Study [24]. There, a standard baseline model is used at initial visit (the pooled cohort equations) and relative risk adjustments applied based on subsequent interventions and changes in risk factors. This is similar in to the Two-Component Model, but our approach has several key novelties. First, rather than using a standard prediction model for the baseline prediction model, we additionally account for treatment drop-in to ensure that baseline risk under current treatment regime is well-defined [3,4,10]. Second, in the intervention component, rather than using relative risk reductions, we propose building a DAG to explicitly capture ‘knock-on’ effects (such as weight loss leading to a reduction in SBP). Third, we allow for multiple arbitrary interventions.

### 4.3. Strengths and Limitations

The approaches described have several strengths. First, bringing causal formality allows us to be more explicit about the targeted estimands and assumptions than is usual in prediction modelling. Second, despite the causal clarity, the approach has a pragmatic flavour in that risks can be estimated from limited data, such as is often encountered in healthcare, and notably in electronic health records. Third, we have provided several different model structures that allow for different trade-offs to be made. For example, including the baseline predictive model (following the Two Component Model approach) is expected to increase predictive performance, but at the price of making a distinction between what happened with risk factors before the first recorded visit (modelled in a flexible predictive way) and changes after this visit (modelled causally).

The approaches have several limitations – some of these represent opportunities for further research, while some are inevitable trade-offs. First, the accuracy of developed models depends upon the quality of the causal estimates that are used as input. Causal estimates must be based on the correct estimands – particularly to avoid non-collapsibility issues affecting (for example) hazard ratios and odds ratios. All offset-based approaches require conditional effect estimates corresponding to the predictor set conditioned on [8], but external evidence often provides marginal effect estimates. If marginal estimates are used, estimates are biased. In our example we were limited to available causal estimates, so many of those we used are vulnerable to bias through non-collapsibility and linearity assumptions. van Amsterdam and Ranganath [8] propose a marginal constraint that recovers conditional effects from reported marginal ones – implementing this is future work. Concerning linearity, we took the pragmatic step of thresholding risk factors such that reductions below 120 (for SBP), 25 (for BMI), and 2.6 (for non-HDL) led to no further risk reduction. This is in line with primary prevention targets in the UK, but evidence is mixed on whether further reduction leads to further improvement. This thresholding step is particularly limiting for the fit of the Modifiable Risk Factor model, and further investigation is warranted. The transportability assumption between the sources of the causal effects and the target population is also strong. Currently, the causal estimates are input as point estimates allowing no propagation of their uncertainty or bias. A fully Bayesian extension would provide a natural way to incorporate this uncertainty, and this is an important future direction. It was also necessary to change causal effect scales between relative risk ratios, odds ratios and hazard ratios – we did so in a bias-minimising way [25], but this will still introduce bias. Moreover, the implications of deviations from these assumptions need further investigation. In the methods we have indicated where the correct counterfactual risk distributions should be recovered, but further work is needed to underpin this. In future work we hope to follow a more integrated approach to estimating the required causal effects to ensure better consistency, by estimating the causal effects from individual patient data, but there are clearly major practical barriers to overcome in this.

Second, the price of flexibility to arbitrary interventions is the requirement to make a consistency assumption, which also relates to research concerning ambiguous interventions [26] and the importance of well-defined interventions [27]. Further investigations are required both to empirically investigate the plausibility of the consistency assumption used in this context, and to connect with active research concerning estimation for ambiguous interventions.

Third, the need to identify an 'initial visit' before which risk factors are modelled non-causally, and after which subsequent changes are modelled causally, leads to undesirable features and raises complex questions. The Two-Component Model is anchored to this first visit. In many primary prevention settings, this first visit is arbitrary, and this approach is affording it a special status that seems undesirable, and also imposes a requirement to 'remember' the levels of risk factors at this initial visit. If data from the initial visit is lost, the sequential predictions would need to be re-anchored to a new initial visit. This may be an issue if a patient moves GP practices, especially if the GPs use different software – but previous data typically moves with the patient. It is also an issue if the underlying model is updated (e.g., based on new training data). Trading off causal consistency and maintenance of performance requires further investigation. Related to this is the assumption that any change in risk factors after initial visits is related to intervention, and 'natural' progression of risk factors under no intervention is disallowed. Relaxing this requires a model for risk factor progressions under no intervention, for which it would be challenging to obtain appropriate data. Furthermore, history and other longitudinal information concerning risk factors and past interventions could be added to the models.

Fourth, in our models we addressed treatment drop-in (treatment changes during follow-up) for binary treatment variables including statin and anti-hypertensive use. However, it is less clear how this should be done for interventions that are represented in non-binary variables. In particular, the ‘do nothing’ risk that we wish to be able to calculate is not necessarily well defined: while it is clear from the perspective of not initiating e.g., statins during follow-up, it is less clear what it would mean, and indeed how to measure, the absence of lifestyle interventions.

Fifth, our primary validations focused on evaluating predictive performance assuming that the estimated causal effects are correct. Evaluation based on artificial censoring and inverse probability weighting [12] avoids this dependence but substitutes a different and, in our setting, demanding set of assumptions — positivity with respect to sustained treatment, and conditional exchangeability in a prevalent-user cohort — neither of which was supported by the data in our example. While we mitigated this through various sensitivity analyses, non-circular evaluation of sequential PUI models remains challenging, and more thorough studies evaluating this are required. A distinct literature addresses evaluation of models where the target of interest is the treatment effects themselves, e.g., conditional average treatment effects (CATE) $E[Y(a_1) - Y(a_0)|X]$. This includes model-based extensions to discrimination and calibration for individualised treatment effects [13], and the reporting guidance is set out in the PATH statement [14]. Our evaluation focused on absolute counterfactual risk accuracy; combining these two forms of evaluation is an interesting future direction.

### 4.4. Conclusion

We offer pragmatic approaches for PUI in settings where many interventions might be available, and risk is evaluated repeatedly over time. While the methods presented are a logical progression, in any given setting all approaches could be used. The modelling approaches presented here can be used to build PUI models in practice, though careful validation of assumptions and outputs is required.

## 5. Acknowledgements

This research was funded by The National Institute for Health Research (NIHR) School for Primary Care Research (SPCR) (reference: NIHR SPCR-2021-2026, grant number 648) and Endeavour Health Charitable Trust. We acknowledge support of the UKRI AI programme, and the Engineering and Physical Sciences Research Council, for CHAI - Causality in Healthcare AI Hub [grant number EP/Y028856/1]. The views expressed are those of the authors and not necessarily those of the NIHR, the Department of Health and Social Care, or Endeavour Health. The authors are grateful to the two anonymous reviewers and editor whose comments substantially improved the final paper.

## 6. Ethics

CPRD has ethics approval from the Health Research Authority to support research using anonymised patient data. CHARIOT application (protocol 22_002333) was reviewed via the CPRD Research Data Governance (RDG) Process to ensure that the proposed research is of benefit to patients and public health.

## 7. Data and Code Sharing

Code for running all examples in the paper are provided on GitHub (https://github.com/manchester-predictive-healthcare-group/CHI-CHARIOT/tree/main/project-1-pui-methods). See Supplementary file 1 Sections 1 and 2 for further information.

The data used in this study were obtained from the Clinical Practice Research Datalink (CPRD). CPRD is a research service that provides anonymized primary care data for public health research. Due to ethical and legal restrictions related to the Data Protection Act and the terms of the license, the authors are not permitted to share the raw data.

Researchers may apply for access to CPRD data by submitting a protocol for review to the Research Data Governance (RDG) process. Details on how to apply for data access and the costs involved can be found at www.cprd.com.

## 8. Competing Interests

No authors had conflicts of interest to declare.

# Supplementary File 1 – technical appendix

**Associated manuscript:** A flexible approach to sequential prediction under intervention

**Authors:** Matthew Sperrin, Bowen Jiang. Yun-Ting Huang, Pedro Da Costa Avelar, Niels Peek, Alexander Pate

**Table of Contents**

# 1. Code for CPRD Aurum data extraction

The code and supporting documentation for the CPRD Aurum data extraction is available on GitHub (see folder generic-data-etraction). This includes operational definitions for each variable extracted, the code lists used to define each variable, name and description of all code used for data extraction, and information on the structure of the code. The code used for data extraction in this study formed the basis for development of an R package, rCPRD,[1] 'to simplify the extraction and processing of CPRD Aurum data, and creating analysis-ready datasets'. If you are reading this data extraction document to better understand how to extract and work with CPRD Aurum data, you are better off looking at the package documentation. Specifically, the rCPRD package vignette: https://alexpate30.github.io/rcprd/articles/rcprd.html, is a step-by-step guide on how to use this package to work with CPRD AURUM data, and will be kept up to date. The code and associated information used for the data extraction in this project is available in the generic data extraction GitHub repo,[2] including operation definitions of all variables and codelists.

# 2. Code for running analyses

## 2.1. General overview

All analyses were run in R version 4.4.2. The code for running the analysis is available on the GitHub page for this manuscript.[2] All this information is also available on the GitHub repository.

In order to re-use the code, the working directory should be set to the folder containing the 'code' subfolder. This must be set using setwd() at the top of every .R file. Sadly, we could not find a way to get the 'here' package working with the computational facility and data storage requirements we had.

In order to re-use the code, there should also be a 'data', 'R, and 'figures' folders in the same directory.

Inside 'code', there are then 2 subfolders, which we detail the contents of here:

**p1_data_prep:** data preparation and imputation.

**p2_analyses:** Run the analyses

Within each of these folders, .R files are all prefixed with 'p1', 'p2', …., and are designed to be run in order. Often (but not always), subsequent .R files are dependent on objects created in previous .R files.

Finally, the code has been written on the assumption that the CPRD Aurum data has already been extracted, formatted and imputed into an analysis-ready dataset. The process for doing do in detail in section 2 of this manuscript.

## 2.2. Dictionary of files

Programs p0_run_all.sh are batch scripts to submit all the R scripts in the right order, with the correct command line inputs (i.e., p1_fit_model.R is used to fit all five models, with the model and sex defined as command line inputs).

| **Program name** | **Utility** |
|---|---|
| **Directory location: code/p1_cohort_definition** | |
| P1_create_cohort.R | Create baseline cohort |
| P2_create_variables_objects.R | Create vectors of variables to be used in models, imputation, etc. |
| P3_impute_cohort.R | Multiple imputation (1 chain) |
| P4_create_split_sample.R | Create a split sample out of the data |
| P5_assess_imputations.R | Assess imputation procedure |
| P6.1_extract_prescriptions.R | Extract prescription data from SQLite database |
| P6.2_extract_medication_status.R | Create interval censored data for models 1 and 2, with respect to statin and antihypertensive use. |
| P6.3_layer_medication_status.R | Combine the intervals from program 6.2. |
| P7.1_augment_medication_status.R | Augment the interval censored data, so the statin/antihypertensive status is defined relative to value at baseline, required for models 3 and 4. |
| P7.2_layer_augmented_medication_status.R | Combine the intervals from program 7.2. |
| P8_calculate_treatment_offsets.R | Define causal effects for modelling. |
| **Directory location: code/p2_cohort_baseline** | |
| P1_fit_model.R | Fit models |
| P2_calculate_cf_surv_times.R | Calculate the counterfactual survival times required for model evaluation |
| P3_calibration_ph.R | Model evaluation – calibration |
| P4_discrimination.R | Model evaluation – discrimination |
| P5_worked_example.Rmd | Worked example for manuscript |
| **Directory location: R** **Note, these functions have been superseded by the functions in the rCPRD package: [https://alexpate30.github.io/rCPRD/index.html](https://alexpate30.github.io/rCPRD/index.html) .** | |
| Db_query.R | A function to be used for querying sqlite database |
| Functions.R | All functions used through the program scripts. |

# 3. Choice of predictors in the model

Three changes were made from the predictors included in QRISK4, which is the basis for the predictors included in the initial risk estimation layer. Non-HDL cholesterol was chosen rather than total cholesterol/HDL ratio because this is a variable which we plan to intervene on, but there is a lack of evidence in the literature about the effect of intervening on total cholesterol/HDL ratio on cardiovascular risk. Furthermore, there is a lack of evidence on the effect of our interventions (such as statins and antihypertensives) on total cholesterol/HDL ratio. There was evidence around the effect of non-HDL cholesterol on cardiovascular risk, and its effects with respect to the other modifiable risk factors, and therefore we chose to include this. For the same reasons, systolic blood pressure variability was excluded as a predictor. This is a variable we would have been intervening on, but there was a lack of evidence about the effect reducing systolic blood pressure variability on cardiovascular risk. It would therefore not possible to include these in the intervention layer and the DAG. Finally, calendar time was included as a predictor due to the secular trend in cardiovascular disease observed in the data (see supplementary data file 2). When implementing the model in practice, we would recommend making all predicts at the maximum follow-up time date, which was 01/03/2020. The numeric value for this is 5537 (number of days from 01/01/2005).

# 4. Modelling approach

## 4.1. Summary

We start by introducing each model broadly. We then define some variables and present the modelling formula.

The **Non-causal** model is a cox proportional hazards regression model with no offset terms. Antihypertensive use at baseline is included as a covariate.

The **treatment offset model** uses the approach of Xu et al.,[3] with some adjustments to this setting. In the referenced study, they create a time-varying variable which is split after an individual initiates statin treatment, and fix the coefficient of this variable to be the effect of initiating statins. In this study, we have a scenario where individuals may be on or off antihypertensive treatment at baseline, and may move on and off treatment multiple times. A time-varying variable is therefore created to represent this process. The estimand of this model is the "risk under the treatment strategy of no antihypertensive", or "risk under the treatment strategy of antihypertensive treatment".

The **unexposed mediator model** uses the same approach as the treatment offset model, except that systolic blood pressure is defined differently, only extracting values which were not recorded after antihypertensive treatment, otherwise systolic blood pressure is treated as missing. No new time-varying variables need to be defined. The estimand remains the same.

The **modifiable risk factor model** uses a slightly different approach. This model is introduced because we want to deal with multiple interventions (in this example statins, antihypertensives, or changes to diet and exercise) which may impact any of the modifiable risk factors (in this example SBP, BMI or Non-HDL cholesterol). We fix the coefficient for each of the modifiable risk factors at baseline to their direct causal effect, and adjust for any changes to intervention status during follow-up (note, this is possible for interventions like statin and antihypertensive use, but not for changes in diet and exercise during follow-up). This means when defining our time-varying variables to adjust for treatment drop-in, they are defined relative to each individual's intervention status at baseline.

The reason for this is to avoid double counting the effect of interventions. Suppose we kept statin and antihypertensive use as time-varying variables which are simply "on" or "off" treatment, like in the treatment offset model. Now consider two identical individuals (A and B) with SBP of 140, where individual B has received antihypertensives to reduce this SBP value to 120. The effect of the antihypertensives is modelled using the offset term on SBP at baseline. We do not want to also apply the offset term for the fact individual B is on antihypertensives at baseline, and individual A is not. We therefore only adjust for changes in treatment use during follow-up.

Importantly, this means the Estimand of this model is the risk of staying on the current intervention strategy (i.e. individuals intervention status remains unchanged), rather than the risk under the treatment strategy of not receiving (or receiving) the intervention.

Finally, the **two-component model**. We ignore the intervention layer of the two-component model here, focusing on the initial risk estimation layer, which involves the modelling. The approach is similar to the modifiable risk factor model, in that we adjust for changes in intervention status during follow-up, not differences in intervention status at baseline. This means the Estimand of this model is again the risk of staying on the current intervention strategy. The key difference from the modifiable risk factor model is that the modifiable risk factors at baseline are modelled freely with splines (like in the Non-causal model)

## 4.2. Definitions and formulas

### 4.2.1. Time-varying intervention offsets

We define two variables. Statin use at baseline ($A_{stat} = 0/1$). Antihypertensive use at baseline ($A_{ah} = 0/1$).

We then define two-time varying variables, which are the individuals treatment status during follow-up. Statin use at time $t$ ($A_{stat}(t) = 0/1$). Antihypertensive use at time $t$ ($A_{ah}(t) = 0/1$).

We then define two more time-varying variables, which are the individual's treatment status during follow-up, relative to their baseline value (the R in the variable name AR represents "relative").

$$AR_{stat}(t) = A_{stat}(t) - A_{stat}$$

$$AR_{ah}(t) = A_{ah}(t) - A_{ah}$$

Finally, we define four new time-varying variables, which are the four intervention status variables defined above, multiplied by the log of the total effect of that intervention ($B$), in the form of a log-hazard ratio.

$$B_{stat}(t) = A_{stat}(t) * B_{stat}$$

$$B_{ah}(t) = A_{ah}(t) * B_{ah}$$

$$BR_{stat}(t) = AR_{stat}(t) * B_{stat}$$

$$BR_{ah}(t) = AR_{ah}(t) * B_{ah}$$

Where $B_{stat} = -0.3102413$, $B_{ah} = -0.3245535$.

The variables $B_{stat}(t)$ and $B_{ah}(t)$ are used as offsets in the **treatment offset** and **unexposed mediator** models. The variables $BR_{stat}(t)$ and $BR_{ah}(t)$ are used as offsets in the **modifiable risk factor** and **two-component** models.

#### 4.2.2. Modifiable risk factor offset terms

In the **modifiable risk factor** model, we also need to include offset terms for the modifiable risk factors. These are defined relative to some target values, $SBP_{target} = 120$, $BMI_{target} = 25$, $NonHDL_{target} = 2.6$, below which no benefit is gained.

We then define:

$$sbp_{adj} = \max(0, sbp - sbp_{target})$$

$$bmi_{adj} = \max(0, bmi - bmi_{target})$$

$$nonhdl = \max(0, nonhdl - nonhdl_{target})$$

And

$$B_{sbp_{adj}} = sbp_{adj} * B_{sbp}$$

$$B_{bmi_{adj}} = bmi_{adj} * B_{bmi}$$

$$B_{nonhdl_{adj}} = nonhdl_{adj} * B_{nonhdl}$$

Where $B_{sbp} = 0.02433163$, $B_{bmi} = 0.02173449$ and $B_{nonhdl} = 0.1936187$ are the log-hazard ratios of the direct effects of each of SBP, BMI and NonHDL on CVD, as calculated in the Supplemental file 5.

### 4.3. Formulas

Let $Z$ be the set of non-modifiable risk factors, excluding antihypertensive use (either at baseline or time-varying), statin use (either at baseline or time-varying), systolic blood pressure, body mass index or non-HDL cholesterol. IMD (which takes integer values between 1 and 20) is modelled as a continuous variable using restricted cubic splines with knots at 1, 10 and 20. All variables are interacted with age. $sbp_{unexposed}$ is systolic blood pressure values extracted with no antihypertensive treatment in the preceding 6 months.

Non-causal model

$$f(X;\ \beta) = age + Z + rcs(sbp, 4) + rcs(bmi, 4) + rcs(nonhdl, 4) +$$
$$age * (Z + \ rcs(sbp, 4) + rcs(bmi, 4) + rcs(nonhdl, 4))$$

Treatment offset model

$$f(X, A;\ \beta) = age + Z + rcs(sbp, 4) + rcs(bmi, 4) + rcs(nonhdl, 4) +$$
$$age * (Z + \ rcs(sbp, 4) + rcs(bmi, 4) + rcs(nonhdl, 4)) +$$
$$offset(B_{ah}(t))$$

Unexposed mediator model

$$f(X, A;\ \beta) = age + Z + rcs\left(sbp_{unexposed}, 4\right) + rcs(bmi, 4) + rcs(nonhdl, 4) +$$
$$age * \left(Z + \ rcs(sbp_{unexposed}, 4) + rcs(bmi, 4) + rcs(nonhdl, 4)\right) +$$
$$offset(B_{ah}(t))$$

Modifiable risk factor model

$$f(X, A;\ \beta) = age + Z + age * Z +$$
$$offset\left(B_{sbp_{adj}}\right) + offset\left(B_{bmi_{adj}}\right) + offset\left(B_{nonhdl_{adj}}\right) +$$
$$offset(BR_{ah}(t)) + offset(BR_{stat}(t))$$

Two-component model; baseline component:

$$f(X, A;\ \beta) = age + Z + rcs(sbp, 4) + rcs(bmi, 4) + rcs(nonhdl, 4) +$$
$$age * (X + \ rcs(sbp, 4) + rcs(bmi, 4) + rcs(nonhdl, 4)) +$$
$$offset(BR_{ah}(t)) + offset(BR_{stat}(t))$$

## 5. Model evaluation

We use a similar process to that proposed by Xu et al.[3] In the referenced study, they obtain treatment naive counter factual survival times (survival times if the individual had not received treatment) by adjusting the cumulative baseline hazard after an individual initiates statin treatment, to get the cumulative hazard that would have been observed if the individual had not initiated statins. A new counterfactual survival time is then estimated based off this augmented cumulative baseline hazard. Once counterfactual survival times have been calculated, model performance can be assessed using the same methodology used to evaluate a non-counterfactual clinical prediction model predicting a survival outcome. In section 5.1, we detail the process for estimating counterfactual survival times in our setting, which has a few key differences from Xu et al.[3] In section 5.2, we detail the approach for estimating calibration and discrimination. Note that the treatment offset and unexposed mediator models have a different estimand (risk under the untreated treatment strategy) compared to the modifiable risk factor model and the two component model (risk under treatment strategy of 'do nothing'). This means the process of estimating the counterfactual survival times under these different treatment strategies is also different. However, once the counterfactual survival times have been calculated, the process for model evaluation with respect to calibration and discrimination is identical.

## 5.1. Methodology for estimation of counterfactual survival times

In this study, we have a scenario where individuals may be on or off treatment at baseline, and may move on and off multiple treatments more than once. There is also a different estimand for models 1 and 2, compared to models 3 and 4, which will alter how the counterfactual survival times are estimated. The estimand of models 1 and 2 is the risk under the treatment strategy of "no statins or antihypertensives over the prediction period". The estimand of models 3 and 4 is the risk under the treatment strategy of "continue on whatever treatment strategy you are currently on over the prediction period", which could involve remaining on statins or antihypertensives. staying on the current intervention strategy (with respect to statins, antihypertensives and an individual's smoking status). The process to augment the process of Xu et al., into this setting is as follows. We re-use some of the notation introduced in section 5.2.

Let $t_i, i \in \{1,2, \dots, n\}$ be all the times at which an individual changes status with respect to statin or antihypertensive use. Let $t_{event}$ be the time at which an individual either has a cardiovascular outcome event or is censored. The cumulative hazard is calculated at each time point $t_i$: $H_0(t_i)$.

For models 1 and 2, the counterfactual cumulative hazard at time $t$ is then estimated as:

$$H_{cf}(t) = H(t_1) + \sum_{i=1}^{n} \exp\big(B_{stat}(t_i) + B_{ah}(t_i)\big) * H(t_{i+1}) - \exp\big(B_{stat}(t_i) + B_{ah}(t_i)\big) * H(t_i)$$

For models 3 and 4, the counterfactual cumulative hazard at time $t$ is estimated as:

$$H_{cf}(t) = H(t_1) + \sum_{i=1}^{n} \exp\big(BR_{stat}(t_i) + BR_{ah}(t_i)\big) * H(t_{i+1}) - \exp\big(BR_{stat}(t_i) + BR_{ah}(t_i)\big) * H(t_i)$$

Where $t_{n+1} = t_{event}$ is the time at individual is either censored or has a cardiovascular outcome event. In Plain English Terms, the cumulative hazard is broken up into intervals, defined by when an individual changes treatment status.

For models 1 and 2, if an individual is on statins or antihypertensives during an interval, the amount of cumulative hazard in that interval is adjusted to the amount of cumulative hazard that would have been observed if the individual had not been on statins or antihypertensives.

For models 3 and 4, if the treatment status during an interval is different from what it was at zero, the amount of cumulative hazard in that interval is adjusted to the amount of hazard if the treatment status had been the same as at baseline.

Note, the values of time-varying variables at the times $t_i$, $B_{stat}(t_i)$, $B_{ah}(t_i)$, $BR_{stat}(t_i)$ and $BR_{ah}(t_i)$ are the value they take for the upcoming interval $[t_i, t_{i+1})$.

The counterfactual survival time, $t_{cf}$, is then estimated in the same way as Xu et al.,[3] by finding the minimum time point at which $H(t) = H_{cf}(t_{event})$. For individuals where $t_{cf}$ is bigger than the maximum follow-up, we set $t_{cf}$ to be the value at maximum follow-up, and the event indicator is set to zero. Absolute risk at the prediction horizon is obtained from the fitted Cox model where the baseline cumulative hazard is estimated by the Breslow estimator and evaluated at 10 years.

### 5.2. Methodology for evaluating calibration and discrimination

Once the counterfactual survival times had been calculated, we followed well-established methods for evaluating model performance. Both calibration and discrimination were assessed for predictions at time $t = 10$.

Calibration was assessed by fitting a Cox proportional hazards model with the counterfactual survival time and indicator as the outcome in the model, and the conditional log-log of the predicted risks as the sole predictor in the model. The conditional log-log of the predicted risks were modelled using restricted cubic splines with 5 knots (located at $5^{th}$, $27.5^{th}$, $50^{th}$, $72.5^{th}$ and $95^{th}$ percentiles). This is an approach detailed by Austin et al.[4]

Also taken from Austin et al.,[4] we report the integrated calibration index (ICI): the mean magnitude of the difference between the calibration curve and the 1:1 line; E50: the median magnitude of the difference between the calibration curve and the 1:1 line; and E90: the $90^{th}$ percentile of the magnitude of the difference between the calibration curve and the 1:1 line. This provides quantifiable summary statistics of the calibration curve.

For discrimination we report Harrell's C-index,[5] which is the proportional of all pairs of comparable individuals, where the individual with the higher risk has the event first. A comparable pair is a pair where it can be verified who had the event first (i.e. pairs of individuals who are both censored before having an event are not deemed comparable).

# Supplementary file 2 - results

## 1 Prelim

This document presents results for both the female and male cohorts, as well as modifiable risk factor models where we fixed the effect the modifiable risk factors to their causal value one at a time.

The models are as follows:

- Model 0 = Non-causal model
- Model 1 = Treatment offset model
- Model 2 = Unexposed mediator model
- Model 3 = Modifiable risk factor model
- Model 4 = Two-component model
- Model 5 = Modifiable risk factor model (fix effect for SBP only)
- Model 6 = Modifiable risk factor model (fix effect for BMI only)
- Model 7 = Modifiable risk factor model (fix effect for Non-HDL cholesterol only)

# 2 Calibration

## 2.1 Female

Model 0: Non-causal model; t = 10 years; Estimand: Non-causal estimate

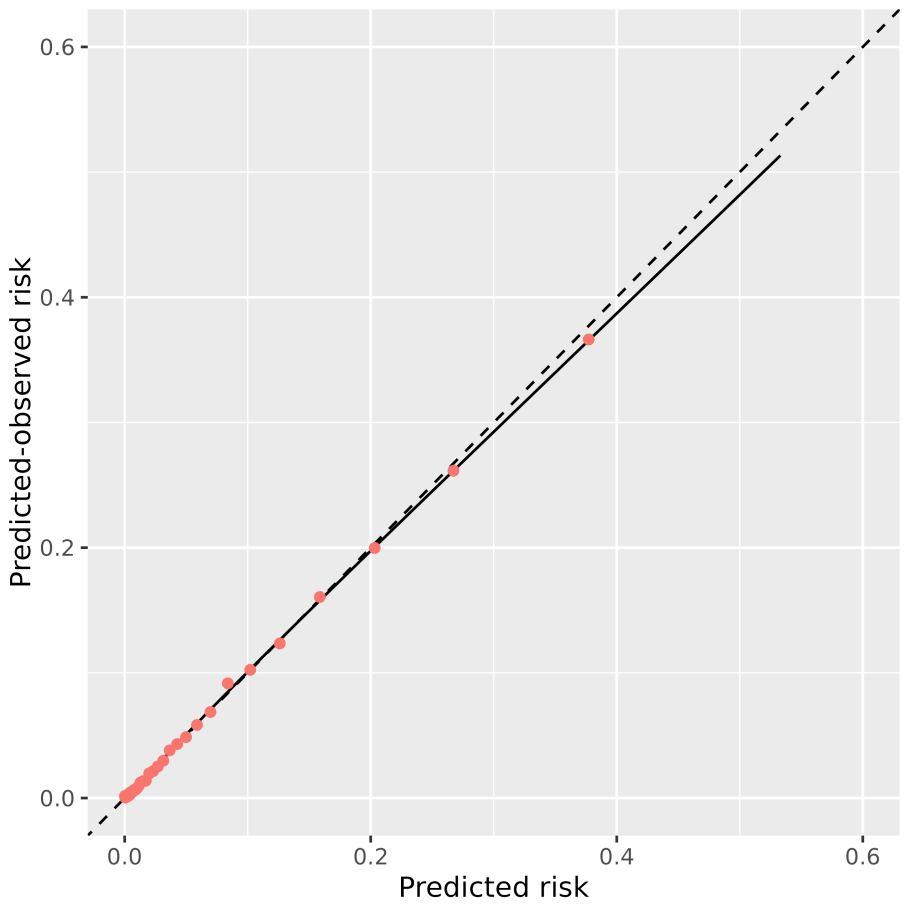

Model 1: Treatment offset model; t = 10 years; Estimand: Risk under no antihypertensives

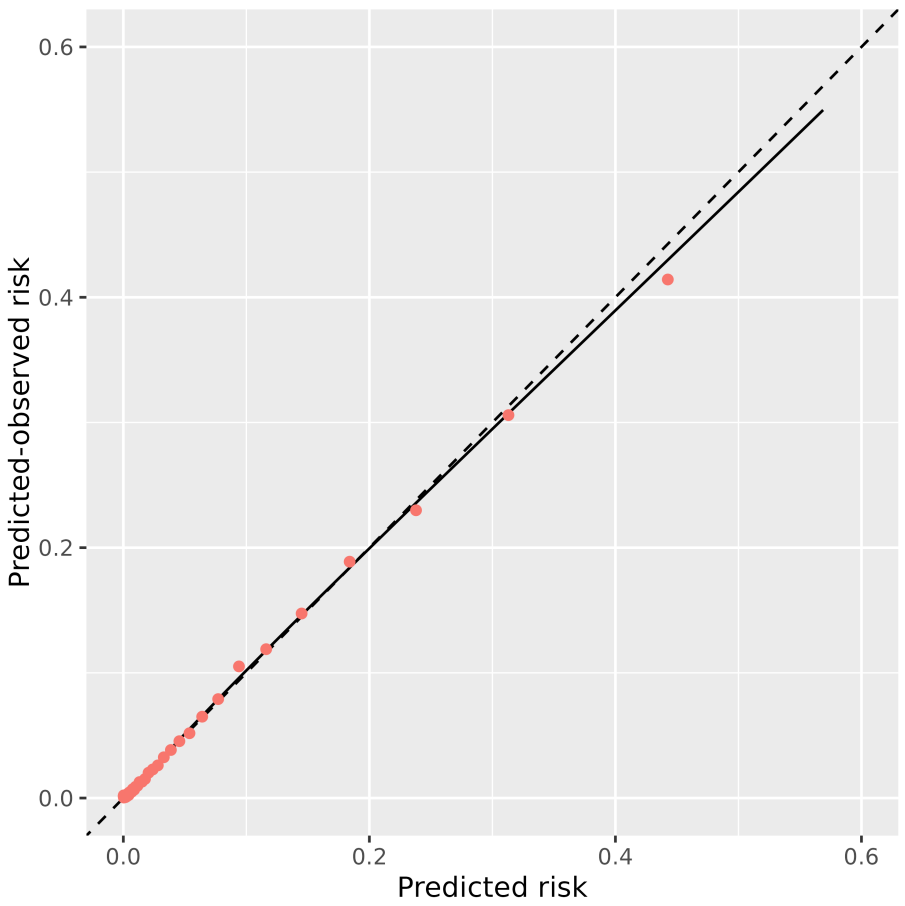

Model 2: Unexposed mediator model; t = 10 years; Estimand: Risk under no antihypertensives

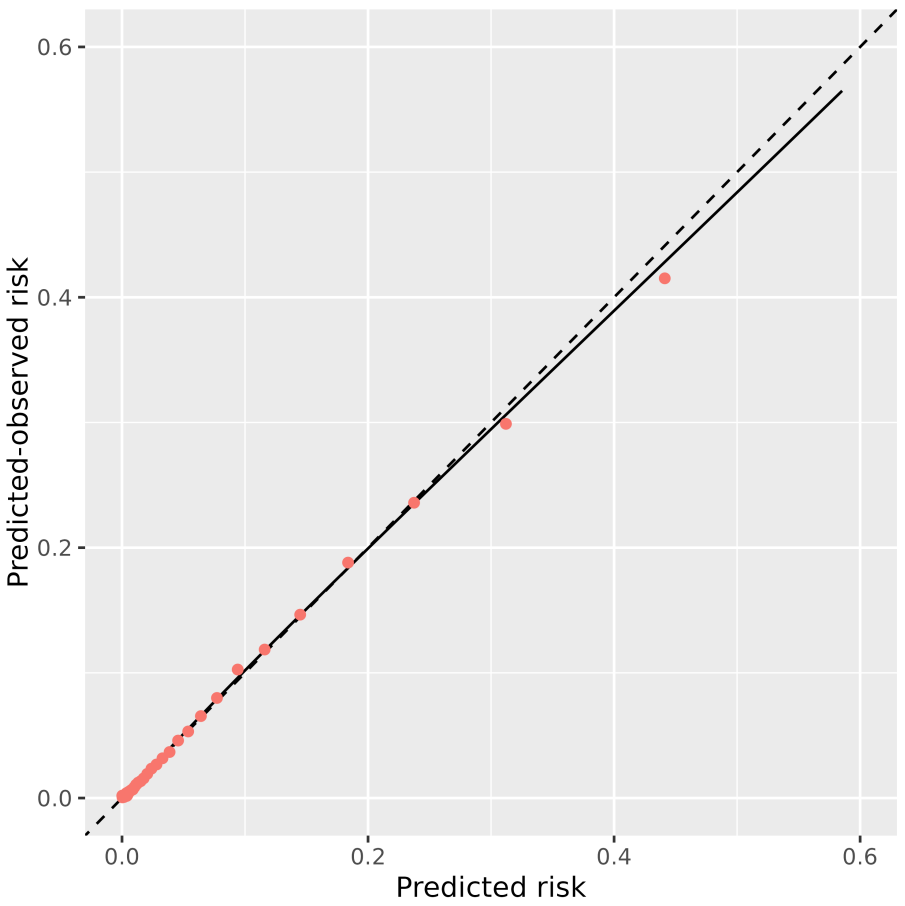

Model 3: Modifiable risk factor model; t = 10 years; Estimand: Risk under current intervention strategy

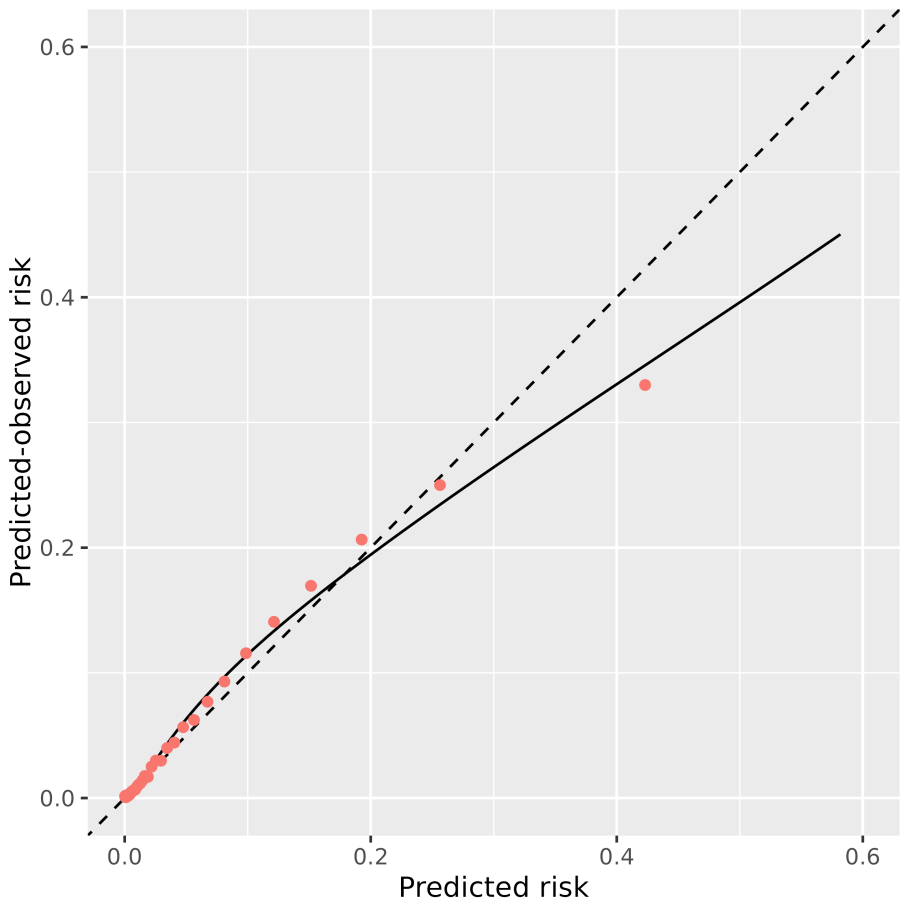

Model 4: Two-component model; Estimand: Risk under current intervention strategy

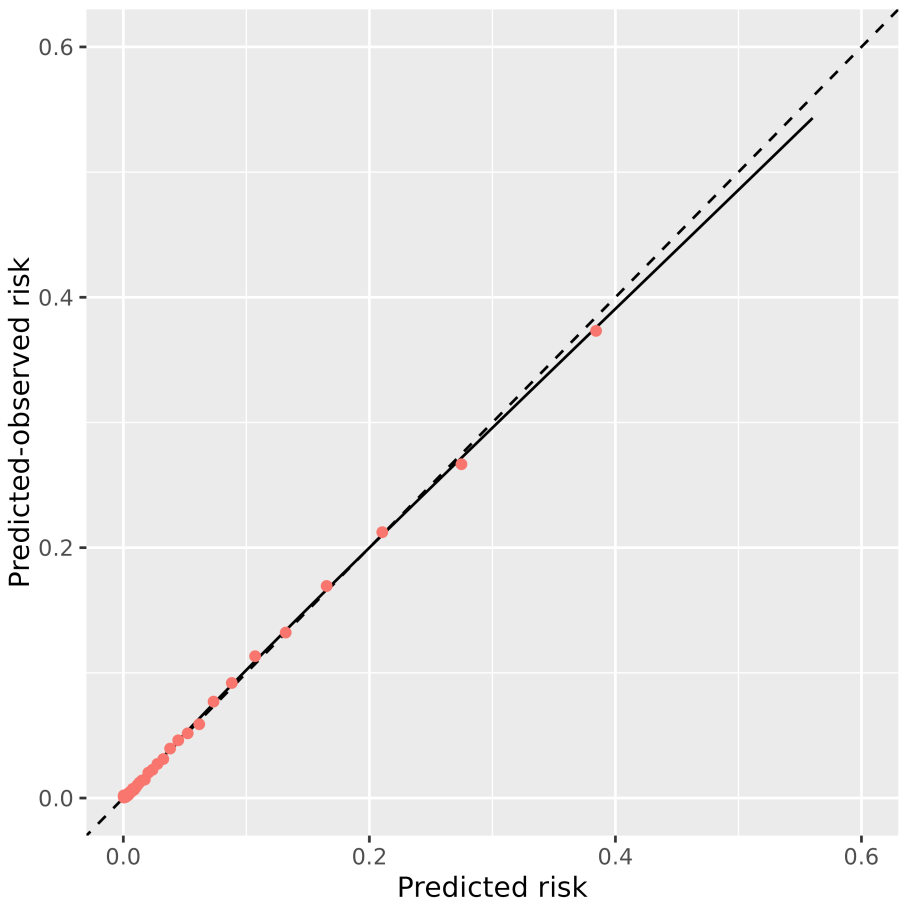

Model 5: Modifiable risk factor model (SBP only); t = 10 years; Estimand: Risk under current intervention strategy

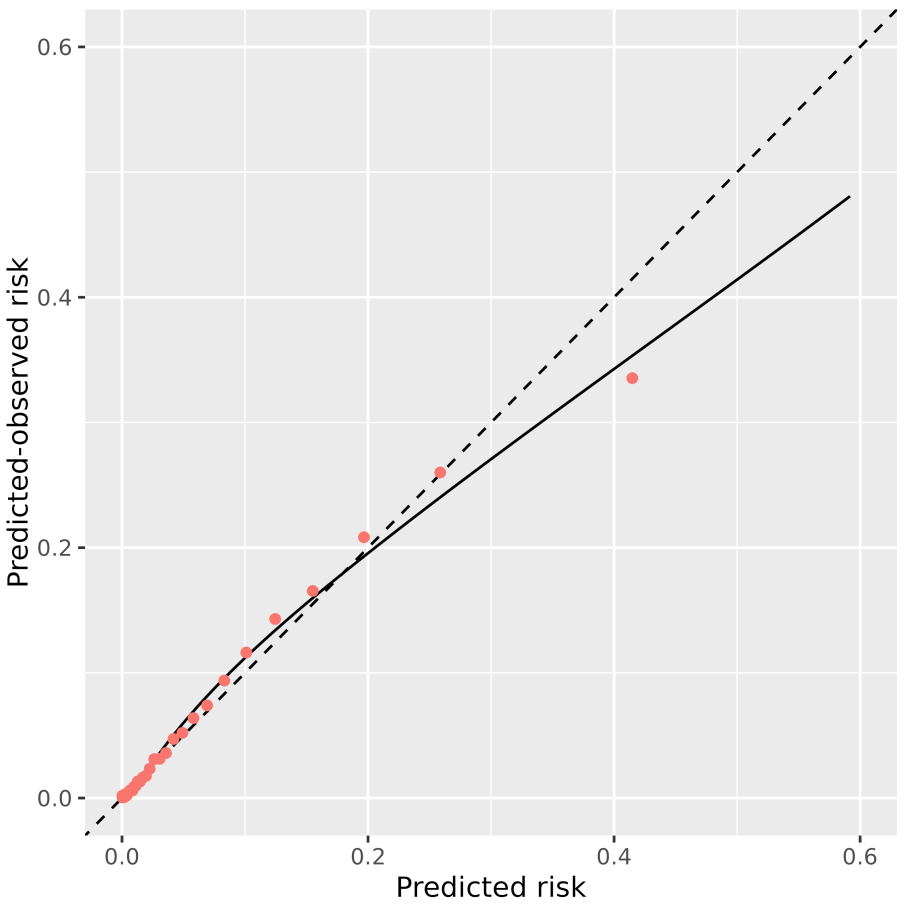

Model 6: Modifiable risk factor model (BMI only); t = 10 years; Estimand: Risk under current intervention strategy

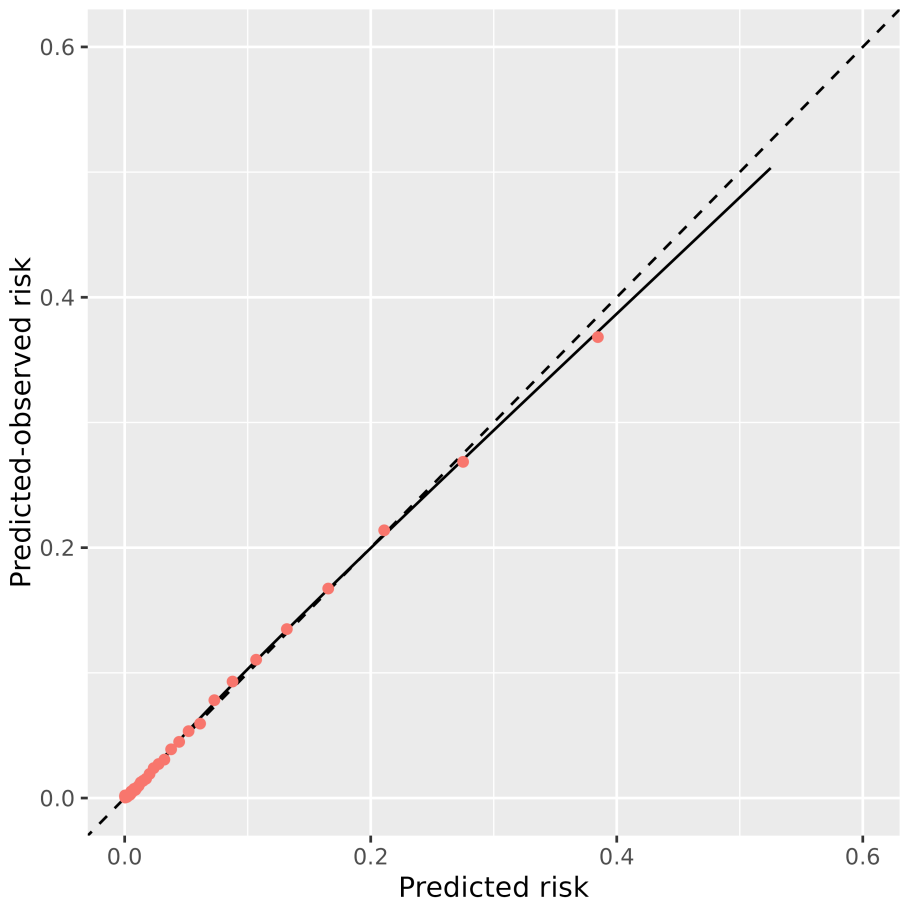

Model 7: Modifiable risk factor model (non-HDL cholesterol only); t = 10 years; Estimand: Risk under current intervention strategy

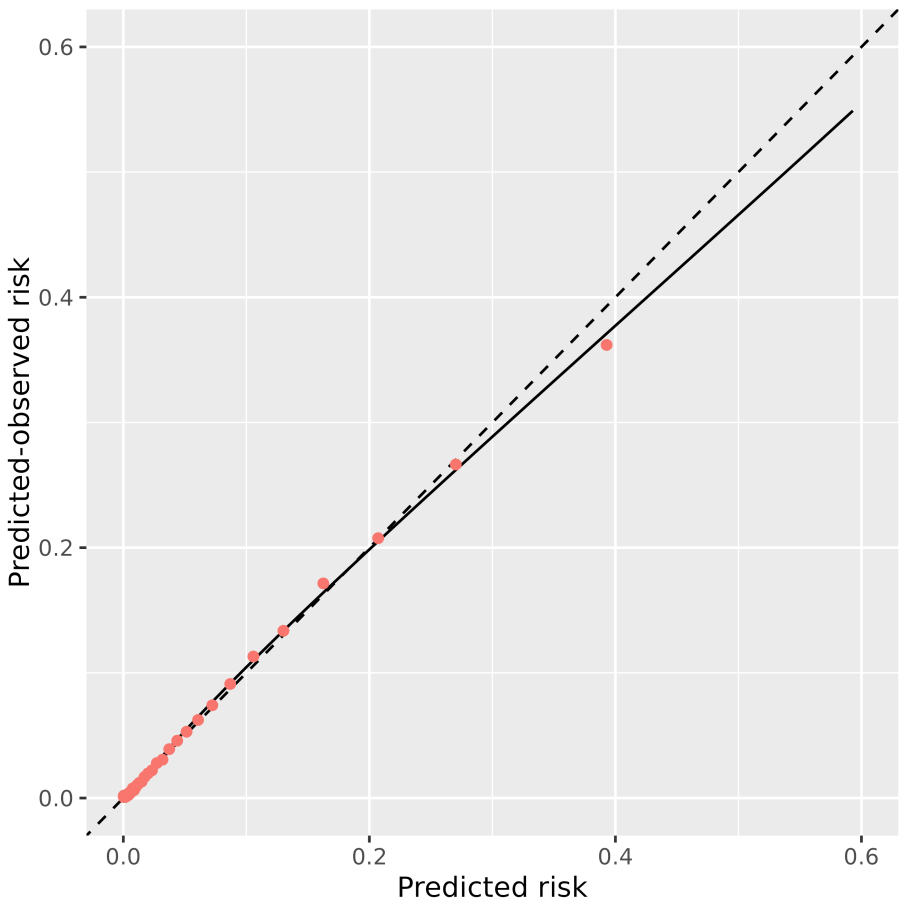

| Model | Estimand | ICI | E50 | E90 |
|---|---|---|---|---|
| 0 | Non-causal estimate | 0.001 (0.001, 0.001) | 0.001 (0, 0.001) | 0.001 (0.001, 0.002) |
| 1 | No antihypertensive use | 0.001 (0.001, 0.001) | 0.001 (0, 0.001) | 0.002 (0.001, 0.003) |
| 2 | No antihypertensive use | 0.001 (0.001, 0.001) | 0.001 (0, 0.001) | 0.002 (0.001, 0.003) |
| 3 | Continue on current interventions | 0.005 (0.005, 0.005) | 0.001 (0.001, 0.001) | 0.014 (0.012, 0.014) |
| 4 | Continue on current interventions | 0.001 (0.001, 0.001) | 0.001 (0, 0.001) | 0.002 (0.001, 0.003) |
| 5 | Continue on current interventions | 0.004 (0.004, 0.004) | 0.001 (0.001, 0.001) | 0.011 (0.009, 0.011) |
| 6 | Continue on current interventions | 0.001 (0.001, 0.001) | 0.001 (0, 0.001) | 0.003 (0.001, 0.004) |
| 7 | Continue on current interventions | 0.002 (0.001, 0.002) | 0.001 (0.001, 0.001) | 0.004 (0.003, 0.005) |

## 2.2 Male

Model 0: Non-causal model; t = 10 years; Estimand: Non-causal estimate

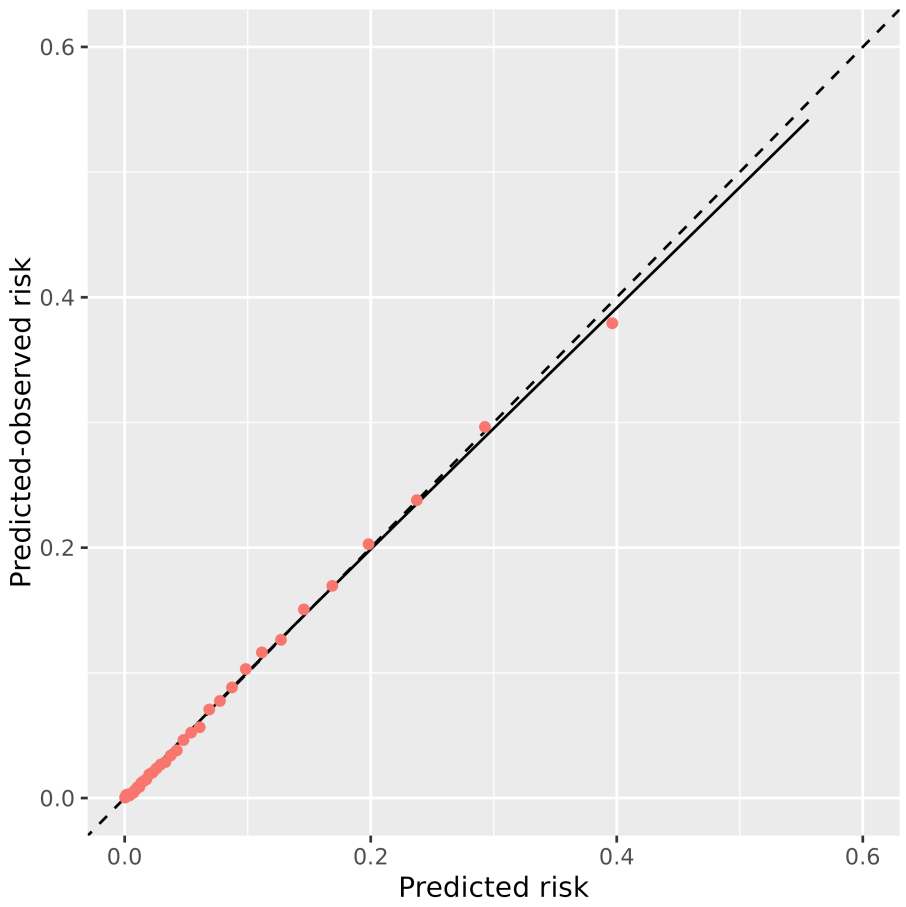

Model 1: Treatment offset model; t = 10 years; Estimand: Risk under no antihypertensives

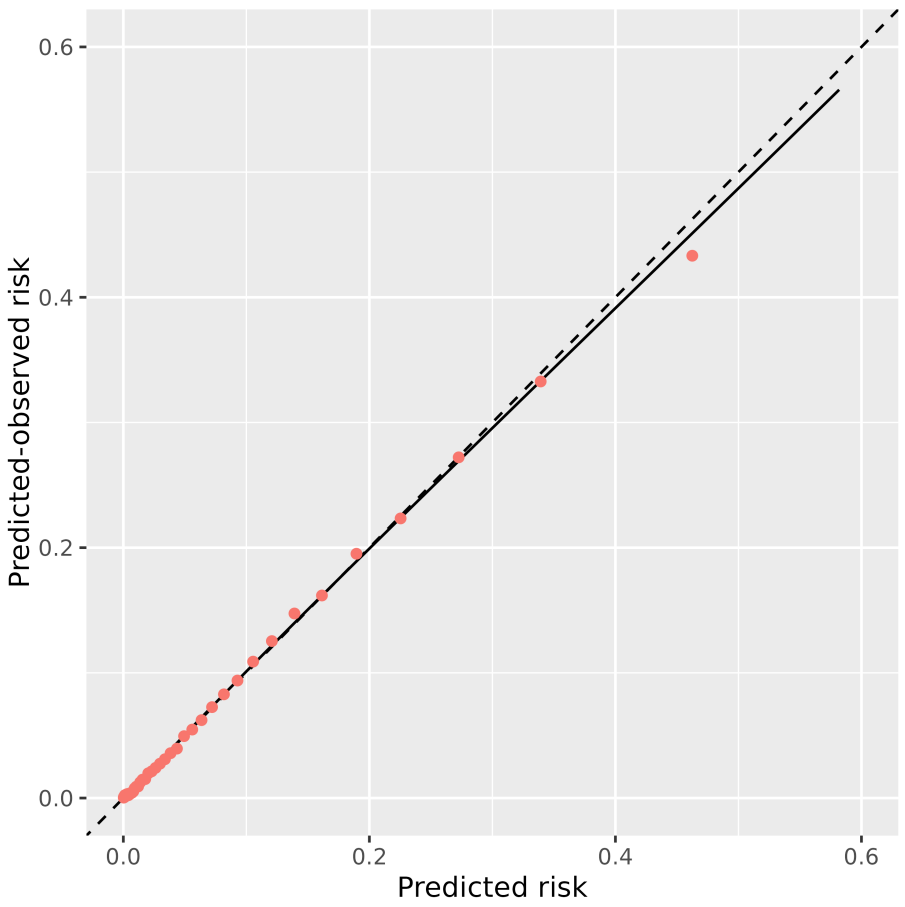

Model 2: Unexposed mediator model; t = 10 years; Estimand: Risk under no antihypertensives

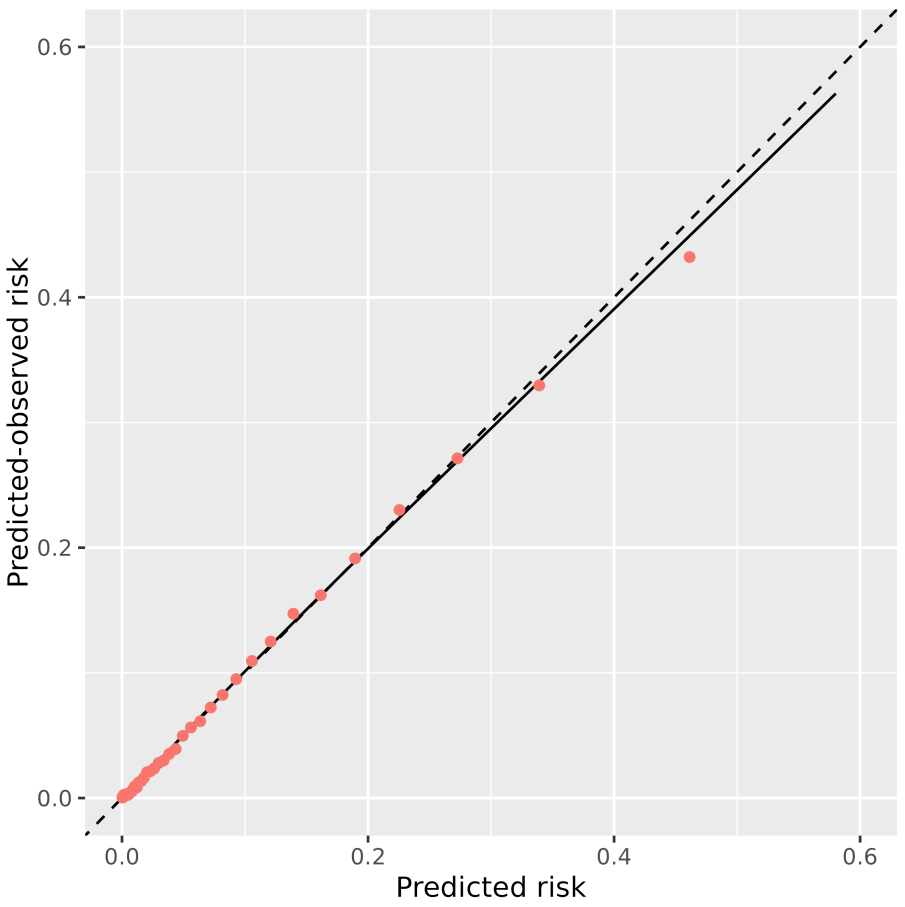

Model 3: Modifiable risk factor model; t = 10 years; Estimand: Risk under current intervention strategy

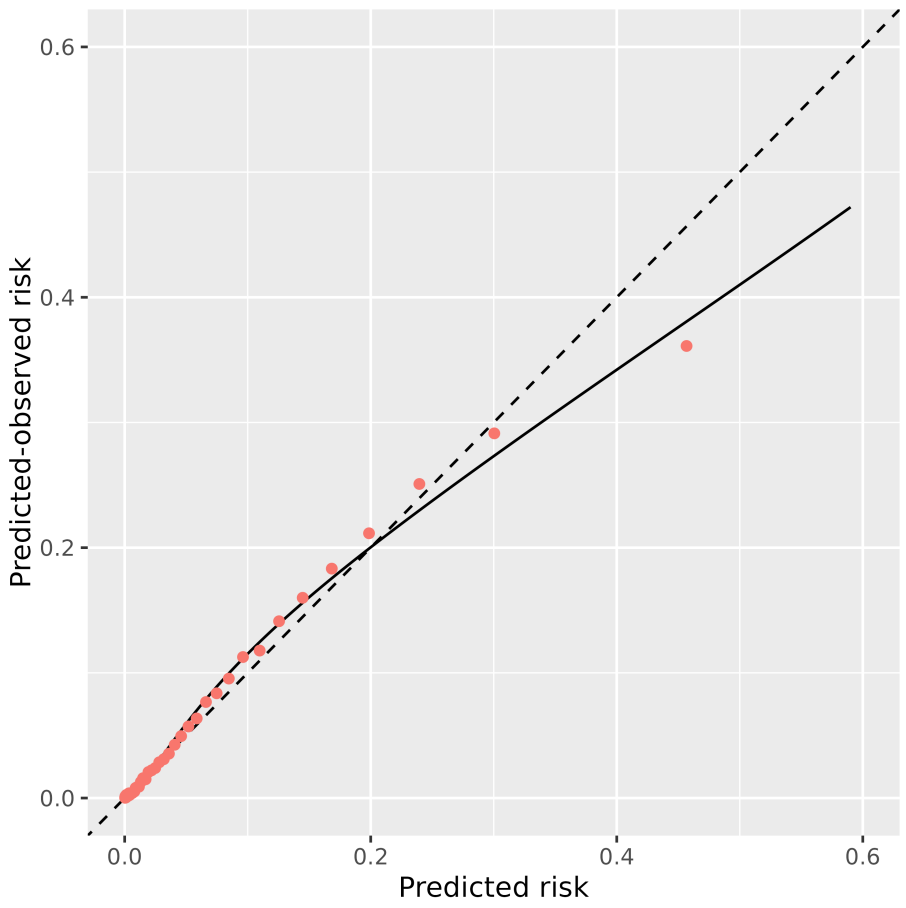

Model 4: Two-component model; Estimand: Risk under current intervention strategy

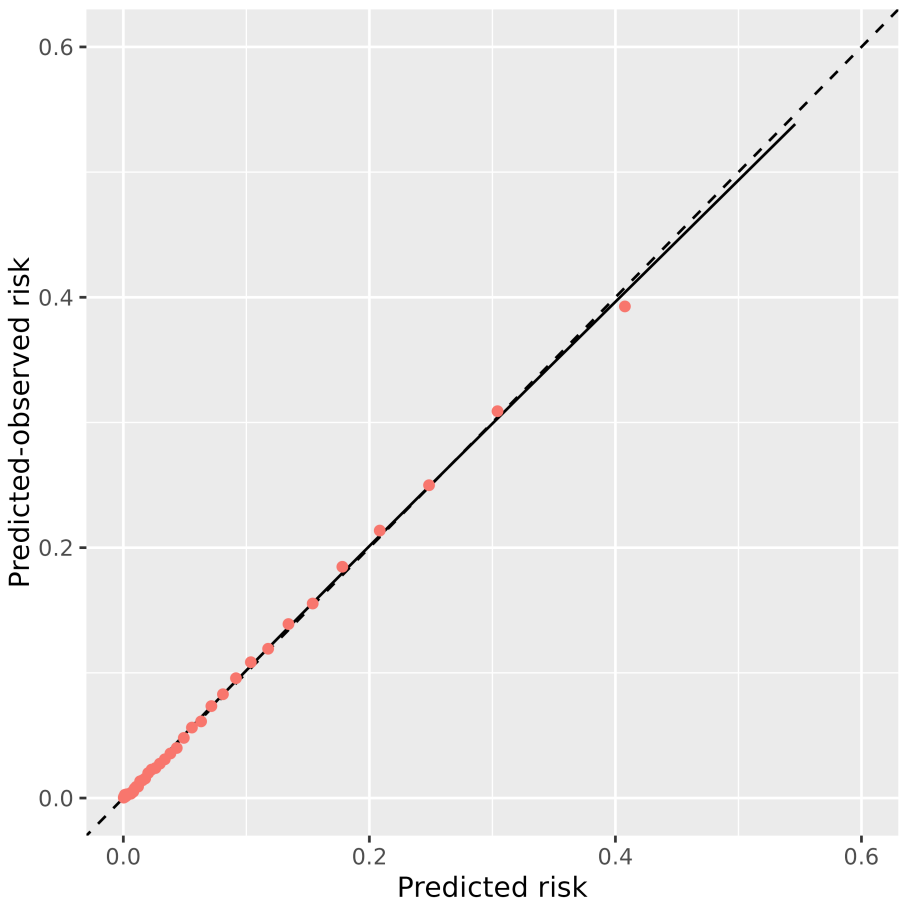

Model 5: Modifiable risk factor model (SBP only); t = 10 years; Estimand: Risk under current intervention strategy

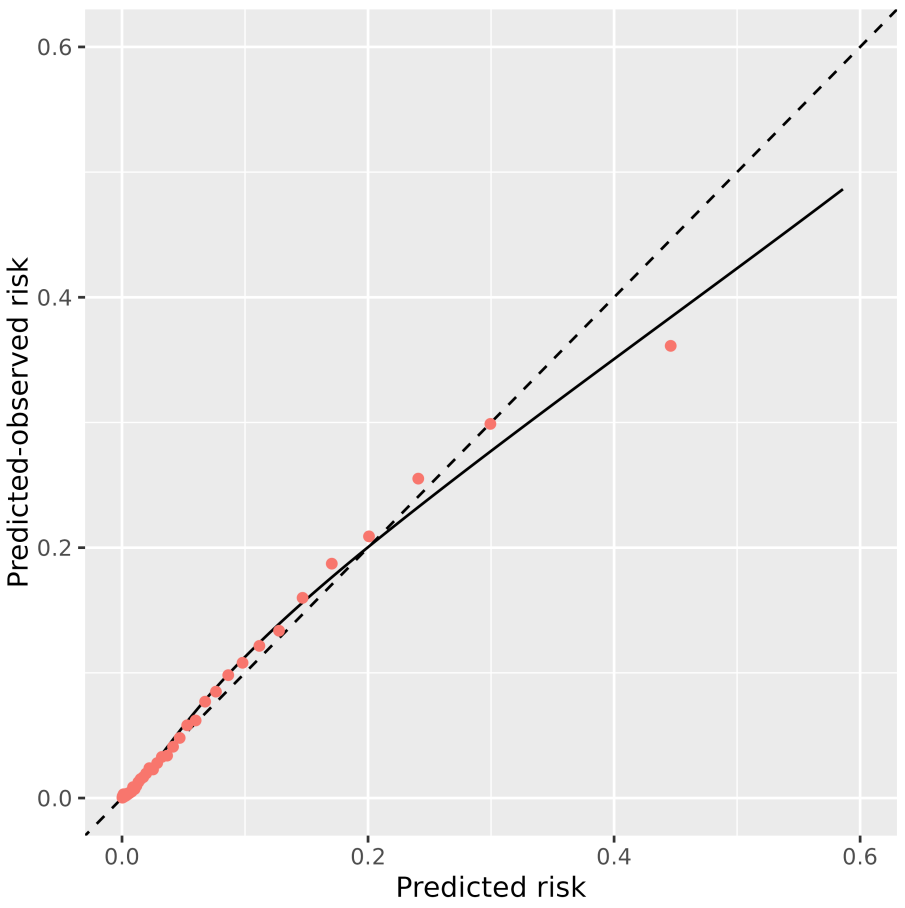

Model 6: Modifiable risk factor model (BMI only); t = 10 years; Estimand: Risk under current intervention strategy

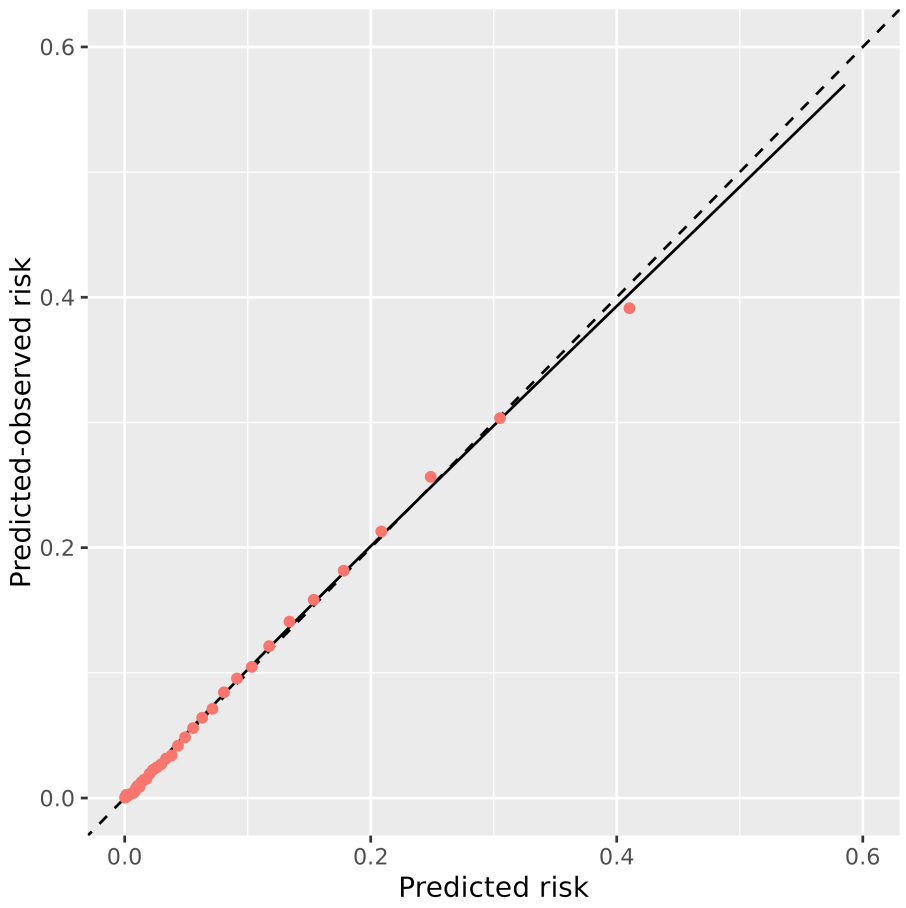

Model 7: Modifiable risk factor model (non-HDL cholesterol only); t = 10 years; Estimand: Risk under current intervention strategy

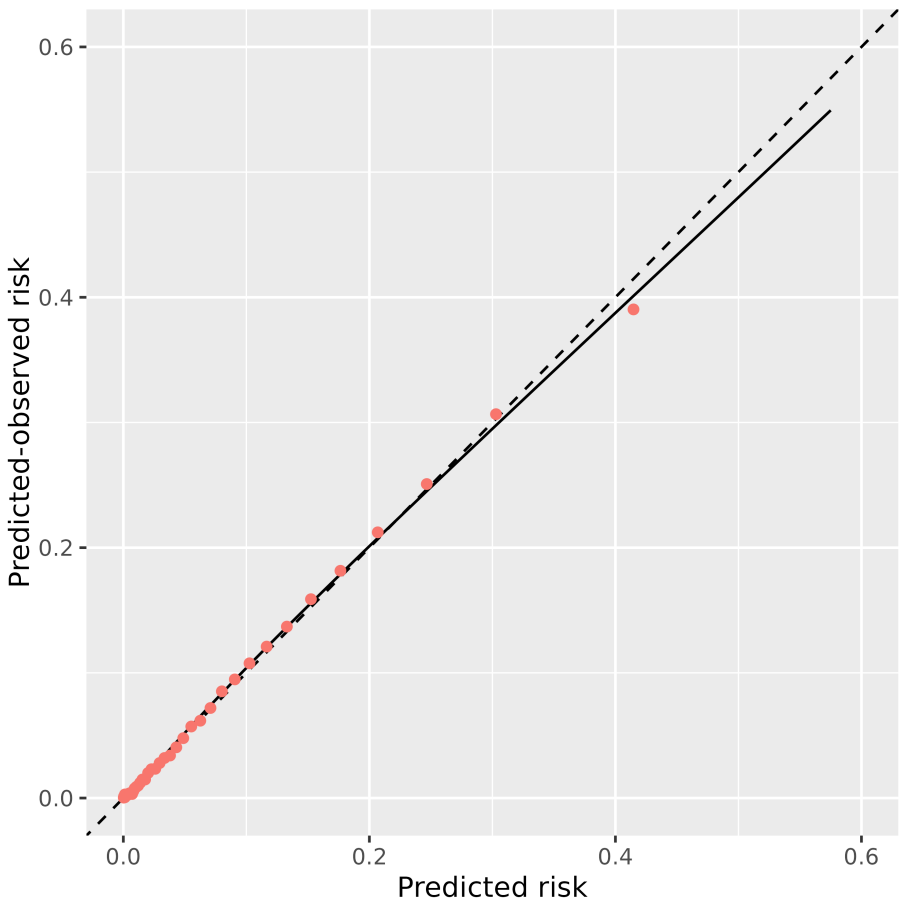

| Model | Estimand | ICI | E50 | E90 |
|---|---|---|---|---|
| 0 | Non-causal estimate | 0.001 (0.001, 0.001) | 0.001 (0, 0.001) | 0.001 (0.001, 0.003) |
| 1 | No antihypertensive use | 0.001 (0.001, 0.002) | 0.001 (0, 0.001) | 0.001 (0.001, 0.003) |
| 2 | No antihypertensive use | 0.001 (0.001, 0.002) | 0.001 (0, 0.001) | 0.001 (0.001, 0.003) |
| 3 | Continue on current interventions | 0.006 (0.005, 0.006) | 0.001 (0.001, 0.002) | 0.015 (0.012, 0.015) |
| 4 | Continue on current interventions | 0.001 (0.001, 0.001) | 0.001 (0, 0.001) | 0.002 (0.001, 0.004) |
| 5 | Continue on current interventions | 0.005 (0.004, 0.005) | 0.001 (0.001, 0.002) | 0.012 (0.01, 0.012) |
| 6 | Continue on current interventions | 0.001 (0.001, 0.002) | 0.001 (0.001, 0.001) | 0.003 (0.002, 0.005) |
| 7 | Continue on current interventions | 0.002 (0.001, 0.002) | 0.001 (0.001, 0.001) | 0.004 (0.003, 0.005) |

# 3 Discrimination

## 3.1 Female

| Model | Estimand | index | index_lower | index_upper |
|---|---|---|---|---|
| 0 | Non-causal estimate | 0.869 | 0.869 | 0.869 |
| 1 | No antihypertensive use | 0.880 | 0.880 | 0.880 |
| 2 | No antihypertensive use | 0.880 | 0.880 | 0.880 |
| 3 | Continue on current interventions | 0.868 | 0.868 | 0.868 |
| 4 | Continue on current interventions | 0.871 | 0.871 | 0.871 |
| 5 | Continue on current interventions | 0.869 | 0.869 | 0.869 |
| 6 | Continue on current interventions | 0.871 | 0.871 | 0.871 |
| 7 | Continue on current interventions | 0.871 | 0.871 | 0.871 |

## 3.2 Male

| Model | Estimand | index | index_lower | index_upper |
|---|---|---|---|---|
| 0 | Non-causal estimate | 0.851 | 0.851 | 0.851 |
| 1 | No antihypertensive use | 0.861 | 0.861 | 0.861 |
| 2 | No antihypertensive use | 0.861 | 0.861 | 0.861 |
| 3 | Continue on current interventions | 0.851 | 0.851 | 0.851 |
| 4 | Continue on current interventions | 0.854 | 0.854 | 0.854 |
| 5 | Continue on current interventions | 0.852 | 0.852 | 0.852 |
| 6 | Continue on current interventions | 0.854 | 0.854 | 0.854 |
| 7 | Continue on current interventions | 0.854 | 0.854 | 0.854 |

# Derivation of the DAG used to drive the intervention component in the CHARIOT project

**Associated manuscript:** A flexible approach to sequential prediction under intervention

**Authors:** Matthew Sperrin, Bowen Jiang. Yun-Ting Huang, Pedro Da Costa Avelar, Niels Peek, Alexander Pate

## Preface

This document outlines the process for estimating the DAG used to drive the intervention layer in the CHARIOT project. This includes the structure of the DAG, the direct effects of modifiable risk factors (MFRs) on cardiovascular disease risk, and the total effect of interventions on cardiovascular disease risk.

In **Section 1**, we introduce the DAG, terminology used through the manuscripts, and goals of this document.

In **Section 2**, we estimate the direct effects of modifiable risk factors (MFRs) on cardiovascular diseases (CVD), and the total effect of each MFR on the other MFRs.

In **Section 3**, we give some examples of how to use to values derived in section 2 to apply the intervention layer of the CHARIOT model.

In **Section 4**, we conduct a sense check/validation, by comparing the total effect of statins/antihypertensives estimated through our DAG, with values from systematic reviews of trials.

In **Section 5**, we clarify where the total effects (presented in Table 3 in the main manuscript) of statins and antihypertensives are obtained from.

## 1. Preliminaries

### 1.1. The causal structure and terminology

The DAG is presented in Figure S1**.** The variables depicted in the DAG are either modifiable risk factors for cardiovascular disease, represented by rectangles, or interventions which should change these modifiable risk factors, represented by diamonds. We are not using this DAG to identify an adjustment set,[1] instead using it to present our assumptions about the causal relationship between the interventions and modifiable risk factors being considered in CHARIOT. We appreciate that normally, “omitting an arc between two variables implies that there is precisely no causal effect of one on the other”.[1] However in this DAG omitted arrows, whether between MFRs or in the reverse direction of two directly connected MFRs, are due to a lack of supporting evidence in the literature for a causal effect in that direction. For example, there is no evidence suggesting that changes in systolic blood pressure (SBP) has a causal effect on body mass index (BMI) or Non-high density lipoprotein (Non-HDL) cholesterol. We assume there are no direct effects of statins or antihypertensives and therefore these arrows have been removed from the DAG (in comparison to the DAG from the main manuscript). Literature which supported the inclusion of an arrow is discussed in the following sections, and the structure of the DAG follows from there.

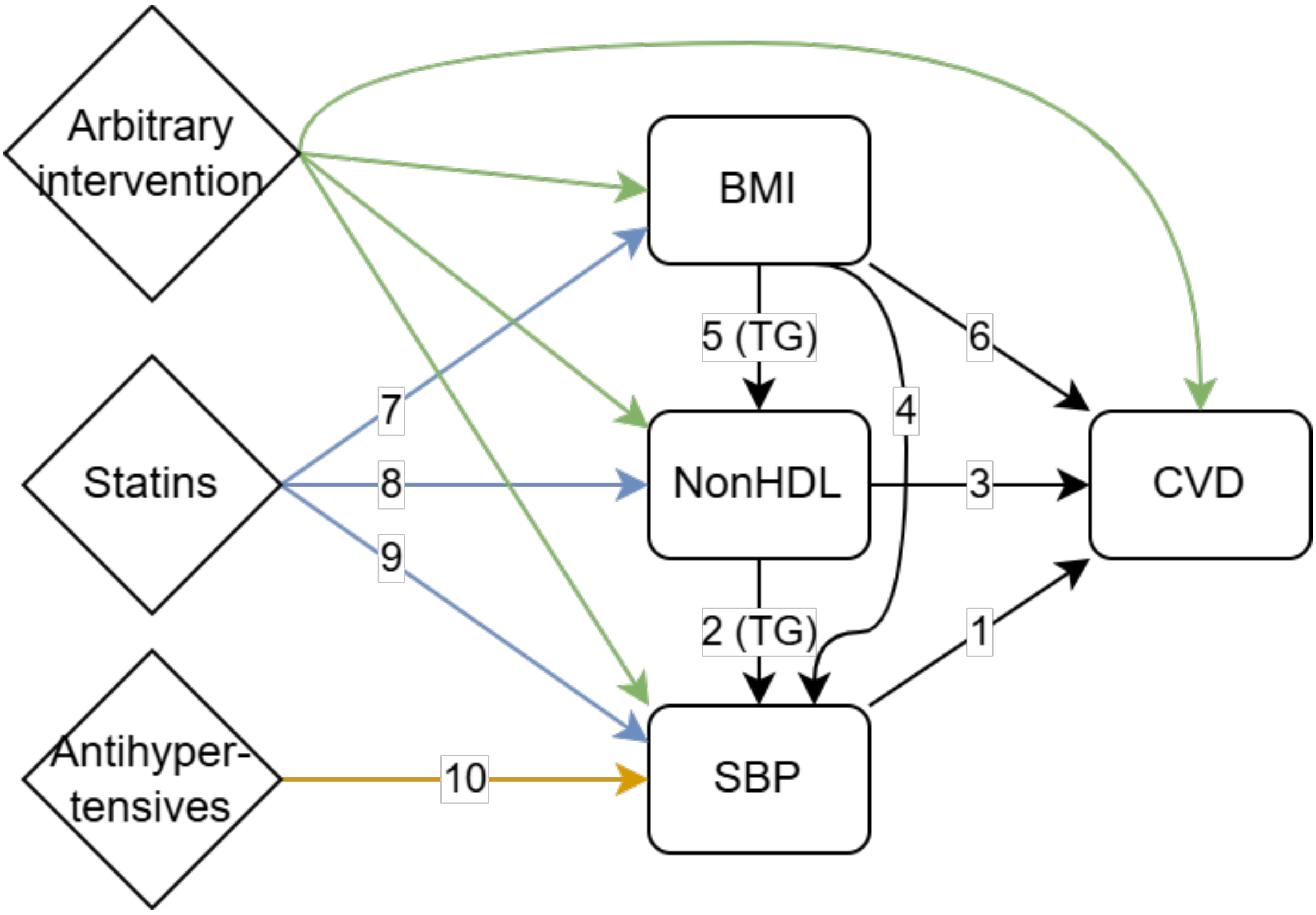

*Figure S3.1: DAG*

Direct effects refer to those that can be represented by a single arrow in the causal structure shown in Figure 1. For example, the direct effect of SBP on CVD is represented by arrow number 1, or the direct effect of BMI on CVD is represented by arrow 6. Indirect effects are represented by a series of arrows. For example, the effect of BMI on CVD that is mediated by non-HDL cholesterol alone is the arrows 5 -> 3. When necessary, these indirect effects will be discussed later. The total effect of a variable on another variable or CVD, is the sum of all direct and indirect effects between the two nodes. For example, the total effect of BMI on CVD is the sum of 6, 5 -> 3, 5 -> 2 -> 1 and 4 -> 1.

## 1.2. Lipid equations

Below are the formulas used for lipid conversion. LDL = low-density lipoproteins. TC = total cholesterol. The "-" in "Non-HDL" is removed when writing formulas due avoid confusion with the deduct symbol, and NonHDL is used through the rest of this document for brevity.

(a) $LDL = TC - HDL - TG/2.2$ (mmol/L)
(b) $NonHDL = TC - HDL$
(c) $NonHDL = LDL + TG/2.2$
(d) $NonHDL = 1.24 * LDL$
(e) $NonHDL = 2.348 * TG$

Formulas (a), the Friedwald equation, is taken from the GP notebook.[2] Formula (b) is by definition. Formula (c) comes from plugging (b) into (a). Formula (d) is taken from the Joint British Societies' consensus recommendations for the prevention of cardiovascular disease.[3] Formula (e) follows from substituting (d) into (c).

It is worth nothing here that although NonHDL includes two lipid components, LDL and TG, the TG labels on directed edges 2 and 5 indicate that triglycerides (TG) is the only component contributing to these causal pathways. This claim is based on the following Mendelian randomisation studies by Xu et al.,[4] "*There was no evidence for a causal relation between BMI and LDL-Cholesterol and TG mediated the effect between BMI and lipids.*" and Liu et al.,[5] "*TG was the most likely casual exposure among major lipids to increase SBP".*

## 1.3. Our aim/what is required for CHARIOT

Our first aim is to estimate the direct effect of the MFRs on CVD, that is, the direct effect of each of SBP, NonHDL and BMI on CVD (arrows 1, 3, 6). CHARIOT uses these direct effects to estimate the risk under some intervention which modifies these modifiable risk factors. The process for doing this is illustrated in section 3. Of note, the process works by converting an individual's risk score on to the hazard scale and applying the effect of the intervention through a hazard ratio, before converting back to a risk. This ensures the resulting risk score is between 0 and 1, and ensures scale consistency between the two components. This means relative risks (RRs) and odds ratios (ORs) reported in the literature will be converted to hazard ratios (HRs) where necessary for the DAG. Our second aim is to estimate the total effect of each of the MFRs on the other MFRs, and the total effect of an intervention on each of the MFRs. CHARIOT uses these to estimate the total risk reduction in CVD from an intervention by its effect on each of the MFRs.

In some cases, the direct effects were readily available in the literature. In other cases, these were estimated by decomposing total effects into sum of the direct and indirect effects. The method for estimating these effects is imperfect, and at times requires crude averaging. This is somewhat a limitation of the available data in terms of the reported effect estimates. While effect estimates are taken from systematic reviews and meta-analyses where possible, given the nature of this work, estimates are also taken from a number of mendelian randomization studies which may be considered a weaker form of evidence. For this reason, attempts are made to validate our process in section 4. For example, we compare the total effect of an intervention on CVD reported directly in the literature, with the total effect of the intervention on CVD as calculated by the sum of the expected change in each of the MFRs in the DAG.

In this iteration of CHARIOT we do not consider heterogeneity of the transitions in the DAG (e.g. by gender, or medical history). This is primarily because there was a lack of scenarios in the literature where we found significant effect heterogeneity, but could be considered in future iterations.

The derivation of these values in replicated in code on GitHub[6] (see file code/p1_data_prep/p8_calculate_treatment_offsets.R).

*Table S3.1: The direct effects of MFRs on CVD*

| **Pathway** | **Study type** | **Extra analyses required** | **Direct effect on CVD (HR)** | **Section derived** |
|---|---|---|---|---|
| (1) SBP -> CVD | Meta-analysis of trials.[7] | Direct from study | 1 mmHg increase: 1.02463 | Section 2.1 |
| (3) NonHDL -> CVD<br>- 1 mmol/L increase in NonHDL | Meta-analysis of trials of LDL.[3] | Total effect reported in study. Direct effect estimated by deducting the indirect effect through pathway:<br>NonHDL -> SBP -> CVD. | 1mmol/L increase: 1.21371 | Section 2.3 |
| (6) BMI -> CVD<br>Per unit BMI reduction | Causal Mediation. Lu et al., 2015[8] | Calculation required no use of other sources, but effect was not directly reported in that study. Some calculations and assumptions were required based on the available data in that study. | $1\mathrm{kg/m^2}$ Increase: 1.021503 | Section 2.6 |

*Table S3.2: The total effects between MFRs*

| **Effect** | **Study type** | **Extra analyses required** | **Value** | **Section derived** |
|---|---|---|---|---|
| Total effect of NonHDL on SBP (per 1 mmol/L reduction in NonHDL)<br><br>Paths: 2 | Mendelian Randomisation[5] | Effect of TG ->SBP taken directly from a study, this is converted to NonHDL -> SBP using formula (e). | 0.2749577 mmHg reduction in SBP | Section 2.2 |
| Total effect of BMI on SBP (per unit reduction in BMI)<br><br>Paths: 4; 5 -> 2 | Mendelian Randomisation[9] | Direct from study | 0.7 mmHg reduction in SBP | Section 2.4 |
| Total effect of BMI on NonHDL (per unit reduction in BMI)<br><br>Paths: 5 | Mendelian Randomisation[4] | Estimate of BMI -> TG is taken directly from study, this is then converted into effect on NonHDL based on formula (e) | 0.1043451 mmol/L decrease | Section 2.5 |
| Total effect of statins on BMI<br><br>Paths: 7 | Mendelian Randomisation[10] | Direct from study | Increase 0.33 BMI unit | Section 2.7 |
| Total effect of statins on NonHDL cholesterol<br><br>Paths: 8; 7 -> 5. | Meta-analysis of randomised controlled trials[11] | Study reported the reduction for LDL cholesterol, then we convert for the reduction for NonHDL | Decrease 1.3268 mmol/L | Section 2.8 |
| Total effect of statins on SBP<br><br>Paths: 9; 8 -> 2, 7 -> 5 -> 2; 7 -> 4. | Meta-analysis of randomised controlled trials[12] | Direct from study | Decrease 2.62 mmHg | Section 2.9 |

# 2. Derivation of the direct effects of each MFR on CVD, and the total effect of each MFR on the other MFRs

In this section, we explain the process to estimate the direct effects of the MFRs on CVD (Table 1). This is arrows 1, 3 and 6, as well as the total effect of changes in the MFRs and some interventions on the other MFRs (Table 2). We start as far downstream the DAG and possible and work back up.

## 2.1. The direct effect of SBP on CVD (path 1)

As assumed in the DAGs, there are no MFRs downstream of SBP. This implies that a change in SBP does not result in any indirect effects. In other words, the total effect of SBP is equivalent to its direct effect. From Ettehad et al.,[7] a relative risk (RR) 0.8 was reported for 10 mmHg reduction in SBP. This is equivalent to an RR of $0.8^{(1/10)} = 0.9779328$ per 1 mmHg reduction. The RR was then converted to an hazard ratio (HR) of 0.975962, or 1.02463 per 1mmHg increase.

There is not a 1:1 conversion between RR and HR, as the conversion is dependent on the baseline risk of the individual (prior to applying the HR or RR). We therefore used an optimal minimax conversion.[13] Note, this conversion formula requires an upper and lower risk score, for which we chose 30% and 0% for all conversions made throughout this manuscript. While some individuals have risks higher than 30%, we considered these individuals high risk outliers, who may have specific needs or bespoke treatment plans. Including these individuals would make the conversion less accurate for pool of individuals with a risk between 0% and 30%, and therefore we set the upper limit at 30%, rather than the actual max risk of the population.

We present values to many digits to avoid any rounding error confusion. The code for running these calculations is provided on the project GitHub page.

## 2.2. The total effect of NonHDL on SBP

Liu et al.,[5] reports that per one SD decrease in TG (1.0244 mmol/L), this results in an average 0.6615 mmHg decrease in SBP.

In the formula (e) NonHDL=2.348*TG, we can say that when we have a 1 mmol/L decrease in NonHDL, it causes a decrease in TG by 0.4258 mmol/L.

The estimated reduction in SBP from a 0.4258 mmol/L reduction in TG is:

$$\frac{0.4258}{1.0244} * 0.6615 = 0.2749577 \text{ mmHg.}$$

Therefore, a 1mmol/L reduction in NonHDL results in a 0.2749577 reduction in SBP.

## 2.3. The direct effect of Non-HDL cholesterol on CVD (path 3)

Taken from the Joint British Societies consensus recommendations for the prevention of cardiovascular disease,[3] a meta-analysis on LDL cholesterol trials[14] report a HR of 0.8184262 per 1 mmol/L reduction in nonHDL, or 1.221857 per 1 mmol/L increase.

We assume the total effect can be decomposed into the direct effect, and indirect effect via SBP, which corresponds to the combined effect of causal paths 2->1 and 3. To estimate the direct effect of NonHDL on CVD (path 3), the indirect effect must be subtracted from the total effect. Given the effect estimates are hazard ratios, which are multiplicative, this is done by dividing the total effect by the indirect effect.

As derived in Section 2.2, a 1 mmol/L increase in Non-HDL cholesterol leads to a 0.2749577 mmHg increase in SBP. As derived in section 2.1, a 1mmHg increase in SBP increases the hazard of CVD by 1.02463. The indirect effect of a 1mmol/L increase in NonHDL is therefore the equivalent of the effect of a 0.2749 mmHg increase in SBP, calculated by:

$$1.02463^{0.2749577} = 1.006713.$$

The direct effect can then be calculated as:

$$direct\ effect = \frac{total\ effect}{indirect\ effect} = \frac{1.221857}{1.006713} = 1.21371$$

## 2.4. The total effect of BMI on SBP

As reported by Holmes et al.,[9] each unit increase in BMI results in a 0.7 mmHg increase in SBP. This is the combined effect of causal paths 4, and 5->2.

## 2.5. The total effect of BMI on NonHDL

As reported by Xu et al.,[4] an increase in BMI of 4.5 results in a 0.2 mmol/ L increase in TG. Therefore a unit increase in BMI results in a $0.2 \times \frac{1}{4.5} = 0.04444$ mmol/L increase in TG.

Applying formula (e), a unit increase in BMI results in a $2.348 \times 0.044 = 0.1043451$ mmol/L increase in NonHDL.

## 2.6. The direct effect of BMI on CVD (path 6)

A study by Lu et al.,[8] found the direct effect of normal weight vs overweight to be a hazard ratio of 1.12, and normal weight vs obese to be 1.22. We applied the following steps to derive the direct effect of a unit change in BMI on CVD risk:

**Step 1:** Define the BMI median points for each category. Normal weight: 22.5; Overweight 27.5; Obesity: 32.5.

**Step 2:** Calculate the median BMI difference between overweight and obese groups with the normal group.

Overweight – Normal: ΔBMI = 27.5 - 22.5 = 5.0

Obesity – Normal: ΔBMI = 32.5 - 22.5 = 10

**Step 3:** Convert the HRs for each group vs the normal group to a per-unit BMI on the log-hazard scale:

Overweight vs Normal: $\ln(1.12)/_{5} = 0.02266$

Obesity vs Normal: $\ln(1.22)/_{10} = 0.01989$

Take the average: (0.02266+0.01989)/2 = 0.021275

**Step 4:** Convert back to the hazard scale: $e^{0.021275} = 1.0215$

## 2.7. The total effect of statin initiation on BMI

A mendelian randomization study by Li and Schooling(Li & Schooling, 2022) found the total effect of statins on BMI was an increase of 0.33.

## 2.8. The total effect of statin initiation on NonHDL

From a meta-analysis of trials on the effects of lowering LDL cholesterol with statin therapy, the weighted mean difference at 1 year was a -1.07 mmol/L reduction.[11]

Using equation (d), this can be converted to give an estimate for the effect on statins on NonHDL cholesterol to be a 1.24*-1.07= -1.3268 mmol/L reduction.

## 2.9. The total effect of statin initiation on SBP

From a meta-analysis of prospective controlled studies on the antihypertensive effects of statins,[12] the total effect of statins on SBP was found to be a reduction of -2.62 mmHg.

## 2.10. The total effect of antihypertensives on SBP

Unable to define total effect. See section 4.2.

# 3. An exemplar of how to use these estimates to implement the intervention layer

Here we showcase how these estimates are used to implement the intervention layer, aligning this with the exemplar use-case from the main manuscript. The risk from the initial risk estimation layer in the exemplar use case is 16.88%.

## 3.1. Arbitrary intervention that acts on all MFRs

First, we consider a situation where the aim is to estimate the risk under some arbitrary intervention, where the effect of this intervention on each of the MFRs can be hypothesized. This aligns with the 'best case scenario' situation in the exemplar use case, where we hypothisize through a combination of changes to diet and exercise, the individual may reduce their BMI by 5kg/m$^2$, non-HDL cholesterol by 1mmol/L and systolic blood pressure by 7.5mmHg. Utilising the hazard ratios for the direct effects from Table S3.1, the hazard ratio for a reduction of the MFR by the specified amount is then estimated:

HR (effect of change in BMI) = $(1.021503)^{-5}$

HR (effect of change in NonHDL) = $(1.21371)^{-1}$

HR (effect of change in SBP) = $(1.02463)^{-7.5}$

The combined HR is the product of these:

$$= 0.617211.$$

The risk score is converted onto the hazard scale, then multiplied by the combined hazard ratio, and then converted back to a risk score to give 10.78%. See the `convert_risk_HR` in the p5_worked_example.R program, or the process is implemented manually in p8_calculate_treatment_effects.R.

## 3.2. Statin initiation

When applying an arbitrary intervention like in section 3.1, we hypothesize its total effect on each of the MFRs. This means we only need to work with the direct effects, and the magnitude of the change in each of the MFRs. For a pre-specified intervention, such as statin therapy or antihypertensive therapy, we use the DAG to propagate through the expected effect of the intervention on each of the MFRs.

Consider statin initiation. Based on Table S3.2, we assume that statins results in a 1.3268mmol/L decrease in nonHDL cholesterol, a 2.62mmHg decrease in SBP, and a 0.33 increase in BMI.

The hazard ratios induced by the change in each MFR are therefore:

HR (effect of change in BMI) = $(1.021503)^{0.33}$

HR (effect of change in NonHDL) = $(1.21371)^{-1.3268}$

HR (effect of change in SBP) = $(1.02463)^{-2.62}$

The combined HR is the product of these = $0.7307345$, is the total effect of statin initiation estimated via our DAG. As was done previously for the arbitrary intervention, the risk score is converted to the hazard scale, multiplied by the combined HR, and then converted back to a risk score, to give 12.64%.

Note, the total effects of interventions could also be taken directly from literature. This could be applied to the risk score estimated from the initial risk estimation layer instead, rather than calculating the total effect as the sum of the indirect effects mediated through the MFRs. The advantage of latter method is that it very explicit about what changes are expected in each of the MFRs. At follow-up visits, this can help facilitate further discussions around whether desired levels of the MFRs have been achieved. For example, if an individual has been adherent to their statin regime, and had a small increase in BMI, this is in-line with what was expected and therefore would not be discouraging to the individual. Furthermore, if an individual has reduced their nonHDL by 2 mmol/L (as opposed to the expected 1.3268), this indicates the intervention has been particularly effective in this individual, and the risk reduction would be bigger than initially predicted. On the contrary, if the treatment has been ineffective and resulted in no reduction in nonHDL, this would be reflected in the estimated risk under intervention, which would be smaller than initially predicted at visit 0.

The inclusion of a direct effect of statin treatment is considered in supplementary file 7.

## 4. External consistency validation

As an attempt to validate the effect estimates from the DAG, we compare total effects of statins/antihypertensives as estimated from the DAG, with estimates of the total effect published from systematic reviews in literature.

### *4.1.* External consistency validation for statins

The goal of this is to calculate the total effect of statins on CVD through the DAG developed in section 2, and compare with the average total effect of statins reported from a 2016 overview of systematic reviews of trials by Karmali et al.,[15] considered to be high quality evidence. This review found statins reduced the risk of ASCVD events with a relative risk of 0.75.

Alternatively calculating through the DAG (see sections 2.7, 2.8 and 2.9) we expect statin therapy to result in an increase in BMI of 0.33, a reduction in NonHDL cholesterol of 1.3268 mmol/L, and a reduction in SBP of 2.62 mmHg. As derived in section 3.2, the combined HR is the product of these = 0.7307345. We assume there is no direct effect of statins on cardiovascular disease, and it works solely through modification of these risk factors. Therefore this is the total effect of statin initiation estimated via our DAG. Converting this to a RR, for comparison, results in RR of 0.7475701.

The comparison is therefore 0.75 and 0.7475701. This gives strong agreement and supports the derivation of the direct effects of each of the MFRs. Note, any difference between these two could also be considered to be the direct effect of statins.

### *4.2.* External consistency validation for anti-hypertensive

The goal of this is to calculate the total effect of antihypertensives on CVD through the DAG developed in section 2, and compare with the average total effect of antihypertensives reported from a 2016 overview of systematic reviews of trials by Karmali et al.,[15] considered to be high quality evidence. This review found compared with placebo, BP-lowering therapy reduced the risk for coronary heart disease (RR, 0.84; 95% CI, 0.79-0.90) and stroke (RR, 0.64; 95% CI, 0.56-0.73).

The direct effect of SBP on CVD used in our DAG is taken from a 2016 systematic review and meta-analysis on blood pressure lowering trials for the prevention of cardiovascular disease and death, by Ettehad et al.[7] They reported that a 10 mmHg reduction in SBP reduced risk of major cardiovascular event by 0.8 (RR). They also report a 10mmHg reduction in the risk of coronary heart disease by 0.83, and stroke by 0.73.

Given antihypertensives (according to our DAG) operates solely through SBP, and the effect estimate for SBP on CVD is reported as a relative risk in the literature, and the comparator is also reported as a relative risk in the literature, for this validation we will work solely with relative risks, and not convert to HRs.

To estimate the total effect of antihypertensive therapy, we also need the average reduction in SBP achieved with treatment. A 2022 IPD meta-analysis by Canoy et al.,[16] reported that the mean SBP difference between active treatment and placebo, for measurements taken after more than 12 months, was –5.1 mmHg. They also reported the difference between more and less intense blood pressure lowering treatment resulted in a difference of -11.1 mmHg. In an older meta-analysis from 2009, Law et al.,[17] reported the average SBP reduction was estimated at 5.99 mmHg.

We can then standardize the relative risk reductions in CHD and stroke for a 10mmHg change in SBP reported in the Ettehad study,[7] based on the different expected changes in SBP.

For CHD:

$$(0.83)^{5.1/10} = 0.9093 \text{ [Canoy et al., 2022, placebo]}$$

$$(0.83)^{5.9/10} = 0.8959 \text{ [Law et al., 2009, placebo]}$$

$$(0.83)^{11.1/10} = 0.8131 \text{ [Canoy et al., 2022, intensity trials]}$$

For stroke:

$$(0.73)^{5.1/10} = 0.8517 \text{ [Canoy et al., 2022, placebo]}$$

$$(0.73)^{5.9/10} = 0.8305 \text{ [Law et al., 2009, placebo]}$$

$$(0.73)^{11.1/10} = 0.7052 \text{ [Canoy et al., 2022, intensity trials]}$$

The comparator, is the estimate of BP-lowering therapy from the overview of systematic review of trials,[15] which found the risk for coronary heart disease (RR, 0.84; 95% CI, 0.79-0.90) and stroke (RR, 0.64; 95% CI, 0.56-0.73). For CHD, the comparator lies somewhere in the middle of the different estimates based on the DAG. For stroke, the estimates from the DAG underestimate the benefit across the board (RRs are too high).

The systematic reviews indicated a lot of variability in the effect of an antihypertensives, and also a big difference on the expected change in SBP depending on if an individual is initiating antihypertensives, or increasing anti-hypertensive intensity. We believe it is therefore difficult to estimate an average risk reduction from taking antihypertensives, given the expected benefit is so dependent on the strength of the antihypertensives, and how much reduction in SBP is achieved.

Given these inconsistencies, this is why we have opted to not allow for an “antihypertensive” intervention, like is done for statins. Instead, the user is forced to specify the expected/desired reduction in SBP. The effect estimate for the reduction in CVD risk based on a specified reduction in SBP, taken from Ettehad et al.,[7] is taken from a comprehensive systematic review.

## 5. Total effect of statins treatment, antihypertensive treatment

The total effect of statins (0.75) and antihypertensives (0.74) are used to adjust for changes in treatment use during follow-up during development of each of the models.

For clarity on where these are taken from, the effect of statins (0.75) is taken directly from the overview of systematic reviews by Karmali et al.,[15] which reports the average risk reduction for ASCVD. The effect of antihypertensives (0.74) is the average of the effects for stroke (0.64) and CHD (0.84) reported by Karmali et al.[15] These are converted to hazard ratios using the approach of Van Der Weele.[13]

# 6. An algorithm for automatically calculating DAG direct effects from prior knowledge

To generalise this process, we describe an algorithm for calculating direct effects from a DAG and associated information from the literature about total and direct effects, which we also provide an implementation of in *code/p8_dag_de/dag_solver.py*. Following this more general procedure, will result in the same estimates obtained in section 2 of this document, up to differences in numerical errors.

The algorithm expects two inputs: A DAG and a list of evidence for total and direct effects between nodes in the DAG. We assume the DAG is correct and complete, which means it represents the true underlying process behind the interventions, MRFs and outcomes, and that there are no unobserved edges or confounders. As for the list of evidence, we assume that it respects the graph's topology, that is, that it does not contain direct effects between nodes not connected by a direct edge, and that there are no evidence for total effect for nodes that have no path between them. Furthermore, we assume that no duplicate information exists, as that could imply two different estimates for the same direct effect. If such information did exist, a choice could be made as to which comes from a higher quality of evidence base. It might not be possible to recover all direct effects in the DAG, but the algorithm will nonetheless attempt to recover as many direct effects using the total effects available. Each total effect constitutes a subgraph between its source and target node, and such subgraph is solvable if, in all simple paths between the source and target nodes, only a single direct effect is unknown. The algorithm then proceeds as follows:

1. Populate the DAG with the direct effects present in the evidence.
2. Select only total effects from the evidence. Each total effect defines a subgraph between the source and target nodes, which is used in steps 3.1.1-3.1.4 below.
3. Do:
   - 3.1. For each total effect evidence do:
     - 3.1.1. List all paths (both direct and indirect) between the evidence's source and target nodes.
     - 3.1.2. If there is more than one edge with a missing direct effect, skip this subgraph as unsolvable for now. Otherwise:
     - 3.1.3. Accumulate a list of indirect effects over all complete paths' $I$ (i.e., paths with no missing direct effects). That is, for each possible path between the source and target node, accumulate the direct effects through that path.
     - 3.1.4. The path with a missing direct effect will have its indirect effect be the total effect discounting the cumulative indirect effects over the other paths. In the case of a single-edge path (i.e., a direct path), $DE = \frac{TE}{\prod_i^I E_i}$, if we're calculating an Odds Ratio or Hazard Ratio effect, or $DE = \frac{TE}{\prod_i^i E_i}$
     - 3.1.5. We then remove this evidence from the pending total effects to avoid re-calculating it in case the do-while loop repeats.
4. While there was at least one solvable evidence subgraph

Running this algorithm on our DAG in Figure S3.1 and on the evidence provided in tables S3.1 and S3.2, would give us the resulting direct effects in Table S6.1 below. Note that the algorithm calculates the direct effects for every single evidence, while the manual version only calculated the direct effects between MFRs and the outcome, as it could simply use the total effect between the interventions and MFRs instead. Direct effects that were not calculated are displayed as N/C.

*Table S6.1: Direct effects calculated Manually and through the Algorithm*

| Source | Target | Algorithm DE | Manual DE | Effect type |
|---|---|---|---|---|
| NonHDL | CVD | 1.21137745189621 | 1.21371 | HR |
| BMI | SBP | 0.67130951129773 | N/C | Linear |
| Statin | BMI | 0.33 | 0.33 | Linear |
| Statin | NonHDL | -1.361233883 | N/C | Linear |
| Statin | SBP | -2.47671826236825 | N/C | Linear |

For didactive purposes, we provide an example run of the algorithm to repeat the steps taken in section 2. We present plots generated from running the algorithm, which plots the DAG using the `networkx` python library, and then show the subgraph being considered at each step, where edges with known (or already deduced) direct effects will have their effects labeled, with a * meaning that the effect is a Hazard Ratio.

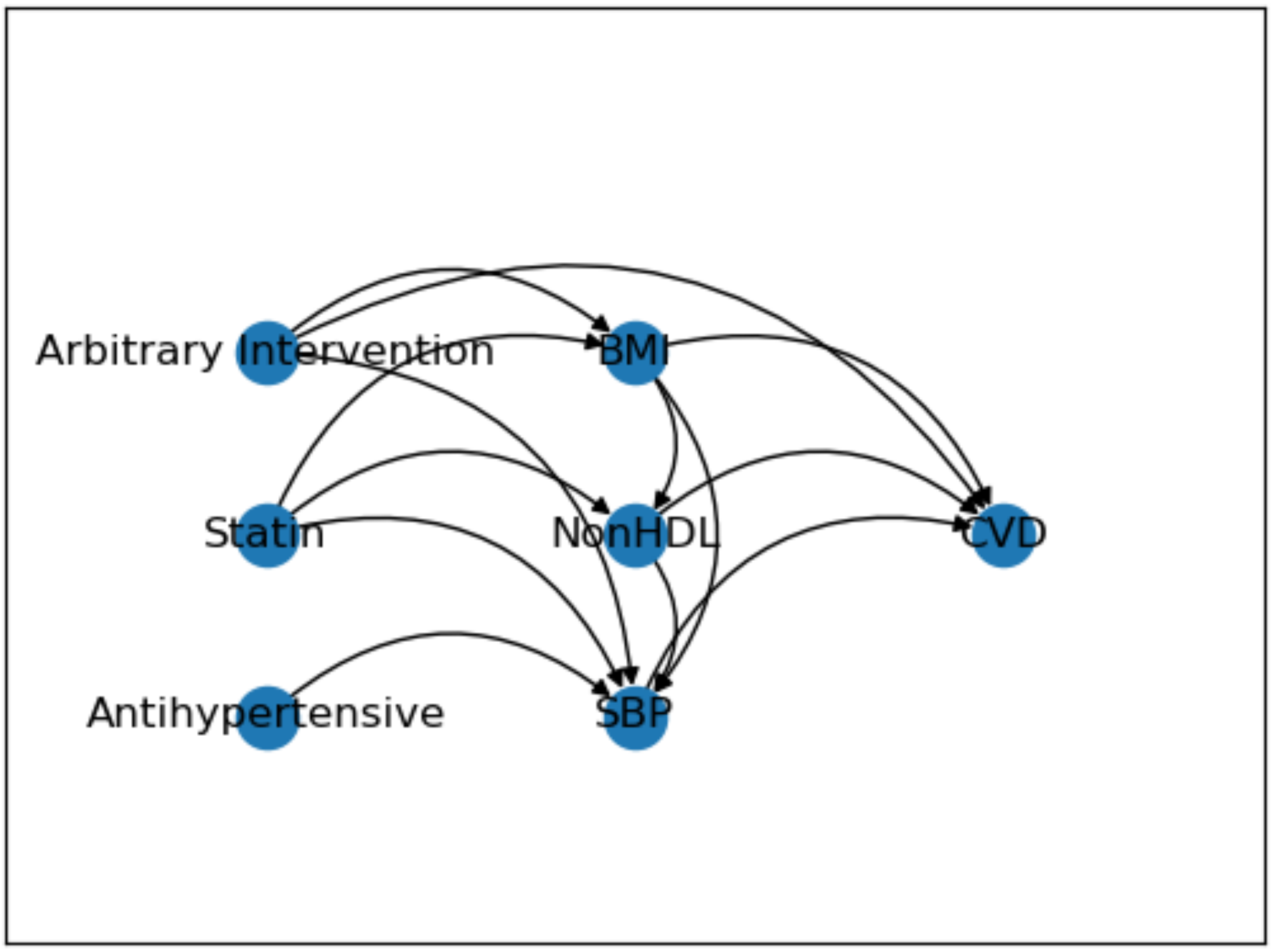

(Step 1): The Direct effects available in the literature were:

- (Direct Effect) SBP [mmHg] → CVD 1.02464* (Hazard Ratio);
- (Direct Effect) NonHDL [mmol/L] → SBP [mmHg] 0.2749577;
- (Direct Effect) BMI [kg/m2] → NonHDL [mmol/L] 0.1043451;
- (Direct Effect) BMI [kg/m2] → CVD 1.0215* (Hazard Ratio);

Which are used to populate the graph's edges initially as below:

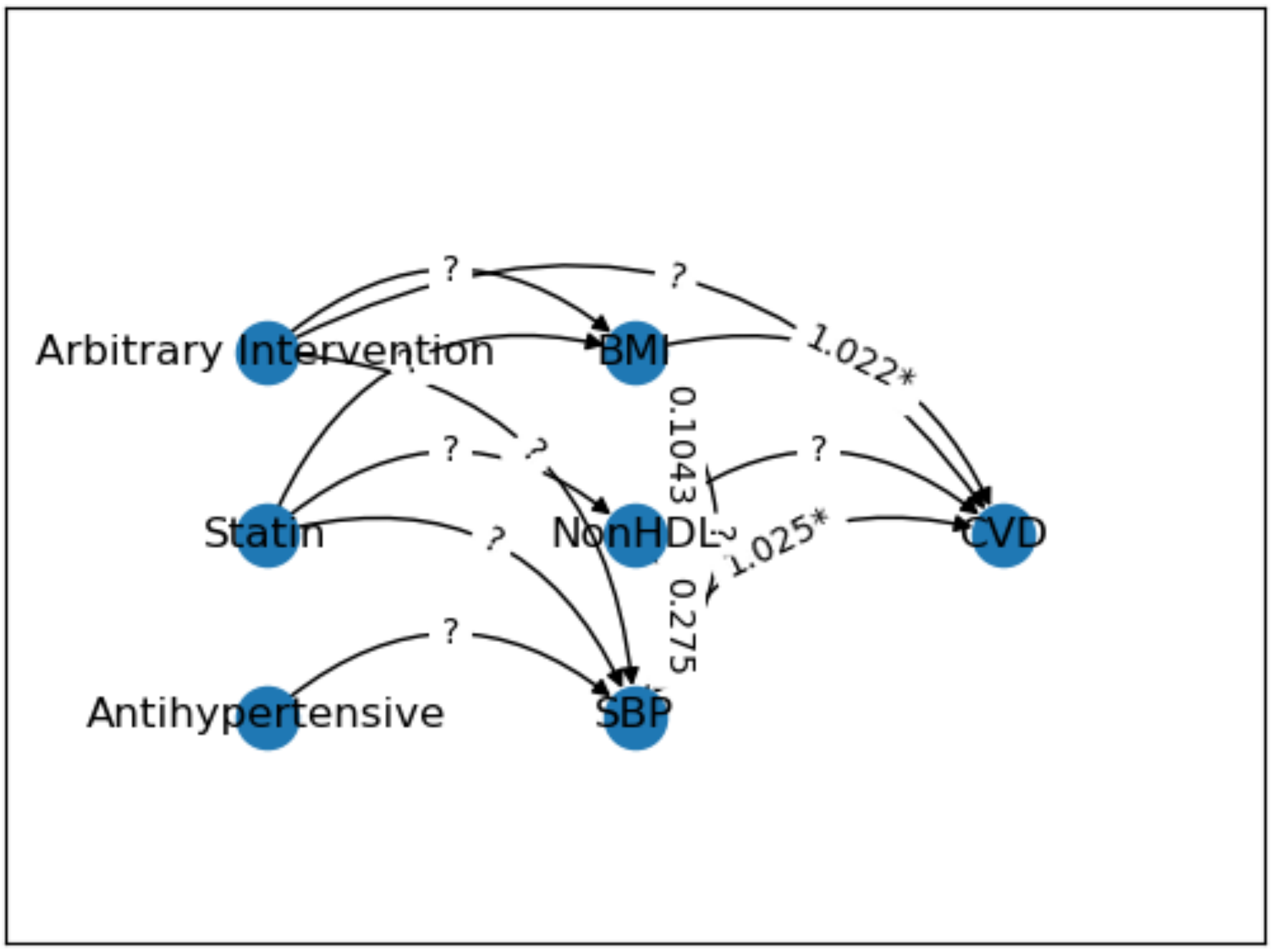

(Step 2) We then collate the following total effects from the evidence:

- (Total Effect) NonHDL [mmol/L] → CVD of 1.219512195* (Hazard Ratio)
- (Total Effect) BMI [kg/m2] → SBP [mmHg] of 0.7
- (Total Effect) Statin → BMI [kg/m2] of 0.33
- (Total Effect) Statin → NonHDL [mmol/L] of -1.3268 [mmol/L]
- (Total Effect) Statin → SBP [mmHg] of -2.62

(Steps 3-4, do-while scan 1) Then we go through each total effect one by one:

(Step 3.1, iteration 1) Using the Total Effect of NonHDL → CVD of 1.219512195:

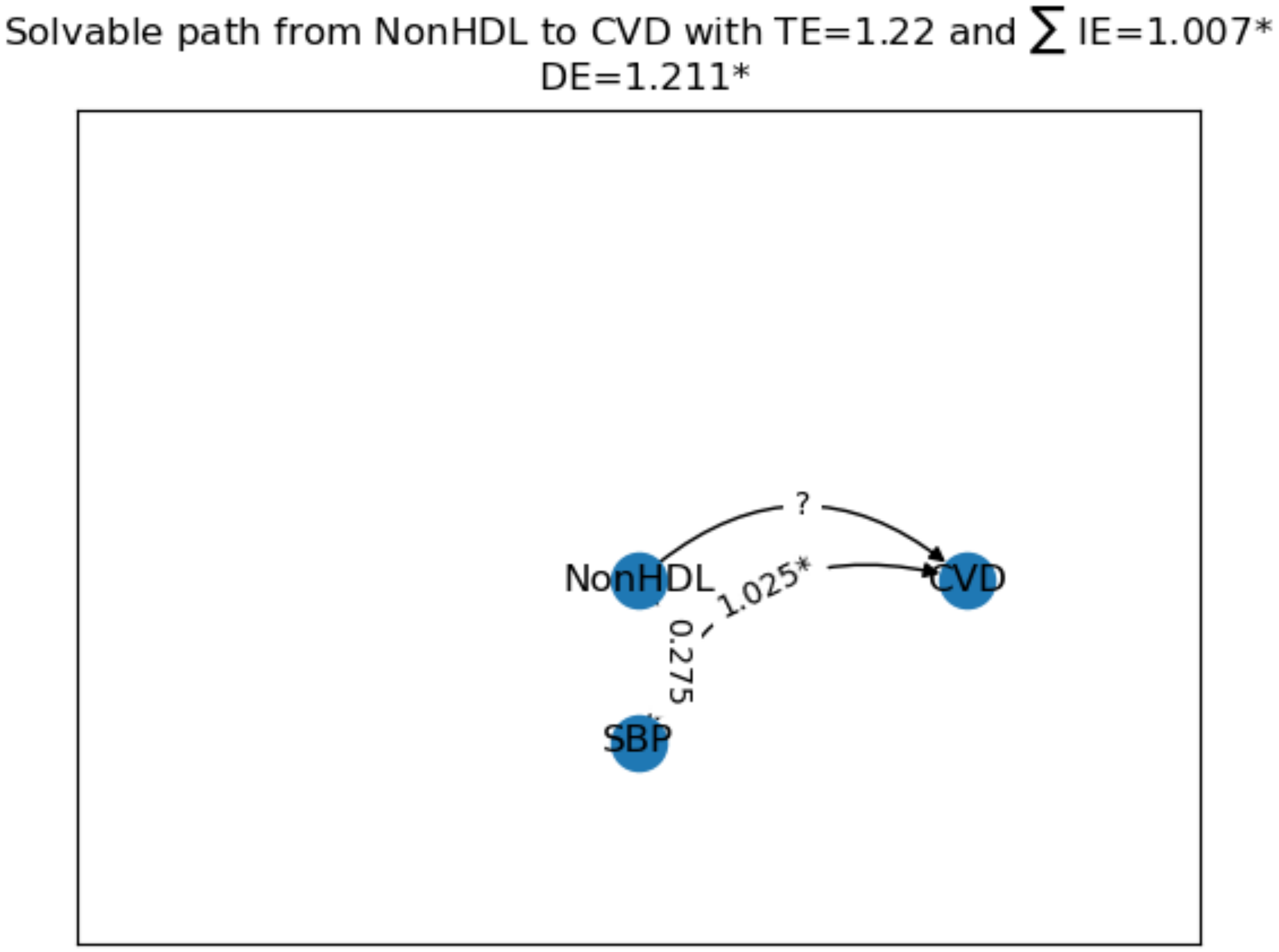

(Step 3.1.1) There are two paths, NonHDL → CVD and NonHDL → SBP → CVD.

(Steo 3.1.2) In all these paths only the NonHDL → CVD direct effect is missing, so the subgraph is solvable.

(Step 3.1.3) The indirect effects of NonHDL on CVD through NonHDL → SBP → CVD is calculated as 1.007.

(Step 3.1.4) The total effect then has this indirect effect discounted giving 1.231 as the missing NonHDL → CVD direct effect.

(Step 3.1, iteration 2) Using the Total Effect of BMI → SBP of 0.7:

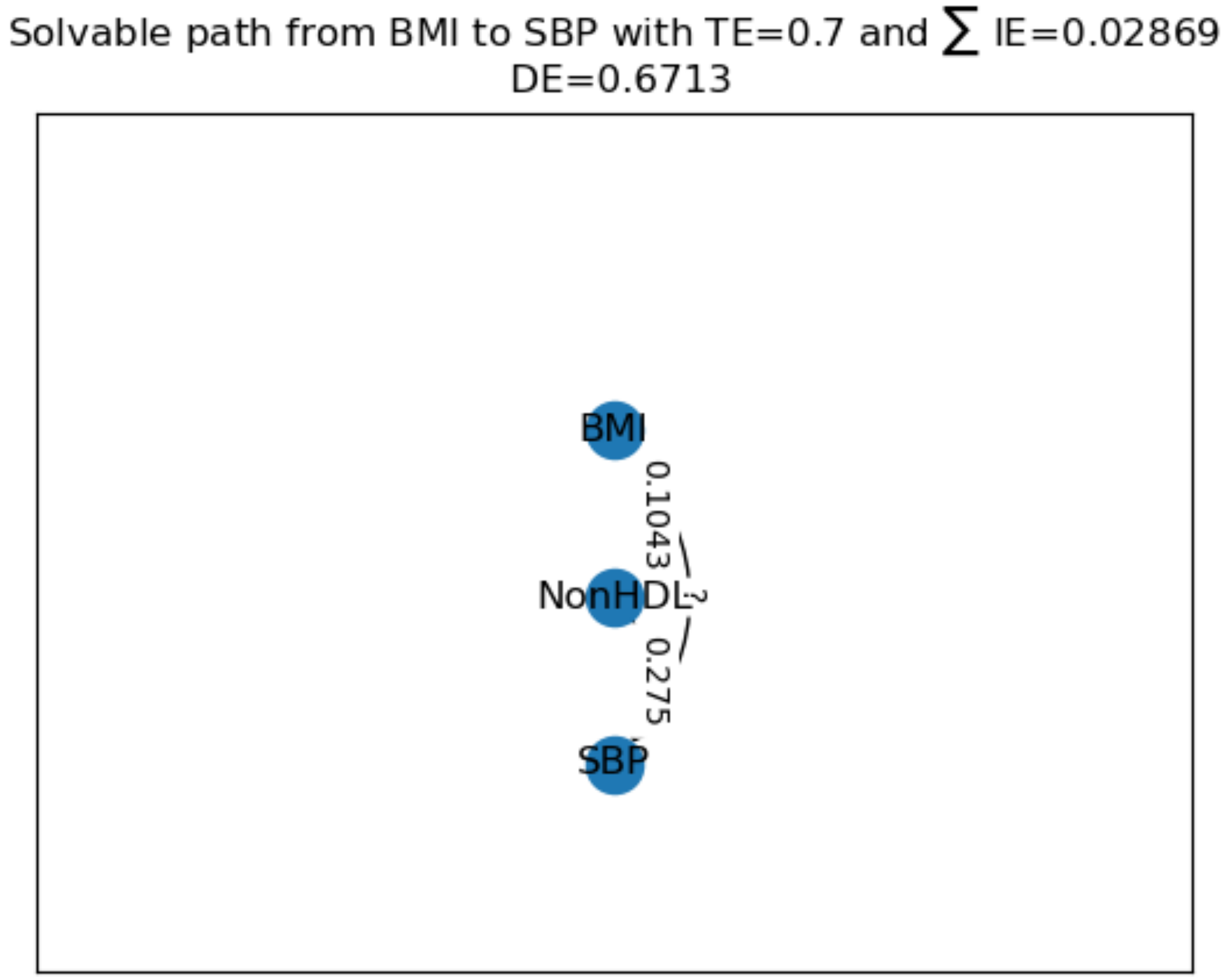

(Step 3.1.1) There are two paths, BMI → NonHDL → SBP and BMI → SBP.

(Steo 3.1.2) In all these paths only the BMI → SBP direct effect is missing, so the subgraph is solvable.

(Step 3.1.3) The indirect effect of BMI into SBP through BMI → NonHDL → SBP is accumulated as 0.02869.

(Step 3.1.4) The total effect has this indirect effect discounted giving 0.6713 as the missing BMI → SBP direct effect.

(Step 3.1, iteration 3) Using the Total Effect of Statin → BMI of 0.33:

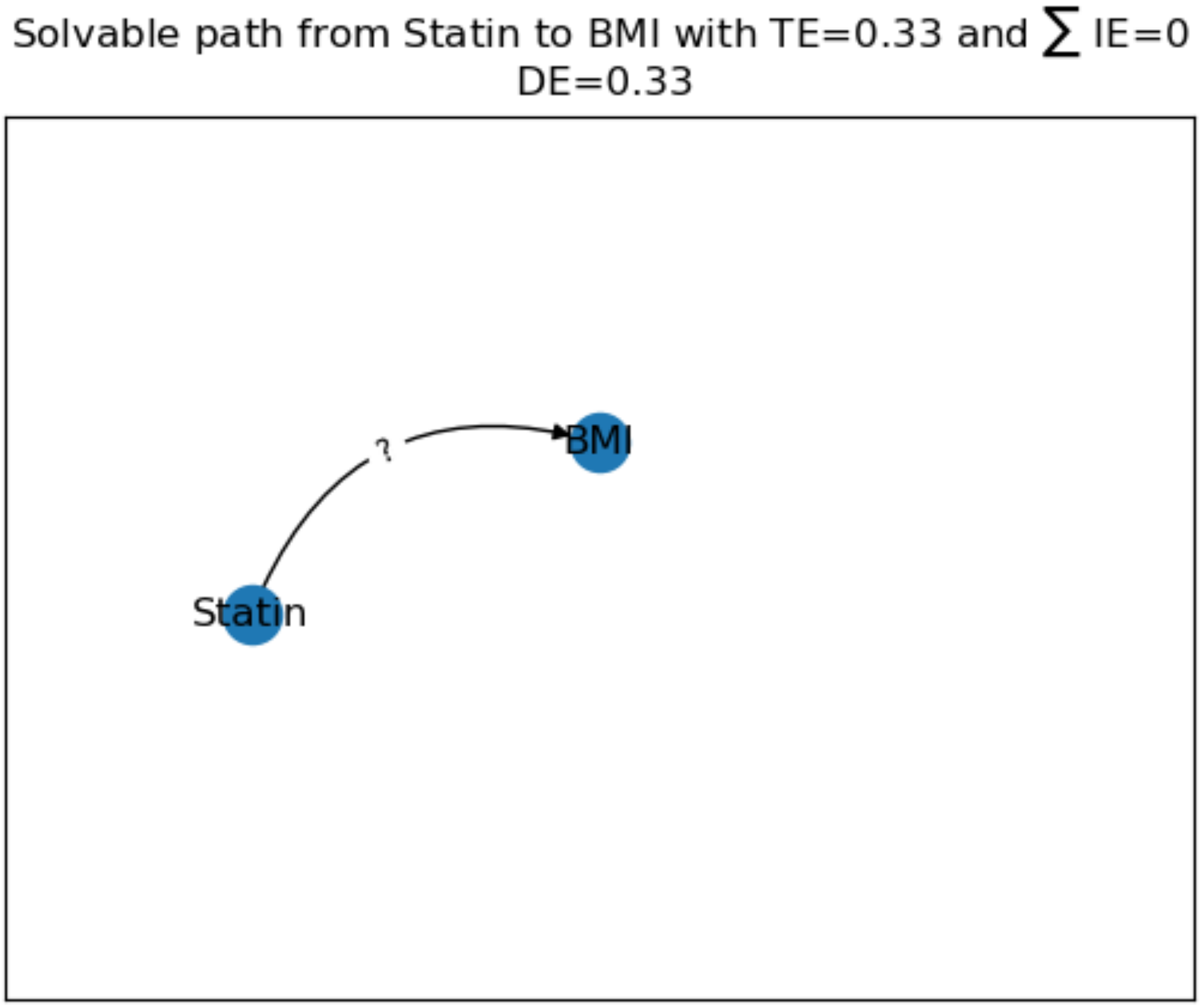

(Step 3.1.1) There is only one path, Statin → BMI.

(Steo 3.1.2) In this path only the Statin → BMI direct effect is missing, so the subgraph is solvable.

(Step 3.1.3) Since there is only one path with only a single edge, there are no indirect effects to accumulate.

(Step 3.1.4) Since there are no other indirect effects to assume, we solve this path's direct effect as 0.33 (Note that in this subgraph we use the assumption of the completeness of the DAG to assume that the total effect of Statin on BMI as a direct effect.)

(Step 3.1, iteration 4) Using the Total Effect of Statin → NonHDL of -1.3268:

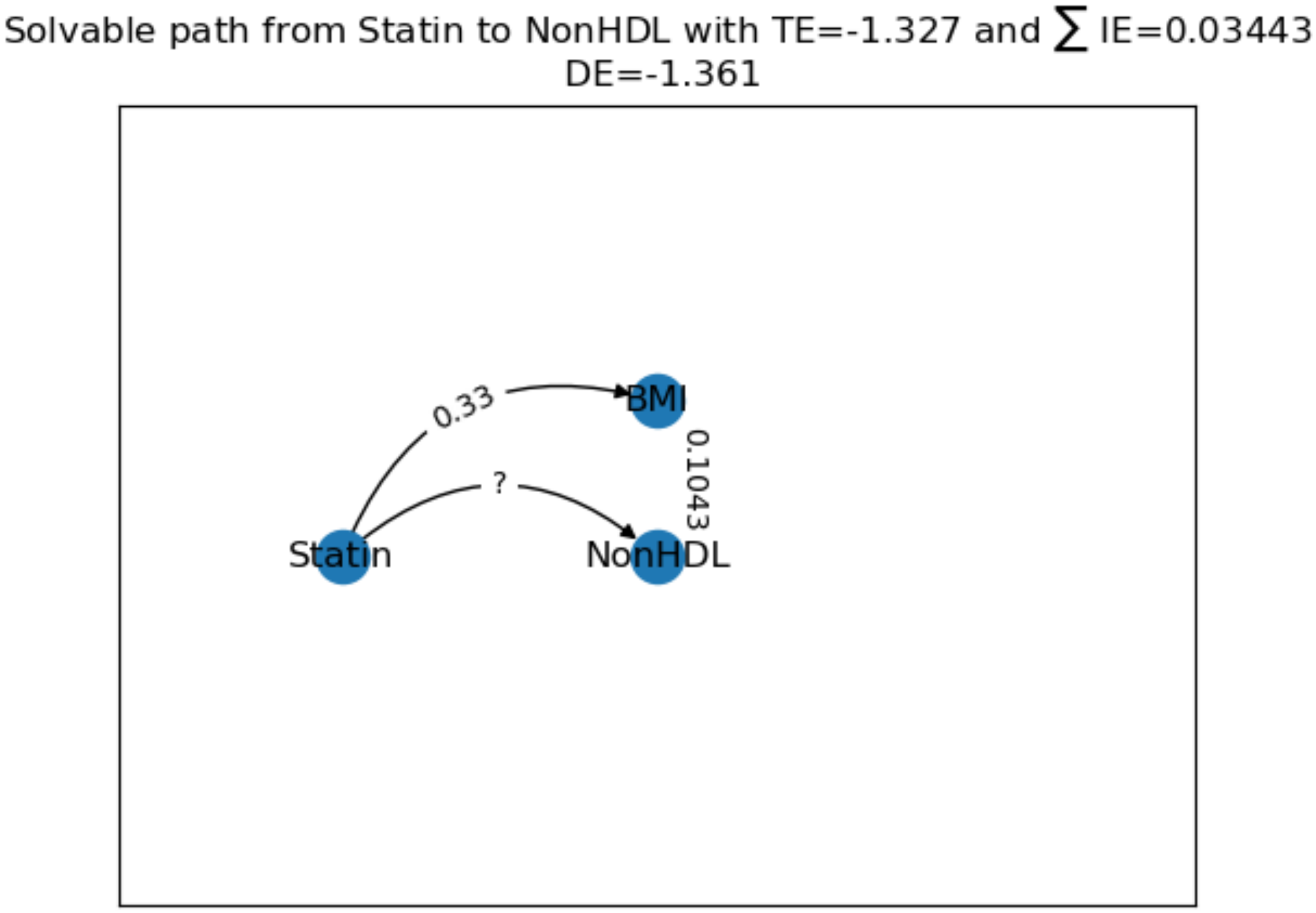

(Step 3.1.1) There are two paths, Statin → BMI → NonHDL and Statin → NonHDL.

(Steo 3.1.2) In all these paths only the Statin → NonHDL direct effect is missing, so the subgraph is solvable.

(Step 3.1.3) The indirect effect of Statin into NonHDL through Statin → BMI → NonHDL is accumulated as 0.03443.

(Step 3.1.4) The total effect has this indirect effect discounted giving 0.6713 as the missing Statin → NonHDL direct effect.

(Step 3.1, iteration 5) Using the Total Effect of Statin → SBP of -2.62:

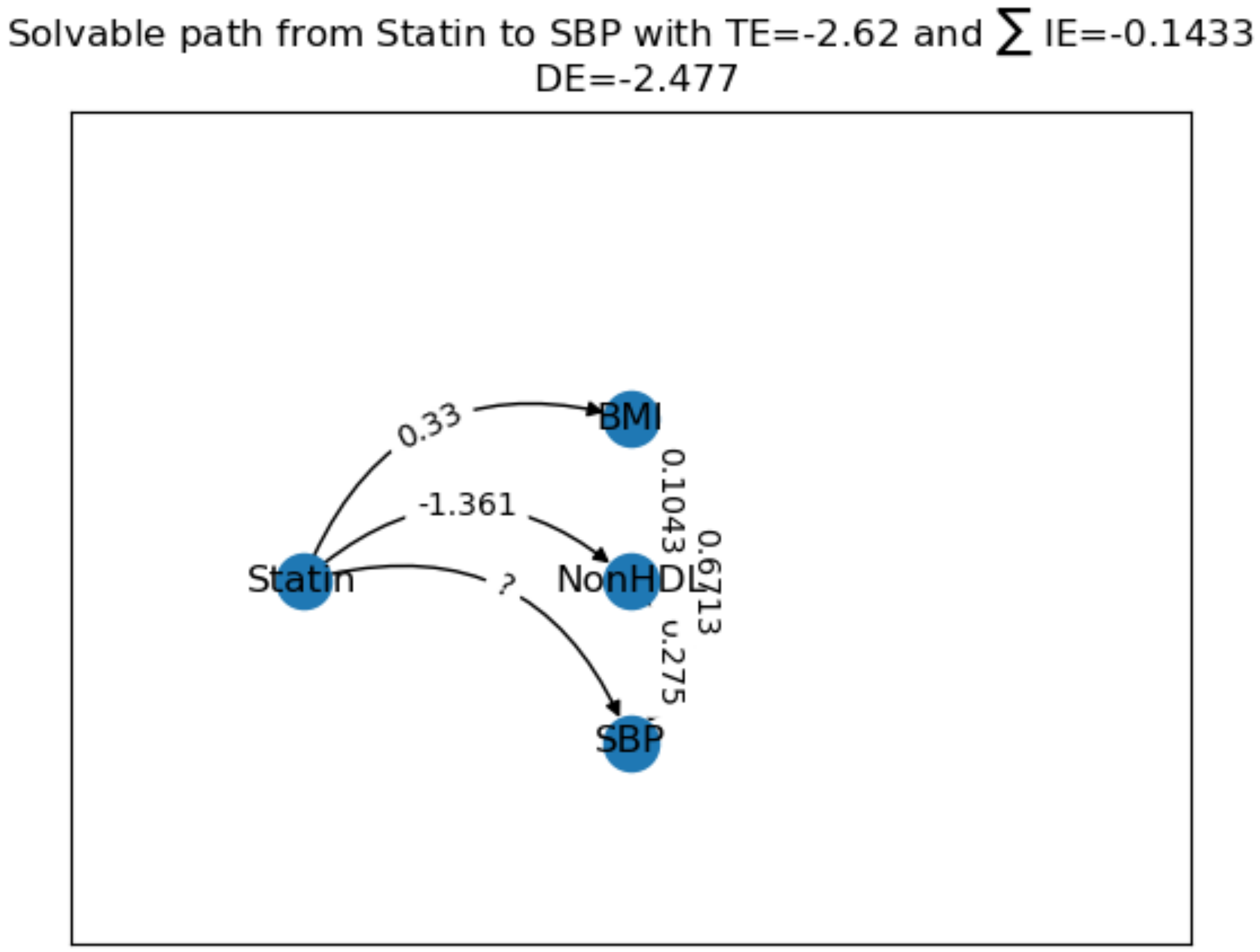

(Step 3.1.1) There are four paths, Statin → BMI → SBP, Statin → BMI → NonHDL → SBP, Statin → NonHDL → SBP and Statin → SBP.

(Steo 3.1.2) In all these paths only the Statin → SBP direct effect is missing, so the subgraph is solvable.

(Step 3.1.3) The indirect effect of Statin into SBP through Statin → BMI → SBP is accumulated as 0.221529, through Statin → BMI → NonHDL → SBP as 0.009465225, through Statin → NonHDL → SBP as -0.374275, which total as –0.1433

(Step 3.1.4) The total effect has this indirect effect discounted giving –2.477 as the missing Statin → NonHDL direct effect.

Note that this subgraph is was only solvable due to iterations 3 and 4, and that the indirect effects happen through multiple overlapping paths.

The effects for the other interventions would follow similarly with their given evidence. When no more paths are identified as solvable, and the algorithm stops. The algorithm then recovers the same values as calculated in section 2 of this document.

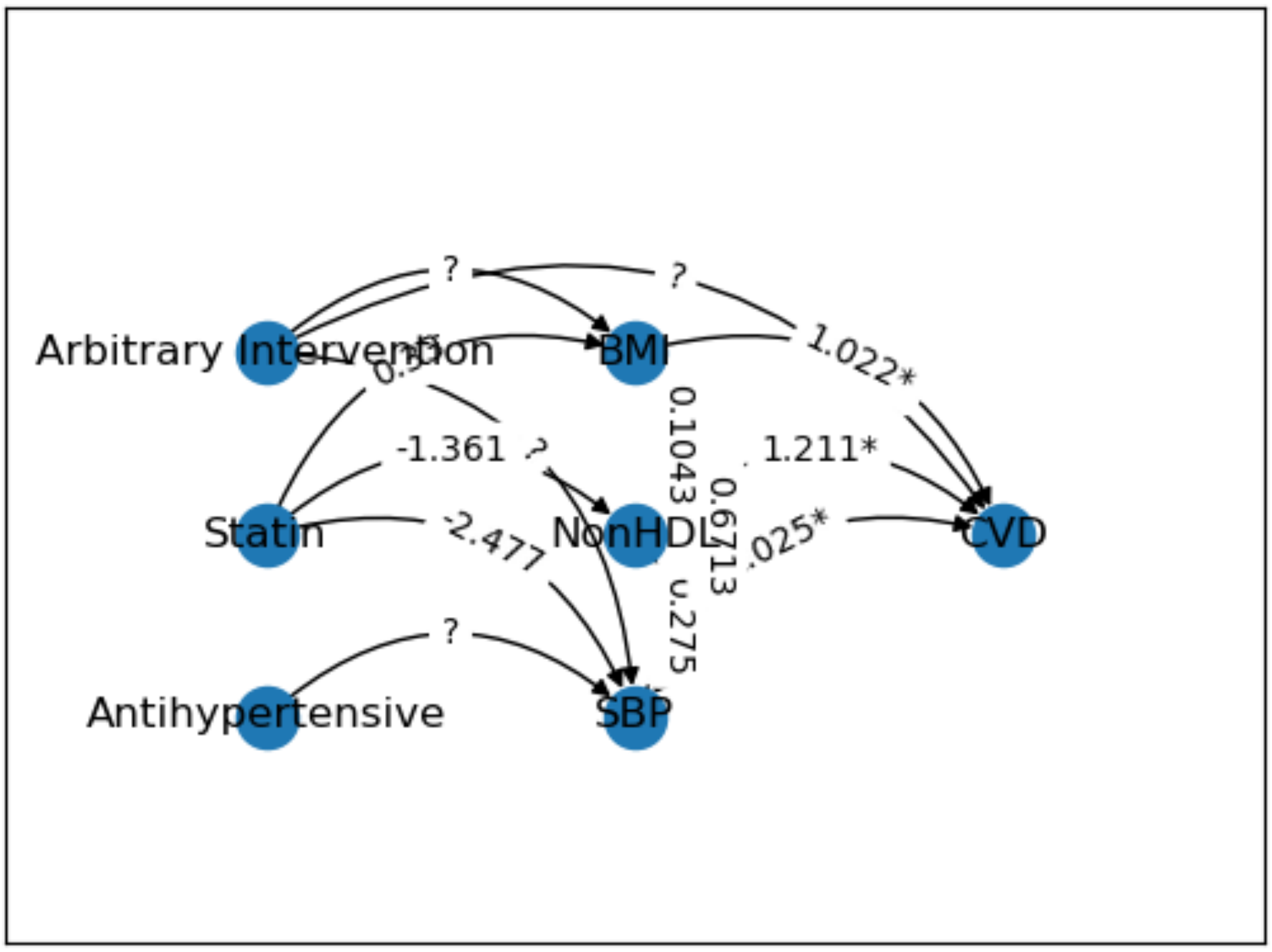

# Assessment of missingness and imputation

26 August 2026

**Associated manuscript: A flexible approach to sequential prediction under intervention**

We start with a summary of missing data stratified by antihypertensive use at baseline, and an exploration of whether the assumptions underpinning the imputation mechanism are met, with a particular focus on the variable *Systolic blood pressure unexposed to antihypertensive use (unexposed SBP)*. This is followed by repeating validation of model 2, stratified within subgroups defined by those with missing unexposed SBP, and those with recorded unexposed SBP.

# 1 Summary of missingess stratified by antihypertensive use at baseline

Missing data in the analyses is imputed using multiple imputation, which assumes that the data is missing at random. This is when the probability of the variable being missing is dependent on observed variables in the dataset, including other predictors and the outcome. If this assumption is not met, this is called missing not at random (MNAR). This is when the probability of the varaible being missing is dependent on the value of the variable itself, even after adjusting for observed variables.

In the unexposed mediator model, there is missingness in the variable "Systolic blood pressure (SBP) unexposed to antihypertensives; *unexposed SBP*". For some individuals, this is because they have no SBP recorded in the five years prior to index date (standard SBP will then also be missing). However, for some individuals, they may have SBP measurements (SBP recorded), but these are only within 180-days of an antihypertensives prescription. In this case, missingness is related to the primary intervention variable of interest. This raises the possibility that the missing at random assumption underlying the multiple imputation may be violated.

To mitigate the risk of MNAR, antihypertensive use at baseline is a predictor in the imputation model for *unexposed SBP* (it is a predictor in the imputation model for all missing variables). We also included *SBP* (not restricted to unexposed) as a predictor in the imputation model. This helps to address the concern that missingness is strongly dependent on antihypertensive use, by ensuring that this driver of missingness is included in the imputation model.

However, if all individuals receiving antihypertensives at baseline had missing *unexposed SBP*, this assumption would be highly likely to be violated. This would be because there would be insufficient non-missing *unexposed SBP* values, in order to predict the missing *unexposed SBP* values among those on antihypertensives. More formally, there would be a lack of overlap between individuals with and without observed values, requiring extrapolation of the imputation model to regions of the covariate space where no data are observed. We therefore present an analysis of missing patterns by treatment status at baseline.

In the case that there is sufficient non-missing *unexposed SBP* among people both on/off antihypertensives at baseline, we still acknowledge that this does not guarantee that the missing at random assumption holds, as missingness may still depend on missing values of *unexposed SBP*, as is the case for all variables with missingness.

We refrain from a complete case analysis, which is highly likely to be biased.

| gender | treatment | receiving | variable | N | Missing (%) |
|---|---|---|---|---|---|
| Female | antihypertensives | no | IMD | 1821231 | 0.12% |
| Female | antihypertensives | yes | IMD | 178769 | 0.11% |
| Female | antihypertensives | no | bmi | 1821231 | 37.03% |
| Female | antihypertensives | yes | bmi | 178769 | 27.44% |
| Female | antihypertensives | no | ethnicity | 1821231 | 12.41% |
| Female | antihypertensives | yes | ethnicity | 178769 | 5.8% |
| Female | antihypertensives | no | nonhdl | 1821231 | 87.07% |
| Female | antihypertensives | yes | nonhdl | 178769 | 40.59% |
| Female | antihypertensives | no | sbp | 1821231 | 22.45% |
| Female | antihypertensives | yes | sbp | 178769 | 1.85% |
| Female | antihypertensives | no | sbp_unexposed | 1821231 | 22.64% |
| Female | antihypertensives | yes | sbp_unexposed | 178769 | 37.43% |
| Female | antihypertensives | no | smoking | 1821231 | 8.86% |
| Female | antihypertensives | yes | smoking | 178769 | 2.55% |
| Male | antihypertensives | no | IMD | 1855349 | 0.13% |
| Male | antihypertensives | yes | IMD | 144651 | 0.11% |
| Male | antihypertensives | no | bmi | 1855349 | 48.81% |
| Male | antihypertensives | yes | bmi | 144651 | 24.15% |
| Male | antihypertensives | no | ethnicity | 1855349 | 22.11% |
| Male | antihypertensives | yes | ethnicity | 144651 | 7.23% |
| Male | antihypertensives | no | nonhdl | 1855349 | 86.81% |
| Male | antihypertensives | yes | nonhdl | 144651 | 32.52% |
| Male | antihypertensives | no | sbp | 1855349 | 44.81% |
| Male | antihypertensives | yes | sbp | 144651 | 2.33% |
| Male | antihypertensives | no | sbp_unexposed | 1855349 | 44.95% |
| Male | antihypertensives | yes | sbp_unexposed | 144651 | 34.37% |
| Male | antihypertensives | no | smoking | 1855349 | 17.57% |
| Male | antihypertensives | yes | smoking | 144651 | 2.33% |

We can see that among those on antihypertensives at baseline, 37.42% (female) and 34.37% (male) have a missing *unexposed SBP* value. We believe this is small enough that this is not a major problem and there is sufficient information among those on antihypertensive treatment baseline about the variable SBP unexposed to antihypertensive recorded, in order to impute this variable for those with missing data that are on antihypertensive treatment at baseline.

# 2 Stratified validation of model 2

We present model validation for the unexposed mediator model, stratified by whether SBP unexposed to antihypertensives is missing at baseline. Model performance is compared with performance of the model among all individuals, which was reported in the main manuscript.

## 2.1 Calibration curves

There are no major differences in the calibration curves when stratifying by whether unexposed SBP was imputed or recorded. All notable differences occur over 20% risk, which is at the tail of the risk distribution.

### 2.1.1 Women

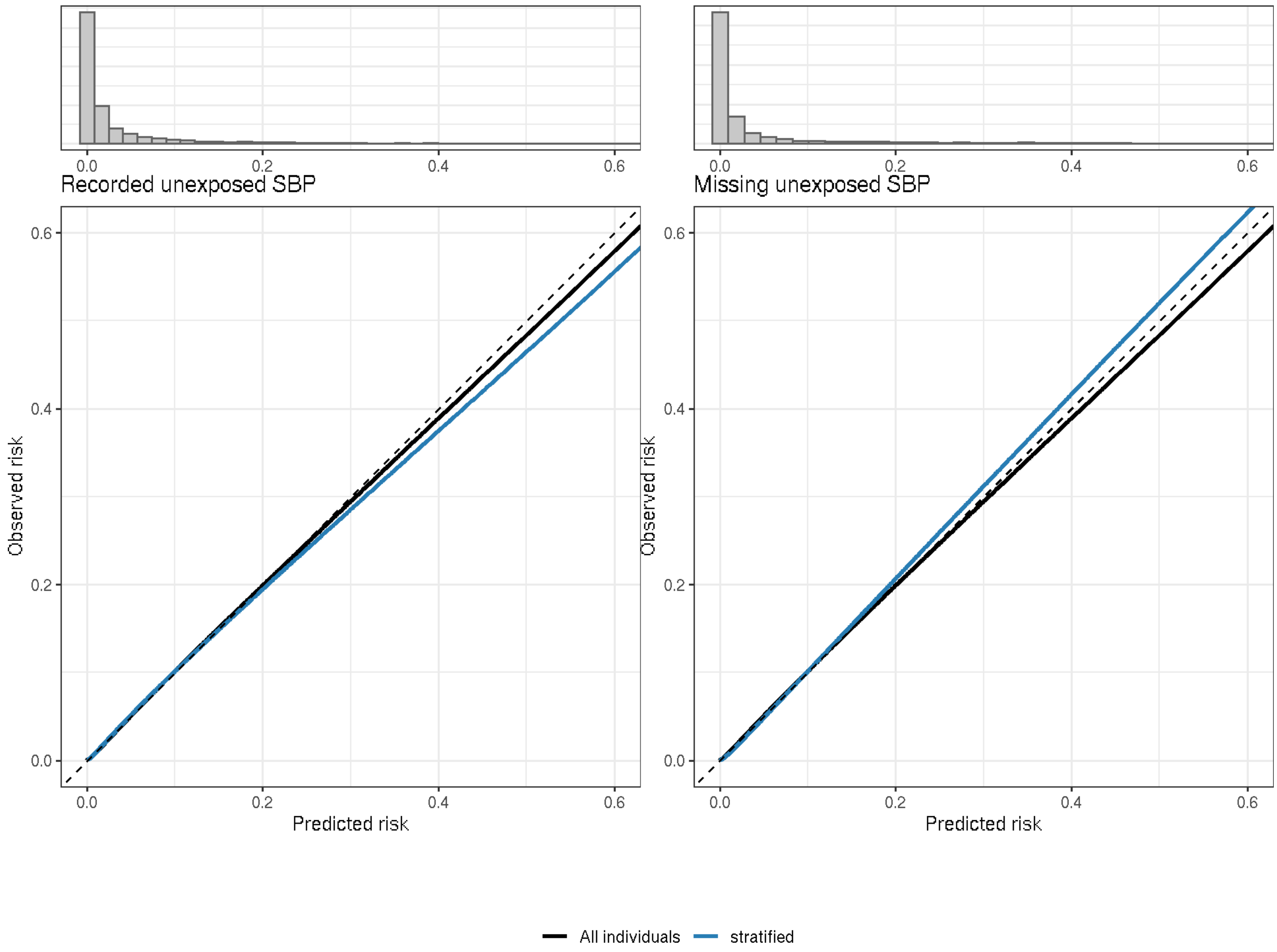

### 2.1.2 Men

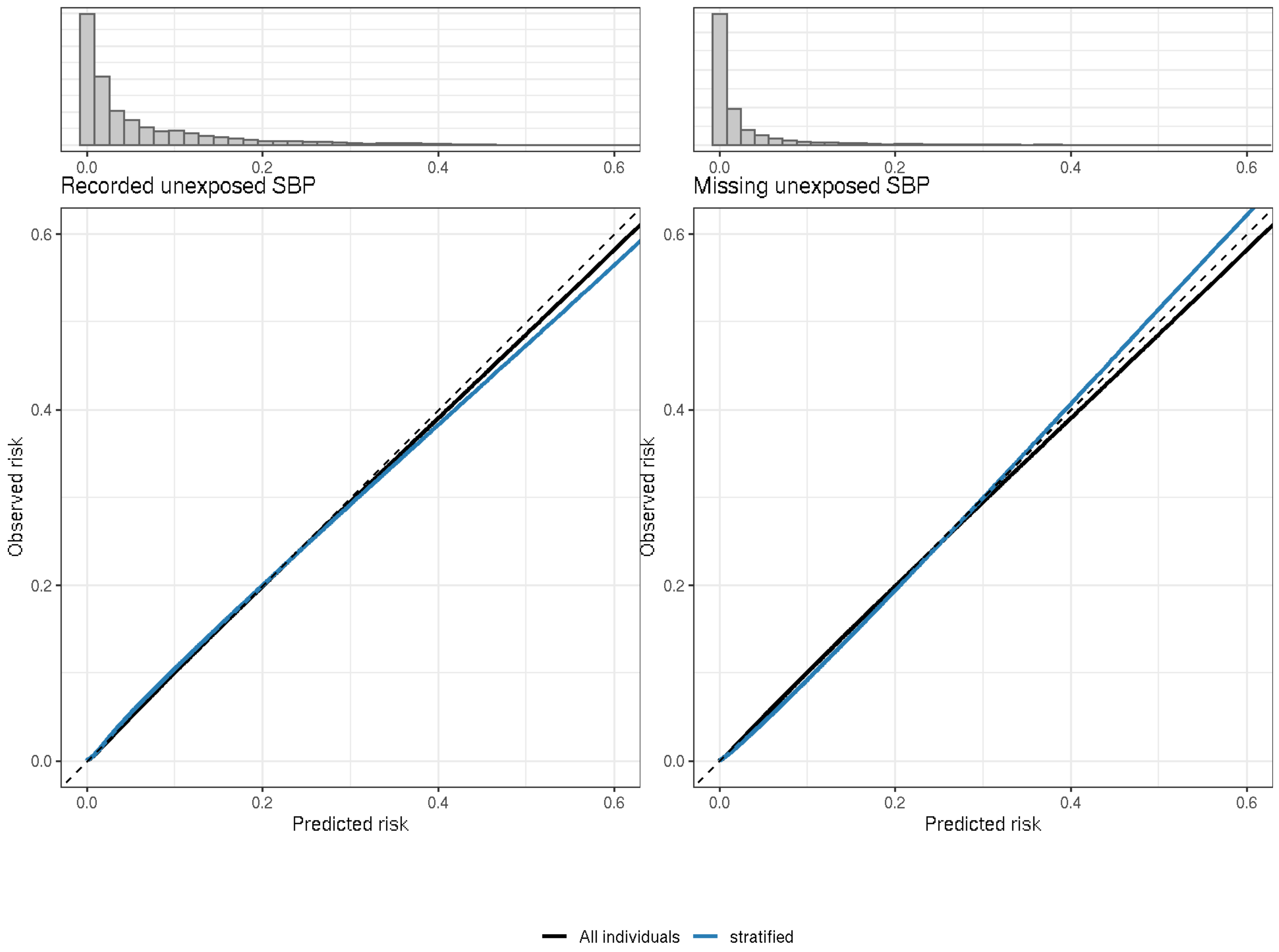

# 2.2 Performance metrics (calibration and discrimination)

Simiarly, there are no major differences in the calibration metrics (ICI, E50 and E90) when stratifying by whether unexposed SBP was imputed or recorded. For men, there is a notable difference in discrimination. There in a drop in discrimination among those with recorded SBP, and an increase in discrimination among those with missing SBP. This is likely due to the fact that a much wider range of individuals will have missing unexposed SBP (i.e. young healthy individuals, or higher risk individuals with no record of SBP without an antihypertensive with the 180 days before it). On the contrary, those with a recorded unexposed SBP, are likely to be a higher risk group with a smaller spread of predicted risk, resulting in a lower discrimination. For women, we see a similar effect, but much less pronounced.

| gender | metric | analysis | estimate | lower CI | upper CI |
|---|---|---|---|---|---|
| Female | ICI | All individuals | 0.001 | 0.001 | 0.001 |
| Female | ICI | Missing unexposed SBP | 0.002 | 0.001 | 0.003 |
| Female | ICI | Recorded unexposed SBP | 0.002 | 0.001 | 0.002 |
| Female | E50 | All individuals | 0.001 | 0.000 | 0.001 |
| Female | E50 | Missing unexposed SBP | 0.001 | 0.001 | 0.001 |
| Female | E50 | Recorded unexposed SBP | 0.001 | 0.000 | 0.001 |

| gender | metric | analysis | estimate | lower CI | upper CI |
|---|---|---|---|---|---|
| Female | E90 | All individuals | 0.002 | 0.001 | 0.003 |
| Female | E90 | Missing unexposed SBP | 0.006 | 0.002 | 0.010 |
| Female | E90 | Recorded unexposed SBP | 0.002 | 0.001 | 0.004 |
| Female | C_index | All individuals | 0.880 | 0.880 | 0.880 |
| Female | C_index | Missing unexposed SBP | 0.893 | 0.893 | 0.893 |
| Female | C_index | Recorded unexposed SBP | 0.873 | 0.873 | 0.873 |
| Male | ICI | All individuals | 0.001 | 0.001 | 0.002 |
| Male | ICI | Missing unexposed SBP | 0.002 | 0.002 | 0.003 |
| Male | ICI | Recorded unexposed SBP | 0.003 | 0.002 | 0.004 |
| Male | E50 | All individuals | 0.001 | 0.000 | 0.001 |
| Male | E50 | Missing unexposed SBP | 0.001 | 0.001 | 0.001 |
| Male | E50 | Recorded unexposed SBP | 0.001 | 0.000 | 0.001 |
| Male | E90 | All individuals | 0.001 | 0.001 | 0.003 |
| Male | E90 | Missing unexposed SBP | 0.007 | 0.005 | 0.009 |
| Male | E90 | Recorded unexposed SBP | 0.006 | 0.004 | 0.008 |
| Male | C_index | All individuals | 0.861 | 0.861 | 0.861 |
| Male | C_index | Missing unexposed SBP | 0.894 | 0.894 | 0.894 |
| Male | C_index | Recorded unexposed SBP | 0.832 | 0.832 | 0.832 |

# Sensitivity analyses looking at misspecification of the effect of statins and antihypertensives

26 August 2026

**Associated manuscript: A flexible approach to sequential prediction under intervention**

These supplementary analyses validate the model under the assumption that the treatment effect of statins and antihypertensives in the validation dataset are 10% and 20% bigger/smaller than those assumed in the development dataset when building the model. We refer to this as a misspecification of the treatment effect (i.e., the treatment effect in the development dataset has been misspecified compared to the true effect in the validation dataset).

We multiply the effect of statins or antihypertensives (on the hazard ratio scale) by values of 0.8, 0.9, 1.1 and 1.2. Given that the hazard ratios for statin and antihypertensive use are < 1, the multipliers 0.8 and 0.9 correspond to a bigger effect in the validation dataset, and multipliers 1.1 and 1.2 correspond to a smaller effect in the validation dataset.

In order to assess the impact of misspecification, we estimate counterfactual survival times in the validation dataset (see supplementary material for this process) using these miss-specified treatment effects. After this, calibration and discrimination are assessed in the way as described in the main manuscript, and supplementary file 1 (methodology).

In section 1, we report calibration and discrimination in the entire validation dataset, for various levels of misspecification of the treatment effect of statins/antihypertensives. For calibration, we report calibration curves, and the confidence interval for the calibration curve (note, the confidence intervals in section 1 are very small and so are not visible), and summary metrics ICI, E50 and E90. For discrimination, Harrell's C-index is reported. In section 2, we repeat the analyses within subgroups defined by treatment use (either statins or antihypertensives) at baseline. This is because we expect the biggest effect to be among individuals who are on treatment at baseline. The cohort of individuals off treatment at baseline contains a large number of individuals who never receive any treatment, and are therefore unaffected by the misspecification.

Calibration and discrimination are always compared with an analysis assuming there is no misspecification of the treatment effects, aligned with the primary analysis from the main manuscript.

# 1 Sensitivity analysis in entire validation cohort

In the following tables and plots (section 1), 'no misspecification' is the same value from the primary analysis, assuming no misspecification of the treatment effect. 'Difference' is the difference between the sensitivity analysis value, and the 'no misspecification' value.

When the treatment effects are miss-specified by 10%/20% we see negligible changes to model performance. This will be for two reasons. Firstly, the calibration and discrimination metrics are dominated by low risk individuals who have a small probability of receiving treatment, among whom the impact of misspecification of the treatment is small. Although, it should be noted that misspecification seems to have minimal effect among high risk individuals too. This could be partially due to the second reason, which is that the cohort is a mix of individuals on and off treatment at baseline. If equal numbers of individuals start/stop taking treatment during follow-up, the effect of the misspecification would cancel out. This is why we look at stratified analyses within those on/off treatment at baseline (see section 2).

# 1.1 Women

## 1.1.1 Calibration curves

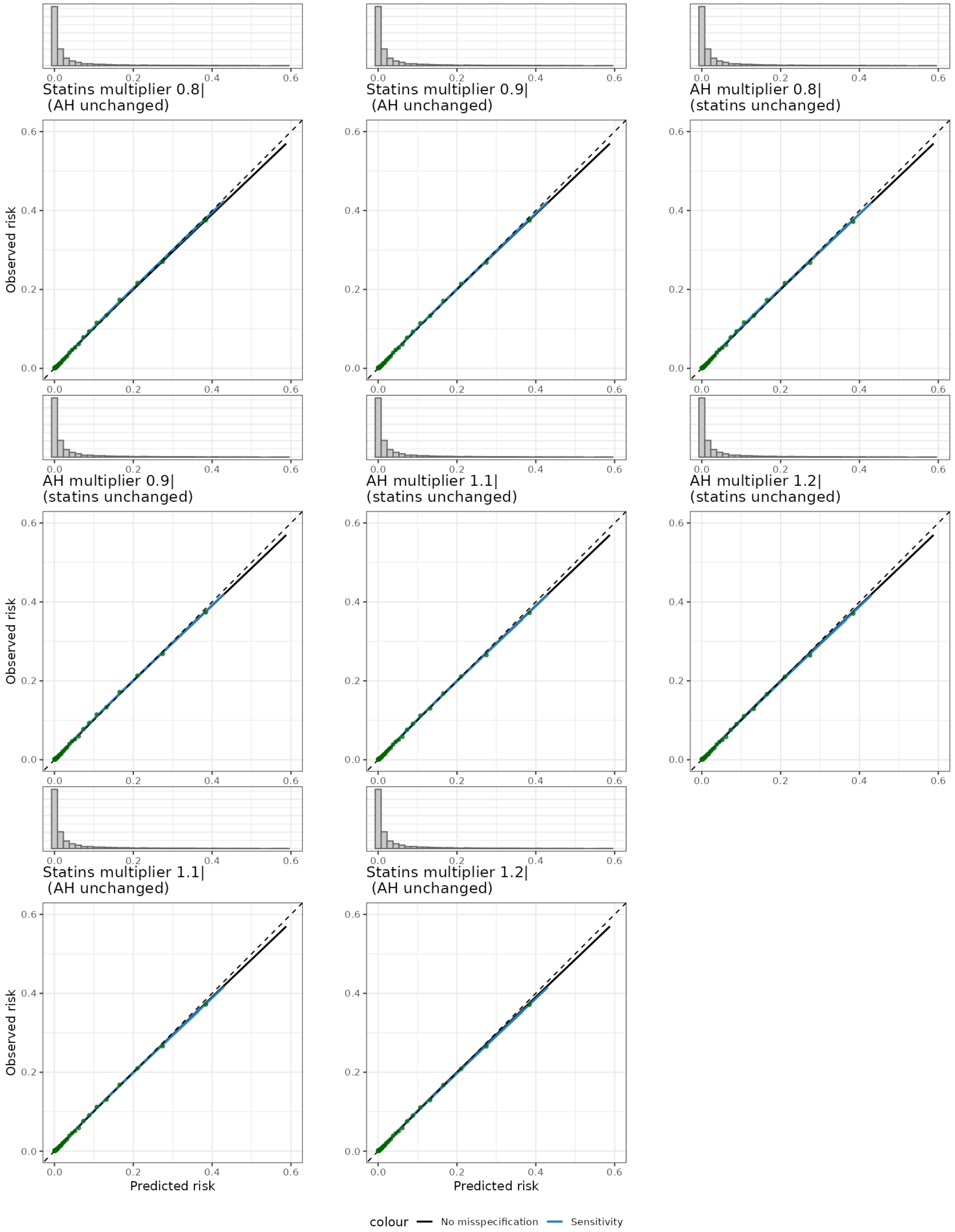

# 1.1.2 Calibration Metrics (ICI, E50, E90)

Calibration metrics for: Women. 'No misspecification' is the value from the main analysis, assuming no misspecification of the treatment effect

| scenario | metric | value | lower_ci | upper_ci | no misspecification | difference |
|---|---|---|---|---|---|---|
| AH multiplier 0.8 (statins unchanged) | ICI | 0.001 | 0.001 | 0.002 | 0.001 | 0.000 |
| AH multiplier 0.9 (statins unchanged) | ICI | 0.001 | 0.001 | 0.001 | 0.001 | 0.000 |
| AH multiplier 1.1 (statins unchanged) | ICI | 0.001 | 0.001 | 0.001 | 0.001 | 0.000 |
| AH multiplier 1.2 (statins unchanged) | ICI | 0.001 | 0.001 | 0.001 | 0.001 | 0.000 |
| Statins multiplier 0.8 (AH unchanged) | ICI | 0.001 | 0.001 | 0.002 | 0.001 | 0.000 |
| Statins multiplier 0.9 (AH unchanged) | ICI | 0.001 | 0.001 | 0.001 | 0.001 | 0.000 |
| Statins multiplier 1.1 (AH unchanged) | ICI | 0.001 | 0.001 | 0.001 | 0.001 | 0.000 |
| Statins multiplier 1.2 (AH unchanged) | ICI | 0.001 | 0.001 | 0.001 | 0.001 | 0.000 |
| AH multiplier 0.8 (statins unchanged) | E50 | 0.001 | 0.000 | 0.001 | 0.001 | 0.000 |
| AH multiplier 0.9 (statins unchanged) | E50 | 0.001 | 0.000 | 0.001 | 0.001 | 0.000 |
| AH multiplier 1.1 (statins unchanged) | E50 | 0.001 | 0.000 | 0.001 | 0.001 | 0.000 |
| AH multiplier 1.2 (statins unchanged) | E50 | 0.000 | 0.000 | 0.001 | 0.001 | 0.000 |
| Statins multiplier 0.8 (AH unchanged) | E50 | 0.001 | 0.000 | 0.001 | 0.001 | 0.000 |
| Statins multiplier 0.9 (AH unchanged) | E50 | 0.001 | 0.000 | 0.001 | 0.001 | 0.000 |
| Statins multiplier 1.1 (AH unchanged) | E50 | 0.001 | 0.000 | 0.001 | 0.001 | 0.000 |
| Statins multiplier 1.2 (AH unchanged) | E50 | 0.001 | 0.000 | 0.001 | 0.001 | 0.000 |
| AH multiplier 0.8 (statins unchanged) | E90 | 0.004 | 0.002 | 0.005 | 0.002 | 0.002 |

| scenario | metric | value | lower_ci | upper_ci | no misspecification | difference |
|---|---|---|---|---|---|---|
| AH multiplier 0.9 (statins unchanged) | E90 | 0.003 | 0.002 | 0.004 | 0.002 | 0.001 |
| AH multiplier 1.1 (statins unchanged) | E90 | 0.001 | 0.001 | 0.003 | 0.002 | -0.001 |
| AH multiplier 1.2 (statins unchanged) | E90 | 0.001 | 0.001 | 0.002 | 0.002 | -0.001 |
| Statins multiplier 0.8 (AH unchanged) | E90 | 0.004 | 0.002 | 0.005 | 0.002 | 0.002 |
| Statins multiplier 0.9 (AH unchanged) | E90 | 0.003 | 0.001 | 0.004 | 0.002 | 0.001 |
| Statins multiplier 1.1 (AH unchanged) | E90 | 0.001 | 0.001 | 0.003 | 0.002 | -0.001 |
| Statins multiplier 1.2 (AH unchanged) | E90 | 0.001 | 0.001 | 0.002 | 0.002 | -0.001 |

# 1.1.3 Discrimination (C-index)

Discrimination results for: Women.

| scenario | c_index | lower_ci | upper_ci | no misspecification | difference |
|---|---|---|---|---|---|
| AH multiplier 0.8 (statins unchanged) | 0.872 | 0.872 | 0.872 | 0.871 | 0.000 |
| AH multiplier 0.9 (statins unchanged) | 0.871 | 0.871 | 0.871 | 0.871 | 0.000 |
| AH multiplier 1.1 (statins unchanged) | 0.871 | 0.871 | 0.871 | 0.871 | 0.000 |
| AH multiplier 1.2 (statins unchanged) | 0.871 | 0.871 | 0.871 | 0.871 | 0.000 |
| Statins multiplier 0.8 (AH unchanged) | 0.872 | 0.872 | 0.872 | 0.871 | 0.001 |
| Statins multiplier 0.9 (AH unchanged) | 0.872 | 0.872 | 0.872 | 0.871 | 0.000 |
| Statins multiplier 1.1 (AH unchanged) | 0.871 | 0.871 | 0.871 | 0.871 | 0.000 |
| Statins multiplier 1.2 (AH unchanged) | 0.870 | 0.870 | 0.870 | 0.871 | -0.001 |

# 1.2 Men

## 1.2.1 Calibration curves

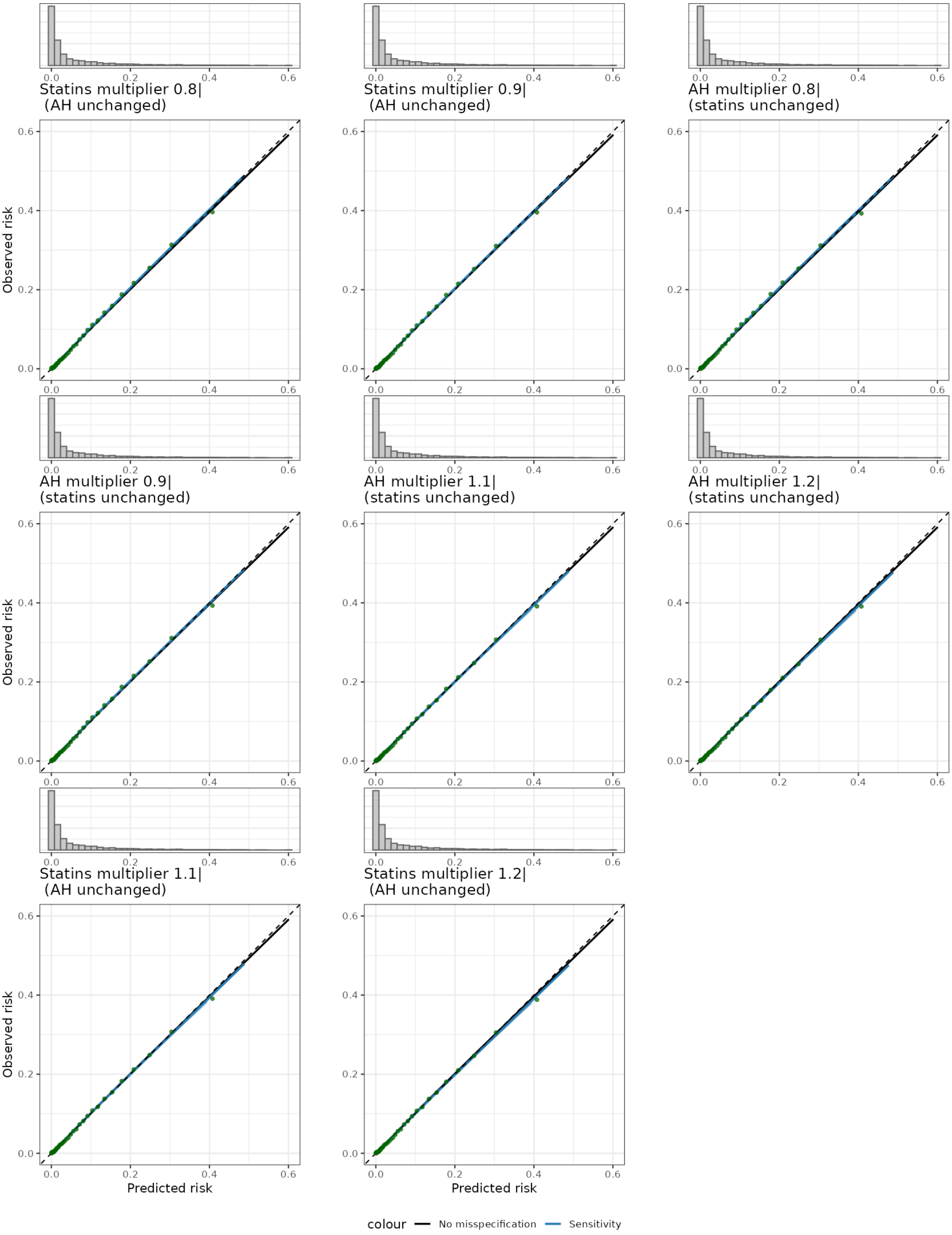

### 1.2.2 Calibration Metrics (ICI, E50, E90)

Calibration metrics for: Men. 'No misspecification' is the value from the main analysis, assuming no misspecification of the treatment effect

| scenario | metric | value | lower_ci | upper_ci | no misspecification | difference |
|---|---|---|---|---|---|---|
| AH multiplier 0.8 (statins unchanged) | ICI | 0.002 | 0.001 | 0.002 | 0.001 | 0.001 |
| AH multiplier 0.9 (statins unchanged) | ICI | 0.001 | 0.001 | 0.002 | 0.001 | 0.000 |
| AH multiplier 1.1 (statins unchanged) | ICI | 0.001 | 0.001 | 0.001 | 0.001 | 0.000 |
| AH multiplier 1.2 (statins unchanged) | ICI | 0.001 | 0.001 | 0.001 | 0.001 | 0.000 |
| Statins multiplier 0.8 (AH unchanged) | ICI | 0.002 | 0.001 | 0.002 | 0.001 | 0.001 |
| Statins multiplier 0.9 (AH unchanged) | ICI | 0.001 | 0.001 | 0.002 | 0.001 | 0.000 |
| Statins multiplier 1.1 (AH unchanged) | ICI | 0.001 | 0.001 | 0.001 | 0.001 | 0.000 |
| Statins multiplier 1.2 (AH unchanged) | ICI | 0.001 | 0.001 | 0.001 | 0.001 | 0.000 |
| AH multiplier 0.8 (statins unchanged) | E50 | 0.001 | 0.000 | 0.001 | 0.001 | 0.000 |
| AH multiplier 0.9 (statins unchanged) | E50 | 0.001 | 0.000 | 0.001 | 0.001 | 0.000 |
| AH multiplier 1.1 (statins unchanged) | E50 | 0.001 | 0.000 | 0.001 | 0.001 | 0.000 |
| AH multiplier 1.2 (statins unchanged) | E50 | 0.000 | 0.000 | 0.001 | 0.001 | 0.000 |
| Statins multiplier 0.8 (AH unchanged) | E50 | 0.001 | 0.001 | 0.001 | 0.001 | 0.000 |
| Statins multiplier 0.9 (AH unchanged) | E50 | 0.001 | 0.000 | 0.001 | 0.001 | 0.000 |
| Statins multiplier 1.1 (AH unchanged) | E50 | 0.001 | 0.000 | 0.001 | 0.001 | 0.000 |
| Statins multiplier 1.2 (AH unchanged) | E50 | 0.001 | 0.000 | 0.001 | 0.001 | 0.000 |
| AH multiplier 0.8 (statins unchanged) | E90 | 0.005 | 0.004 | 0.007 | 0.002 | 0.003 |

| scenario | metric | value | lower_ci | upper_ci | no misspecification | difference |
|---|---|---|---|---|---|---|
| AH multiplier 0.9 (statins unchanged) | E90 | 0.003 | 0.003 | 0.006 | 0.002 | 0.002 |
| AH multiplier 1.1 (statins unchanged) | E90 | 0.001 | 0.001 | 0.003 | 0.002 | -0.001 |
| AH multiplier 1.2 (statins unchanged) | E90 | 0.001 | 0.001 | 0.002 | 0.002 | -0.001 |
| Statins multiplier 0.8 (AH unchanged) | E90 | 0.004 | 0.003 | 0.007 | 0.002 | 0.003 |
| Statins multiplier 0.9 (AH unchanged) | E90 | 0.003 | 0.002 | 0.005 | 0.002 | 0.001 |
| Statins multiplier 1.1 (AH unchanged) | E90 | 0.001 | 0.001 | 0.003 | 0.002 | -0.001 |
| Statins multiplier 1.2 (AH unchanged) | E90 | 0.001 | 0.001 | 0.003 | 0.002 | -0.001 |

### 1.2.3 Discrimination (C-index)

Discrimination results for: Men.

| scenario | c_index | lower_ci | upper_ci | no misspecification | difference |
|---|---|---|---|---|---|
| AH multiplier 0.8 (statins unchanged) | 0.855 | 0.855 | 0.855 | 0.854 | 0.001 |
| AH multiplier 0.9 (statins unchanged) | 0.855 | 0.855 | 0.855 | 0.854 | 0.000 |
| AH multiplier 1.1 (statins unchanged) | 0.854 | 0.854 | 0.854 | 0.854 | 0.000 |
| AH multiplier 1.2 (statins unchanged) | 0.853 | 0.853 | 0.853 | 0.854 | -0.001 |
| Statins multiplier 0.8 (AH unchanged) | 0.855 | 0.855 | 0.855 | 0.854 | 0.001 |
| Statins multiplier 0.9 (AH unchanged) | 0.855 | 0.855 | 0.855 | 0.854 | 0.001 |
| Statins multiplier 1.1 (AH unchanged) | 0.854 | 0.854 | 0.854 | 0.854 | 0.000 |
| Statins multiplier 1.2 (AH unchanged) | 0.853 | 0.853 | 0.853 | 0.854 | -0.001 |

# 2 Treatment-stratified sensitivity analysis

In the following tables and plots (section 2), 'no misspecification' is an estimate assuming no misspecification of the treatment effect (i.e. aligned with the primary analysis), but stratified to the same subgroup. 'Difference' is the difference between the sensitivity analysis value, and the 'no misspecification' value.

As expected, misspecification of the effects of either statins/antihypertensives has a bigger impact on calibration and discrimination among those on the respective treatment at baseline.

For example, consider section 2.2.1, or 2.6.1. When the effect of statins is bigger in the validation cohort (statins multiplier = 0.8 or 0.9), the calibration curve indicates more over-prediction than the calibration curve under no misspecification. Similarly, when the effect of statins is smaller in the validation cohort (statins

multiplier = 1.1 or 1.2), the calibration curve indicates less over-prediction. Despite this, the magnitude of the difference in the calibration curves is not alarming, and is only visible among higher risk individuals.

Note, the calibration within subgroups defined by treatment use at baseline is imperfect, even when there is no misspecification of the treatment effect between development and validation. This will largely be due to the fact that statin and antihypertensive use is a strong predictor of incident cardiovascular disease, which have not been used as predictors in the model. This effect is much stronger for the statin users group (sections 2.2 and 2.6) compared to the antihypertensive users group (sections 2.4 and 2.8), where the model is reasonably well calibrated across the entire distribution of predicted risk. This trade off between predictive performance in the classical sense, and the ability to use the model for prediction under these interventions, is discussed in the main manuscript.

# 2.1 Women; Statins; not on treatment at baseline

## 2.1.1 Calibration curves

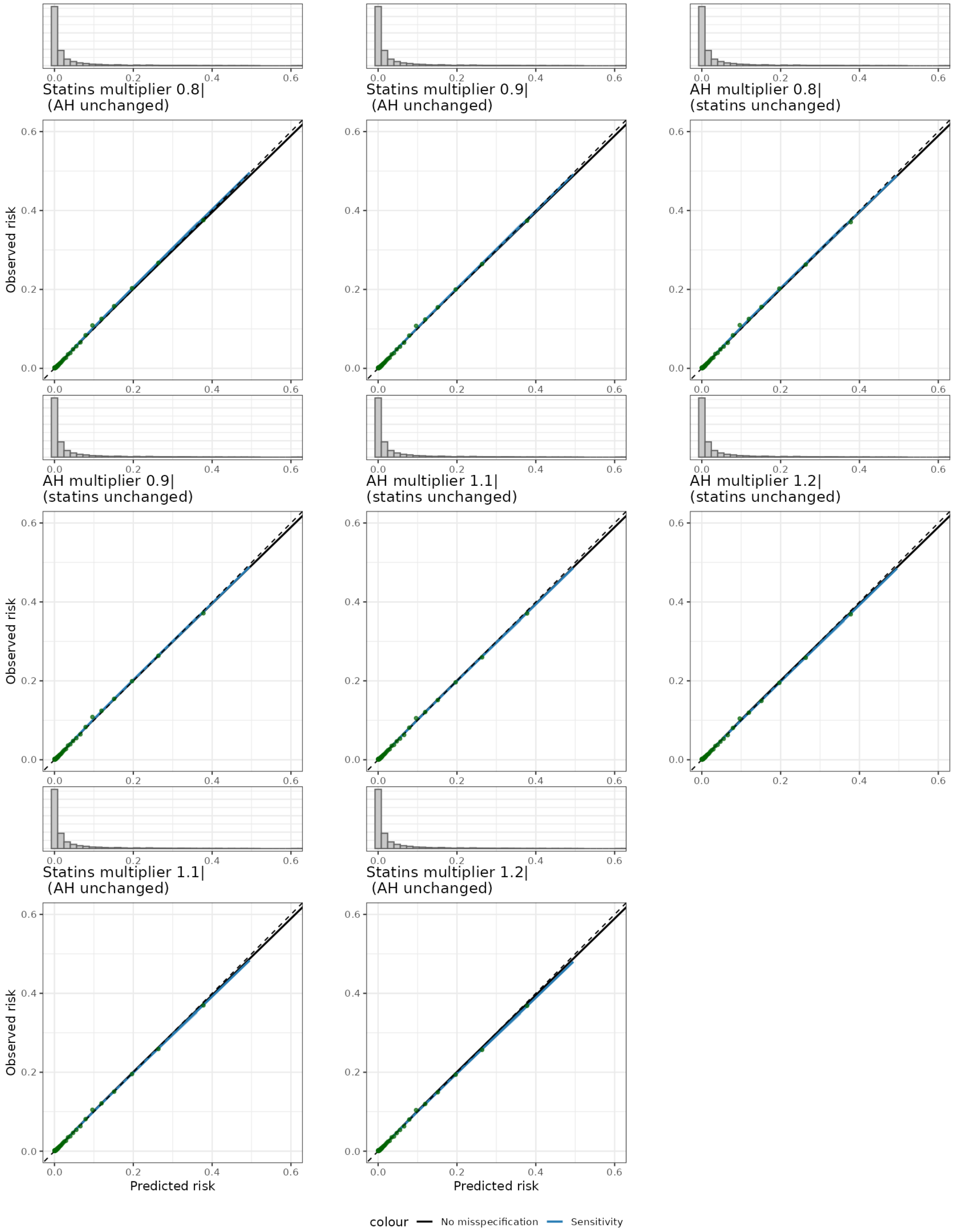

## 2.1.2 Calibration Metrics (ICI, E50, E90)

Calibration metrics for: Women; Statins; not on treatment at baseline.

| scenario | metric | value | lower_ci | upper_ci | no misspecification | difference |
|---|---|---|---|---|---|---|
| AH multiplier 0.8 (statins unchanged) | ICI | 0.001 | 0.001 | 0.001 | 0.001 | 0.000 |
| AH multiplier 0.9 (statins unchanged) | ICI | 0.001 | 0.001 | 0.001 | 0.001 | 0.000 |
| AH multiplier 1.1 (statins unchanged) | ICI | 0.001 | 0.000 | 0.001 | 0.001 | 0.000 |
| AH multiplier 1.2 (statins unchanged) | ICI | 0.001 | 0.000 | 0.001 | 0.001 | 0.000 |
| Statins multiplier 0.8 (AH unchanged) | ICI | 0.001 | 0.001 | 0.002 | 0.001 | 0.000 |
| Statins multiplier 0.9 (AH unchanged) | ICI | 0.001 | 0.001 | 0.001 | 0.001 | 0.000 |
| Statins multiplier 1.1 (AH unchanged) | ICI | 0.001 | 0.000 | 0.001 | 0.001 | 0.000 |
| Statins multiplier 1.2 (AH unchanged) | ICI | 0.001 | 0.001 | 0.001 | 0.001 | 0.000 |
| AH multiplier 0.8 (statins unchanged) | E50 | 0.001 | 0.000 | 0.001 | 0.001 | 0.000 |
| AH multiplier 0.9 (statins unchanged) | E50 | 0.001 | 0.000 | 0.001 | 0.001 | 0.000 |
| AH multiplier 1.1 (statins unchanged) | E50 | 0.001 | 0.000 | 0.001 | 0.001 | 0.000 |
| AH multiplier 1.2 (statins unchanged) | E50 | 0.000 | 0.000 | 0.001 | 0.001 | 0.000 |
| Statins multiplier 0.8 (AH unchanged) | E50 | 0.001 | 0.000 | 0.001 | 0.001 | 0.000 |
| Statins multiplier 0.9 (AH unchanged) | E50 | 0.001 | 0.000 | 0.001 | 0.001 | 0.000 |
| Statins multiplier 1.1 (AH unchanged) | E50 | 0.001 | 0.000 | 0.001 | 0.001 | 0.000 |
| Statins multiplier 1.2 (AH unchanged) | E50 | 0.000 | 0.000 | 0.001 | 0.001 | 0.000 |
| AH multiplier 0.8 (statins unchanged) | E90 | 0.003 | 0.002 | 0.005 | 0.002 | 0.002 |
| AH multiplier 0.9 (statins unchanged) | E90 | 0.003 | 0.001 | 0.004 | 0.002 | 0.001 |
| AH multiplier 1.1 (statins unchanged) | E90 | 0.001 | 0.001 | 0.002 | 0.002 | -0.001 |

| scenario | metric | value | lower_ci | upper_ci | no misspecification | difference |
|---|---|---|---|---|---|---|
| AH multiplier 1.2 (statins unchanged) | E90 | 0.001 | 0.001 | 0.002 | 0.002 | -0.001 |
| Statins multiplier 0.8 (AH unchanged) | E90 | 0.004 | 0.002 | 0.006 | 0.002 | 0.003 |
| Statins multiplier 0.9 (AH unchanged) | E90 | 0.003 | 0.001 | 0.004 | 0.002 | 0.001 |
| Statins multiplier 1.1 (AH unchanged) | E90 | 0.001 | 0.001 | 0.002 | 0.002 | -0.001 |
| Statins multiplier 1.2 (AH unchanged) | E90 | 0.001 | 0.001 | 0.002 | 0.002 | -0.001 |

## 2.1.3 Discrimination (C-index)

Discrimination results for: Women; Statins; not on treatment at baseline. 'No misspecification' is the value from the main analysis, assuming no misspecification of the treatment effect, but stratified to the same subgroup.

| scenario | c_index | lower_ci | upper_ci | no misspecification | difference |
|---|---|---|---|---|---|
| AH multiplier 0.8 (statins unchanged) | 0.874 | 0.874 | 0.874 | 0.873 | 0.001 |
| AH multiplier 0.9 (statins unchanged) | 0.874 | 0.874 | 0.874 | 0.873 | 0.000 |
| AH multiplier 1.1 (statins unchanged) | 0.873 | 0.873 | 0.873 | 0.873 | 0.000 |
| AH multiplier 1.2 (statins unchanged) | 0.873 | 0.873 | 0.873 | 0.873 | -0.001 |
| Statins multiplier 0.8 (AH unchanged) | 0.875 | 0.875 | 0.875 | 0.873 | 0.001 |
| Statins multiplier 0.9 (AH unchanged) | 0.874 | 0.874 | 0.874 | 0.873 | 0.001 |
| Statins multiplier 1.1 (AH unchanged) | 0.873 | 0.873 | 0.873 | 0.873 | -0.001 |
| Statins multiplier 1.2 (AH unchanged) | 0.872 | 0.872 | 0.872 | 0.873 | -0.001 |

# 2.2 Women; Statins; on treatment at baseline

## 2.2.1 Calibration curves

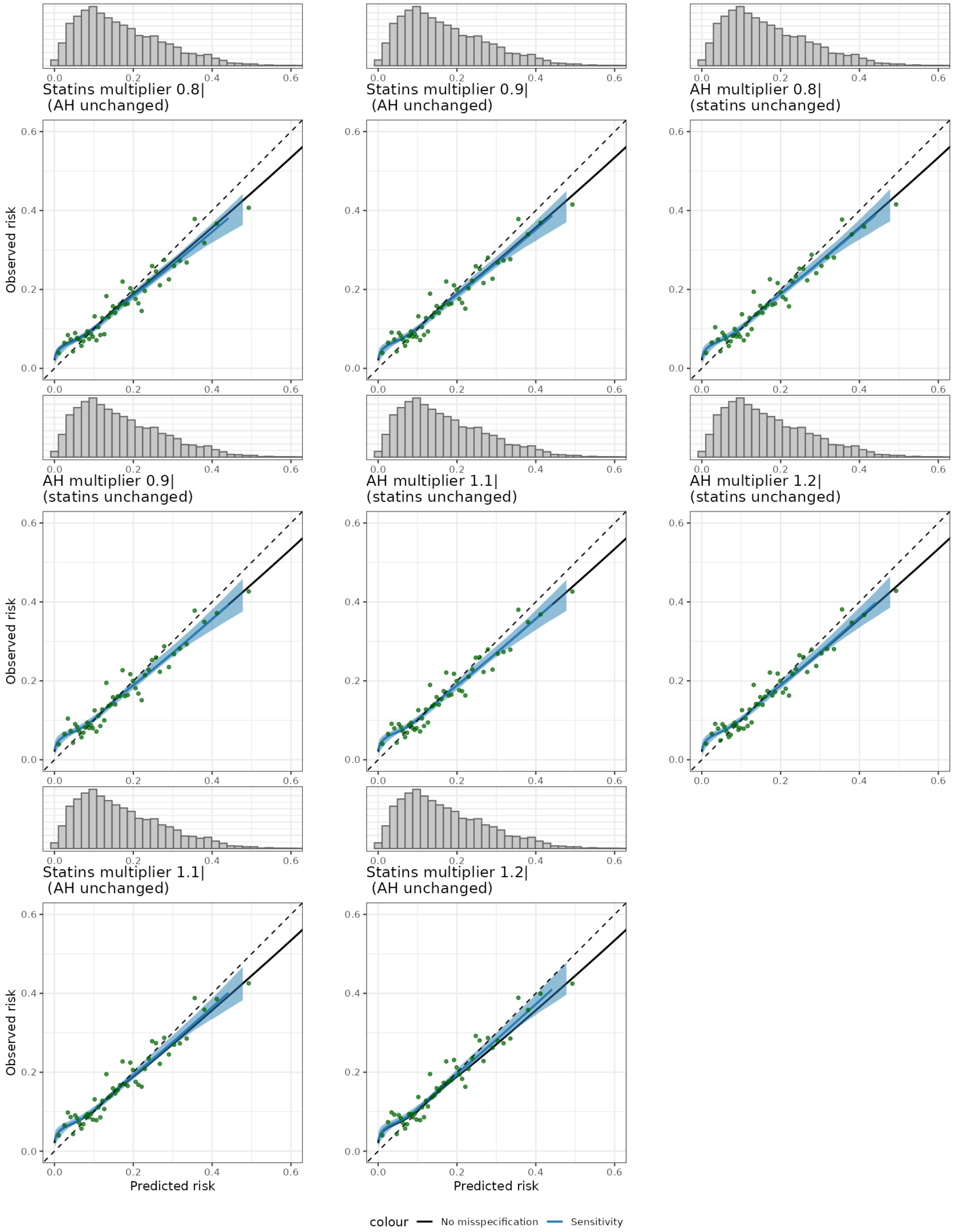

## 2.2.2 Calibration Metrics (ICI, E50, E90)

Calibration metrics for: Women; Statins; on treatment at baseline.

| scenario | metric | value | lower_ci | upper_ci | no misspecification | difference |
|---|---|---|---|---|---|---|
| AH multiplier 0.8 (statins unchanged) | ICI | 0.016 | 0.011 | 0.023 | 0.016 | 0.001 |
| AH multiplier 0.9 (statins unchanged) | ICI | 0.016 | 0.011 | 0.022 | 0.016 | 0.000 |
| AH multiplier 1.1 (statins unchanged) | ICI | 0.016 | 0.011 | 0.022 | 0.016 | 0.000 |
| AH multiplier 1.2 (statins unchanged) | ICI | 0.015 | 0.010 | 0.021 | 0.016 | -0.001 |
| Statins multiplier 0.8 (AH unchanged) | ICI | 0.019 | 0.014 | 0.026 | 0.016 | 0.003 |
| Statins multiplier 0.9 (AH unchanged) | ICI | 0.017 | 0.012 | 0.024 | 0.016 | 0.001 |
| Statins multiplier 1.1 (AH unchanged) | ICI | 0.014 | 0.009 | 0.020 | 0.016 | -0.002 |
| Statins multiplier 1.2 (AH unchanged) | ICI | 0.013 | 0.009 | 0.019 | 0.016 | -0.003 |
| AH multiplier 0.8 (statins unchanged) | E50 | 0.012 | 0.005 | 0.021 | 0.012 | 0.000 |
| AH multiplier 0.9 (statins unchanged) | E50 | 0.012 | 0.005 | 0.020 | 0.012 | 0.000 |
| AH multiplier 1.1 (statins unchanged) | E50 | 0.012 | 0.005 | 0.021 | 0.012 | 0.000 |
| AH multiplier 1.2 (statins unchanged) | E50 | 0.012 | 0.004 | 0.020 | 0.012 | -0.001 |
| Statins multiplier 0.8 (AH unchanged) | E50 | 0.014 | 0.008 | 0.022 | 0.012 | 0.002 |
| Statins multiplier 0.9 (AH unchanged) | E50 | 0.013 | 0.006 | 0.022 | 0.012 | 0.001 |
| Statins multiplier 1.1 (AH unchanged) | E50 | 0.011 | 0.004 | 0.019 | 0.012 | -0.002 |
| Statins multiplier 1.2 (AH unchanged) | E50 | 0.010 | 0.003 | 0.017 | 0.012 | -0.003 |
| AH multiplier 0.8 (statins unchanged) | E90 | 0.037 | 0.025 | 0.048 | 0.036 | 0.001 |
| AH multiplier 0.9 (statins unchanged) | E90 | 0.036 | 0.024 | 0.047 | 0.036 | 0.000 |
| AH multiplier 1.1 (statins unchanged) | E90 | 0.035 | 0.023 | 0.047 | 0.036 | 0.000 |

| scenario | metric | value | lower_ci | upper_ci | no misspecification | difference |
|---|---|---|---|---|---|---|
| AH multiplier 1.2 (statins unchanged) | E90 | 0.034 | 0.023 | 0.043 | 0.036 | -0.002 |
| Statins multiplier 0.8 (AH unchanged) | E90 | 0.042 | 0.028 | 0.055 | 0.036 | 0.006 |
| Statins multiplier 0.9 (AH unchanged) | E90 | 0.037 | 0.026 | 0.051 | 0.036 | 0.002 |
| Statins multiplier 1.1 (AH unchanged) | E90 | 0.032 | 0.022 | 0.043 | 0.036 | -0.003 |
| Statins multiplier 1.2 (AH unchanged) | E90 | 0.029 | 0.023 | 0.040 | 0.036 | -0.007 |

### 2.2.3 Discrimination (C-index)

Discrimination results for: Women; Statins; on treatment at baseline. 'No misspecification' is the value from the main analysis, assuming no misspecification of the treatment effect, but stratified to the same subgroup.

| scenario | c_index | lower_ci | upper_ci | no misspecification | difference |
|---|---|---|---|---|---|
| AH multiplier 0.8 (statins unchanged) | 0.659 | 0.658 | 0.659 | 0.66 | -0.001 |
| AH multiplier 0.9 (statins unchanged) | 0.659 | 0.659 | 0.659 | 0.66 | -0.001 |
| AH multiplier 1.1 (statins unchanged) | 0.660 | 0.660 | 0.660 | 0.66 | 0.000 |
| AH multiplier 1.2 (statins unchanged) | 0.661 | 0.660 | 0.661 | 0.66 | 0.001 |
| Statins multiplier 0.8 (AH unchanged) | 0.660 | 0.660 | 0.660 | 0.66 | 0.000 |
| Statins multiplier 0.9 (AH unchanged) | 0.660 | 0.660 | 0.660 | 0.66 | 0.000 |
| Statins multiplier 1.1 (AH unchanged) | 0.659 | 0.659 | 0.660 | 0.66 | 0.000 |
| Statins multiplier 1.2 (AH unchanged) | 0.659 | 0.659 | 0.659 | 0.66 | 0.000 |

## 2.3 Women; antihypertensives; not on treatment at

# baseline

## 2.3.1 Calibration curves

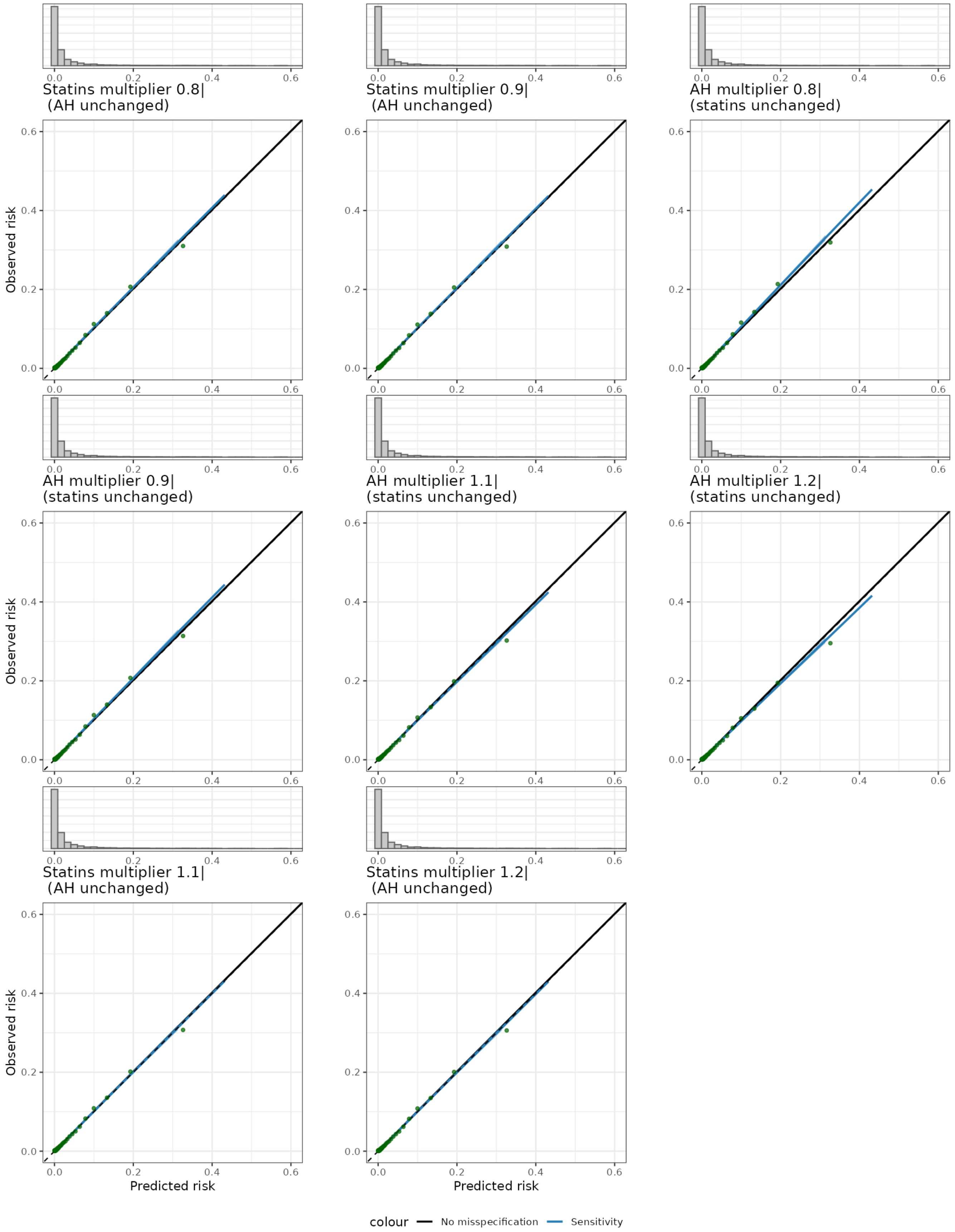

## 2.3.2 Calibration Metrics (ICI, E50, E90)

Calibration metrics for: Women; antihypertensives; not on treatment at baseline.

| scenario | metric | value | lower_ci | upper_ci | no misspecification | difference |
|---|---|---|---|---|---|---|
| AH multiplier 0.8 (statins unchanged) | ICI | 0.001 | 0.001 | 0.002 | 0.001 | 0.001 |
| AH multiplier 0.9 (statins unchanged) | ICI | 0.001 | 0.001 | 0.001 | 0.001 | 0.000 |
| AH multiplier 1.1 (statins unchanged) | ICI | 0.001 | 0.000 | 0.001 | 0.001 | 0.000 |
| AH multiplier 1.2 (statins unchanged) | ICI | 0.001 | 0.001 | 0.001 | 0.001 | 0.000 |
| Statins multiplier 0.8 (AH unchanged) | ICI | 0.001 | 0.001 | 0.001 | 0.001 | 0.000 |
| Statins multiplier 0.9 (AH unchanged) | ICI | 0.001 | 0.000 | 0.001 | 0.001 | 0.000 |
| Statins multiplier 1.1 (AH unchanged) | ICI | 0.000 | 0.000 | 0.001 | 0.001 | 0.000 |
| Statins multiplier 1.2 (AH unchanged) | ICI | 0.000 | 0.000 | 0.001 | 0.001 | 0.000 |
| AH multiplier 0.8 (statins unchanged) | E50 | 0.001 | 0.000 | 0.001 | 0.001 | 0.000 |
| AH multiplier 0.9 (statins unchanged) | E50 | 0.001 | 0.000 | 0.001 | 0.001 | 0.000 |
| AH multiplier 1.1 (statins unchanged) | E50 | 0.001 | 0.000 | 0.001 | 0.001 | 0.000 |
| AH multiplier 1.2 (statins unchanged) | E50 | 0.001 | 0.000 | 0.001 | 0.001 | 0.000 |
| Statins multiplier 0.8 (AH unchanged) | E50 | 0.001 | 0.000 | 0.001 | 0.001 | 0.000 |
| Statins multiplier 0.9 (AH unchanged) | E50 | 0.001 | 0.000 | 0.001 | 0.001 | 0.000 |
| Statins multiplier 1.1 (AH unchanged) | E50 | 0.000 | 0.000 | 0.001 | 0.001 | 0.000 |
| Statins multiplier 1.2 (AH unchanged) | E50 | 0.000 | 0.000 | 0.001 | 0.001 | 0.000 |
| AH multiplier 0.8 (statins unchanged) | E90 | 0.004 | 0.002 | 0.005 | 0.001 | 0.003 |
| AH multiplier 0.9 (statins unchanged) | E90 | 0.002 | 0.001 | 0.004 | 0.001 | 0.001 |
| AH multiplier 1.1 (statins unchanged) | E90 | 0.001 | 0.001 | 0.002 | 0.001 | 0.000 |

| scenario | metric | value | lower_ci | upper_ci | no misspecification | difference |
|---|---|---|---|---|---|---|
| AH multiplier 1.2 (statins unchanged) | E90 | 0.001 | 0.001 | 0.003 | 0.001 | 0.001 |
| Statins multiplier 0.8 (AH unchanged) | E90 | 0.002 | 0.001 | 0.004 | 0.001 | 0.001 |
| Statins multiplier 0.9 (AH unchanged) | E90 | 0.002 | 0.001 | 0.003 | 0.001 | 0.001 |
| Statins multiplier 1.1 (AH unchanged) | E90 | 0.001 | 0.001 | 0.002 | 0.001 | 0.000 |
| Statins multiplier 1.2 (AH unchanged) | E90 | 0.001 | 0.001 | 0.002 | 0.001 | 0.000 |

### 2.3.3 Discrimination (C-index)

Discrimination results for: Women; antihypertensives; not on treatment at baseline. 'No misspecification' is the value from the main analysis, assuming no misspecification of the treatment effect, but stratified to the same subgroup.

| scenario | c_index | lower_ci | upper_ci | no misspecification | difference |
|---|---|---|---|---|---|
| AH multiplier 0.8 (statins unchanged) | 0.876 | 0.876 | 0.876 | 0.873 | 0.002 |
| AH multiplier 0.9 (statins unchanged) | 0.875 | 0.875 | 0.875 | 0.873 | 0.001 |
| AH multiplier 1.1 (statins unchanged) | 0.872 | 0.872 | 0.872 | 0.873 | -0.001 |
| AH multiplier 1.2 (statins unchanged) | 0.871 | 0.871 | 0.871 | 0.873 | -0.002 |
| Statins multiplier 0.8 (AH unchanged) | 0.874 | 0.874 | 0.874 | 0.873 | 0.001 |
| Statins multiplier 0.9 (AH unchanged) | 0.874 | 0.874 | 0.874 | 0.873 | 0.000 |
| Statins multiplier 1.1 (AH unchanged) | 0.873 | 0.873 | 0.873 | 0.873 | 0.000 |
| Statins multiplier 1.2 (AH unchanged) | 0.873 | 0.873 | 0.873 | 0.873 | -0.001 |

## 2.4 Women; antihypertensives; on treatment at

# baseline

## 2.4.1 Calibration curves

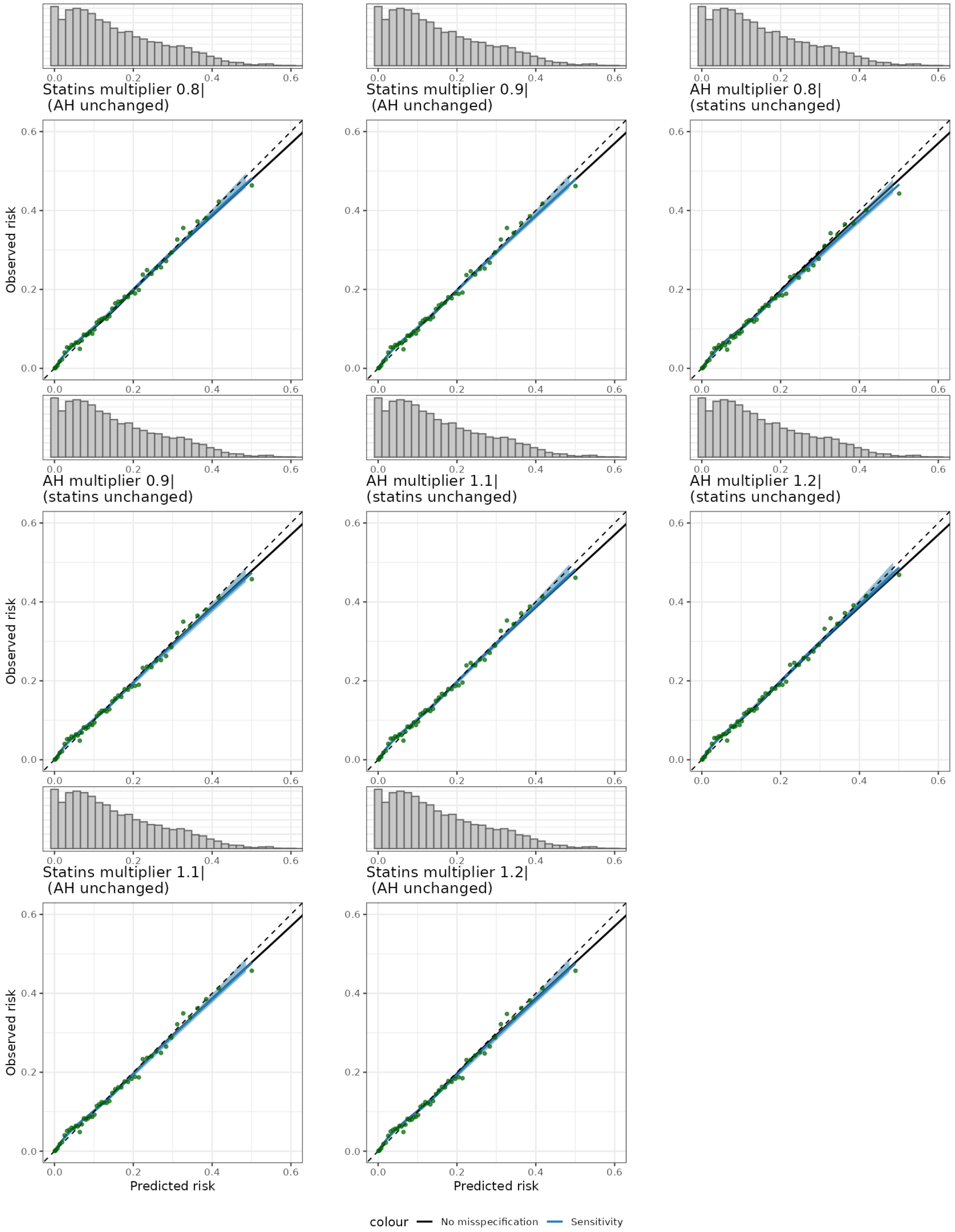

## 2.4.2 Calibration Metrics (ICI, E50, E90)

Calibration metrics for: Women; antihypertensives; on treatment at baseline.

| scenario | metric | value | lower_ci | upper_ci | no misspecification | difference |
|---|---|---|---|---|---|---|
| AH multiplier 0.8 (statins unchanged) | ICI | 0.008 | 0.006 | 0.011 | 0.006 | 0.003 |
| AH multiplier 0.9 (statins unchanged) | ICI | 0.007 | 0.004 | 0.009 | 0.006 | 0.001 |
| AH multiplier 1.1 (statins unchanged) | ICI | 0.005 | 0.003 | 0.008 | 0.006 | -0.001 |
| AH multiplier 1.2 (statins unchanged) | ICI | 0.004 | 0.003 | 0.007 | 0.006 | -0.001 |
| Statins multiplier 0.8 (AH unchanged) | ICI | 0.005 | 0.004 | 0.008 | 0.006 | -0.001 |
| Statins multiplier 0.9 (AH unchanged) | ICI | 0.005 | 0.004 | 0.008 | 0.006 | -0.001 |
| Statins multiplier 1.1 (AH unchanged) | ICI | 0.006 | 0.004 | 0.009 | 0.006 | 0.001 |
| Statins multiplier 1.2 (AH unchanged) | ICI | 0.007 | 0.004 | 0.010 | 0.006 | 0.002 |
| AH multiplier 0.8 (statins unchanged) | E50 | 0.007 | 0.004 | 0.008 | 0.004 | 0.004 |
| AH multiplier 0.9 (statins unchanged) | E50 | 0.005 | 0.004 | 0.008 | 0.004 | 0.001 |
| AH multiplier 1.1 (statins unchanged) | E50 | 0.003 | 0.003 | 0.008 | 0.004 | -0.001 |
| AH multiplier 1.2 (statins unchanged) | E50 | 0.002 | 0.003 | 0.008 | 0.004 | -0.002 |
| Statins multiplier 0.8 (AH unchanged) | E50 | 0.003 | 0.003 | 0.008 | 0.004 | -0.001 |
| Statins multiplier 0.9 (AH unchanged) | E50 | 0.003 | 0.003 | 0.008 | 0.004 | -0.001 |
| Statins multiplier 1.1 (AH unchanged) | E50 | 0.005 | 0.004 | 0.008 | 0.004 | 0.002 |
| Statins multiplier 1.2 (AH unchanged) | E50 | 0.007 | 0.004 | 0.009 | 0.004 | 0.003 |
| AH multiplier 0.8 (statins unchanged) | E90 | 0.016 | 0.011 | 0.025 | 0.010 | 0.006 |
| AH multiplier 0.9 (statins unchanged) | E90 | 0.011 | 0.007 | 0.021 | 0.010 | 0.001 |
| AH multiplier 1.1 (statins unchanged) | E90 | 0.010 | 0.007 | 0.014 | 0.010 | 0.000 |

| scenario | metric | value | lower_ci | upper_ci | no misspecification | difference |
|---|---|---|---|---|---|---|
| AH multiplier 1.2 (statins unchanged) | E90 | 0.011 | 0.007 | 0.013 | 0.010 | 0.001 |
| Statins multiplier 0.8 (AH unchanged) | E90 | 0.011 | 0.007 | 0.014 | 0.010 | 0.001 |
| Statins multiplier 0.9 (AH unchanged) | E90 | 0.010 | 0.007 | 0.016 | 0.010 | 0.000 |
| Statins multiplier 1.1 (AH unchanged) | E90 | 0.010 | 0.007 | 0.019 | 0.010 | 0.000 |
| Statins multiplier 1.2 (AH unchanged) | E90 | 0.012 | 0.008 | 0.021 | 0.010 | 0.002 |

## 2.4.3 Discrimination (C-index)

Discrimination results for: Women; antihypertensives; on treatment at baseline. 'No misspecification' is the value from the main analysis, assuming no misspecification of the treatment effect, but stratified to the same subgroup.

| scenario | c_index | lower_ci | upper_ci | no misspecification | difference |
|---|---|---|---|---|---|
| AH multiplier 0.8 (statins unchanged) | 0.713 | 0.713 | 0.713 | 0.711 | 0.001 |
| AH multiplier 0.9 (statins unchanged) | 0.712 | 0.712 | 0.712 | 0.711 | 0.001 |
| AH multiplier 1.1 (statins unchanged) | 0.710 | 0.710 | 0.710 | 0.711 | -0.001 |
| AH multiplier 1.2 (statins unchanged) | 0.710 | 0.710 | 0.710 | 0.711 | -0.001 |
| Statins multiplier 0.8 (AH unchanged) | 0.711 | 0.711 | 0.711 | 0.711 | 0.000 |
| Statins multiplier 0.9 (AH unchanged) | 0.711 | 0.711 | 0.711 | 0.711 | 0.000 |
| Statins multiplier 1.1 (AH unchanged) | 0.711 | 0.711 | 0.711 | 0.711 | 0.000 |
| Statins multiplier 1.2 (AH unchanged) | 0.711 | 0.711 | 0.711 | 0.711 | 0.000 |

# 2.5 Men; Statins; not on treatment at baseline

## 2.5.1 Calibration curves

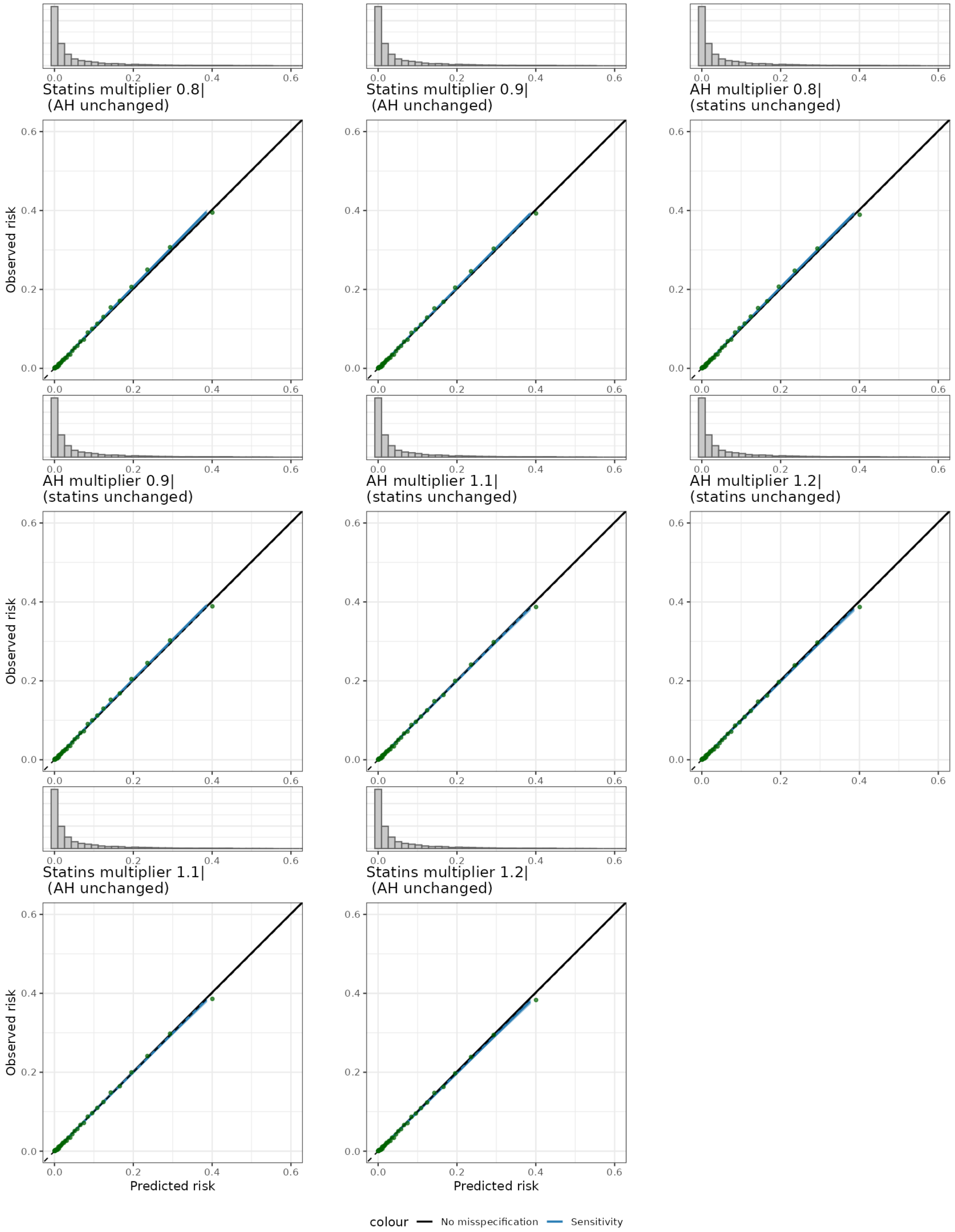

## 2.5.2 Calibration Metrics (ICI, E50, E90)

Calibration metrics for: Men; Statins; not on treatment at baseline.

| scenario | metric | value | lower_ci | upper_ci | no misspecification | difference |
|---|---|---|---|---|---|---|
| AH multiplier 0.8 (statins unchanged) | ICI | 0.002 | 0.001 | 0.002 | 0.001 | 0.001 |
| AH multiplier 0.9 (statins unchanged) | ICI | 0.001 | 0.001 | 0.002 | 0.001 | 0.000 |
| AH multiplier 1.1 (statins unchanged) | ICI | 0.000 | 0.000 | 0.001 | 0.001 | 0.000 |
| AH multiplier 1.2 (statins unchanged) | ICI | 0.001 | 0.001 | 0.001 | 0.001 | 0.000 |
| Statins multiplier 0.8 (AH unchanged) | ICI | 0.002 | 0.001 | 0.002 | 0.001 | 0.001 |
| Statins multiplier 0.9 (AH unchanged) | ICI | 0.001 | 0.001 | 0.002 | 0.001 | 0.000 |
| Statins multiplier 1.1 (AH unchanged) | ICI | 0.000 | 0.001 | 0.001 | 0.001 | 0.000 |
| Statins multiplier 1.2 (AH unchanged) | ICI | 0.001 | 0.001 | 0.001 | 0.001 | 0.000 |
| AH multiplier 0.8 (statins unchanged) | E50 | 0.001 | 0.000 | 0.001 | 0.001 | 0.000 |
| AH multiplier 0.9 (statins unchanged) | E50 | 0.001 | 0.000 | 0.001 | 0.001 | 0.000 |
| AH multiplier 1.1 (statins unchanged) | E50 | 0.000 | 0.000 | 0.001 | 0.001 | 0.000 |
| AH multiplier 1.2 (statins unchanged) | E50 | 0.001 | 0.000 | 0.001 | 0.001 | 0.000 |
| Statins multiplier 0.8 (AH unchanged) | E50 | 0.001 | 0.000 | 0.001 | 0.001 | 0.000 |
| Statins multiplier 0.9 (AH unchanged) | E50 | 0.001 | 0.000 | 0.001 | 0.001 | 0.000 |
| Statins multiplier 1.1 (AH unchanged) | E50 | 0.000 | 0.000 | 0.001 | 0.001 | 0.000 |
| Statins multiplier 1.2 (AH unchanged) | E50 | 0.001 | 0.000 | 0.001 | 0.001 | 0.000 |
| AH multiplier 0.8 (statins unchanged) | E90 | 0.005 | 0.003 | 0.008 | 0.002 | 0.003 |
| AH multiplier 0.9 (statins unchanged) | E90 | 0.003 | 0.002 | 0.006 | 0.002 | 0.002 |
| AH multiplier 1.1 (statins unchanged) | E90 | 0.001 | 0.001 | 0.002 | 0.002 | -0.001 |

| scenario | metric | value | lower_ci | upper_ci | no misspecification | difference |
|---|---|---|---|---|---|---|
| AH multiplier 1.2 (statins unchanged) | E90 | 0.001 | 0.001 | 0.002 | 0.002 | 0.000 |
| Statins multiplier 0.8 (AH unchanged) | E90 | 0.005 | 0.004 | 0.008 | 0.002 | 0.004 |
| Statins multiplier 0.9 (AH unchanged) | E90 | 0.003 | 0.002 | 0.006 | 0.002 | 0.002 |
| Statins multiplier 1.1 (AH unchanged) | E90 | 0.001 | 0.001 | 0.002 | 0.002 | -0.001 |
| Statins multiplier 1.2 (AH unchanged) | E90 | 0.002 | 0.001 | 0.002 | 0.002 | 0.000 |

## 2.5.3 Discrimination (C-index)

Discrimination results for: Men; Statins; not on treatment at baseline. 'No misspecification' is the value from the main analysis, assuming no misspecification of the treatment effect, but stratified to the same subgroup.

| scenario | c_index | lower_ci | upper_ci | no misspecification | difference |
|---|---|---|---|---|---|
| AH multiplier 0.8 (statins unchanged) | 0.858 | 0.858 | 0.858 | 0.857 | 0.001 |
| AH multiplier 0.9 (statins unchanged) | 0.857 | 0.857 | 0.857 | 0.857 | 0.001 |
| AH multiplier 1.1 (statins unchanged) | 0.856 | 0.856 | 0.856 | 0.857 | -0.001 |
| AH multiplier 1.2 (statins unchanged) | 0.856 | 0.856 | 0.856 | 0.857 | -0.001 |
| Statins multiplier 0.8 (AH unchanged) | 0.858 | 0.858 | 0.858 | 0.857 | 0.001 |
| Statins multiplier 0.9 (AH unchanged) | 0.857 | 0.857 | 0.857 | 0.857 | 0.001 |
| Statins multiplier 1.1 (AH unchanged) | 0.856 | 0.856 | 0.856 | 0.857 | -0.001 |
| Statins multiplier 1.2 (AH unchanged) | 0.855 | 0.855 | 0.855 | 0.857 | -0.001 |

# 2.6 Men; Statins; on treatment at baseline

## 2.6.1 Calibration curves

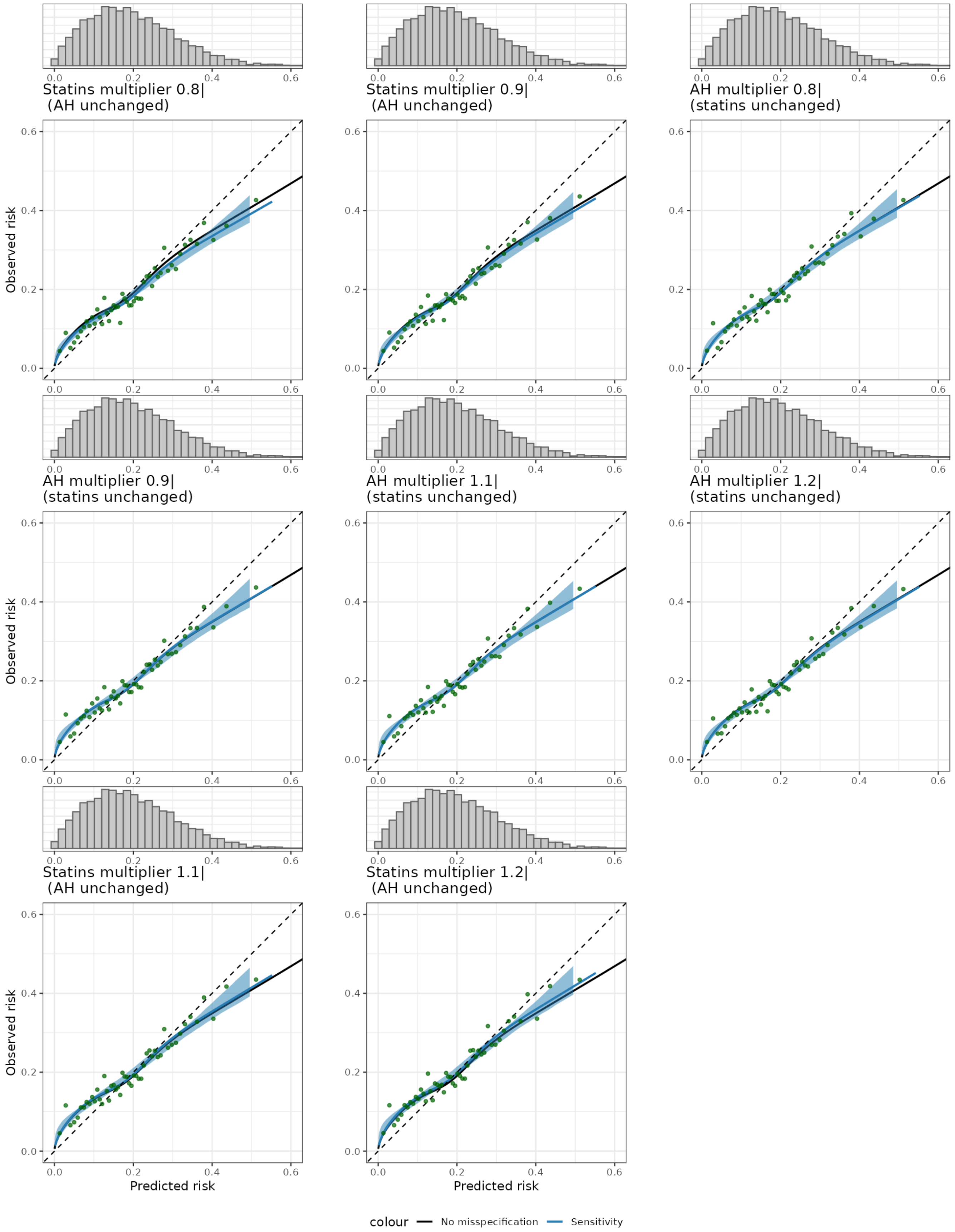

## 2.6.2 Calibration Metrics (ICI, E50, E90)

Calibration metrics for: Men; Statins; on treatment at baseline.

| scenario | metric | value | lower_ci | upper_ci | no misspecification | difference |
|---|---|---|---|---|---|---|
| AH multiplier 0.8 (statins unchanged) | ICI | 0.023 | 0.018 | 0.029 | 0.023 | 0.000 |
| AH multiplier 0.9 (statins unchanged) | ICI | 0.022 | 0.018 | 0.029 | 0.023 | 0.000 |
| AH multiplier 1.1 (statins unchanged) | ICI | 0.023 | 0.018 | 0.028 | 0.023 | 0.000 |
| AH multiplier 1.2 (statins unchanged) | ICI | 0.023 | 0.018 | 0.029 | 0.023 | 0.001 |
| Statins multiplier 0.8 (AH unchanged) | ICI | 0.026 | 0.020 | 0.031 | 0.023 | 0.004 |
| Statins multiplier 0.9 (AH unchanged) | ICI | 0.025 | 0.019 | 0.030 | 0.023 | 0.002 |
| Statins multiplier 1.1 (AH unchanged) | ICI | 0.022 | 0.017 | 0.028 | 0.023 | -0.001 |
| Statins multiplier 1.2 (AH unchanged) | ICI | 0.021 | 0.017 | 0.028 | 0.023 | -0.002 |
| AH multiplier 0.8 (statins unchanged) | E50 | 0.016 | 0.013 | 0.028 | 0.016 | 0.000 |
| AH multiplier 0.9 (statins unchanged) | E50 | 0.016 | 0.013 | 0.029 | 0.016 | 0.000 |
| AH multiplier 1.1 (statins unchanged) | E50 | 0.016 | 0.013 | 0.028 | 0.016 | 0.000 |
| AH multiplier 1.2 (statins unchanged) | E50 | 0.017 | 0.013 | 0.029 | 0.016 | 0.001 |
| Statins multiplier 0.8 (AH unchanged) | E50 | 0.021 | 0.015 | 0.029 | 0.016 | 0.005 |
| Statins multiplier 0.9 (AH unchanged) | E50 | 0.018 | 0.014 | 0.029 | 0.016 | 0.002 |
| Statins multiplier 1.1 (AH unchanged) | E50 | 0.015 | 0.013 | 0.028 | 0.016 | -0.001 |
| Statins multiplier 1.2 (AH unchanged) | E50 | 0.014 | 0.013 | 0.028 | 0.016 | -0.002 |
| AH multiplier 0.8 (statins unchanged) | E90 | 0.042 | 0.035 | 0.054 | 0.042 | 0.001 |
| AH multiplier 0.9 (statins unchanged) | E90 | 0.042 | 0.035 | 0.054 | 0.042 | 0.001 |
| AH multiplier 1.1 (statins unchanged) | E90 | 0.041 | 0.034 | 0.055 | 0.042 | 0.000 |

| scenario | metric | value | lower_ci | upper_ci | no misspecification | difference |
|---|---|---|---|---|---|---|
| AH multiplier 1.2 (statins unchanged) | E90 | 0.040 | 0.034 | 0.056 | 0.042 | -0.001 |
| Statins multiplier 0.8 (AH unchanged) | E90 | 0.046 | 0.039 | 0.065 | 0.042 | 0.004 |
| Statins multiplier 0.9 (AH unchanged) | E90 | 0.040 | 0.037 | 0.061 | 0.042 | -0.002 |
| Statins multiplier 1.1 (AH unchanged) | E90 | 0.043 | 0.034 | 0.053 | 0.042 | 0.002 |
| Statins multiplier 1.2 (AH unchanged) | E90 | 0.045 | 0.035 | 0.053 | 0.042 | 0.004 |

### 2.6.3 Discrimination (C-index)

Discrimination results for: Men; Statins; on treatment at baseline. 'No misspecification' is the value from the main analysis, assuming no misspecification of the treatment effect, but stratified to the same subgroup.

| scenario | c_index | lower_ci | upper_ci | no misspecification | difference |
|---|---|---|---|---|---|
| AH multiplier 0.8 (statins unchanged) | 0.630 | 0.630 | 0.630 | 0.631 | -0.001 |
| AH multiplier 0.9 (statins unchanged) | 0.630 | 0.630 | 0.631 | 0.631 | 0.000 |
| AH multiplier 1.1 (statins unchanged) | 0.631 | 0.631 | 0.631 | 0.631 | 0.000 |
| AH multiplier 1.2 (statins unchanged) | 0.631 | 0.631 | 0.631 | 0.631 | 0.001 |
| Statins multiplier 0.8 (AH unchanged) | 0.632 | 0.632 | 0.632 | 0.631 | 0.001 |
| Statins multiplier 0.9 (AH unchanged) | 0.631 | 0.631 | 0.632 | 0.631 | 0.001 |
| Statins multiplier 1.1 (AH unchanged) | 0.630 | 0.630 | 0.630 | 0.631 | -0.001 |
| Statins multiplier 1.2 (AH unchanged) | 0.630 | 0.629 | 0.630 | 0.631 | -0.001 |

## 2.7 Men; antihypertensives; not on treatment at

# baseline

## 2.7.1 Calibration curves

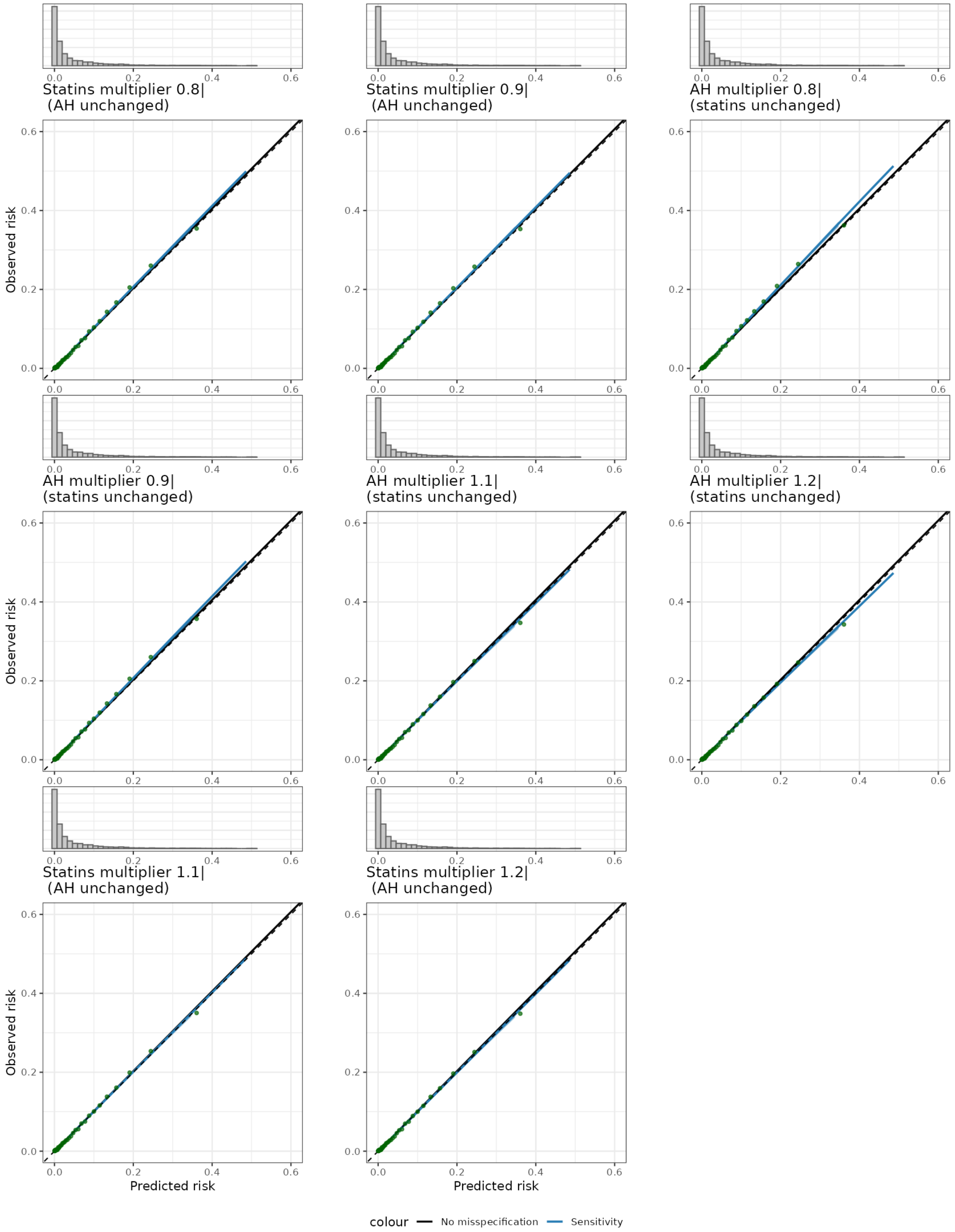

## 2.7.2 Calibration Metrics (ICI, E50, E90)

Calibration metrics for: Men; antihypertensives; not on treatment at baseline.

| scenario | metric | value | lower_ci | upper_ci | no misspecification | difference |
|---|---|---|---|---|---|---|
| AH multiplier 0.8 (statins unchanged) | ICI | 0.002 | 0.002 | 0.003 | 0.001 | 0.001 |
| AH multiplier 0.9 (statins unchanged) | ICI | 0.002 | 0.001 | 0.002 | 0.001 | 0.001 |
| AH multiplier 1.1 (statins unchanged) | ICI | 0.001 | 0.001 | 0.001 | 0.001 | 0.000 |
| AH multiplier 1.2 (statins unchanged) | ICI | 0.001 | 0.001 | 0.002 | 0.001 | 0.000 |
| Statins multiplier 0.8 (AH unchanged) | ICI | 0.001 | 0.001 | 0.002 | 0.001 | 0.001 |
| Statins multiplier 0.9 (AH unchanged) | ICI | 0.001 | 0.001 | 0.002 | 0.001 | 0.000 |
| Statins multiplier 1.1 (AH unchanged) | ICI | 0.001 | 0.001 | 0.001 | 0.001 | 0.000 |
| Statins multiplier 1.2 (AH unchanged) | ICI | 0.001 | 0.001 | 0.001 | 0.001 | 0.000 |
| AH multiplier 0.8 (statins unchanged) | E50 | 0.001 | 0.001 | 0.001 | 0.001 | 0.000 |
| AH multiplier 0.9 (statins unchanged) | E50 | 0.001 | 0.001 | 0.001 | 0.001 | 0.000 |
| AH multiplier 1.1 (statins unchanged) | E50 | 0.001 | 0.000 | 0.001 | 0.001 | 0.000 |
| AH multiplier 1.2 (statins unchanged) | E50 | 0.001 | 0.000 | 0.002 | 0.001 | 0.000 |
| Statins multiplier 0.8 (AH unchanged) | E50 | 0.001 | 0.001 | 0.001 | 0.001 | 0.000 |
| Statins multiplier 0.9 (AH unchanged) | E50 | 0.001 | 0.000 | 0.001 | 0.001 | 0.000 |
| Statins multiplier 1.1 (AH unchanged) | E50 | 0.001 | 0.000 | 0.001 | 0.001 | 0.000 |
| Statins multiplier 1.2 (AH unchanged) | E50 | 0.001 | 0.000 | 0.001 | 0.001 | 0.000 |
| AH multiplier 0.8 (statins unchanged) | E90 | 0.006 | 0.005 | 0.008 | 0.002 | 0.004 |
| AH multiplier 0.9 (statins unchanged) | E90 | 0.004 | 0.003 | 0.006 | 0.002 | 0.002 |
| AH multiplier 1.1 (statins unchanged) | E90 | 0.001 | 0.001 | 0.002 | 0.002 | -0.001 |

| scenario | metric | value | lower_ci | upper_ci | no misspecification | difference |
|---|---|---|---|---|---|---|
| AH multiplier 1.2 (statins unchanged) | E90 | 0.002 | 0.001 | 0.003 | 0.002 | 0.000 |
| Statins multiplier 0.8 (AH unchanged) | E90 | 0.004 | 0.003 | 0.006 | 0.002 | 0.002 |
| Statins multiplier 0.9 (AH unchanged) | E90 | 0.003 | 0.002 | 0.005 | 0.002 | 0.001 |
| Statins multiplier 1.1 (AH unchanged) | E90 | 0.001 | 0.001 | 0.003 | 0.002 | -0.001 |
| Statins multiplier 1.2 (AH unchanged) | E90 | 0.001 | 0.001 | 0.002 | 0.002 | -0.001 |

### 2.7.3 Discrimination (C-index)

Discrimination results for: Men; antihypertensives; not on treatment at baseline. 'No misspecification' is the value from the main analysis, assuming no misspecification of the treatment effect, but stratified to the same subgroup.

| scenario | c_index | lower_ci | upper_ci | no misspecification | difference |
|---|---|---|---|---|---|
| AH multiplier 0.8 (statins unchanged) | 0.859 | 0.859 | 0.859 | 0.857 | 0.002 |
| AH multiplier 0.9 (statins unchanged) | 0.858 | 0.858 | 0.858 | 0.857 | 0.001 |
| AH multiplier 1.1 (statins unchanged) | 0.855 | 0.855 | 0.855 | 0.857 | -0.001 |
| AH multiplier 1.2 (statins unchanged) | 0.854 | 0.854 | 0.854 | 0.857 | -0.002 |
| Statins multiplier 0.8 (AH unchanged) | 0.858 | 0.858 | 0.858 | 0.857 | 0.001 |
| Statins multiplier 0.9 (AH unchanged) | 0.857 | 0.857 | 0.857 | 0.857 | 0.000 |
| Statins multiplier 1.1 (AH unchanged) | 0.856 | 0.856 | 0.856 | 0.857 | 0.000 |
| Statins multiplier 1.2 (AH unchanged) | 0.856 | 0.856 | 0.856 | 0.857 | -0.001 |

## 2.8 Men; antihypertensives; on treatment at

# baseline

## 2.8.1 Calibration curves

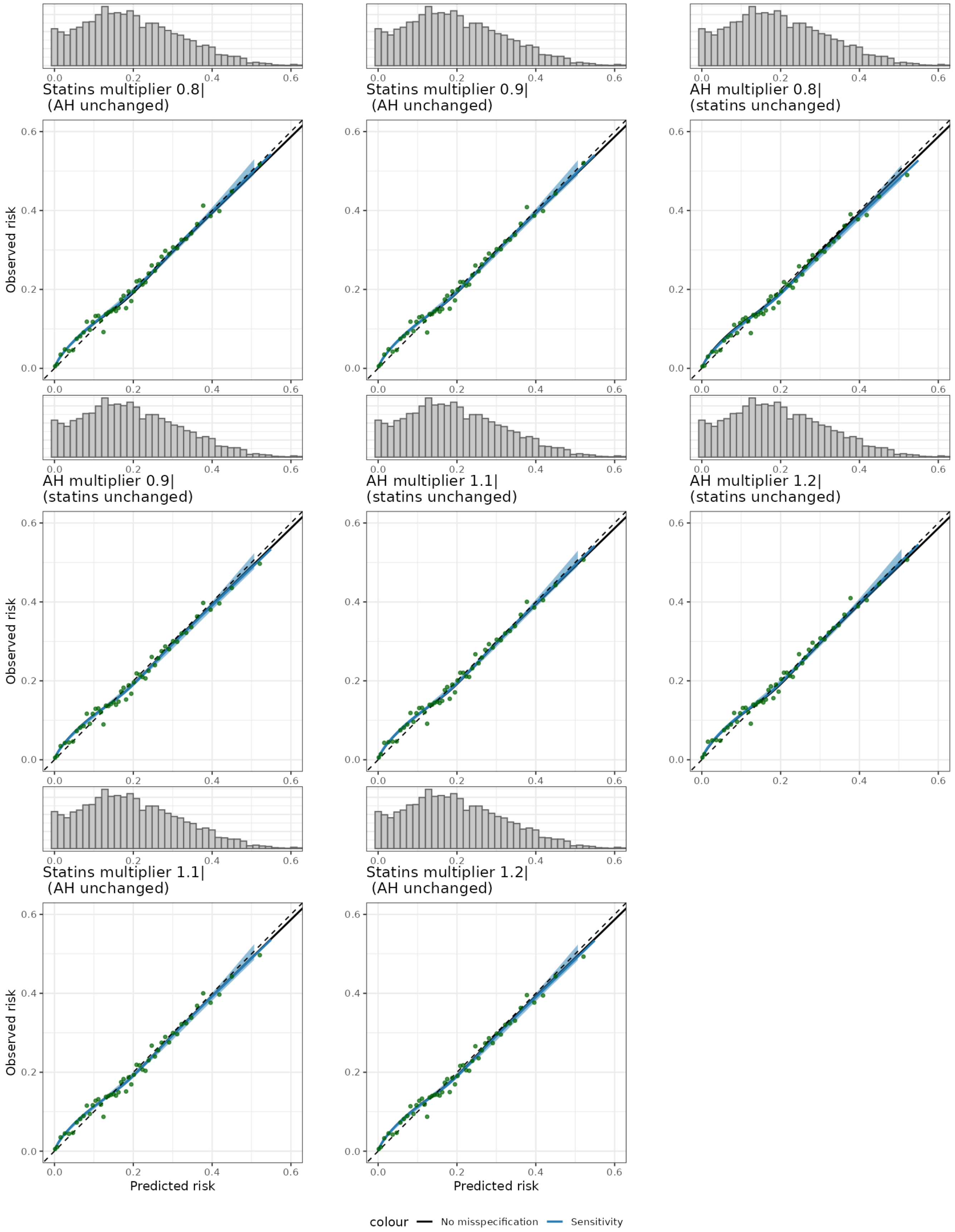

## 2.8.2 Calibration Metrics (ICI, E50, E90)

Calibration metrics for: Men; antihypertensives; on treatment at baseline.

| scenario | metric | value | lower_ci | upper_ci | no misspecification | difference |
|---|---|---|---|---|---|---|
| AH multiplier 0.8 (statins unchanged) | ICI | 0.011 | 0.007 | 0.013 | 0.009 | 0.002 |
| AH multiplier 0.9 (statins unchanged) | ICI | 0.009 | 0.006 | 0.012 | 0.009 | 0.001 |
| AH multiplier 1.1 (statins unchanged) | ICI | 0.008 | 0.006 | 0.011 | 0.009 | -0.001 |
| AH multiplier 1.2 (statins unchanged) | ICI | 0.007 | 0.006 | 0.011 | 0.009 | -0.002 |
| Statins multiplier 0.8 (AH unchanged) | ICI | 0.007 | 0.005 | 0.010 | 0.009 | -0.002 |
| Statins multiplier 0.9 (AH unchanged) | ICI | 0.008 | 0.006 | 0.011 | 0.009 | -0.001 |
| Statins multiplier 1.1 (AH unchanged) | ICI | 0.009 | 0.007 | 0.012 | 0.009 | 0.001 |
| Statins multiplier 1.2 (AH unchanged) | ICI | 0.010 | 0.007 | 0.013 | 0.009 | 0.001 |
| AH multiplier 0.8 (statins unchanged) | E50 | 0.011 | 0.007 | 0.014 | 0.007 | 0.004 |
| AH multiplier 0.9 (statins unchanged) | E50 | 0.009 | 0.005 | 0.014 | 0.007 | 0.002 |
| AH multiplier 1.1 (statins unchanged) | E50 | 0.006 | 0.003 | 0.012 | 0.007 | -0.002 |
| AH multiplier 1.2 (statins unchanged) | E50 | 0.005 | 0.003 | 0.011 | 0.007 | -0.003 |
| Statins multiplier 0.8 (AH unchanged) | E50 | 0.004 | 0.003 | 0.010 | 0.007 | -0.003 |
| Statins multiplier 0.9 (AH unchanged) | E50 | 0.006 | 0.003 | 0.012 | 0.007 | -0.002 |
| Statins multiplier 1.1 (AH unchanged) | E50 | 0.008 | 0.005 | 0.013 | 0.007 | 0.001 |
| Statins multiplier 1.2 (AH unchanged) | E50 | 0.010 | 0.006 | 0.014 | 0.007 | 0.002 |
| AH multiplier 0.8 (statins unchanged) | E90 | 0.017 | 0.012 | 0.023 | 0.018 | -0.001 |
| AH multiplier 0.9 (statins unchanged) | E90 | 0.017 | 0.012 | 0.020 | 0.018 | -0.001 |
| AH multiplier 1.1 (statins unchanged) | E90 | 0.019 | 0.013 | 0.021 | 0.018 | 0.001 |

| scenario | metric | value | lower_ci | upper_ci | no misspecification | difference |
|---|---|---|---|---|---|---|
| AH multiplier 1.2 (statins unchanged) | E90 | 0.020 | 0.014 | 0.022 | 0.018 | 0.002 |
| Statins multiplier 0.8 (AH unchanged) | E90 | 0.019 | 0.013 | 0.021 | 0.018 | 0.001 |
| Statins multiplier 0.9 (AH unchanged) | E90 | 0.019 | 0.013 | 0.020 | 0.018 | 0.001 |
| Statins multiplier 1.1 (AH unchanged) | E90 | 0.018 | 0.012 | 0.020 | 0.018 | 0.000 |
| Statins multiplier 1.2 (AH unchanged) | E90 | 0.017 | 0.012 | 0.020 | 0.018 | -0.001 |

## 2.8.3 Discrimination (C-index)

Discrimination results for: Men; antihypertensives; on treatment at baseline. 'No misspecification' is the value from the main analysis, assuming no misspecification of the treatment effect, but stratified to the same subgroup.

| scenario | c_index | lower_ci | upper_ci | no misspecification | difference |
|---|---|---|---|---|---|
| AH multiplier 0.8 (statins unchanged) | 0.678 | 0.678 | 0.678 | 0.677 | 0.002 |
| AH multiplier 0.9 (statins unchanged) | 0.678 | 0.677 | 0.678 | 0.677 | 0.001 |
| AH multiplier 1.1 (statins unchanged) | 0.676 | 0.676 | 0.676 | 0.677 | -0.001 |
| AH multiplier 1.2 (statins unchanged) | 0.675 | 0.675 | 0.676 | 0.677 | -0.001 |
| Statins multiplier 0.8 (AH unchanged) | 0.677 | 0.677 | 0.677 | 0.677 | 0.000 |
| Statins multiplier 0.9 (AH unchanged) | 0.677 | 0.677 | 0.677 | 0.677 | 0.000 |
| Statins multiplier 1.1 (AH unchanged) | 0.677 | 0.677 | 0.677 | 0.677 | 0.000 |
| Statins multiplier 1.2 (AH unchanged) | 0.676 | 0.676 | 0.677 | 0.677 | 0.000 |

# Validation of the two component model using artificial censoring

26 August 2026

**Associated manuscript: A flexible approach to sequential prediction under intervention**

This document contains a number of analyses pertaining to validation of the two-component model using, as outlined by Keogh and Van Geloven.(Keogh and Van Geloven 2024) This is in contrast to validating the model with a process aligned with Xu et al, (Xu et al. 2021) which was utilised in the main analysis. This document in split into three sections. Section 1 explains our choice of methodology. Section 2 implements a simulation to ensure application of the methodology of Keogh and Van Geloven(Keogh and Van Geloven 2024) is being applied appropriately, given the data/setting in this study is slightly different from the referenced work. Section 3 presents a validation of the two-component model using these approaches. Section 4 presents an analysis to estimate the causal effect of statin and antihypertensive use on cardiovascular risk, using artificial censoring weights. Given the causal effect of these interventions are broadly known, we can assess whether the artificial censoring weights framework allows correct recovery statin and antihypertensive use. Failure to do so may explain any differences in the validation results from section 2, and the validation results from the main paper.

# 1 Choice of methodology when evaluating model performance using artificial censoring

## 1.1 Motivation

The work of (Keogh and Van Geloven 2024) applies artificial censoring in the validation cohort when anyone deviates from the treatment strategy under which predictions are being made. Inverse probability of artificial censoring weights (IPACWs) are then estimated, and used in a weighted Kaplan-Meier estimator to estimate observed risk in the validation cohort. This observed risk is in a pseudo-population where everybody followed the assigned treatment strategy. This is then compared with the mean predicted risk under intervention. This analysis is also executed within subgroups defined by predicted risk. When using artificial censoring to assess the validation of the two-component model (see section 2), we do not follow this approach. There are a few reasons for this.

Firstly, our data is continuously observed, and 10-year risk scores contain 3652 different possible event times. Given the size of the cohort, the majority of these event times are observed. A weighted Kaplan-Meier requires time-varying weights at every possible event time using interval censored data, which presents a computational problem. One option would be to bin the data at 30/90/180 intervals, and apply weighted Kaplan-Meier estimator to this data. However, the impact of this choice, and magnitude of the bias induced by doing so, is unclear. There are alternative approaches (detailed subsequently in this document) which do not suffer this issue which we prefer.

Secondly, our non-artificial censoring mechanism is not independent from the outcome mechanism. This is primarily due to individuals at high risk of death (censoring event), are generally older and are also more likely to experience a cardiovascular event. Assessment of mean calibration through Kaplan-Meier observed risk vs mean predicted risk will be biased, as the Kaplan-Meier estimate will be biased. This is protected somewhat by looking at weighted Kaplan-Meier vs Mean Predicted risk within groups of individuals with similar risk (i.e. deciles). (Keogh and Van Geloven 2024) state that "*some estimators for evaluating predictive performance in the (non interventiuonal) prediction setting apply inverse probability of censoring weights to account for standard right censoring*". When using such estimators, and independence between the censoring mechanisms, combined weights can be calculated as the product of the inverse probability of censoring weights (IPCWs) and the IPACWs. These combined weights could be used to weight the Kaplan-Meier estimator, however, this feels like an overly complicated solution.

Thirdly, comparing Kaplan-Meier observed risk with mean predicted risk cannot be easily extended to produce calibration curves. Doing so within subgroups defined by predicted risk may often be sufficient, but has it's limitations (primarily results dependent on number of arbitrarily defined groups).(Riley et al. 2024) Given the aforementioned computational issues, and that we must use combined weights due to a dependent non-artificial censoring mechanism, and we prefer to use alternative approaches which do not require estimating both time-varying IPCWs and IPACWs, and that can be easily extended to produce calibration curves.

The methods we choose to use are pseudo-values, and cross sectional approaches looking at who has and hasn't had the event by the time point of interest. Both must be weighted using IPCW's and IPACW's (specifically the product of these, the combined weights). We demonstrate approaches to estimate mean calibration, and produce smoothed calibration curves.

# 1.2 Proposed methodology

We first introduce methodology than can be used to estimate calibration in a validation cohort with right censoring, using IPCWs.

## 1.2.1 Mean calibration

The first suggestion is to use a Horvitz-Thompson Estimator to estimate the mean observed risk, and compare this with mean predicted risk. An outcome indicator is defined for who has $(Y_i(t_{eval}) = 1)$ and hasn't $(Y_i(t_{eval}) = 0)$ had the event by $t_{eval}$. The indicator itself is multiplied by the weights, and then we take the mean.

$$\widehat{R}_{HT}(t_{eval}) = \frac{1}{N}\sum_{i=1}^{N} w_i Y_i(t_{eval}) = \frac{\sum_{i=1}^{N} w_i Y_i(t_{eval})}{N}.$$

The second suggestion is to use a weighted mean estimator in the subset of uncensored individuals at 10 years. This is a Hajek-style estimator. This includes those who have had the event prior to 10 years, or who have not had the event and are uncensored at 10 years. Rather than multiplying the outcome indicator $Y_i(t_{eval})$ by the weight and taking the mean across the whole validation cohort (Horvitz-Thompson), we take a weighted mean in the subset of uncensored individuals $U$:

$$\widehat{R}_{Hajek}(t_{eval}) = \frac{\sum_{i \in U} w_i Y_i(t_{eval})}{\sum_{i \in U} w_i}.$$

The difference between the Hajek style and Horvitz-Thompson estimator, is that the denominator when calculating the mean for the Hajek style estimator is 1/sum(weights), rather than 1/n. More generally, the Hajek style estimator is a stabilised version of the Horvitz-Thompson estimator.

A third approach is the use of weighted pseudo-values. See (Overgaard, Parner, and Pedersen 2019; Binder, Gerds, and Andersen 2014). The mean of the weighted pseudo-values would be compared against the mean predicted risk.

## 1.2.2 Moderate calibration

For the first approach, we extended the Horvitz-Thompson estimator approach for mean calibration to produce calibration curves. We multiply the outcome indicator $(Y_{eval})$ by the weights, and then apply calibration methods that would be used for a linear model (regression weighted outcome on predictions and then estimate fitted values as a function of the predictions). This is done within the entire validation cohort (not just those that are uncensored at $t_{eval}$). We refer to this as smoothed HT approach.

For the second approach, we look in the cohort of individuals uncensored at time $t_{eval}$, and apply methods for producing calibration curves for a binary logistic regression model, with the addition of IPCWs. We refer to this as the BLR-IPCW approach.

The third approach uses the weighted pseudo-values, again appling calibration methods that would be used for a linear model (weighted pseudo-values regressed on predictions and then estimate fitted values as a function of the predictions).

We also consider the use of proportional hazards regression to estimate the calibration curve, following the work of (Austin, Harrell, and Klaveren 2020). Given that this approach regresses the observed survival outcome on the predicted risks using a cox proportional hazards model, the implementation with respect to the IP weights is less clear. This is therefore implemented as a comparator, rather than a proposed analysis when working in an artificially censored cohort.

Finally, I will try and use the PH regression approach of (Austin, Harrell, and Klaveren 2020). Given that this approach regresses the observed survival outcome on the predicted risks using a cox proportional hazards model, the implementation with respect to the weights is less clear. I believe the correct thing to do would be to use the IPACWs, rather than the product of the IPCWs and IPACWs, however I am unsure. Furthermore, these weights should be time-varying.

We report the calibration plot and integrated calibration index (mean magnitude of bias of the calibration curve), E50 (median magnitude of bias), and E90 (90th percentile of the magnitude of bias) for each method.

### 1.2.3 Extension to assess validation in artificially censored cohorts

All these methods can be extended to the setting validating a model in a validation cohort with artificial censoring, in order to evaluate the performance of a prediction under intervention model.

The IPCWs are estimated by fitting a model in the non-artificially censored cohort, to predict the risk of being censored. The IPACWs are estimated by fitting a model in the artificially censored cohort, to predict the risk of being artificially censored. Both weights are calculated by evaluating the probability of being uncensored at $min\{t_{eval}, t_{event}, t_{censored}, t_{artifcensored}\}$. The combined weights are defined as $IPCW * IPACW$, as detailed by (Keogh and Van Geloven 2024).

Note, it may feel natural to estimate the IPCWs at $mint_{eval}, t_{event}, t_{censored}$, which is the weight that would be estimated in a cohort without artificial censoring. I have tried this and this does not impact results. This minimum of these only changes for individuals who are artificially censored, and these individuals will always contribute zero to the outcome model, and $0 * weight = 0$. Therefore either is fine.

# 2 Simulation

We implement a simulation to ensure application of the methodology of (Keogh and Van Geloven 2024) is being applied appropriately, given the metrics for evaluating model performance, data and setting is different from the referenced work.

# 2.1 Simulation setup

This simulation now applies two censoring mechanisms. The first reflects the regular censoring mechanisms, and the second represents the artificial censoring mechanism applied in the K&VG work. Only the artificial censoring mechanism is dependent with the outcome. This is to ensure that when applying non time-varying weights (for the IPACWs) to the Kaplan-Meier estimator, we have no bias induced by the regular censoring mechanism. In section 3, we have a simulation where both the censoring mechanisms are dependent.

We simulate survival times, censoring times, and artificial censoring times. All have exponential hazard. We simulate a cohort of size $N = 500000$.

The regular censoring time and survival time share a single predictor (z) with a log-hazard of 1, proportional hazards. The artificial censoring time and survival time share a single predictor (x) with a log-hazard of 1, proportional hazards. This means both censoring mechanisms are dependent with the outcome. A perfectly specified prediction model is developed in the non-artificially censored cohort. Validation is then assessed in the artificially censored cohort. Weights are estimated following the process defined in the previous section, and compared with weights derived directly from the data generating mechanism, to ensure they are correct. The weights derived directly from the data generating mechanism are used when assessing validation.

# 2.2 Simulation results

We know the prediction model is perfectly specified and risks are correct. The assessment of calibration should therefore indicate near perfect calibration.

## 2.2.1 Mean calibration

Horvitz-Thompson (HT) Estimator of observed risk indicates the model is well calibrated:

```
## Observed HT: 0.418
##  Predicted: 0.413
##  Predicted/observed ratio: 0.987
```

The weighted mean estimator of observed risk in subset of uncensored individuals (Hajek style) also indicates the model is well calibrated:

```
## Observed Hajek: 0.417
##  Predicted: 0.413
##  Predicted/observed ratio: 0.991
```

The weighted pseudo-value estimator of observed risk also indicates the model is well calibrated:

## 2.2.2 Moderate calibration

Calibration plot and ICI for the smoothed-HT approach indicates the model is well calibrated:

```
## ICI: 0.005778362
```

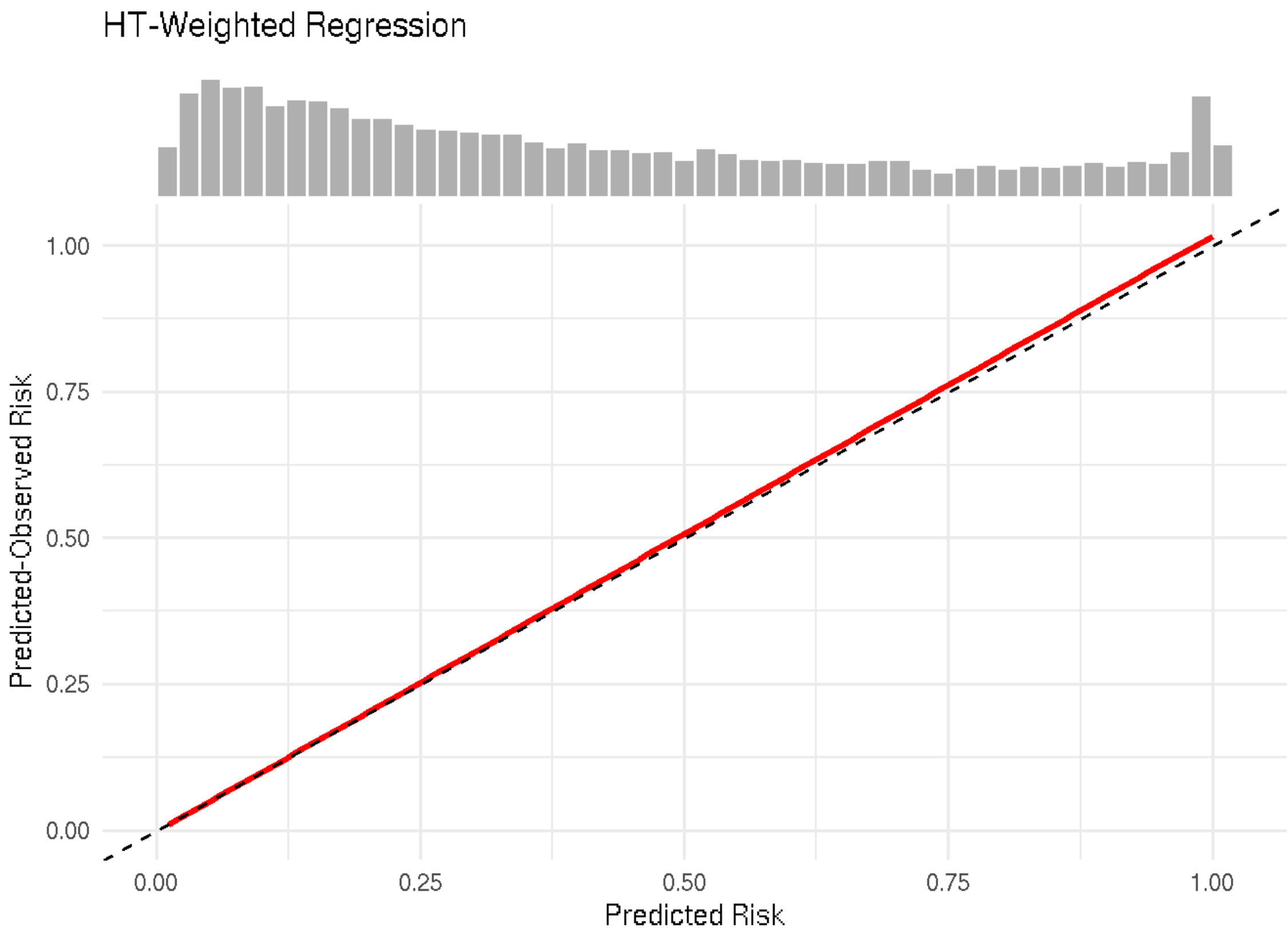

Calibration plot and ICI for the BLR-IPCW approach indicates the model is well calibrated:

```
## ICI: 0.00316427
```

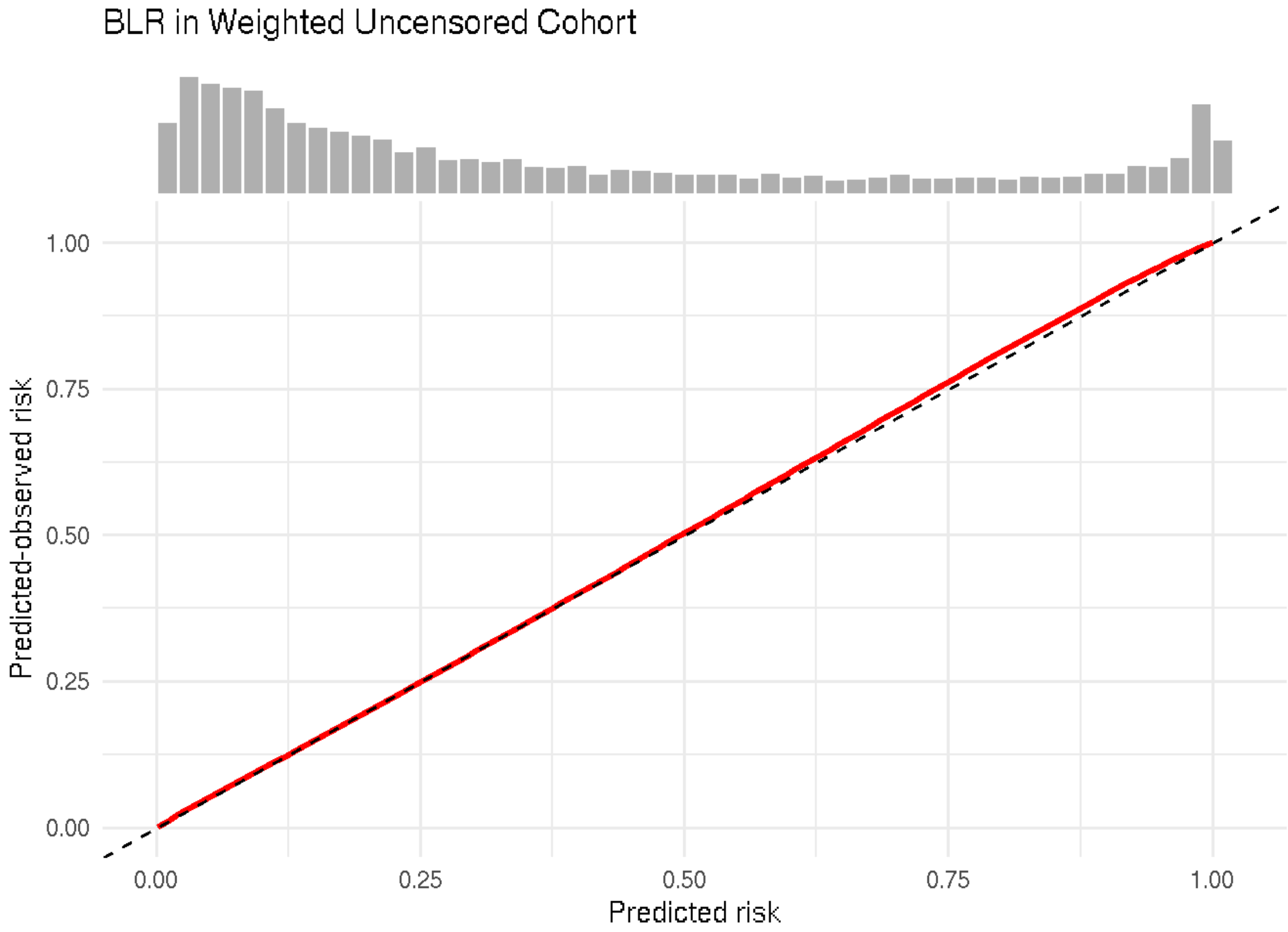

Calibration plot and ICI for the weighted pseudo-value approach indicates the model is well calibrated:

```
## ICI: 0.0058583
```

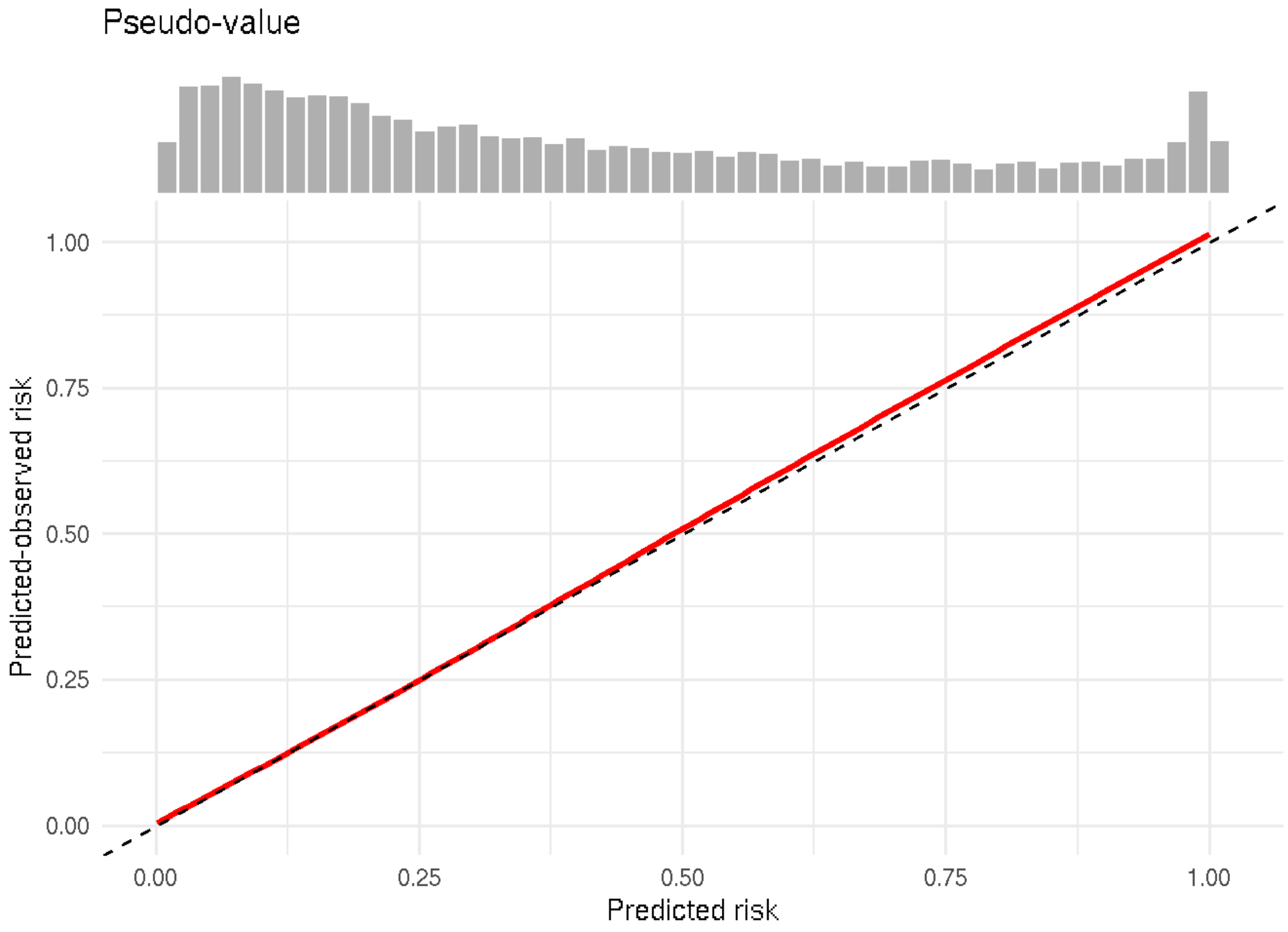

Finally, we will use the proportional hazards regression approach for assessing calibration.(Austin, Harrell, and Klaveren 2020) Given that this approach regresses the observed survival outcome on the predicted risks using a Cox proportional hazards model, the implementation with respect to the weights is less clear. I believe the correct thing to do would be to use the IPACWs, rather than the product of the IPACW and IPCW weights, given the Cox already makes assumptions about the censoring mechanism, however I am unsure about this. Furthermore, these weights should be time-varying.

We implement the method using combined weights (not time-varying), and as expected, this is not appropriate:

```
## ICI: 0.115728
```

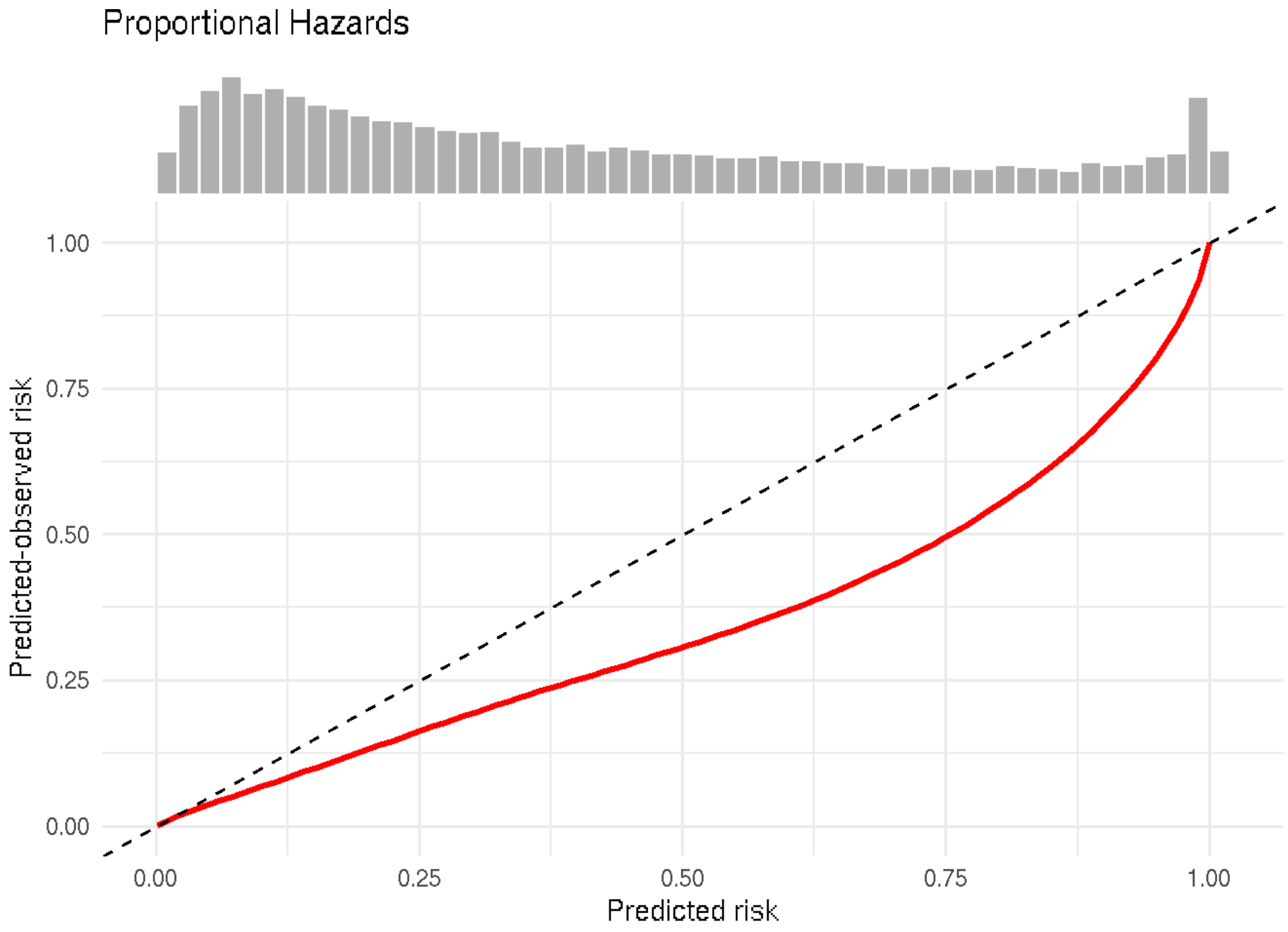

We also implement using just the IPACWs (not time-varying), and find similar results:

```
## ICI: 0.06777245
```

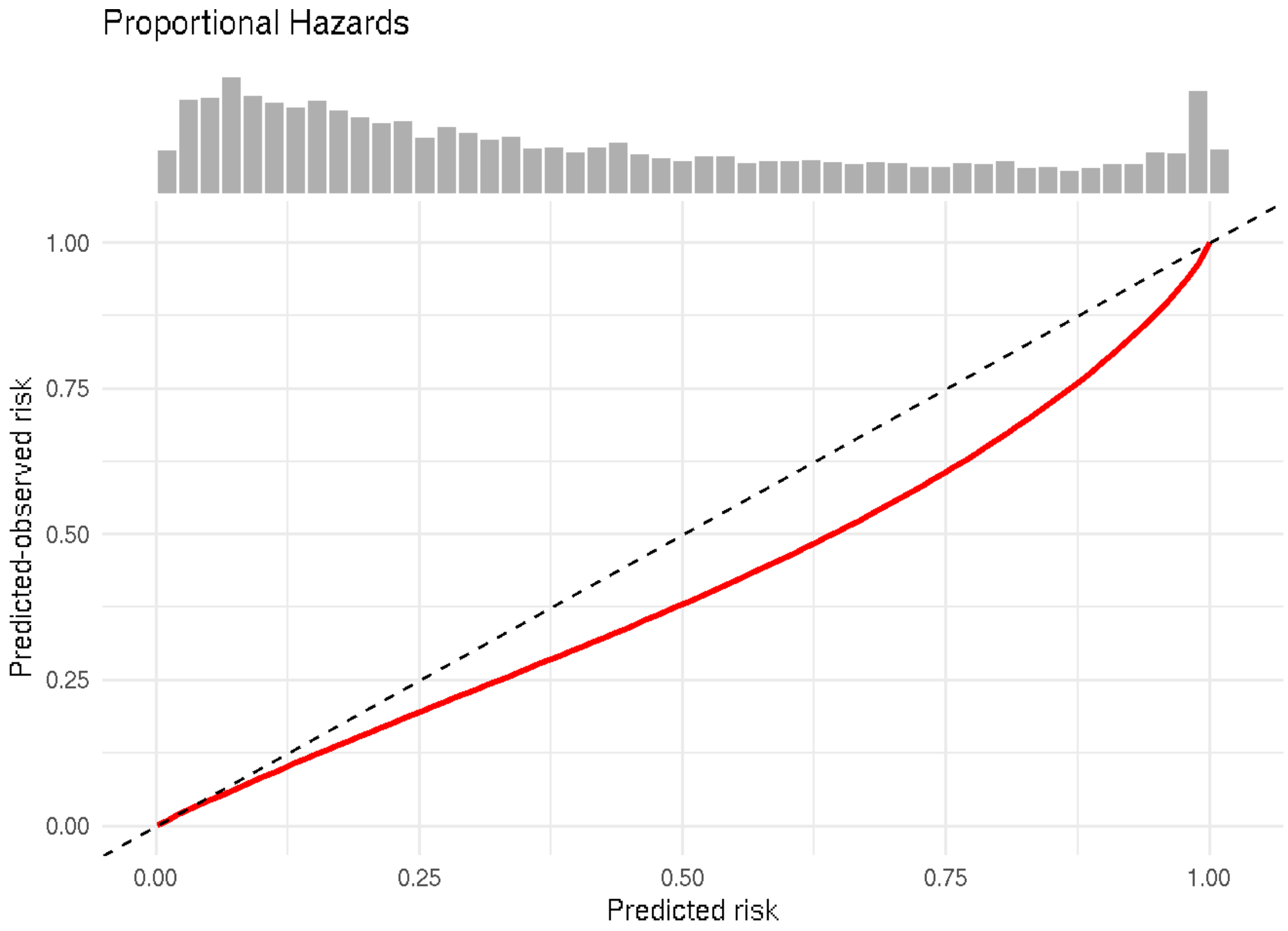

Interestingly though, if we use no weights at all, this approach indicates the model is well calibrated:

```
## ICI: 0.001316977
```

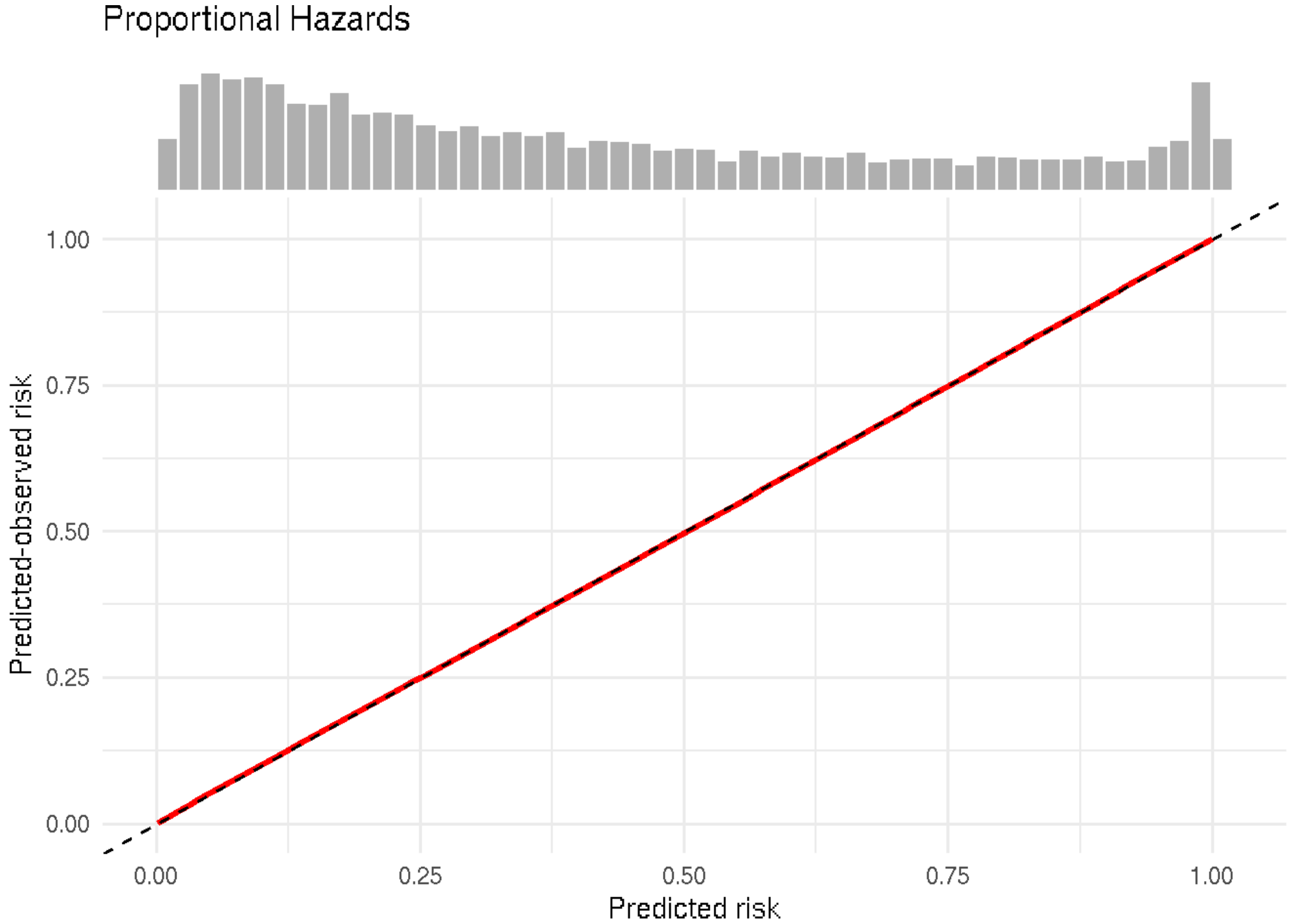

We believe this is because the prediction model is perfectly specified, and therefore the censoring mechanism (combined normal and artificial) is conditionally independent from the outcome in the validation dataset, given the predicted risks. When we start weighting the cohort, this no longer holds. If the prediction model was not perfectly specified, this would also violate this assumption, and the assessment of calibration would be biased. We will therefore avoid this approach in the real data examples.

## 2.3 A note on time-varying confounding

We have ignored time-varying confounding in these simulations. The goal of these simulations was to show how to assess calibration (mean or moderate) using approaches avoiding the Kaplan-Meier estimator. We assumed that we can accurately estimate or know the weights. If there is time-dependent confounding between treatment initiation (artificial censoring) and the outcome, this would need to be reflected in the process for estimating the weights. This could either be a pooled logistic regression model, or an interval censored survival model. However, this does induce extra considerations, as the IPCWs and IPACWs are estimated at $min\{t_{eval}, t_{event}, t_{censored}, t_{artifcensored}\}$.

Currently, our models for estimating the weights (cox proportional hazard models) allow us to estimate probabilities of censoring at these exact times. If we use a pooled logistic regression models in order to estimate the weights, then we will lose the ability to do this. This is based on the assumption that the pooled logistic regression model would be implemented within intervals of say 30/90/180/365 days. While a pooled logistic regression model could be implemented with 1 day intervals, again this introduces a large computational burden, and potentially insufficient sample sizes/event rates for the individual logistic regression models.

An alternative is to use interval censored survival models to estimate the weights. This would allow us to both adjustment for time-varying confounding (at say 1 year intervals), and still estimate weights at the exact times of $min\{t_{eval}, t_{event}, t_{censored}, t_{artifcensored}\}$. Once estimated, these weights (estimated with adjustment for time-varying confounding), can then be used in the same way as the weights have been throughout this manuscript, in conjunction with the Hajek style/Horvitz Thompson estimators of mean calibration, and the BLR-IPCW, smoothed-HT and weighted pseudo-values approaches for estimating calibration curves. We prefer these approaches over the Kaplan-Meier approach for computational reasons, and more explicit handling of the non-artificial censoring mechanism.

# 3 Validation of the two-component model using inverse probability of artificial censoring weights

# 3.1 Calibration results

We apply IPACW weights as soon as anyone deviates from their treatment at baseline, with respect to statin use, or antihypertensive use. IPCWs are estimated by modelling the entire validation cohort together. IPACWs are estimated using stratified models for those off treatment at baseline, antihypertensives at baseline, statins at baseline, or on both treatments at baseline. Predictors are the same for each model, but due to smaller sample sizes for the latter three cohorts, we do not include interaction terms between age and other predictors in these models. Sample size criteria of (Riley et al. 2024) are adhered to, to minimise risk of overfitting. This is all implemented in program p1_create_validation_data_and_assess_calibration.R. These are the results from the three methods.

**Female cohort:**

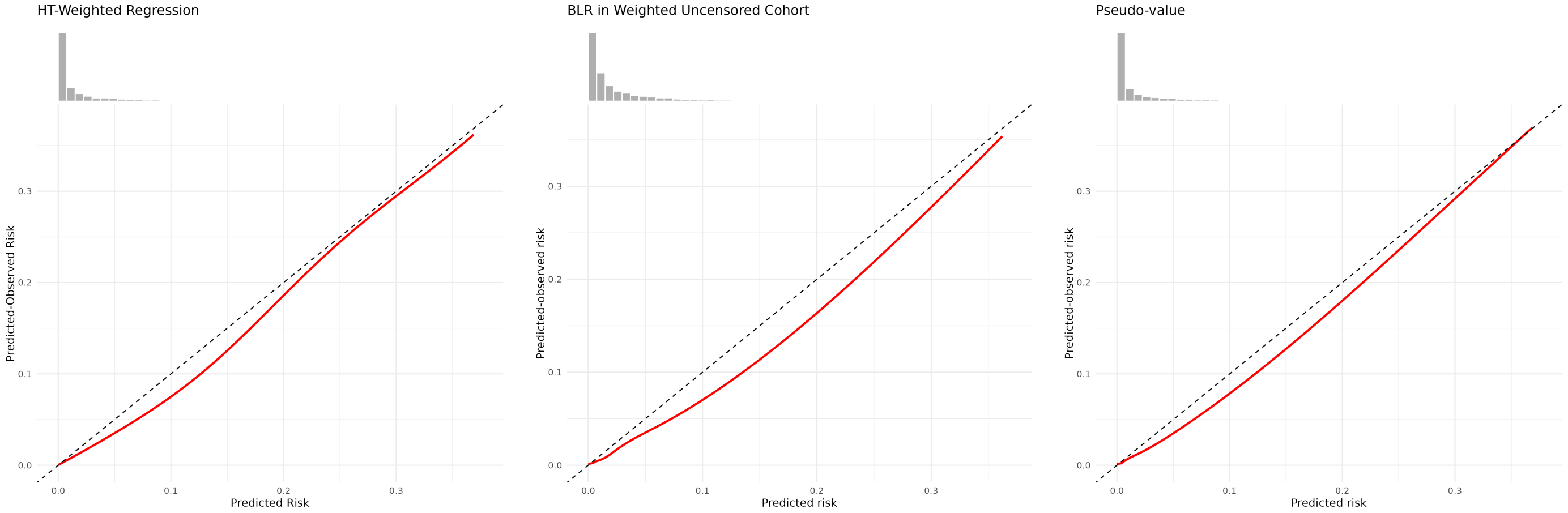

Smoothed-HT Hajek-style Weighted pseudo-value

**Male cohort:**

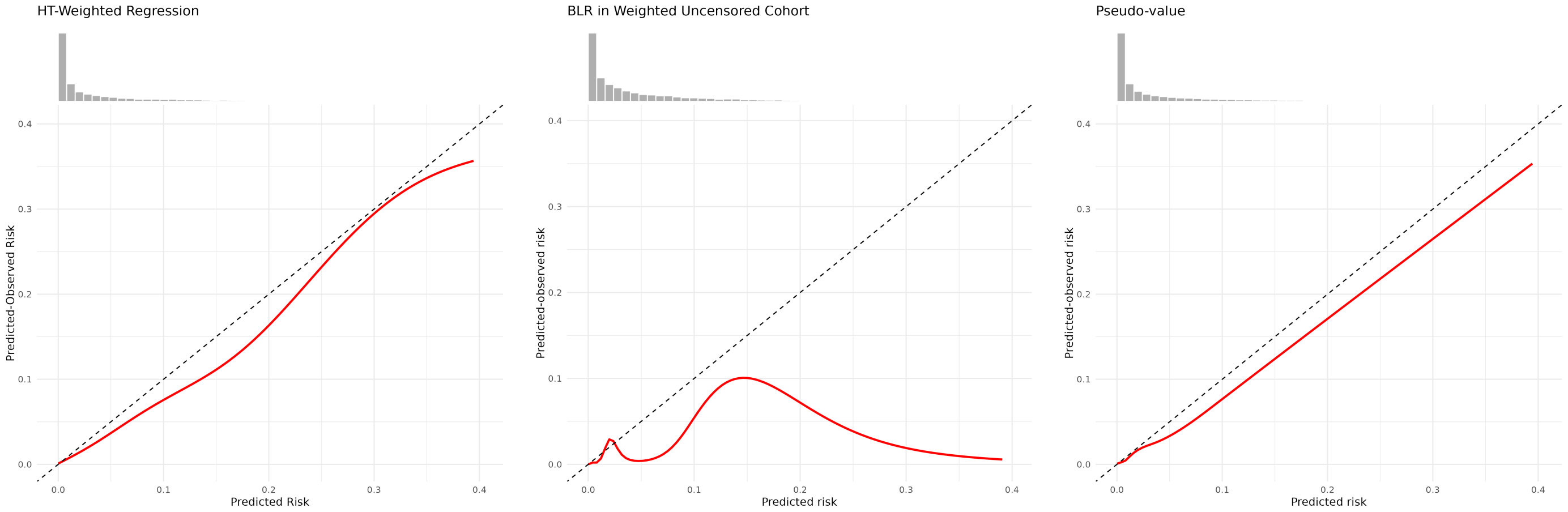

Smoothed-HT Hajek-style Weighted pseudo-value

Table of calibration metrics:

| Cohort | Method | ICI | E50 | E90 |
|---|---|---|---|---|
| Female | BLR | 0.009 | 0.006 | 0.028 |
| Female | Smoothed HT | 0.005 | 0.002 | 0.019 |
| Female | Pseudo-value | 0.006 | 0.002 | 0.019 |
| Male | BLR | 0.039 | 0.008 | 0.068 |
| Male | Smoothed HT | 0.010 | 0.004 | 0.030 |
| Male | Pseudo-value | 0.010 | 0.003 | 0.027 |

We find the model to be miscalibrated for all methods in the same direction for both female and male cohorts. There appears to be a clear issue with the Hajek-style estimator in the male cohort, but the rest present a similar pattern.

It's impossible to say whether the model is really miscalibrated, or there is just a difference between the average conditional treatment effect taken from the literature, vs what can be recovered from the data using the IP weights. The weighting could be subject to a number of assumption violations. In particular, the positivity assumption (we need support to effectively estimate the effect of sustaining treatment for 10 years), as well as conditional exchangeability between the artificial censoring mechanism and the outcome.

A small exploration/summary of key variables in the data follows.

# 3.2 Exploration of the data

We inspect the distribution of the weights. There are some extremely large combined weights (IPCW*IPACW) indicating positivity violations. This is driven primarily by the artificial censoring mechanism, indicating we do not have support for estimating calibration of predictions under sustained treatment for 10 years.

| Cohort | Weights | 50% | 90% | 95% | 97.5% | 99% | 99.9% | 99.99% | 100% |
|---|---|---|---|---|---|---|---|---|---|
| Female | IPCW | 1.80 | 4.47 | 6.23 | 8.11 | 10.52 | 19.10 | 67.29 | 1390.62 |
| Female | IPACW | 1.08 | 1.54 | 1.92 | 2.37 | 3.11 | 9.75 | 77.83 | 21576.13 |
| Female | Combined | 2.22 | 5.61 | 7.71 | 9.96 | 12.99 | 37.05 | 330.72 | 49144.48 |
| Male | IPCW | 1.84 | 4.36 | 5.44 | 6.29 | 7.48 | 12.21 | 22.92 | 271.75 |
| Male | IPACW | 1.05 | 1.48 | 1.82 | 2.24 | 3.01 | 8.45 | 69.48 | 1545462.00 |
| Male | Combined | 2.19 | 5.20 | 6.28 | 7.40 | 9.66 | 26.71 | 224.09 | 3346185.96 |

We next present a summary of how the number of outcome events and follow-up time changes after artificial censoring is applied. Variables are presented in the original cohort (orig), artificially censored (ac) cohorts. There are also columns for the ratio of number of events and follow-up time in the artificially censored cohort, compared to the original cohort (i.e. what proportion of follow-up time and number of events remains in artificially censored cohort). This is stratified by treatment status at baseline:

| Cohort | on statins | on antihyp | Prop events orig | N events orig | Followup time orig | Prop events ac | N events ac | Followup time ac | Followup time ratio (ac/orig) | N events ratio (ac/orig) |
|---|---|---|---|---|---|---|---|---|---|---|
| Male | 0 | 0 | 0.0311914 | 28679 | 5778.51291 | 0.0167948 | 15442 | 5069.95060 | 0.8773798 | 0.5384428 |
| Male | 0 | 1 | 0.1581179 | 9584 | 435.73580 | 0.0811707 | 4920 | 234.89620 | 0.5390794 | 0.5133556 |
| Male | 1 | 0 | 0.1319863 | 1117 | 56.25386 | 0.0450195 | 381 | 25.63067 | 0.4556252 | 0.3410922 |
| Male | 1 | 1 | 0.1766757 | 2027 | 77.43268 | 0.1135710 | 1303 | 53.77249 | 0.6944418 | 0.6428219 |
| Female | 0 | 0 | 0.0195045 | 17622 | 5458.34513 | 0.0096548 | 8723 | 4655.77264 | 0.8529641 | 0.4950062 |
| Female | 0 | 1 | 0.1277524 | 10049 | 604.36603 | 0.0651538 | 5125 | 323.13226 | 0.5346632 | 0.5100010 |
| Female | 1 | 0 | 0.1093318 | 751 | 48.71951 | 0.0366866 | 252 | 21.27467 | 0.4366767 | 0.3355526 |
| Female | 1 | 1 | 0.1641940 | 1804 | 81.18179 | 0.0946573 | 1040 | 53.32090 | 0.6568087 | 0.5764967 |

We see that in the untreated at baseline group, that after applying artificial censoring, there is a ~50% reduction in the number of events, however only a ~15% reduction in total follow-up time/at-risk time. This is highlighting that untreated individuals who are at high risk of CVD events, are much more likely to initiate treatment than those at low risk of CVD events. This is reflected in the calibration plot, that after applying artificial censoring, the model appears to under predict, due to the large number of outcome events being artificially censored. This is logical sense, and shows statins/antihypertensives are being initiated in the right people. However, the level of this dependence is extremely strong, and it is likely there may be unmeasured confounding between the outcome and censoring mechanism, which would result in biased estimation of the weights as conditional exchangeability assumption will not hold.

Furthermore, and most likely, there could be positivity issues, indicated by large weights. We are predicting the risk of a cardiovascular event if sustaining treatment for 10 years. There are a very small percentage of individuals who sustain treatment for this length of time, and do not have an event, meaning the risk set gets very small. We present data on the proportion of individuals who sustain treatment and are followed up for 10 years.

| Cohort | on statins | on antihyp | N | N sustained treatment | Proportion sustained treatment |
|---|---|---|---|---|---|
| Male | 0 | 0 | 919451 | 185817 | 0.2020956 |
| Male | 0 | 1 | 60613 | 7115 | 0.1173841 |

| Cohort | on statins | on antihyp | N | N sustained treatment | Proportion sustained treatment |
|---|---|---|---|---|---|
| Male | 1 | 0 | 8463 | 551 | 0.0651069 |
| Male | 1 | 1 | 11473 | 1827 | 0.1592434 |
| Female | 0 | 0 | 903484 | 171654 | 0.1899912 |
| Female | 0 | 1 | 78660 | 10675 | 0.1357107 |
| Female | 1 | 0 | 6869 | 456 | 0.0663852 |
| Female | 1 | 1 | 10987 | 1838 | 0.1672886 |

Truncating the weights to a maximum of 100, or 10, does not have a major impact on the calibration plots, beyond the male plot using the Hajek method now aligning with the others. We still see considerable over-prediction in both the female and male cohorts.

**Female cohort, truncated at 100:**

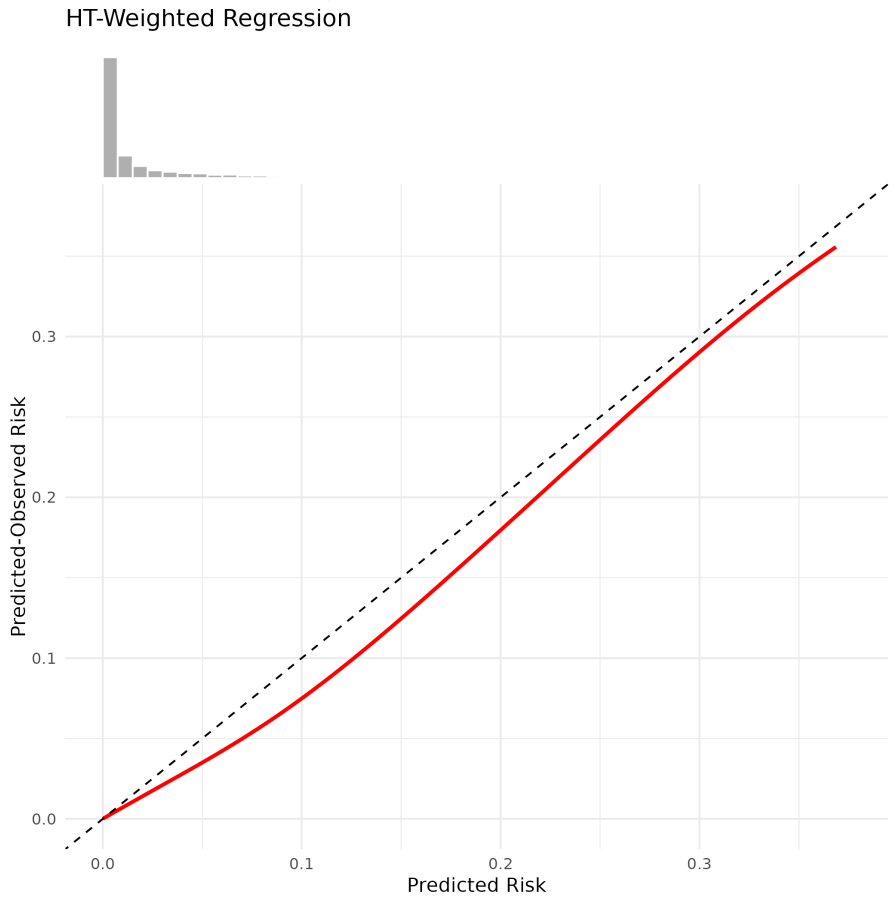

Smoothed-HT

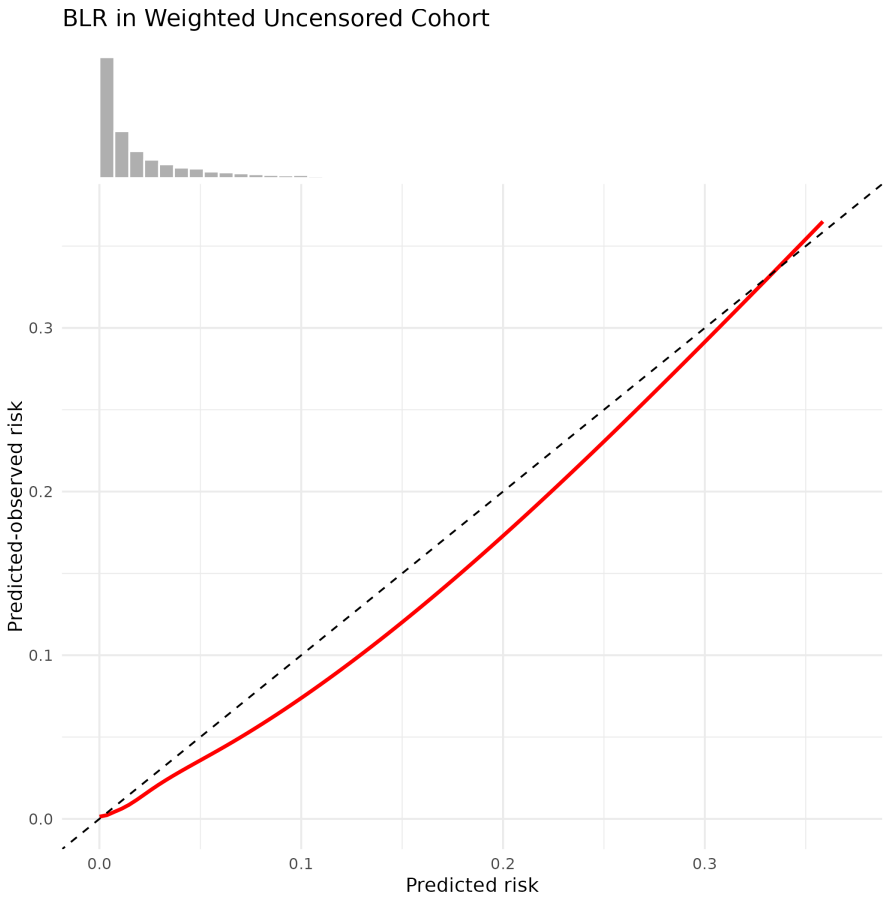

Hajek-style

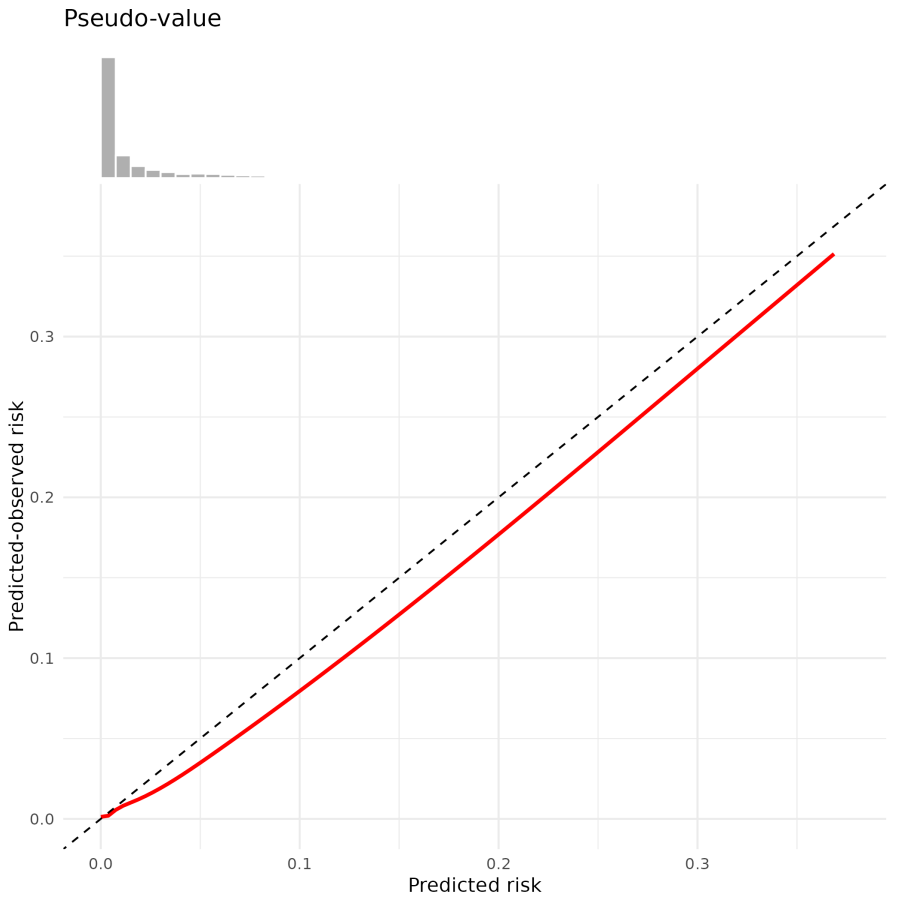

Weighted pseudo-value

**Male cohort, truncated at 100:**

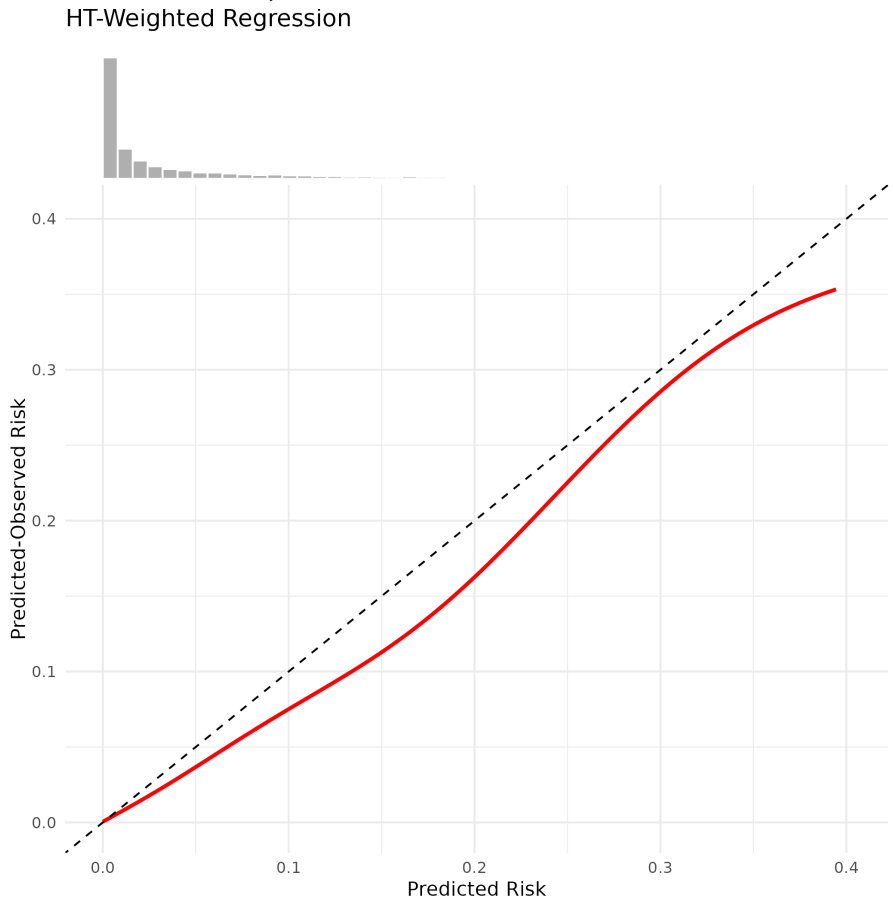

Smoothed-HT

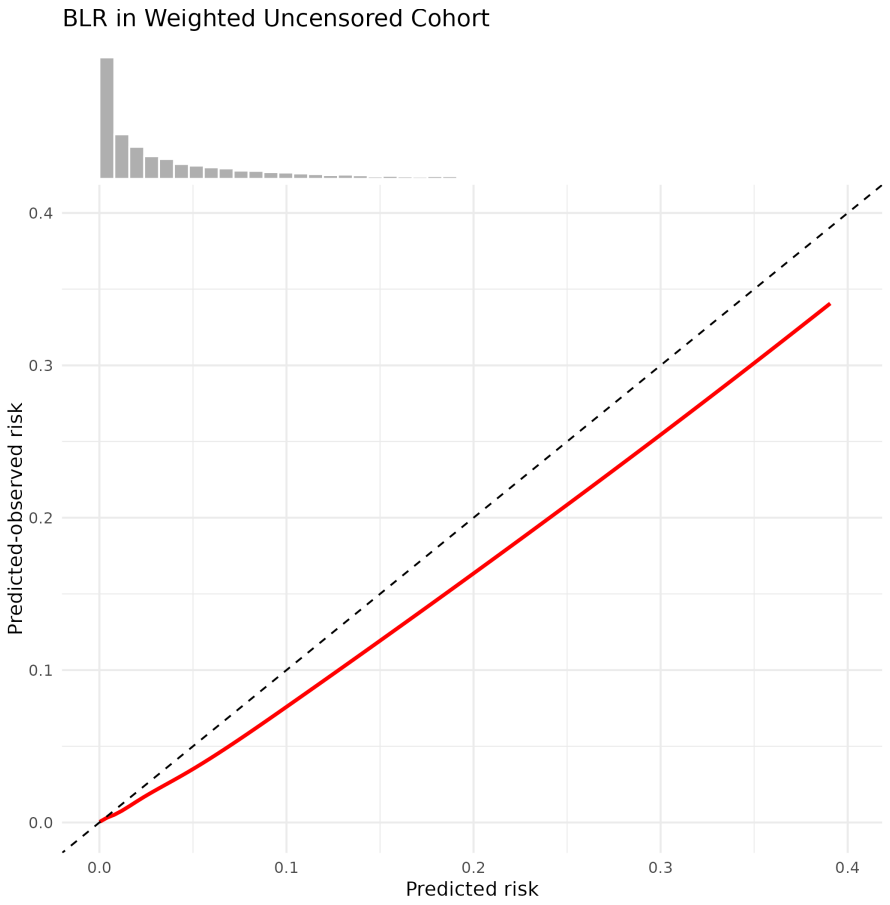

Hajek-style

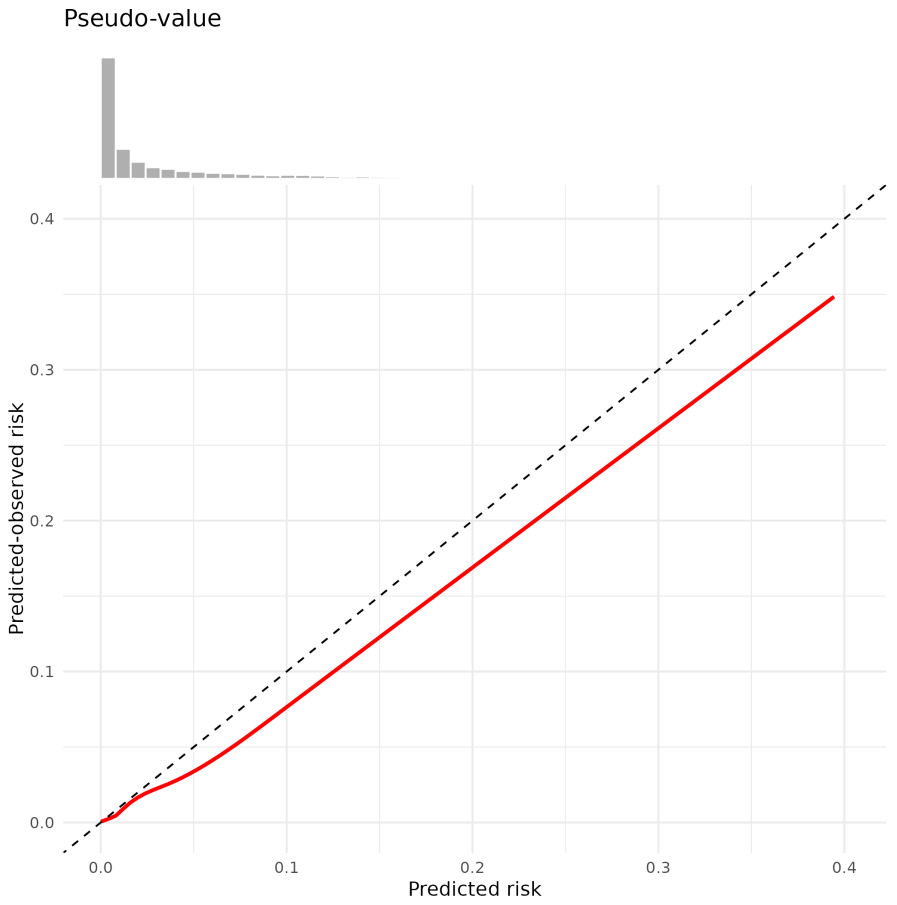

Weighted pseudo-value

**Female cohort, truncated at 10:**

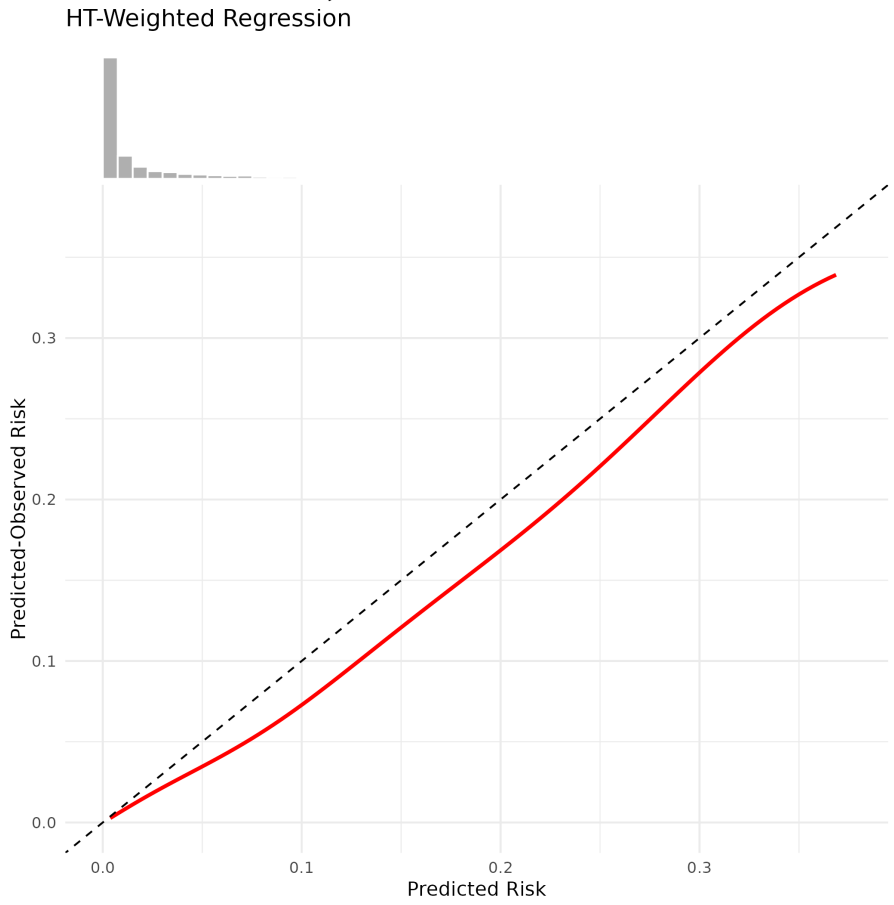

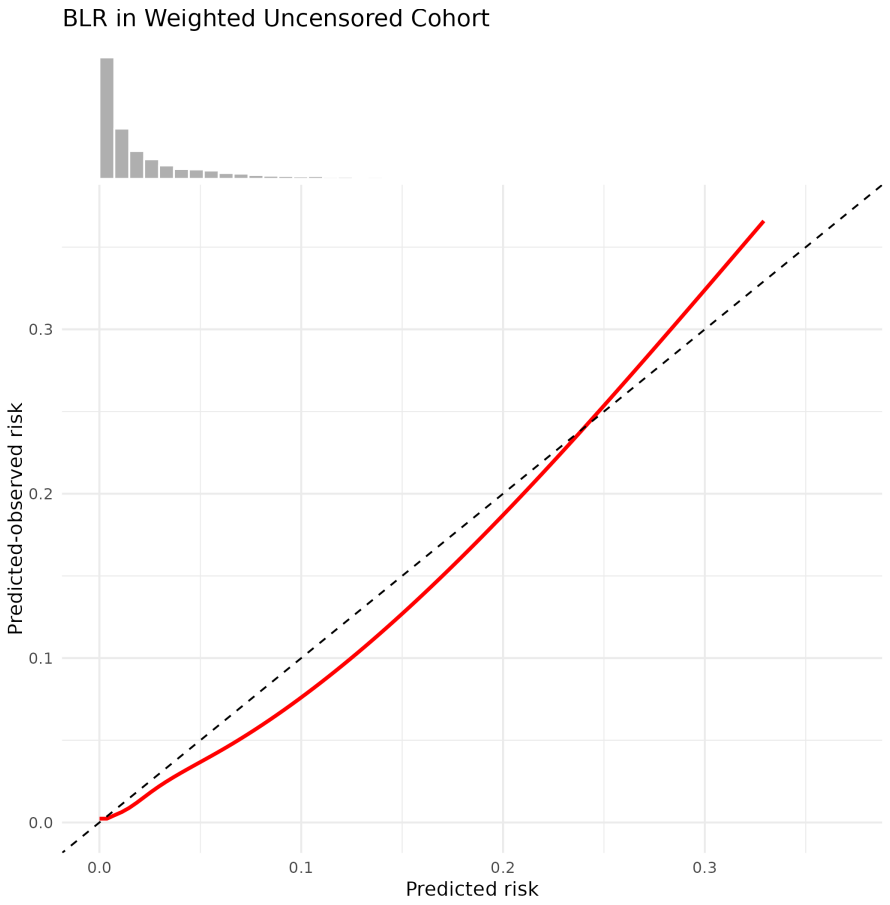

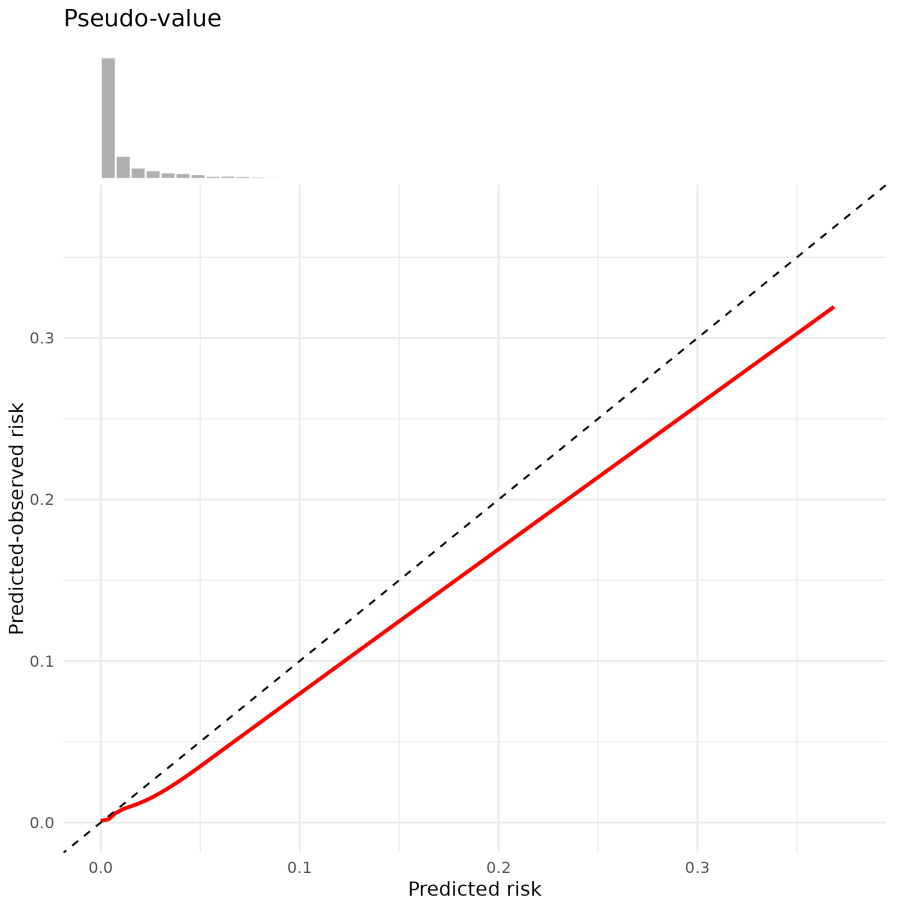

Smoothed-HT | Hajek-style | Weighted pseudo-value

**Male cohort, truncated at 10:**

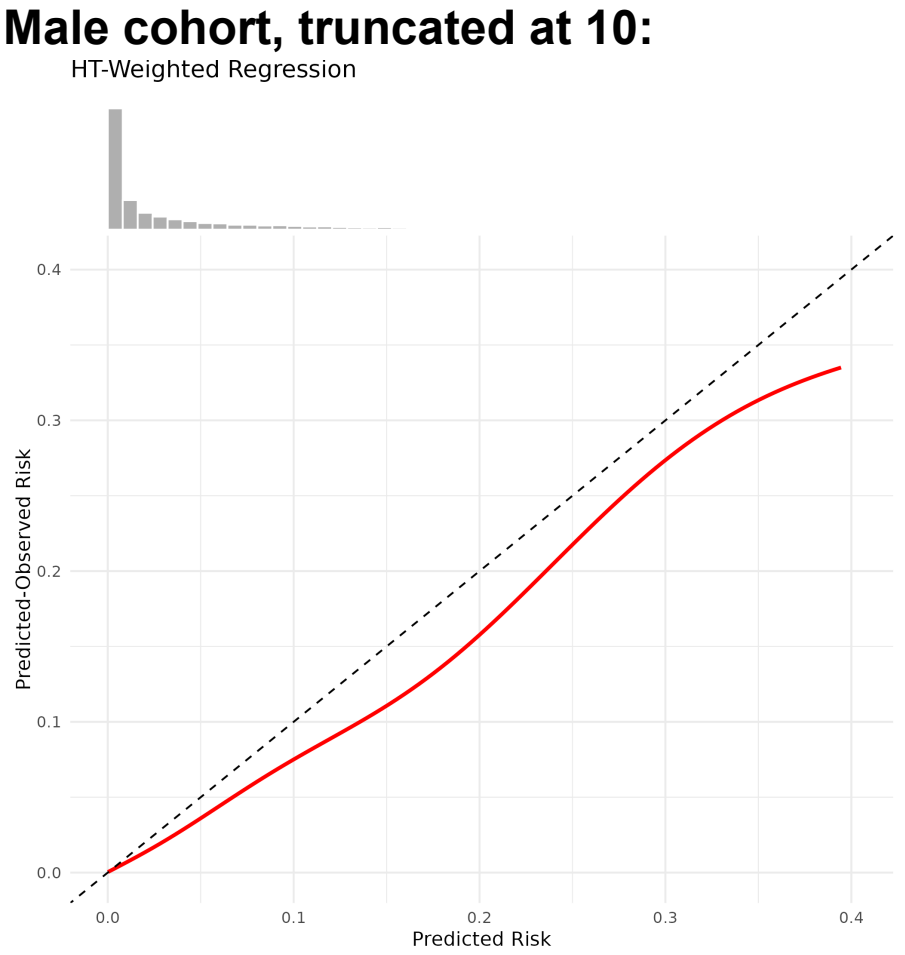

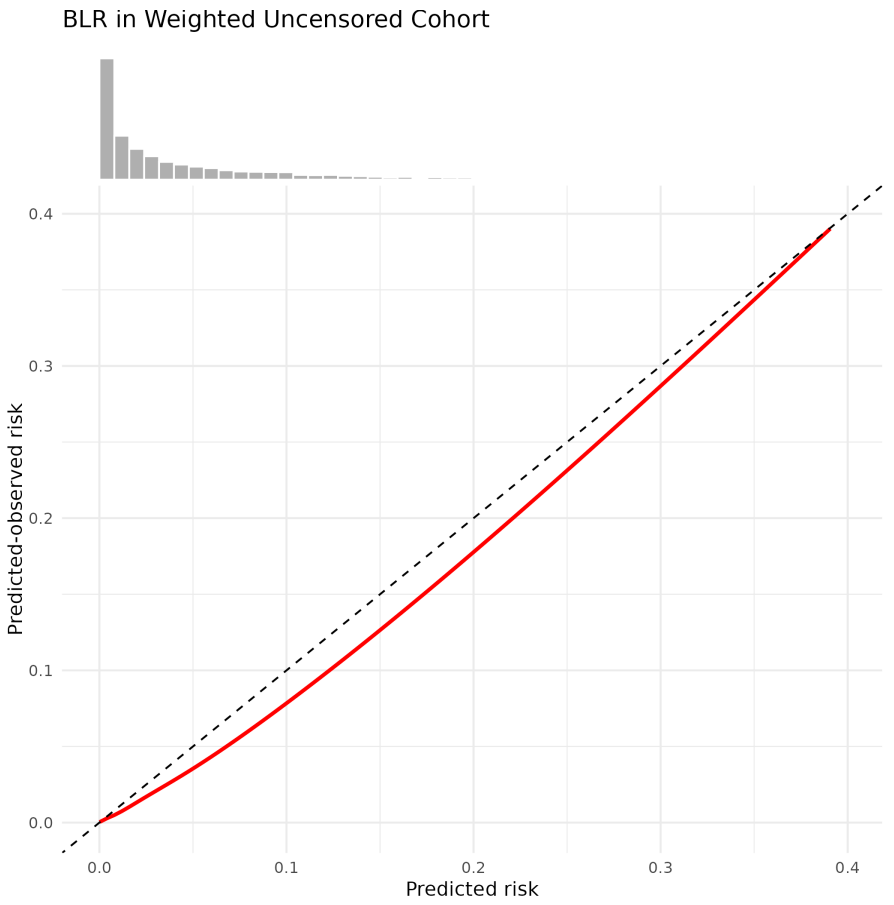

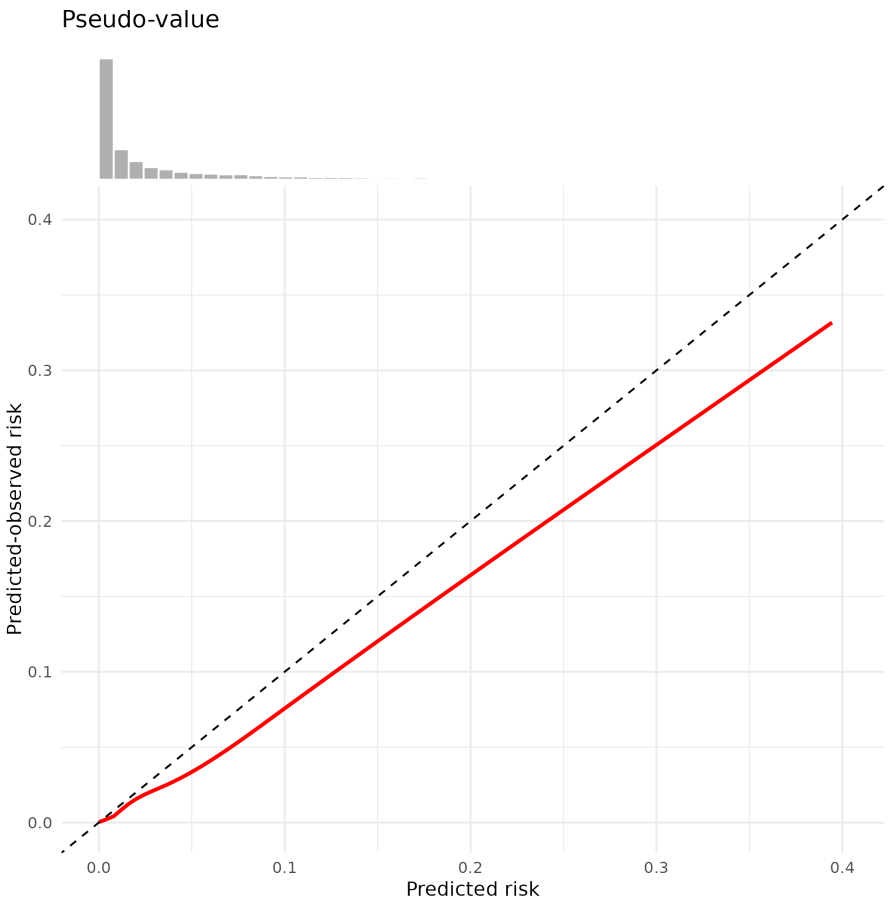

Smoothed-HT | Hajek-style | Weighted pseudo-value

To investigate further, we exploit a feature of the artificial censoring approach: in estimating observed risk under a sustained treatment strategy, it implicitly recovers the effect of that strategy through the censoring weights. If those weights were adequate, it would be possible to recover the effects of statin and antihypertensive use, obtaining estimates broadly consistent with the trial evidence. If they are not, the same weights cannot support a calibration estimate either. We therefore attempt this explicitly, at the ten-year horizon and at shorter intervals where positivity violations induced by artificial censoring are less severe. The analysis is reported in Appendix A. The results indicate that there is not sufficient support to estimate the effect of sustaining statin or antihypertensive treatment for 10 years. Relative risk estimates indicate statins and antihypertensive use leading to higher risk of cardiovascular outcomes, in the opposite direction to the trial evidence.

# 4 Summary

In summary, evaluation of calibration using the artificial censoring approach of Keogh and van Geloven indicates a miscalibrated model. There are two possible explanations: either the model is miscalibrated, or the assumptions underpinning the analysis are violated and the evaluation of calibration is not correct. The diagnostics we have examined point to clear violations in positivity, and the recovered causal effects are qualitatively reversed compared with the trial estimates. Because the assumptions required for this evaluation are not met, the exercise cannot inform us about the calibration of the model.

# 5 References

# 6 Appendix

This section summarises the results of an analysis to estimate the causal effect of sustained statin and antihypertensive treatment on cardiovascular disease risk. Analyses were conducted separately for statins and antihypertensives. We use artificial censoring in combination with IP weighting to estimate the effects. The process of artificial censoring and estimation of weights (IPCW and IPACW) is similar to the analysis used to estimate the calibration (Keogh and Van Geloven 2024). Given the effect of statins and antihypertensives are broadly known, this allows us to compare the estimates to a ground truth, unlike when assessing the calibration of the prediction models. This provides further evidence about the validity of estimation and usage of such weights in this cohort.

The entire cohort (development + validation) were combined. We aimed to estimate the risk under no-intervention (p0) and risk under sustained treatment (p1). The relative risk reduction is estimated as the ratio of these (p1/p0). Values below 1 indicate a protective effect of treatment. We therefore created two cohorts. One where individuals were artificially censored as soon as they initiated treatment (instantly censoring those on treatment at baseline), and a second where individuals were artificially censored as soon as they stopped receiving treatment (instantly censoring those off treatment at baseline). We estimate the risk at 1/3/5/10 years follow-up in these two cohorts using weights via a Horvitz-Thompson style estimator, and a Hajek-style estimator.

The weights used to estimate p0 and p1 were combined weights IPCW × IPACW × IPTW. The IPCWs and IPACWs were estimated as defined previously. The IPTWs are inverse probability of treatment weights when estimating p1, or inverse probability of untreated if calculating p0, and are the inverse of the probability of the receiving treatment which individuals were on at baseline. Under these weights, we can create pseudo-populations as if the whole population had either been treated, or untreated, over the follow-up horizon of interest.

Treatment use was defined in the same way as in the main analysis with a 180 wash-out period. We used all the same predictors for estimating the weights and included splines on continuous variables. Two weighting models were used, one with no interaction terms, and another where all predictors were interacted with age.

As a result of finding highly unstable relative risk estimates and some really extreme weights/small effective sample sizes, indicative of violations of the positivity assumption, we also implement these analyses in overlap cohorts. This is where the cohort is restricted to individuals who had a reasonable probability of receiving either treatment (our threshold is 1%). We create two overlap cohorts. The first ( `Overlap cohort 1` ) restricts to individuals with sufficient probability of receiving either treatment at baseline (> 1%). The second ( `Overlap cohort 2` ) applies this restriction, but also restricts to those with probability of sustaining the treatment they did receive for the follow-up interval, i.e. probability of not being artificially censored over prediction interval > 1%; and probability of being observed for follow-up interval, i.e. not being censored from standard censoring mechanism > 1%. While these analyses alter the target population, we assume the effect of statins and antihypertensives to be reasonably homogeneous, the estimand should remain similar, but this should reduce violations of the positivity assumption, leading to more stable estimates.

We prefer estimation in overlap cohorts as opposed to truncating the weights. Extreme weights indicate there is little information in the data about this treatment strategy for the highly weighted individuals. Truncation will bias the estimator in order to improve variance, but the issue of positivity remains. Overlap restriction removes observations with little or no support before estimation, dealing with the cause of extreme weights, whereas truncation addresses the consequences of extreme weights. Both do have value, but we have opted to create overlap cohorts.

# 6.1 Relative Risk Estimates

Green RR values indicate a protective treatment effect (RR < 1). Red values indicate the estimate is in the wrong direction, which may signal positivity violations or other issues (see diagnostics sections below).

The **effective sample size (ESS)** for each treatment group is also included in the Table. This is calculated using the Kish formula:

$$\text{ESS} = \frac{(\sum w_i)^2}{\sum w_i^2}$$

The ESS equals N when all weights are equal, and falls below N when weights are variable. A low ESS (expressed as a % of N) indicates that a small number of highly-weighted individuals are driving the analysis, which undermines the reliability of the estimates regardless of the nominal sample size. The table below shows the ESS at the final combined weighting stage (IPCW × IPACW × IPTW) alongside the actual sample size, for both cohorts. ESS% values below 10% are shown in red; below 25% in orange.

Relative risk estimates and ESS by cohort, medication, gender, follow-up, and model

| Cohort | Med | Gender | Follow-up (years) | Age interactions | p0_hajek | p1_hajek | RR_hajek | p0_ht | p1_ht | RR_ht | N (untreated) | ESS (untre |
|---|---|---|---|---|---|---|---|---|---|---|---|---|

| Cohort | Med | Gender | Follow-up (years) | Age interactions | p0_hajek | p1_hajek | RR_hajek | p0_ht | p1_ht | RR_ht | N (untreated) | ESS (untre |
|---|---|---|---|---|---|---|---|---|---|---|---|---|
| Entire cohort | statins | Female | 1 | No | 0.0031 | 0.0057 | **1.797** | 0.0032 | 0.0056 | **1.775** | 9545467 | 73388 |
| | | | 1 | Yes | 0.0030 | 0.0052 | **1.748** | 0.0030 | 0.0054 | **1.827** | 9545467 | 18007 |
| | | | 3 | No | 0.0093 | 0.0146 | **1.566** | 0.0098 | 0.0158 | **1.611** | 9545467 | 1828 |
| | | | 3 | Yes | 0.0091 | 0.0146 | **1.611** | 0.0091 | 0.0167 | **1.828** | 9545467 | 99747 |
| | | | 5 | No | 0.0141 | 0.0221 | **1.568** | 0.0175 | 0.0267 | **1.526** | 9545467 | 121 |
| | | | 5 | Yes | 0.0146 | 0.0200 | **1.371** | 0.0159 | 0.0258 | **1.624** | 9545467 | 972 |
| | | | 10 | No | 0.0006 | 0.0185 | **28.531** | 0.0678 | 0.0525 | **0.775** | 9545467 | 1 |
| | | | 10 | Yes | 0.0124 | 0.0347 | **2.795** | 0.1536 | 0.0495 | **0.322** | 9545467 | 1 |
| | | Male | 1 | No | 0.0031 | 0.0057 | **1.796** | 0.0032 | 0.0056 | **1.774** | 9545467 | 73413 |
| | | | 1 | Yes | 0.0030 | 0.0051 | **1.742** | 0.0030 | 0.0054 | **1.832** | 9545467 | 17894 |
| | | | 3 | No | 0.0093 | 0.0146 | **1.567** | 0.0098 | 0.0158 | **1.611** | 9545467 | 1846 |
| | | | 3 | Yes | 0.0091 | 0.0146 | **1.614** | 0.0091 | 0.0167 | **1.828** | 9545467 | 99810 |
| | | | 5 | No | 0.0141 | 0.0221 | **1.570** | 0.0175 | 0.0267 | **1.526** | 9545467 | 120 |
| | | | 5 | Yes | 0.0146 | 0.0200 | **1.369** | 0.0159 | 0.0258 | **1.622** | 9545467 | 1090 |
| | | | 10 | No | 0.0007 | 0.0185 | **27.930** | 0.0684 | 0.0525 | **0.768** | 9545467 | 1 |
| | | | 10 | Yes | 0.0125 | 0.0351 | **2.798** | 0.1533 | 0.0500 | **0.326** | 9545467 | 1 |
| | Antihyp | Female | 1 | No | 0.0040 | 0.0073 | **1.803** | 0.0041 | 0.0055 | **1.323** | 9015065 | 11295 |
| | | | 1 | Yes | 0.0032 | 0.0064 | **1.989** | 0.0032 | 0.0052 | **1.594** | 9015065 | 11211 |
| | | | 3 | No | 0.0122 | 0.0145 | **1.188** | 0.0128 | 0.0155 | **1.214** | 9015065 | 74955 |
| | | | 3 | Yes | 0.0092 | 0.0123 | **1.327** | 0.0093 | 0.0145 | **1.558** | 9015065 | 46345 |
| | | | 5 | No | 0.0219 | 0.0150 | **0.687** | 0.0283 | 0.0258 | **0.912** | 9015065 | 164 |
| | | | 5 | Yes | 0.0176 | 0.0141 | **0.801** | 0.0182 | 0.0248 | **1.362** | 9015065 | 14929 |
| | | | 10 | No | 0.0421 | 0.0067 | **0.159** | 0.1059 | 0.0618 | **0.583** | 9015065 | 1 |
| | | | 10 | Yes | 0.0213 | 0.0117 | **0.549** | 0.0427 | 0.0694 | **1.627** | 9015065 | 127 |
| | | Male | 1 | No | 0.0040 | 0.0073 | **1.810** | 0.0041 | 0.0055 | **1.327** | 9015065 | 11829 |
| | | | 1 | Yes | 0.0032 | 0.0064 | **1.994** | 0.0032 | 0.0052 | **1.592** | 9015065 | 11210 |
| | | | 3 | No | 0.0121 | 0.0145 | **1.198** | 0.0127 | 0.0155 | **1.220** | 9015065 | 68228 |
| | | | 3 | Yes | 0.0092 | 0.0123 | **1.328** | 0.0093 | 0.0145 | **1.556** | 9015065 | 58085 |
| | | | 5 | No | 0.0521 | 0.0151 | **0.289** | 0.0691 | 0.0258 | **0.373** | 9015065 | 172 |
| | | | 5 | Yes | 0.0177 | 0.0143 | **0.811** | 0.0182 | 0.0248 | **1.359** | 9015065 | 16174 |
| | | | 10 | No | 0.0581 | 0.0067 | **0.116** | 0.1491 | 0.0618 | **0.415** | 9015065 | 1 |
| | | | 10 | Yes | 0.0238 | 0.0119 | **0.501** | 0.0426 | 0.0690 | **1.618** | 9015065 | 198 |

| Cohort | Med | Gender | Follow-up (years) | Age interactions | p0_hajek | p1_hajek | RR_hajek | p0_ht | p1_ht | RR_ht | N (untreated) | ESS (untre |
|---|---|---|---|---|---|---|---|---|---|---|---|---|
| Overlap cohort 1 | statins | Female | 1 | No | 0.0094 | 0.0161 | **1.718** | 0.0095 | 0.0140 | **1.476** | 2707256 | 16888 |
| | | | 1 | Yes | 0.0096 | 0.0161 | **1.679** | 0.0097 | 0.0146 | **1.506** | 2488539 | 16149 |
| | | | 3 | No | 0.0281 | 0.0427 | **1.520** | 0.0287 | 0.0372 | **1.295** | 2707256 | 91439 |
| | | | 3 | Yes | 0.0289 | 0.0428 | **1.479** | 0.0294 | 0.0389 | **1.324** | 2488539 | 65443 |
| | | | 5 | No | 0.0477 | 0.0684 | **1.434** | 0.0498 | 0.0601 | **1.205** | 2707256 | 25446 |
| | | | 5 | Yes | 0.0474 | 0.0680 | **1.434** | 0.0504 | 0.0626 | **1.241** | 2488539 | 9338 |
| | | | 10 | No | 0.0927 | 0.1295 | **1.397** | 0.1234 | 0.1194 | **0.967** | 2707256 | 1217 |
| | | | 10 | Yes | 0.0688 | 0.1279 | **1.860** | 0.1621 | 0.1255 | **0.774** | 2488539 | 2 |
| | | Male | 1 | No | 0.0094 | 0.0161 | **1.718** | 0.0095 | 0.0140 | **1.476** | 2707273 | 16890 |
| | | | 1 | Yes | 0.0096 | 0.0161 | **1.679** | 0.0097 | 0.0146 | **1.506** | 2488491 | 16156 |
| | | | 3 | No | 0.0281 | 0.0427 | **1.519** | 0.0287 | 0.0372 | **1.295** | 2707273 | 91385 |
| | | | 3 | Yes | 0.0289 | 0.0427 | **1.479** | 0.0294 | 0.0389 | **1.324** | 2488491 | 65860 |
| | | | 5 | No | 0.0477 | 0.0684 | **1.434** | 0.0499 | 0.0601 | **1.205** | 2707273 | 25398 |
| | | | 5 | Yes | 0.0474 | 0.0680 | **1.433** | 0.0504 | 0.0626 | **1.241** | 2488491 | 9551 |
| | | | 10 | No | 0.0927 | 0.1295 | **1.397** | 0.1234 | 0.1195 | **0.968** | 2707273 | 1216 |
| | | | 10 | Yes | 0.0688 | 0.1282 | **1.863** | 0.1623 | 0.1259 | **0.776** | 2488491 | 2 |
| | Antihyp | Female | 1 | No | 0.0051 | 0.0096 | **1.866** | 0.0053 | 0.0072 | **1.372** | 6439280 | 18626 |
| | | | 1 | Yes | 0.0049 | 0.0095 | **1.935** | 0.0050 | 0.0078 | **1.559** | 5533489 | 14966 |
| | | | 3 | No | 0.0144 | 0.0222 | **1.540** | 0.0150 | 0.0204 | **1.363** | 6439280 | 39300 |
| | | | 3 | Yes | 0.0140 | 0.0227 | **1.622** | 0.0142 | 0.0219 | **1.542** | 5533489 | 73633 |
| | | | 5 | No | 0.0301 | 0.0246 | **0.816** | 0.0331 | 0.0336 | **1.014** | 6439280 | 8721 |
| | | | 5 | Yes | 0.0255 | 0.0297 | **1.165** | 0.0266 | 0.0361 | **1.358** | 5533489 | 82653 |
| | | | 10 | No | 0.0405 | 0.0060 | **0.148** | 0.1109 | 0.0759 | **0.685** | 6439280 | 3 |
| | | | 10 | Yes | 0.0254 | 0.0295 | **1.161** | 0.0619 | 0.0831 | **1.342** | 5533489 | 53 |
| | | Male | 1 | No | 0.0051 | 0.0096 | **1.869** | 0.0053 | 0.0072 | **1.374** | 6410348 | 18586 |
| | | | 1 | Yes | 0.0050 | 0.0096 | **1.934** | 0.0050 | 0.0078 | **1.556** | 5504677 | 15268 |
| | | | 3 | No | 0.0144 | 0.0223 | **1.546** | 0.0150 | 0.0205 | **1.365** | 6410348 | 40520 |
| | | | 3 | Yes | 0.0141 | 0.0225 | **1.600** | 0.0143 | 0.0219 | **1.537** | 5504677 | 78686 |
| | | | 5 | No | 0.0301 | 0.0248 | **0.826** | 0.0330 | 0.0337 | **1.021** | 6410348 | 9968 |
| | | | 5 | Yes | 0.0257 | 0.0298 | **1.160** | 0.0267 | 0.0362 | **1.354** | 5504677 | 95249 |
| | | | 10 | No | 0.0403 | 0.0062 | **0.154** | 0.1092 | 0.0762 | **0.697** | 6410348 | 3 |
| | | | 10 | Yes | 0.0308 | 0.0303 | **0.984** | 0.0621 | 0.0832 | **1.339** | 5504677 | 95 |

| Cohort | Med | Gender | Follow-up (years) | Age interactions | p0_hajek | p1_hajek | RR_hajek | p0_ht | p1_ht | RR_ht | N (untreated) | ESS (untre |
|---|---|---|---|---|---|---|---|---|---|---|---|---|
| Overlap cohort 2 | statins | Female | 1 | No | 0.0094 | 0.0161 | 1.718 | 0.0095 | 0.0140 | 1.476 | 2707256 | 16888 |
| | | | 1 | Yes | 0.0096 | 0.0161 | 1.679 | 0.0097 | 0.0146 | 1.507 | 2488538 | 16458 |
| | | | 3 | No | 0.0281 | 0.0427 | 1.519 | 0.0287 | 0.0372 | 1.295 | 2707209 | 96058 |
| | | | 3 | Yes | 0.0289 | 0.0426 | 1.474 | 0.0293 | 0.0388 | 1.321 | 2488502 | 89930 |
| | | | 5 | No | 0.0478 | 0.0683 | 1.430 | 0.0498 | 0.0601 | 1.206 | 2706828 | 33912 |
| | | | 5 | Yes | 0.0485 | 0.0679 | 1.401 | 0.0501 | 0.0625 | 1.248 | 2488234 | 36598 |
| | | | 10 | No | 0.1009 | 0.1291 | 1.279 | 0.1134 | 0.1192 | 1.052 | 2702758 | 58511 |
| | | | 10 | Yes | 0.1004 | 0.1266 | 1.261 | 0.1115 | 0.1255 | 1.125 | 2483950 | 59981 |
| | | Male | 1 | No | 0.0094 | 0.0161 | 1.718 | 0.0095 | 0.0140 | 1.476 | 2707273 | 16890 |
| | | | 1 | Yes | 0.0096 | 0.0162 | 1.680 | 0.0097 | 0.0146 | 1.507 | 2488490 | 16459 |
| | | | 3 | No | 0.0281 | 0.0427 | 1.519 | 0.0287 | 0.0372 | 1.295 | 2707227 | 96024 |
| | | | 3 | Yes | 0.0289 | 0.0426 | 1.474 | 0.0293 | 0.0388 | 1.321 | 2488453 | 89966 |
| | | | 5 | No | 0.0478 | 0.0683 | 1.429 | 0.0498 | 0.0600 | 1.206 | 2706845 | 33886 |
| | | | 5 | Yes | 0.0485 | 0.0679 | 1.401 | 0.0501 | 0.0625 | 1.248 | 2488185 | 36541 |
| | | | 10 | No | 0.1008 | 0.1291 | 1.280 | 0.1133 | 0.1193 | 1.053 | 2702773 | 58716 |
| | | | 10 | Yes | 0.1005 | 0.1265 | 1.259 | 0.1115 | 0.1254 | 1.124 | 2483899 | 60249 |
| | Antihyp | Female | 1 | No | 0.0051 | 0.0096 | 1.866 | 0.0053 | 0.0072 | 1.372 | 6439280 | 18626 |
| | | | 1 | Yes | 0.0049 | 0.0095 | 1.934 | 0.0050 | 0.0078 | 1.559 | 5533488 | 22549 |
| | | | 3 | No | 0.0144 | 0.0224 | 1.552 | 0.0150 | 0.0204 | 1.363 | 6439273 | 41314 |
| | | | 3 | Yes | 0.0139 | 0.0227 | 1.628 | 0.0141 | 0.0219 | 1.549 | 5533468 | 79563 |
| | | | 5 | No | 0.0282 | 0.0290 | 1.028 | 0.0303 | 0.0336 | 1.109 | 6439050 | 10838 |
| | | | 5 | Yes | 0.0234 | 0.0301 | 1.285 | 0.0243 | 0.0361 | 1.484 | 5533324 | 96986 |
| | | | 10 | No | 0.0585 | 0.0382 | 0.653 | 0.0679 | 0.0745 | 1.097 | 6434678 | 24971 |
| | | | 10 | Yes | 0.0293 | 0.0389 | 1.331 | 0.0542 | 0.0805 | 1.486 | 5530224 | 20 |
| | | Male | 1 | No | 0.0051 | 0.0096 | 1.869 | 0.0053 | 0.0072 | 1.374 | 6410348 | 18586 |
| | | | 1 | Yes | 0.0050 | 0.0096 | 1.933 | 0.0050 | 0.0078 | 1.556 | 5504676 | 22383 |
| | | | 3 | No | 0.0144 | 0.0225 | 1.557 | 0.0150 | 0.0205 | 1.365 | 6410341 | 42465 |
| | | | 3 | Yes | 0.0140 | 0.0229 | 1.635 | 0.0142 | 0.0219 | 1.545 | 5504655 | 78677 |
| | | | 5 | No | 0.0282 | 0.0290 | 1.029 | 0.0302 | 0.0337 | 1.114 | 6410114 | 10939 |
| | | | 5 | Yes | 0.0236 | 0.0302 | 1.280 | 0.0245 | 0.0362 | 1.479 | 5504505 | 12196 |
| | | | 10 | No | 0.0582 | 0.0396 | 0.682 | 0.0678 | 0.0749 | 1.104 | 6405767 | 22936 |
| | | | 10 | Yes | 0.0344 | 0.0400 | 1.164 | 0.0546 | 0.0807 | 1.478 | 5501399 | 40 |

## 6.2 Diagnostics

We look at effective sample size (ESS), broken down by each weight, and the distribution of weights.

The following tables break down the ESS loss at each weighting stage — IPCW, IPACW, IPTW, and the combined weight. This helps identify which component of the weighting is driving the ESS collapse. ESS% values below 10% are shown in red; below 25% in orange.

**IPCW** (Inverse Probability of Censoring Weights) corrects for informative censoring — individuals who are lost to follow-up for reasons related to their risk. High IPCW weights indicate that some individuals had a very low predicted probability of remaining uncensored, meaning the model is trying to upweight a small group to represent a much larger one.

**IPACW** (Inverse Probability of Adherence to Censoring Weights) corrects for artificial censoring — individuals who deviate from their assigned treatment strategy (sustained treatment or sustained non-treatment) are censored, and IPACW reweights the remaining individuals to represent the full population. High IPACW weights arise when some individuals have a very low predicted probability of sustaining their assigned strategy.

**IPTW** (Inverse Probability of Treatment Weights) balances the treated and untreated groups at baseline on measured confounders. High IPTW weights arise when some individuals have a propensity score close to 0 or 1, indicating near-deterministic treatment assignment.

## 6.2.1 Entire cohort — Statins — Female — Age interactions: No

**Table: ESS diagnostics**

| Follow-up | Stage | ESS (overall) | ESS% (overall) | ESS (untreated) | ESS% (untreated) | ESS (treated) | ESS% (treated) |
|---|---|---|---|---|---|---|---|
| 1 | IPCW | 9821720 | **99.2** | 9471105 | **99.2** | 353023 | **99.7** |
| | IPACW | 9881094 | **99.8** | 9535869 | **99.9** | 351805 | **99.4** |
| | IPCW×IPACW | 9826069 | **99.3** | 9477788 | **99.3** | 349234 | **98.6** |
| | IPTW | 1857246 | **18.8** | 8339922 | **87.4** | 2655 | **0.7** |
| | Combined | 564386 | **5.7** | 7338860 | **76.9** | 1360 | **0.4** |
| 3 | IPCW | 9154962 | **92.5** | 8827894 | **92.5** | 345236 | **97.5** |
| | IPACW | 134607 | **1.4** | 123271 | **1.3** | 344942 | **97.4** |
| | IPCW×IPACW | 221571 | **2.2** | 206096 | **2.2** | 324680 | **91.7** |
| | IPTW | 1857246 | **18.8** | 8339922 | **87.4** | 2655 | **0.7** |
| | Combined | 2021 | **0.0** | 1828 | **0.0** | 538 | **0.2** |
| 5 | IPCW | 8169156 | **82.5** | 7874409 | **82.5** | 331204 | **93.5** |
| | IPACW | 11129 | **0.1** | 10162 | **0.1** | 337625 | **95.3** |
| | IPCW×IPACW | 20125 | **0.2** | 18739 | **0.2** | 284469 | **80.3** |
| | IPTW | 1857246 | **18.8** | 8339922 | **87.4** | 2655 | **0.7** |
| | Combined | 134 | **0.0** | 121 | **0.0** | 354 | **0.1** |
| 10 | IPCW | 5197972 | **52.5** | 5010831 | **52.5** | 249235 | **70.4** |
| | IPACW | 6 | **0.0** | 6 | **0.0** | 309124 | **87.3** |
| | IPCW×IPACW | 6 | **0.0** | 6 | **0.0** | 17621 | **5.0** |
| | IPTW | 1857246 | **18.8** | 8339922 | **87.4** | 2655 | **0.7** |
| | Combined | 1 | **0.0** | 1 | **0.0** | 18 | **0.0** |

**Table: Weight distributions**

| Follow-up (years) | Stage | Group | N | Median | P90 | P95 | P99 | P99.9 | Max |
|---|---|---|---|---|---|---|---|---|---|
| 1 | IPCW | Treated | 354098 | 1.090 | 1.162 | 1.197 | 1.292 | 1.522 | 5.4 |
| | IPACW | | | 1.131 | 1.233 | 1.282 | 1.468 | 1.977 | 6.4 |
| | IPCW x IPACW | | | 1.244 | 1.393 | 1.477 | 1.783 | 2.559 | 8.2 |
| | IPTW | | | 0.141 | 0.881 | 1.569 | 10.052 | 137.162 | 1037.8 |
| | Combined | | | 0.174 | 1.137 | 2.090 | 15.226 | 254.799 | 3273.7 |
| | IPCW | Untreated | 9545467 | 1.193 | 1.320 | 1.359 | 1.479 | 1.634 | 6.2 |
| | IPACW | | | 1.002 | 1.031 | 1.045 | 1.094 | 1.358 | 29.8 |
| | IPCW x IPACW | | | 1.199 | 1.323 | 1.364 | 1.490 | 1.658 | 31.7 |
| | IPTW | | | 0.966 | 1.025 | 1.100 | 1.471 | 6.066 | 87.6 |
| | Combined | | | 1.179 | 1.298 | 1.368 | 1.727 | 8.542 | 537.7 |
| 3 | IPCW | Treated | 354098 | 1.236 | 1.464 | 1.575 | 1.892 | 2.635 | 14.3 |
| | IPACW | | | | 1.452 | 1.552 | 1.919 | 3.015 | 17.2 |
| | IPCW x IPACW | | | 1.576 | 2.016 | 2.245 | 3.077 | 5.760 | 36.2 |
| | IPTW | | | 0.141 | 0.881 | 1.569 | 10.052 | 137.162 | 1037.8 |
| | Combined | | | 0.213 | 1.433 | 2.677 | 20.148 | 346.188 | 22131.1 |
| | IPCW | Untreated | 9545467 | 1.432 | 2.131 | 2.334 | 2.949 | 4.354 | 107.3 |
| | IPACW | | | 1.004 | 1.084 | 1.126 | 1.264 | 2.044 | 27864.4 |
| | IPCW x IPACW | | | 1.485 | 2.143 | 2.347 | 2.993 | 4.522 | 33009.1 |
| | IPTW | | | 0.966 | 1.025 | 1.100 | 1.471 | 6.066 | 87.6 |
| | Combined | | | 1.481 | 2.098 | 2.309 | 3.170 | 13.297 | 368041.9 |
| 5 | IPCW | Treated | 354098 | 1.344 | 1.759 | 1.957 | 2.564 | 4.072 | 30.2 |
| | IPACW | | | 1.294 | 1.606 | 1.751 | 2.248 | 3.808 | 22.0 |
| | IPCW x IPACW | | | 1.827 | 2.648 | 3.067 | 4.570 | 9.385 | 186.9 |
| | IPTW | | | 0.141 | 0.881 | 1.569 | 10.052 | 137.162 | 1037.8 |
| | Combined | | | 0.244 | 1.667 | 3.152 | 23.518 | 405.233 | 24291.2 |
| | IPCW | Untreated | 9545467 | 1.576 | 2.903 | 3.417 | 4.726 | 7.822 | 198.0 |
| | IPACW | | | 1.005 | 1.126 | 1.195 | 1.416 | 2.608 | 72558.2 |
| | IPCW x IPACW | | | 1.696 | 2.933 | 3.448 | 4.841 | 8.474 | 100755.3 |
| | IPTW | | | 0.966 | 1.025 | 1.100 | 1.471 | 6.066 | 87.6 |
| | Combined | | | 1.689 | 2.885 | 3.397 | 5.127 | 17.770 | 1703732.6 |

| Follow-up (years) | Stage | Group | N | Median | P90 | P95 | P99 | P99.9 | Max |
|---|---|---|---|---|---|---|---|---|---|
| 10 | IPCW | Treated | 354098 | 1.435 | 2.390 | 2.813 | 4.169 | 7.934 | 311.7 |
| | IPACW | | | 1.336 | 1.878 | 2.125 | 2.927 | 5.752 | 103.6 |
| | IPCW x IPACW | | | 1.971 | 4.210 | 5.191 | 8.870 | 23.115 | 6264.6 |
| | IPTW | | | 0.141 | 0.881 | 1.569 | 10.052 | 137.162 | 1037.8 |
| | Combined | | | 0.298 | 2.095 | 4.119 | 29.326 | 499.344 | 549617.6 |
| | IPCW | Untreated | 9545467 | 1.781 | 4.435 | 6.170 | 10.707 | 24.654 | 1071.6 |
| | IPACW | | | 1.007 | 1.192 | 1.323 | 1.741 | 3.603 | 10810492.2 |
| | IPCW x IPACW | | | 1.964 | 4.570 | 6.296 | 10.953 | 26.777 | 22489535.7 |
| | IPTW | | | 0.966 | 1.025 | 1.100 | 1.471 | 6.066 | 87.6 |
| | Combined | | | 1.952 | 4.548 | 6.248 | 11.108 | 35.325 | 825785273.0 |

# 6.2.2 Entire cohort — Statins — Female — Age interactions: Yes

**Table: ESS diagnostics**

| Follow-up | Stage | ESS (overall) | ESS% (overall) | ESS (untreated) | ESS% (untreated) | ESS (treated) | ESS% (treated) |
|---|---|---|---|---|---|---|---|
| 1 | IPCW | 1984701 | 20.0 | 1866883 | 19.6 | 350772 | **99.1** |
| | IPACW | 9883127 | **99.8** | 9538153 | **99.9** | 351644 | **99.3** |
| | IPCW×IPACW | 2010275 | 20.3 | 1877564 | 19.7 | 346373 | **97.8** |
| | IPTW | 1266915 | 12.8 | 8738573 | 91.5 | 1620 | 0.5 |
| | Combined | 337903 | 3.4 | 1800737 | 18.9 | 915 | 0.3 |
| 3 | IPCW | 8993381 | **90.8** | 8685522 | **91.0** | 315672 | **89.1** |
| | IPACW | 9633046 | **97.3** | 9306145 | **97.5** | 344247 | **97.2** |
| | IPCW×IPACW | 8819613 | **89.1** | 8534937 | **89.4** | 289071 | **81.6** |
| | IPTW | 1266915 | 12.8 | 8738573 | 91.5 | 1620 | 0.5 |
| | Combined | 113860 | 1.2 | 997476 | 10.4 | 418 | 0.1 |
| 5 | IPCW | 7134413 | **72.1** | 7051499 | **73.9** | 127042 | **35.9** |
| | IPACW | 1025691 | 10.4 | 941080 | 9.9 | 335554 | **94.8** |
| | IPCW×IPACW | 575638 | 5.8 | 539637 | 5.7 | 76065 | 21.5 |
| | IPTW | 1266915 | 12.8 | 8738573 | 91.5 | 1620 | 0.5 |
| | Combined | 1083 | 0.0 | 972 | 0.0 | 252 | 0.1 |
| 10 | IPCW | 3787067 | **38.3** | 4378099 | **45.9** | 14034 | 4.0 |
| | IPACW | 3 | 0.0 | 3 | 0.0 | 291314 | **82.3** |
| | IPCW×IPACW | 2 | 0.0 | 2 | 0.0 | 9364 | 2.6 |
| | IPTW | 1266915 | 12.8 | 8738573 | 91.5 | 1620 | 0.5 |
| | Combined | 1 | 0.0 | 1 | 0.0 | 68 | 0.0 |

**Table: Weight distributions**

| Follow-up (years) | Stage | Group | N | Median | P90 | P95 | P99 | P99.9 | Max |
|---|---|---|---|---|---|---|---|---|---|
| 1 | IPCW | Treated | 354098 | 1.086 | 1.185 | 1.236 | 1.386 | 1.865 | 25.4 |
| | IPACW | | | 1.130 | 1.233 | 1.285 | 1.482 | 2.026 | 6.3 |
| | IPCW x IPACW | | | 1.239 | 1.425 | 1.533 | 1.871 | 2.760 | 31.0 |
| | IPTW | | | 0.127 | 0.809 | 1.408 | 8.690 | 132.387 | 1699.8 |
| | Combined | | | 0.158 | 1.034 | 1.853 | 13.092 | 257.396 | 4100.5 |
| | IPCW | Untreated | 9545467 | 1.196 | 1.321 | 1.348 | 1.451 | 1.624 | 7471.3 |
| | IPACW | | | 1.001 | 1.032 | 1.046 | 1.089 | 1.329 | 12.2 |
| | IPCW x IPACW | | | 1.203 | 1.324 | 1.352 | 1.481 | 1.645 | 7518.7 |
| | IPTW | | | 0.965 | 1.024 | 1.101 | 1.538 | 5.068 | 115.0 |
| | Combined | | | 1.182 | 1.294 | 1.344 | 1.836 | 7.024 | 7580.5 |
| 3 | IPCW | Treated | 354098 | 1.223 | 1.519 | 1.688 | 2.251 | 4.348 | 101.9 |
| | IPACW | | | 1.236 | 1.452 | 1.555 | 1.943 | 3.156 | 21.9 |
| | IPCW x IPACW | | | 1.560 | 2.098 | 2.433 | 3.586 | 7.683 | 163.0 |
| | IPTW | | | 0.127 | 0.809 | 1.408 | 8.690 | 132.387 | 1699.8 |
| | Combined | | | 0.198 | 1.296 | 2.358 | 17.603 | 355.661 | 16229.5 |
| | IPCW | Untreated | 9545467 | 1.438 | 2.147 | 2.335 | 2.709 | 4.498 | 545.2 |
| | IPACW | | | 1.003 | 1.089 | 1.129 | 1.252 | 1.927 | 246.9 |
| | IPCW x IPACW | | | 1.486 | 2.159 | 2.348 | 2.750 | 4.649 | 549.4 |
| | IPTW | | | 0.965 | 1.024 | 1.101 | 1.538 | 5.068 | 115.0 |
| | Combined | | | 1.481 | 2.118 | 2.311 | 3.066 | 10.752 | 12534.6 |
| 5 | IPCW | Treated | 354098 | 1.327 | 1.832 | 2.124 | 3.226 | 7.750 | 905.9 |
| | IPACW | | | 1.294 | 1.604 | 1.754 | 2.289 | 3.941 | 33.1 |
| | IPCW x IPACW | | | 1.809 | 2.748 | 3.363 | 5.632 | 14.194 | 1902.5 |
| | IPTW | | | 0.127 | 0.809 | 1.408 | 8.690 | 132.387 | 1699.8 |
| | Combined | | | 0.229 | 1.507 | 2.778 | 20.889 | 399.230 | 26597.6 |
| | IPCW | Untreated | 9545467 | 1.577 | 2.931 | 3.431 | 4.373 | 8.143 | 1251.0 |
| | IPACW | | | 1.005 | 1.134 | 1.200 | 1.400 | 2.408 | 9012.3 |
| | IPCW x IPACW | | | 1.689 | 2.964 | 3.463 | 4.439 | 8.905 | 23046.8 |
| | IPTW | | | 0.965 | 1.024 | 1.101 | 1.538 | 5.068 | 115.0 |
| | Combined | | | 1.680 | 2.922 | 3.417 | 4.626 | 15.053 | 612917.8 |

| Follow-up (years) | Stage | Group | N | Median | P90 | P95 | P99 | P99.9 | Max |
|---|---|---|---|---|---|---|---|---|---|
| 10 | IPCW | Treated | 354098 | 1.436 | 2.456 | 3.029 | 5.369 | 15.868 | 3051.7 |
| | IPACW | | | 1.335 | 1.874 | 2.125 | 2.965 | 6.029 | 212.9 |
| | IPCW x IPACW | | | 1.982 | 4.283 | 5.555 | 11.389 | 37.173 | 4830.7 |
| | IPTW | | | 0.127 | 0.809 | 1.408 | 8.690 | 132.387 | 1699.8 |
| | Combined | | | 0.280 | 1.901 | 3.668 | 26.636 | 492.607 | 118978.4 |
| | IPCW | Untreated | 9545467 | 1.763 | 4.499 | 6.247 | 10.498 | 19.124 | 2562.9 |
| | IPACW | | | 1.006 | 1.202 | 1.336 | 1.725 | 3.285 | 24292472.0 |
| | IPCW x IPACW | | | 1.956 | 4.650 | 6.388 | 10.714 | 21.160 | 105106497.6 |
| | IPTW | | | 0.965 | 1.024 | 1.101 | 1.538 | 5.068 | 115.0 |
| | Combined | | | 1.942 | 4.627 | 6.345 | 10.784 | 30.025 | 2795257655.2 |

# 6.2.3 Entire cohort — Statins — Male — Age interactions: No

**Table: ESS diagnostics**

| Follow-up | Stage | ESS (overall) | ESS% (overall) | ESS (untreated) | ESS% (untreated) | ESS (treated) | ESS% (treated) |
|---|---|---|---|---|---|---|---|
| 1 | IPCW | 9821622 | **99.2** | 9471009 | **99.2** | 353019 | **99.7** |
| | IPACW | 9881104 | **99.8** | 9535879 | **99.9** | 351805 | **99.4** |
| | IPCW×IPACW | 9825981 | **99.3** | 9477701 | **99.3** | 349234 | **98.6** |
| | IPTW | 1854526 | **18.7** | 8339844 | **87.4** | 2653 | **0.7** |
| | Combined | 565012 | **5.7** | 7341302 | **76.9** | 1362 | **0.4** |
| 3 | IPCW | 9154486 | **92.5** | 8827439 | **92.5** | 345195 | **97.5** |
| | IPACW | 136119 | **1.4** | 124657 | **1.3** | 344950 | **97.4** |
| | IPCW×IPACW | 223918 | **2.3** | 208280 | **2.2** | 324672 | **91.7** |
| | IPTW | 1854526 | **18.7** | 8339844 | **87.4** | 2653 | **0.7** |
| | Combined | 2041 | **0.0** | 1846 | **0.0** | 541 | **0.2** |
| 5 | IPCW | 8168584 | **82.5** | 7873879 | **82.5** | 331098 | **93.5** |
| | IPACW | 11177 | **0.1** | 10206 | **0.1** | 337630 | **95.3** |
| | IPCW×IPACW | 20185 | **0.2** | 18794 | **0.2** | 284430 | **80.3** |
| | IPTW | 1854526 | **18.7** | 8339844 | **87.4** | 2653 | **0.7** |
| | Combined | 134 | **0.0** | 120 | **0.0** | 356 | **0.1** |
| 10 | IPCW | 5194220 | **52.5** | 5007591 | **52.5** | 247439 | **69.9** |
| | IPACW | 6 | **0.0** | 6 | **0.0** | 309290 | **87.3** |
| | IPCW×IPACW | 7 | **0.0** | 6 | **0.0** | 17688 | **5.0** |
| | IPTW | 1854526 | **18.7** | 8339844 | **87.4** | 2653 | **0.7** |
| | Combined | 1 | **0.0** | 1 | **0.0** | 18 | **0.0** |

**Table: Weight distributions**

| Follow-up (years) | Stage | Group | N | Median | P90 | P95 | P99 | P99.9 | Max |
|---|---|---|---|---|---|---|---|---|---|
| 1 | IPCW | Treated | 354098 | 1.090 | 1.163 | 1.197 | 1.292 | 1.522 | 5.4 |
| | IPACW | | | 1.131 | 1.233 | 1.282 | 1.467 | 1.976 | 6.4 |
| | IPCW x IPACW | | | 1.244 | 1.393 | 1.477 | 1.783 | 2.567 | 8.2 |
| | IPTW | | | 0.141 | 0.881 | 1.570 | 10.058 | 137.329 | 1038.3 |
| | Combined | | | 0.174 | 1.137 | 2.090 | 15.236 | 254.853 | 3266.2 |
| | IPCW | Untreated | 9545467 | 1.192 | 1.320 | 1.359 | 1.478 | 1.634 | 6.2 |
| | IPACW | | | 1.002 | 1.031 | 1.045 | 1.094 | 1.358 | 29.7 |
| | IPCW x IPACW | | | 1.199 | 1.323 | 1.364 | 1.489 | 1.657 | 31.6 |
| | IPTW | | | 0.966 | 1.025 | 1.100 | 1.471 | 6.066 | 87.6 |
| | Combined | | | 1.179 | 1.298 | 1.368 | 1.728 | 8.540 | 534.9 |
| 3 | IPCW | Treated | 354098 | 1.236 | 1.465 | 1.575 | 1.894 | 2.644 | 14.3 |
| | IPACW | | | | 1.452 | 1.552 | 1.920 | 3.018 | 17.1 |
| | IPCW x IPACW | | | 1.576 | 2.016 | 2.246 | 3.080 | 5.769 | 36.1 |
| | IPTW | | | 0.141 | 0.881 | 1.570 | 10.058 | 137.329 | 1038.3 |
| | Combined | | | 0.213 | 1.432 | 2.676 | 20.159 | 346.016 | 22020.5 |
| | IPCW | Untreated | 9545467 | 1.432 | 2.133 | 2.335 | 2.948 | 4.350 | 106.3 |
| | IPACW | | | 1.004 | 1.084 | 1.126 | 1.264 | 2.045 | 27706.4 |
| | IPCW x IPACW | | | 1.484 | 2.144 | 2.349 | 2.991 | 4.519 | 32831.7 |
| | IPTW | | | 0.966 | 1.025 | 1.100 | 1.471 | 6.066 | 87.6 |
| | Combined | | | 1.481 | 2.100 | 2.310 | 3.168 | 13.299 | 366224.4 |
| 5 | IPCW | Treated | 354098 | 1.344 | 1.759 | 1.958 | 2.567 | 4.077 | 30.0 |
| | IPACW | | | 1.294 | 1.606 | 1.751 | 2.246 | 3.804 | 21.8 |
| | IPCW x IPACW | | | 1.827 | 2.649 | 3.067 | 4.569 | 9.415 | 186.4 |
| | IPTW | | | 0.141 | 0.881 | 1.570 | 10.058 | 137.329 | 1038.3 |
| | Combined | | | 0.244 | 1.667 | 3.153 | 23.530 | 410.405 | 24168.7 |
| | IPCW | Untreated | 9545467 | 1.576 | 2.905 | 3.420 | 4.726 | 7.815 | 198.3 |
| | IPACW | | | 1.005 | 1.126 | 1.195 | 1.416 | 2.607 | 72109.3 |
| | IPCW x IPACW | | | 1.696 | 2.935 | 3.451 | 4.841 | 8.475 | 100956.8 |
| | IPTW | | | 0.966 | 1.025 | 1.100 | 1.471 | 6.066 | 87.6 |
| | Combined | | | 1.689 | 2.886 | 3.400 | 5.126 | 17.766 | 1708223.3 |

| Follow-up (years) | Stage | Group | N | Median | P90 | P95 | P99 | P99.9 | Max |
|---|---|---|---|---|---|---|---|---|---|
| 10 | IPCW | Treated | 354098 | 1.435 | 2.392 | 2.815 | 4.171 | 7.932 | 309.0 |
| | IPACW | | | 1.336 | 1.878 | 2.125 | 2.927 | 5.738 | 102.9 |
| | IPCW x IPACW | | | 1.971 | 4.211 | 5.192 | 8.875 | 23.132 | 6247.8 |
| | IPTW | | | 0.141 | 0.881 | 1.570 | 10.058 | 137.329 | 1038.3 |
| | Combined | | | 0.298 | 2.095 | 4.120 | 29.358 | 502.410 | 548315.1 |
| | IPCW | Untreated | 9545467 | 1.781 | 4.435 | 6.171 | 10.720 | 24.653 | 1101.3 |
| | IPACW | | | 1.007 | 1.192 | 1.323 | 1.741 | 3.603 | 10641038.8 |
| | IPCW x IPACW | | | 1.964 | 4.569 | 6.297 | 10.966 | 26.755 | 22159748.3 |
| | IPTW | | | 0.966 | 1.025 | 1.100 | 1.471 | 6.066 | 87.6 |
| | Combined | | | 1.952 | 4.548 | 6.250 | 11.119 | 35.305 | 813991723.7 |

## 6.2.4 Entire cohort — Statins — Male — Age interactions: Yes

**Table: ESS diagnostics**

| Follow-up | Stage | ESS (overall) | ESS% (overall) | ESS (untreated) | ESS% (untreated) | ESS (treated) | ESS% (treated) |
|---|---|---|---|---|---|---|---|
| 1 | IPCW | 1971741 | 19.9 | 1854617 | 19.4 | 350779 | **99.1** |
| | IPACW | 9883122 | **99.8** | 9538150 | **99.9** | 351643 | **99.3** |
| | IPCW×IPACW | 1997222 | 20.2 | 1865273 | 19.5 | 346379 | **97.8** |
| | IPTW | 1238722 | 12.5 | 8738161 | 91.5 | 1588 | 0.4 |
| | Combined | 330351 | 3.3 | 1789405 | 18.7 | 899 | 0.3 |
| 3 | IPCW | 8994771 | **90.9** | 8686544 | **91.0** | 316188 | **89.3** |
| | IPACW | 9585453 | **96.8** | 9257597 | **97.0** | 344251 | **97.2** |
| | IPCW×IPACW | 8765735 | **88.5** | 8479731 | **88.8** | 289740 | **81.8** |
| | IPTW | 1238722 | 12.5 | 8738161 | 91.5 | 1588 | 0.4 |
| | Combined | 113269 | 1.1 | 998103 | 10.5 | 418 | 0.1 |
| 5 | IPCW | 7196061 | **72.7** | 7082149 | **74.2** | 141161 | **39.9** |
| | IPACW | 1139810 | 11.5 | 1046512 | 11.0 | 335532 | **94.8** |
| | IPCW×IPACW | 640673 | 6.5 | 600397 | 6.3 | 87717 | 24.8 |
| | IPTW | 1238722 | 12.5 | 8738161 | 91.5 | 1588 | 0.4 |
| | Combined | 1213 | 0.0 | 1090 | 0.0 | 254 | 0.1 |
| 10 | IPCW | 3865932 | **39.1** | 4421410 | **46.3** | 15083 | 4.3 |
| | IPACW | 3 | 0.0 | 3 | 0.0 | 291361 | **82.3** |
| | IPCW×IPACW | 2 | 0.0 | 2 | 0.0 | 9941 | 2.8 |
| | IPTW | 1238722 | 12.5 | 8738161 | 91.5 | 1588 | 0.4 |
| | Combined | 1 | 0.0 | 1 | 0.0 | 68 | 0.0 |

**Table: Weight distributions**

| Follow-up (years) | Stage | Group | N | Median | P90 | P95 | P99 | P99.9 | Max |
|---|---|---|---|---|---|---|---|---|---|
| 1 | IPCW | Treated | 354098 | 1.085 | 1.186 | 1.236 | 1.386 | 1.864 | 25.3 |
| | IPACW | | | 1.130 | 1.233 | 1.285 | 1.482 | 2.029 | 6.2 |
| | IPCW x IPACW | | | 1.239 | 1.425 | 1.533 | 1.871 | 2.753 | 31.1 |
| | IPTW | | | 0.127 | 0.809 | 1.408 | 8.694 | 132.510 | 1697.0 |
| | Combined | | | 0.158 | 1.034 | 1.853 | 13.100 | 258.912 | 4079.3 |
| | IPCW | Untreated | 9545467 | 1.195 | 1.322 | 1.349 | 1.451 | 1.623 | 7502.1 |
| | IPACW | | | 1.001 | 1.032 | 1.046 | 1.089 | 1.329 | 11.9 |
| | IPCW x IPACW | | | 1.203 | 1.324 | 1.352 | 1.481 | 1.645 | 7549.6 |
| | IPTW | | | 0.965 | 1.024 | 1.101 | 1.537 | 5.067 | 115.1 |
| | Combined | | | 1.182 | 1.295 | 1.345 | 1.836 | 7.027 | 7611.7 |
| 3 | IPCW | Treated | 354098 | 1.222 | 1.520 | 1.689 | 2.253 | 4.353 | 101.2 |
| | IPACW | | | 1.236 | 1.452 | 1.555 | 1.943 | 3.168 | 22.0 |
| | IPCW x IPACW | | | 1.560 | 2.099 | 2.434 | 3.588 | 7.684 | 161.8 |
| | IPTW | | | 0.127 | 0.809 | 1.408 | 8.694 | 132.510 | 1697.0 |
| | Combined | | | 0.198 | 1.296 | 2.359 | 17.625 | 356.210 | 16258.8 |
| | IPCW | Untreated | 9545467 | 1.438 | 2.150 | 2.339 | 2.710 | 4.497 | 542.1 |
| | IPACW | | | 1.003 | 1.089 | 1.129 | 1.252 | 1.927 | 306.6 |
| | IPCW x IPACW | | | 1.486 | 2.162 | 2.351 | 2.750 | 4.646 | 604.3 |
| | IPTW | | | 0.965 | 1.024 | 1.101 | 1.537 | 5.067 | 115.1 |
| | Combined | | | 1.481 | 2.121 | 2.314 | 3.066 | 10.751 | 12128.7 |
| 5 | IPCW | Treated | 354098 | 1.327 | 1.832 | 2.126 | 3.230 | 7.814 | 800.6 |
| | IPACW | | | 1.294 | 1.604 | 1.754 | 2.290 | 3.935 | 33.1 |
| | IPCW x IPACW | | | 1.808 | 2.750 | 3.366 | 5.636 | 14.268 | 1680.1 |
| | IPTW | | | 0.127 | 0.809 | 1.408 | 8.694 | 132.510 | 1697.0 |
| | Combined | | | 0.229 | 1.507 | 2.778 | 20.942 | 399.047 | 26419.9 |
| | IPCW | Untreated | 9545467 | 1.577 | 2.936 | 3.440 | 4.380 | 8.150 | 1251.5 |
| | IPACW | | | 1.005 | 1.134 | 1.200 | 1.400 | 2.408 | 8387.8 |
| | IPCW x IPACW | | | 1.689 | 2.969 | 3.471 | 4.445 | 8.910 | 21641.3 |
| | IPTW | | | 0.965 | 1.024 | 1.101 | 1.537 | 5.067 | 115.1 |
| | Combined | | | 1.680 | 2.927 | 3.425 | 4.627 | 15.045 | 577737.6 |

| Follow-up (years) | Stage | Group | N | Median | P90 | P95 | P99 | P99.9 | Max |
|---|---|---|---|---|---|---|---|---|---|
| 10 | IPCW | Treated | 354098 | 1.436 | 2.457 | 3.030 | 5.369 | 15.966 | 2777.5 |
| | IPACW | | | 1.335 | 1.874 | 2.125 | 2.963 | 6.015 | 212.7 |
| | IPCW x IPACW | | | 1.981 | 4.283 | 5.560 | 11.416 | 37.488 | 4828.0 |
| | IPTW | | | 0.127 | 0.809 | 1.408 | 8.694 | 132.510 | 1697.0 |
| | Combined | | | 0.280 | 1.901 | 3.669 | 26.656 | 493.722 | 120484.0 |
| | IPCW | Untreated | 9545467 | 1.763 | 4.491 | 6.258 | 10.546 | 19.137 | 2338.8 |
| | IPACW | | | 1.006 | 1.202 | 1.336 | 1.725 | 3.287 | 21239281.6 |
| | IPCW x IPACW | | | 1.956 | 4.644 | 6.398 | 10.760 | 21.164 | 92939650.6 |
| | IPTW | | | 0.965 | 1.024 | 1.101 | 1.537 | 5.067 | 115.1 |
| | Combined | | | 1.942 | 4.622 | 6.358 | 10.830 | 30.034 | 2481120847.8 |

## 6.2.5 Entire cohort — Antihypertensives — Female — Age interactions: No

**Table: ESS diagnostics**

| Follow-up | Stage | ESS (overall) | ESS% (overall) | ESS (untreated) | ESS% (untreated) | ESS (treated) | ESS% (treated) |
|---|---|---|---|---|---|---|---|
| 1 | IPCW | 9820176 | **99.2** | 8945311 | **99.2** | 882342 | **99.8** |
| | IPACW | 9739971 | **98.4** | 9004383 | **99.9** | 798762 | **90.3** |
| | IPCW×IPACW | 9668304 | **97.7** | 8948565 | **99.3** | 760438 | **86.0** |
| | IPTW | 3554347 | **35.9** | 3591538 | **39.8** | 140684 | **15.9** |
| | Combined | 1063301 | **10.7** | 1129540 | **12.5** | 58091 | **6.6** |
| 3 | IPCW | 9142454 | **92.4** | 8329635 | **92.4** | 867444 | **98.1** |
| | IPACW | 8716744 | **88.1** | 8800105 | **97.6** | 494448 | **55.9** |
| | IPCW×IPACW | 7153162 | **72.3** | 8244650 | **91.5** | 215910 | **24.4** |
| | IPTW | 3554347 | **35.9** | 3591538 | **39.8** | 140684 | **15.9** |
| | Combined | 71239 | **0.7** | 74955 | **0.8** | 4702 | **0.5** |
| 5 | IPCW | 8148910 | **82.3** | 7417457 | **82.3** | 840629 | **95.0** |
| | IPACW | 2001253 | **20.2** | 1837082 | **20.4** | 180339 | **20.4** |
| | IPCW×IPACW | 1126711 | **11.4** | 1761603 | **19.5** | 21122 | **2.4** |
| | IPTW | 3554347 | **35.9** | 3591538 | **39.8** | 140684 | **15.9** |
| | Combined | 228 | **0.0** | 164 | **0.0** | 639 | **0.1** |
| 10 | IPCW | 5177683 | **52.3** | 4695792 | **52.1** | 715348 | **80.9** |
| | IPACW | 3 | **0.0** | 3 | **0.0** | 8703 | **1.0** |
| | IPCW×IPACW | 2 | **0.0** | 2 | **0.0** | 156 | **0.0** |
| | IPTW | 3554347 | **35.9** | 3591538 | **39.8** | 140684 | **15.9** |
| | Combined | 1 | **0.0** | 1 | **0.0** | 23 | **0.0** |

**Table: Weight distributions**

| Follow-up (years) | Stage | Group | N | Median | P90 | P95 | P99 | P99.9 | Max |
|---|---|---|---|---|---|---|---|---|---|
| 1 | IPCW | Treated | 884500 | 1.079 | 1.147 | 1.183 | 1.275 | 1.446 | 4.900000e+00 |
| | IPACW | | | 1.107 | 1.354 | 1.632 | 2.951 | 5.657 | 2.660000e+01 |
| | IPCW x IPACW | | | 1.201 | 1.522 | 1.862 | 3.602 | 7.393 | 3.760000e+01 |
| | IPTW | | | 0.115 | 2.721 | 5.527 | 10.655 | 14.581 | 2.910000e+01 |
| | Combined | | | 0.137 | 3.999 | 9.486 | 27.977 | 82.597 | 6.076000e+02 |
| | IPCW | Untreated | 9015065 | 1.199 | 1.322 | 1.362 | 1.482 | 1.636 | 6.200000e+00 |
| | IPACW | | | 1.016 | 1.044 | 1.062 | 1.149 | 1.420 | 1.090000e+01 |
| | IPCW x IPACW | | | 1.226 | 1.344 | 1.387 | 1.511 | 1.678 | 1.300000e+01 |
| | IPTW | | | 0.924 | 0.965 | 0.988 | 4.187 | 12.779 | 6.494000e+02 |
| | Combined | | | 1.144 | 1.250 | 1.317 | 5.067 | 18.944 | 5.626500e+03 |
| 3 | IPCW | Treated | 884500 | 1.209 | 1.403 | 1.500 | 1.795 | 2.444 | 1.430000e+01 |
| | IPACW | | | 1.190 | 1.585 | 1.917 | 3.993 | 14.190 | 4.405000e+02 |
| | IPCW x IPACW | | | 1.462 | 2.089 | 2.602 | 6.038 | 29.333 | 1.270200e+03 |
| | IPTW | | | 0.115 | 2.721 | 5.527 | 10.655 | 14.581 | 2.910000e+01 |
| | Combined | | | 0.171 | 4.732 | 10.764 | 38.144 | 240.544 | 2.050410e+04 |
| | IPCW | Untreated | 9015065 | 1.443 | 2.138 | 2.340 | 2.966 | 4.383 | 1.073000e+02 |
| | IPACW | | | 1.040 | 1.112 | 1.161 | 1.366 | 2.239 | 1.743000e+02 |
| | IPCW x IPACW | | | 1.549 | 2.239 | 2.446 | 3.110 | 4.631 | 2.735000e+02 |
| | IPTW | | | 0.924 | 0.965 | 0.988 | 4.187 | 12.779 | 6.494000e+02 |
| | Combined | | | 1.463 | 2.118 | 2.373 | 6.127 | 30.233 | 4.450760e+04 |
| 5 | IPCW | Treated | 884500 | 1.313 | 1.655 | 1.818 | 2.331 | 3.629 | 3.020000e+01 |
| | IPACW | | | 1.242 | 1.722 | 2.101 | 4.600 | 20.470 | 1.218100e+03 |
| | IPCW x IPACW | | | 1.675 | 2.601 | 3.316 | 8.075 | 53.305 | 5.933400e+03 |
| | IPTW | | | 0.115 | 2.721 | 5.527 | 10.655 | 14.581 | 2.910000e+01 |
| | Combined | | | 0.202 | 5.238 | 11.424 | 44.041 | 392.839 | 5.969420e+04 |
| | IPCW | Untreated | 9015065 | 1.577 | 2.904 | 3.419 | 4.753 | 7.887 | 1.980000e+02 |
| | IPACW | | | 1.056 | 1.170 | 1.247 | 1.549 | 3.099 | 4.854800e+03 |
| | IPCW x IPACW | | | 1.767 | 3.140 | 3.683 | 5.152 | 8.833 | 1.040550e+04 |
| | IPTW | | | 0.924 | 0.965 | 0.988 | 4.187 | 12.779 | 6.494000e+02 |
| | Combined | | | 1.677 | 3.026 | 3.639 | 7.454 | 41.658 | 1.693098e+06 |

| Follow-up (years) | Stage | Group | N | Median | P90 | P95 | P99 | P99.9 | Max |
|---|---|---|---|---|---|---|---|---|---|
| 10 | IPCW | Treated | 884500 | 1.430 | 2.233 | 2.579 | 3.659 | 6.725 | 3.117000e+02 |
| | IPACW | | | 1.310 | 1.885 | 2.437 | 5.607 | 31.702 | 1.088300e+04 |
| | IPCW x IPACW | | | 1.979 | 3.836 | 5.086 | 12.993 | 120.208 | 1.989618e+05 |
| | IPTW | | | 0.115 | 2.721 | 5.527 | 10.655 | 14.581 | 2.910000e+01 |
| | Combined | | | 0.270 | 6.165 | 12.479 | 53.649 | 774.458 | 2.492237e+06 |
| | IPCW | Untreated | 9015065 | 1.746 | 4.346 | 6.046 | 10.600 | 24.533 | 1.071600e+03 |
| | IPACW | | | 1.064 | 1.287 | 1.428 | 1.952 | 5.116 | 1.895614e+07 |
| | IPCW x IPACW | | | 1.973 | 5.088 | 7.035 | 12.261 | 29.446 | 9.215467e+07 |
| | IPTW | | | 0.924 | 0.965 | 0.988 | 4.187 | 12.779 | 6.494000e+02 |
| | Combined | | | 1.884 | 5.013 | 7.071 | 14.049 | 72.935 | 1.499461e+10 |

## 6.2.6 Entire cohort — Antihypertensives — Female — Age interactions: Yes

**Table: ESS diagnostics**

| Follow-up | Stage | ESS (overall) | ESS% (overall) | ESS (untreated) | ESS% (untreated) | ESS (treated) | ESS% (treated) |
|---|---|---|---|---|---|---|---|
| 1 | IPCW | 1978233 | 20.0 | 1693135 | 18.8 | 880543 | **99.6** |
| | IPACW | 9730707 | **98.3** | 9006488 | **99.9** | 792487 | **89.6** |
| | IPCW×IPACW | 1785948 | 18.0 | 1492304 | 16.6 | 754077 | **85.3** |
| | IPTW | 4580756 | **46.3** | 5329648 | **59.1** | 119496 | 13.5 |
| | Combined | 948143 | 9.6 | 1121163 | 12.4 | 48032 | 5.4 |
| 3 | IPCW | 8984198 | **90.8** | 8189664 | **90.8** | 843510 | **95.4** |
| | IPACW | 8908242 | **90.0** | 8919059 | **98.9** | 523319 | **59.2** |
| | IPCW×IPACW | 7548786 | **76.3** | 8145386 | **90.4** | 276869 | **31.3** |
| | IPTW | 4580756 | **46.3** | 5329648 | **59.1** | 119496 | 13.5 |
| | Combined | 128043 | 1.3 | 463458 | 5.1 | 4102 | 0.5 |
| 5 | IPCW | 7492523 | **75.7** | 6865876 | **76.2** | 670068 | **75.8** |
| | IPACW | 6488898 | **65.5** | 8558213 | **94.9** | 214058 | 24.2 |
| | IPCW×IPACW | 2954601 | **29.8** | 6381420 | **70.8** | 43053 | 4.9 |
| | IPTW | 4580756 | **46.3** | 5329648 | **59.1** | 119496 | 13.5 |
| | Combined | 19148 | 0.2 | 149290 | 1.7 | 789 | 0.1 |
| 10 | IPCW | 4382398 | **44.3** | 4160302 | **46.1** | 229515 | **25.9** |
| | IPACW | 124244 | 1.3 | 103256 | 1.1 | 24934 | 2.8 |
| | IPCW×IPACW | 35722 | 0.4 | 40069 | 0.4 | 1494 | 0.2 |
| | IPTW | 4580756 | **46.3** | 5329648 | **59.1** | 119496 | 13.5 |
| | Combined | 206 | 0.0 | 127 | 0.0 | 103 | 0.0 |

**Table: Weight distributions**

| Follow-up (years) | Stage | Group | N | Median | P90 | P95 | P99 | P99.9 | Max |
|---|---|---|---|---|---|---|---|---|---|
| 1 | IPCW | Treated | 884500 | 1.072 | 1.160 | 1.206 | 1.310 | 1.607 | 25.4 |
| | IPACW | | | 1.107 | 1.334 | 1.662 | 3.086 | 6.017 | 20.6 |
| | IPCW x IPACW | | | 1.193 | 1.531 | 1.909 | 3.762 | 7.891 | 34.4 |
| | IPTW | | | 0.115 | 2.485 | 6.115 | 12.458 | 18.877 | 62.4 |
| | Combined | | | 0.137 | 3.801 | 10.332 | 33.961 | 108.148 | 1083.2 |
| | IPCW | Untreated | 9015065 | 1.202 | 1.323 | 1.350 | 1.463 | 1.625 | 7471.3 |
| | IPACW | | | 1.016 | 1.045 | 1.067 | 1.153 | 1.362 | 5.7 |
| | IPCW x IPACW | | | 1.229 | 1.345 | 1.372 | 1.501 | 1.654 | 8256.8 |
| | IPTW | | | 0.922 | 0.970 | 0.999 | 4.175 | 10.180 | 231.6 |
| | Combined | | | 1.147 | 1.249 | 1.291 | 5.088 | 14.357 | 9051.4 |
| 3 | IPCW | Treated | 884500 | 1.189 | 1.427 | 1.560 | 1.972 | 3.190 | 80.2 |
| | IPACW | | | 1.190 | 1.536 | 1.929 | 4.310 | 14.859 | 239.9 |
| | IPCW x IPACW | | | 1.439 | 2.127 | 2.721 | 6.518 | 29.458 | 576.6 |
| | IPTW | | | 0.115 | 2.485 | 6.115 | 12.458 | 18.877 | 62.4 |
| | Combined | | | 0.168 | 4.540 | 11.902 | 45.721 | 290.840 | 31646.9 |
| | IPCW | Untreated | 9015065 | 1.448 | 2.153 | 2.341 | 2.722 | 4.528 | 545.2 |
| | IPACW | | | 1.039 | 1.115 | 1.171 | 1.376 | 2.058 | 126.8 |
| | IPCW x IPACW | | | 1.551 | 2.256 | 2.445 | 2.845 | 4.699 | 613.2 |
| | IPTW | | | 0.922 | 0.970 | 0.999 | 4.175 | 10.180 | 231.6 |
| | Combined | | | 1.465 | 2.133 | 2.349 | 6.204 | 22.590 | 18073.3 |
| 5 | IPCW | Treated | 884500 | 1.286 | 1.677 | 1.901 | 2.659 | 5.291 | 476.9 |
| | IPACW | | | 1.243 | 1.661 | 2.081 | 4.980 | 21.913 | 1117.2 |
| | IPCW x IPACW | | | 1.641 | 2.653 | 3.495 | 8.969 | 55.848 | 3493.3 |
| | IPTW | | | 0.115 | 2.485 | 6.115 | 12.458 | 18.877 | 62.4 |
| | Combined | | | 0.198 | 4.991 | 12.643 | 52.701 | 461.471 | 79382.6 |
| | IPCW | Untreated | 9015065 | 1.578 | 2.932 | 3.433 | 4.385 | 8.257 | 1042.7 |
| | IPACW | | | 1.056 | 1.174 | 1.261 | 1.567 | 2.817 | 233.3 |
| | IPCW x IPACW | | | 1.758 | 3.177 | 3.698 | 4.705 | 9.086 | 1733.3 |
| | IPTW | | | 0.922 | 0.970 | 0.999 | 4.175 | 10.180 | 231.6 |
| | Combined | | | 1.668 | 3.065 | 3.643 | 7.566 | 31.213 | 19930.0 |

| Follow-up (years) | Stage | Group | N | Median | P90 | P95 | P99 | P99.9 | Max |
|---|---|---|---|---|---|---|---|---|---|
| 10 | IPCW | Treated | 884500 | 1.418 | 2.232 | 2.664 | 4.273 | 10.734 | 1395.1 |
| | IPACW | | | 1.314 | 1.848 | 2.355 | 6.066 | 36.416 | 3613.1 |
| | IPCW x IPACW | | | 1.961 | 3.853 | 5.307 | 15.108 | 138.117 | 26927.0 |
| | IPTW | | | 0.115 | 2.485 | 6.115 | 12.458 | 18.877 | 62.4 |
| | Combined | | | 0.264 | 5.763 | 13.702 | 64.202 | 902.439 | 753549.0 |
| | IPCW | Untreated | 9015065 | 1.732 | 4.402 | 6.116 | 10.387 | 19.066 | 2562.9 |
| | IPACW | | | 1.063 | 1.289 | 1.443 | 1.997 | 4.796 | 29539.7 |
| | IPCW x IPACW | | | 1.974 | 5.172 | 7.148 | 12.019 | 23.041 | 117096.6 |
| | IPTW | | | 0.922 | 0.970 | 0.999 | 4.175 | 10.180 | 231.6 |
| | Combined | | | 1.886 | 5.107 | 7.173 | 13.060 | 60.585 | 1567514.7 |

## 6.2.7 Entire cohort — Antihypertensives — Male — Age interactions: No

**Table: ESS diagnostics**

| Follow-up | Stage | ESS (overall) | ESS% (overall) | ESS (untreated) | ESS% (untreated) | ESS (treated) | ESS% (treated) |
|---|---|---|---|---|---|---|---|
| 1 | IPCW | 9820074 | **99.2** | 8945214 | **99.2** | 882340 | **99.8** |
| | IPACW | 9740199 | **98.4** | 9004367 | **99.9** | 798912 | **90.3** |
| | IPCW×IPACW | 9668833 | **97.7** | 8948557 | **99.3** | 760783 | **86.0** |
| | IPTW | 3599755 | **36.4** | 3646346 | **40.4** | 140424 | **15.9** |
| | Combined | 1099846 | **11.1** | 1182960 | **13.1** | 58051 | **6.6** |
| 3 | IPCW | 9141940 | **92.3** | 8329172 | **92.4** | 867403 | **98.1** |
| | IPACW | 8713465 | **88.0** | 8791774 | **97.5** | 495261 | **56.0** |
| | IPCW×IPACW | 7162230 | **72.3** | 8240238 | **91.4** | 217091 | **24.5** |
| | IPTW | 3599755 | **36.4** | 3646346 | **40.4** | 140424 | **15.9** |
| | Combined | 66774 | **0.7** | 68228 | **0.8** | 4724 | **0.5** |
| 5 | IPCW | 8148213 | **82.3** | 7416850 | **82.3** | 840519 | **95.0** |
| | IPACW | 2099539 | **21.2** | 1942332 | **21.5** | 181052 | **20.5** |
| | IPCW×IPACW | 1156042 | **11.7** | 1834394 | **20.3** | 21311 | **2.4** |
| | IPTW | 3599755 | **36.4** | 3646346 | **40.4** | 140424 | **15.9** |
| | Combined | 238 | **0.0** | 172 | **0.0** | 642 | **0.1** |
| 10 | IPCW | 5173309 | **52.3** | 4691803 | **52.0** | 714947 | **80.8** |
| | IPACW | 3 | **0.0** | 3 | **0.0** | 8772 | **1.0** |
| | IPCW×IPACW | 2 | **0.0** | 2 | **0.0** | 158 | **0.0** |
| | IPTW | 3599755 | **36.4** | 3646346 | **40.4** | 140424 | **15.9** |
| | Combined | 1 | **0.0** | 1 | **0.0** | 23 | **0.0** |

**Table: Weight distributions**

| Follow-up (years) | Stage | Group | N | Median | P90 | P95 | P99 | P99.9 | Max |
|---|---|---|---|---|---|---|---|---|---|
| 1 | IPCW | Treated | 884500 | 1.079 | 1.147 | 1.183 | 1.276 | 1.446 | 4.900000e+00 |
| | IPACW | | | 1.107 | 1.354 | 1.632 | 2.952 | 5.640 | 2.650000e+01 |
| | IPCW x IPACW | | | 1.201 | 1.521 | 1.862 | 3.600 | 7.385 | 3.760000e+01 |
| | IPTW | | | 0.115 | 2.713 | 5.483 | 10.681 | 14.589 | 2.910000e+01 |
| | Combined | | | 0.137 | 3.985 | 9.458 | 27.968 | 82.648 | 6.051000e+02 |
| | IPCW | Untreated | 9015065 | 1.198 | 1.322 | 1.362 | 1.482 | 1.635 | 6.200000e+00 |
| | IPACW | | | 1.016 | 1.044 | 1.062 | 1.149 | 1.420 | 1.070000e+01 |
| | IPCW x IPACW | | | 1.226 | 1.344 | 1.387 | 1.510 | 1.677 | 1.270000e+01 |
| | IPTW | | | 0.924 | 0.965 | 0.988 | 4.189 | 12.745 | 6.360000e+02 |
| | Combined | | | 1.144 | 1.250 | 1.317 | 5.071 | 18.921 | 5.278200e+03 |
| 3 | IPCW | Treated | 884500 | 1.209 | 1.403 | 1.500 | 1.795 | 2.448 | 1.430000e+01 |
| | IPACW | | | 1.190 | 1.585 | 1.917 | 3.991 | 14.148 | 4.394000e+02 |
| | IPCW x IPACW | | | 1.462 | 2.088 | 2.602 | 6.033 | 29.240 | 1.265000e+03 |
| | IPTW | | | 0.115 | 2.713 | 5.483 | 10.681 | 14.589 | 2.910000e+01 |
| | Combined | | | 0.171 | 4.711 | 10.732 | 38.063 | 240.011 | 2.037810e+04 |
| | IPCW | Untreated | 9015065 | 1.442 | 2.139 | 2.342 | 2.965 | 4.380 | 1.063000e+02 |
| | IPACW | | | 1.040 | 1.112 | 1.162 | 1.367 | 2.238 | 1.703000e+02 |
| | IPCW x IPACW | | | 1.549 | 2.240 | 2.447 | 3.109 | 4.626 | 2.677000e+02 |
| | IPTW | | | 0.924 | 0.965 | 0.988 | 4.189 | 12.745 | 6.360000e+02 |
| | Combined | | | 1.463 | 2.119 | 2.374 | 6.129 | 30.221 | 4.338000e+04 |
| 5 | IPCW | Treated | 884500 | 1.313 | 1.655 | 1.819 | 2.332 | 3.624 | 3.000000e+01 |
| | IPACW | | | 1.242 | 1.722 | 2.101 | 4.598 | 20.380 | 1.205000e+03 |
| | IPCW x IPACW | | | 1.675 | 2.601 | 3.315 | 8.070 | 53.150 | 5.889100e+03 |
| | IPTW | | | 0.115 | 2.713 | 5.483 | 10.681 | 14.589 | 2.910000e+01 |
| | Combined | | | 0.202 | 5.212 | 11.384 | 43.948 | 391.673 | 5.943640e+04 |
| | IPCW | Untreated | 9015065 | 1.577 | 2.906 | 3.422 | 4.753 | 7.881 | 1.983000e+02 |
| | IPACW | | | 1.056 | 1.171 | 1.248 | 1.551 | 3.094 | 4.671200e+03 |
| | IPCW x IPACW | | | 1.767 | 3.141 | 3.684 | 5.151 | 8.824 | 1.003130e+04 |
| | IPTW | | | 0.924 | 0.965 | 0.988 | 4.189 | 12.745 | 6.360000e+02 |
| | Combined | | | 1.678 | 3.027 | 3.641 | 7.452 | 41.516 | 1.625741e+06 |

| Follow-up (years) | Stage | Group | N | Median | P90 | P95 | P99 | P99.9 | Max |
|---|---|---|---|---|---|---|---|---|---|
| 10 | IPCW | Treated | 884500 | 1.430 | 2.234 | 2.580 | 3.661 | 6.736 | 3.090000e+02 |
| | IPACW | | | 1.310 | 1.885 | 2.437 | 5.602 | 31.806 | 1.084240e+04 |
| | IPCW x IPACW | | | 1.979 | 3.837 | 5.086 | 12.963 | 119.550 | 1.975701e+05 |
| | IPTW | | | 0.115 | 2.713 | 5.483 | 10.681 | 14.589 | 2.910000e+01 |
| | Combined | | | 0.270 | 6.140 | 12.448 | 53.551 | 770.590 | 2.477797e+06 |
| | IPCW | Untreated | 9015065 | 1.746 | 4.347 | 6.049 | 10.613 | 24.517 | 1.101300e+03 |
| | IPACW | | | 1.064 | 1.288 | 1.429 | 1.954 | 5.121 | 1.755336e+07 |
| | IPCW x IPACW | | | 1.973 | 5.088 | 7.036 | 12.271 | 29.415 | 8.563318e+07 |
| | IPTW | | | 0.924 | 0.965 | 0.988 | 4.189 | 12.745 | 6.360000e+02 |
| | Combined | | | 1.884 | 5.014 | 7.072 | 14.044 | 72.958 | 1.387835e+10 |

## 6.2.8 Entire cohort — Antihypertensives — Male — Age interactions: Yes

**Table: ESS diagnostics**

| Follow-up | Stage | ESS (overall) | ESS% (overall) | ESS (untreated) | ESS% (untreated) | ESS (treated) | ESS% (treated) |
|---|---|---|---|---|---|---|---|
| 1 | IPCW | 1965327 | 19.9 | 1681927 | 18.7 | 880552 | **99.6** |
| | IPACW | 9727385 | **98.3** | 9006449 | **99.9** | 790510 | **89.4** |
| | IPCW×IPACW | 1776803 | 17.9 | 1484559 | 16.5 | 752403 | **85.1** |
| | IPTW | 4585359 | **46.3** | 5331076 | **59.1** | 119254 | 13.5 |
| | Combined | 951328 | 9.6 | 1121051 | 12.4 | 48172 | 5.4 |
| 3 | IPCW | 8985369 | **90.8** | 8190858 | **90.9** | 843435 | **95.4** |
| | IPACW | 8578328 | **86.7** | 8926264 | **99.0** | 442027 | **50.0** |
| | IPCW×IPACW | 7293039 | **73.7** | 8155810 | **90.5** | 240683 | **27.2** |
| | IPTW | 4585359 | **46.3** | 5331076 | **59.1** | 119254 | 13.5 |
| | Combined | 136044 | 1.4 | 580854 | 6.4 | 4169 | 0.5 |
| 5 | IPCW | 7524802 | **76.0** | 6898918 | **76.5** | 666704 | **75.4** |
| | IPACW | 4048843 | **40.9** | 8578962 | **95.2** | 86632 | 9.8 |
| | IPCW×IPACW | 2188142 | 22.1 | 6456920 | 71.6 | 27543 | 3.1 |
| | IPTW | 4585359 | **46.3** | 5331076 | **59.1** | 119254 | 13.5 |
| | Combined | 18778 | 0.2 | 161745 | 1.8 | 766 | 0.1 |
| 10 | IPCW | 4423620 | **44.7** | 4208178 | **46.7** | 225642 | **25.5** |
| | IPACW | 124756 | 1.3 | 104807 | 1.2 | 22086 | 2.5 |
| | IPCW×IPACW | 36547 | 0.4 | 40547 | 0.4 | 1570 | 0.2 |
| | IPTW | 4585359 | **46.3** | 5331076 | **59.1** | 119254 | 13.5 |
| | Combined | 299 | 0.0 | 198 | 0.0 | 106 | 0.0 |

**Table: Weight distributions**

| Follow-up (years) | Stage | Group | N | Median | P90 | P95 | P99 | P99.9 | Max |
|---|---|---|---|---|---|---|---|---|---|
| 1 | IPCW | Treated | 884500 | 1.072 | 1.160 | 1.206 | 1.310 | 1.606 | 25.3 |
| | IPACW | | | 1.107 | 1.335 | 1.662 | 3.090 | 6.033 | 36.6 |
| | IPCW x IPACW | | | 1.193 | 1.532 | 1.908 | 3.762 | 7.890 | 47.3 |
| | IPTW | | | 0.115 | 2.488 | 6.009 | 12.494 | 18.833 | 62.1 |
| | Combined | | | 0.137 | 3.790 | 10.259 | 33.805 | 107.197 | 1063.5 |
| | IPCW | Untreated | 9015065 | 1.201 | 1.323 | 1.350 | 1.463 | 1.625 | 7502.1 |
| | IPACW | | | 1.016 | 1.046 | 1.067 | 1.153 | 1.362 | 5.2 |
| | IPCW x IPACW | | | 1.229 | 1.345 | 1.372 | 1.502 | 1.654 | 8282.8 |
| | IPTW | | | 0.922 | 0.970 | 0.999 | 4.177 | 10.192 | 227.6 |
| | Combined | | | 1.146 | 1.249 | 1.291 | 5.091 | 14.386 | 9070.0 |
| 3 | IPCW | Treated | 884500 | 1.189 | 1.428 | 1.561 | 1.975 | 3.195 | 79.7 |
| | IPACW | | | 1.191 | 1.537 | 1.931 | 4.310 | 14.818 | 571.1 |
| | IPCW x IPACW | | | 1.439 | 2.127 | 2.722 | 6.507 | 29.197 | 944.8 |
| | IPTW | | | 0.115 | 2.488 | 6.009 | 12.494 | 18.833 | 62.1 |
| | Combined | | | 0.168 | 4.520 | 11.828 | 45.450 | 287.709 | 30715.1 |
| | IPCW | Untreated | 9015065 | 1.448 | 2.157 | 2.344 | 2.722 | 4.527 | 542.1 |
| | IPACW | | | 1.039 | 1.116 | 1.172 | 1.376 | 2.061 | 91.2 |
| | IPCW x IPACW | | | 1.551 | 2.259 | 2.447 | 2.845 | 4.697 | 609.0 |
| | IPTW | | | 0.922 | 0.970 | 0.999 | 4.177 | 10.192 | 227.6 |
| | Combined | | | 1.465 | 2.135 | 2.351 | 6.210 | 22.676 | 15544.2 |
| 5 | IPCW | Treated | 884500 | 1.286 | 1.678 | 1.902 | 2.664 | 5.295 | 483.4 |
| | IPACW | | | 1.243 | 1.662 | 2.082 | 4.979 | 22.062 | 3345.3 |
| | IPCW x IPACW | | | 1.641 | 2.654 | 3.496 | 8.952 | 55.417 | 7018.9 |
| | IPTW | | | 0.115 | 2.488 | 6.009 | 12.494 | 18.833 | 62.1 |
| | Combined | | | 0.198 | 4.967 | 12.585 | 52.412 | 457.156 | 76839.6 |
| | IPCW | Untreated | 9015065 | 1.579 | 2.937 | 3.442 | 4.391 | 8.263 | 955.3 |
| | IPACW | | | 1.056 | 1.175 | 1.262 | 1.568 | 2.822 | 199.5 |
| | IPCW x IPACW | | | 1.758 | 3.179 | 3.706 | 4.711 | 9.098 | 1589.3 |
| | IPTW | | | 0.922 | 0.970 | 0.999 | 4.177 | 10.192 | 227.6 |
| | Combined | | | 1.668 | 3.069 | 3.651 | 7.573 | 31.340 | 19969.7 |

| Follow-up (years) | Stage | Group | N | Median | P90 | P95 | P99 | P99.9 | Max |
|---|---|---|---|---|---|---|---|---|---|
| 10 | IPCW | Treated | 884500 | 1.418 | 2.233 | 2.666 | 4.277 | 10.821 | 1422.0 |
| | IPACW | | | 1.313 | 1.847 | 2.356 | 6.071 | 36.263 | 3506.7 |
| | IPCW x IPACW | | | 1.961 | 3.854 | 5.311 | 15.105 | 137.579 | 26171.9 |
| | IPTW | | | 0.115 | 2.488 | 6.009 | 12.494 | 18.833 | 62.1 |
| | Combined | | | 0.264 | 5.733 | 13.643 | 63.965 | 896.031 | 730838.8 |
| | IPCW | Untreated | 9015065 | 1.732 | 4.398 | 6.129 | 10.435 | 19.072 | 2338.8 |
| | IPACW | | | 1.063 | 1.290 | 1.445 | 1.999 | 4.800 | 30207.3 |
| | IPCW x IPACW | | | 1.974 | 5.166 | 7.151 | 12.064 | 23.075 | 119670.2 |
| | IPTW | | | 0.922 | 0.970 | 0.999 | 4.177 | 10.192 | 227.6 |
| | Combined | | | 1.886 | 5.102 | 7.182 | 13.098 | 60.734 | 1085571.0 |

## 6.2.9 Overlap cohort 1 — Statins — Female — Age interactions: No

**Table: ESS diagnostics**

| Follow-up | Stage | ESS (overall) | ESS% (overall) | ESS (untreated) | ESS% (untreated) | ESS (treated) | ESS% (treated) |
|---|---|---|---|---|---|---|---|
| 1 | IPCW | 3044884 | **99.7** | 2700427 | **99.7** | 344466 | **99.6** |
| | IPACW | 3043905 | **99.7** | 2702543 | **99.8** | 344492 | **99.6** |
| | IPCW×IPACW | 3034216 | **99.4** | 2695900 | **99.6** | 342103 | **98.9** |
| | IPTW | 2036883 | **66.7** | 2014411 | **74.4** | 114332 | **33.1** |
| | Combined | 1716498 | **56.2** | 1688837 | **62.4** | 108284 | **31.3** |
| 3 | IPCW | 2980494 | **97.6** | 2647016 | **97.8** | 333554 | **96.4** |
| | IPACW | 2974930 | **97.4** | 2639764 | **97.5** | 340287 | **98.4** |
| | IPCW×IPACW | 2880215 | **94.3** | 2568162 | **94.9** | 320961 | **92.8** |
| | IPTW | 2036883 | **66.7** | 2014411 | **74.4** | 114332 | **33.1** |
| | Combined | 1011249 | **33.1** | 914394 | **33.8** | 99573 | **28.8** |
| 5 | IPCW | 2828620 | **92.6** | 2521271 | **93.1** | 307909 | **89.0** |
| | IPACW | 2726186 | **89.3** | 2391481 | **88.3** | 335465 | **97.0** |
| | IPCW×IPACW | 2418213 | **79.2** | 2144125 | **79.2** | 279510 | **80.8** |
| | IPTW | 2036883 | **66.7** | 2014411 | **74.4** | 114332 | **33.1** |
| | Combined | 308266 | **10.1** | 254461 | **9.4** | 86208 | **24.9** |
| 10 | IPCW | 1037034 | **34.0** | 2072993 | **76.6** | 21792 | **6.3** |
| | IPACW | 39228 | **1.3** | 29944 | **1.1** | 321710 | **93.0** |
| | IPCW×IPACW | 18926 | **0.6** | 14677 | **0.5** | 12300 | **3.6** |
| | IPTW | 2036883 | **66.7** | 2014411 | **74.4** | 114332 | **33.1** |
| | Combined | 1501 | **0.0** | 1217 | **0.0** | 5066 | **1.5** |

**Table: Weight distributions**

| Follow-up (years) | Stage | Group | N | Median | P90 | P95 | P99 | P99.9 | Max |
|---|---|---|---|---|---|---|---|---|---|
| 1 | IPCW | Treated | 345885 | 1.092 | 1.174 | 1.215 | 1.328 | 1.635 | 8.5 |
| | IPACW | | | 1.129 | 1.221 | 1.259 | 1.381 | 1.626 | 3.0 |
| | IPCW x IPACW | | | 1.244 | 1.392 | 1.467 | 1.692 | 2.166 | 10.7 |
| | IPTW | | | 0.422 | 2.252 | 3.375 | 6.689 | 10.439 | 13.1 |
| | Combined | | | 0.522 | 2.881 | 4.399 | 8.876 | 14.602 | 24.8 |
| | IPCW | Untreated | 2707256 | 1.089 | 1.154 | 1.191 | 1.283 | 1.483 | 22.2 |
| | IPACW | | | 1.027 | 1.064 | 1.085 | 1.198 | 1.530 | 4.9 |
| | IPCW x IPACW | | | 1.125 | 1.201 | 1.247 | 1.408 | 1.791 | 22.5 |
| | IPTW | | | 0.925 | 1.084 | 1.232 | 2.689 | 9.003 | 134.0 |
| | Combined | | | 1.049 | 1.262 | 1.450 | 3.587 | 12.897 | 265.4 |
| 3 | IPCW | Treated | 345885 | 1.242 | 1.500 | 1.637 | 2.039 | 3.107 | 29.3 |
| | IPACW | | | 1.233 | 1.433 | 1.518 | 1.774 | 2.363 | 6.0 |
| | IPCW x IPACW | | | 1.582 | 2.031 | 2.280 | 3.039 | 4.948 | 36.1 |
| | IPTW | | | 0.422 | 2.252 | 3.375 | 6.689 | 10.439 | 13.1 |
| | Combined | | | 0.649 | 3.608 | 5.565 | 11.534 | 21.641 | 70.1 |
| | IPCW | Untreated | 2707256 | 1.247 | 1.448 | 1.558 | 1.883 | 2.522 | 88.4 |
| | IPACW | | | 1.070 | 1.178 | 1.238 | 1.487 | 2.846 | 113.0 |
| | IPCW x IPACW | | | 1.370 | 1.606 | 1.757 | 2.281 | 4.173 | 153.5 |
| | IPTW | | | 0.925 | 1.084 | 1.232 | 2.689 | 9.003 | 134.0 |
| | Combined | | | 1.280 | 1.667 | 1.939 | 4.892 | 22.771 | 736.2 |
| 5 | IPCW | Treated | 345885 | 1.348 | 1.817 | 2.070 | 2.884 | 5.184 | 95.7 |
| | IPACW | | | 1.291 | 1.584 | 1.710 | 2.073 | 2.986 | 7.8 |
| | IPCW x IPACW | | | 1.836 | 2.688 | 3.162 | 4.729 | 9.149 | 173.9 |
| | IPTW | | | 0.422 | 2.252 | 3.375 | 6.689 | 10.439 | 13.1 |
| | Combined | | | 0.750 | 4.153 | 6.538 | 14.042 | 29.768 | 306.4 |
| | IPCW | Untreated | 2707256 | 1.380 | 1.739 | 1.926 | 2.561 | 3.824 | 226.5 |
| | IPACW | | | 1.099 | 1.272 | 1.370 | 1.726 | 4.312 | 183.2 |
| | IPCW x IPACW | | | 1.597 | 2.030 | 2.306 | 3.268 | 7.746 | 316.0 |
| | IPTW | | | 0.925 | 1.084 | 1.232 | 2.689 | 9.003 | 134.0 |
| | Combined | | | 1.493 | 2.100 | 2.516 | 5.917 | 36.697 | 2915.2 |

| Follow-up (years) | Stage | Group | N | Median | P90 | P95 | P99 | P99.9 | Max |
|---|---|---|---|---|---|---|---|---|---|
| 10 | IPCW | Treated | 345885 | 1.428 | 2.488 | 3.002 | 4.950 | 11.424 | 2953.2 |
| | IPACW | | | 1.331 | 1.855 | 2.079 | 2.699 | 4.393 | 24.7 |
| | IPCW x IPACW | | | 1.953 | 4.323 | 5.414 | 9.860 | 25.613 | 5938.0 |
| | IPTW | | | 0.422 | 2.252 | 3.375 | 6.689 | 10.439 | 13.1 |
| | Combined | | | 0.922 | 5.195 | 8.562 | 20.856 | 57.632 | 10459.3 |
| | IPCW | Untreated | 2707256 | 1.563 | 2.420 | 2.801 | 4.303 | 8.146 | 514.8 |
| | IPACW | | | 1.131 | 1.450 | 1.624 | 2.260 | 7.660 | 17249.6 |
| | IPCW x IPACW | | | 1.900 | 3.140 | 3.756 | 6.210 | 21.242 | 44003.3 |
| | IPTW | | | 0.925 | 1.084 | 1.232 | 2.689 | 9.003 | 134.0 |
| | Combined | | | 1.842 | 3.177 | 4.037 | 8.616 | 76.790 | 161604.3 |

# 6.2.10 Overlap cohort 1 — Statins — Female — Age interactions: Yes

**Table: ESS diagnostics**

| Follow-up | Stage | ESS (overall) | ESS% (overall) | ESS (untreated) | ESS% (untreated) | ESS (treated) | ESS% (treated) |
|---|---|---|---|---|---|---|---|
| 1 | IPCW | 2737028 | **96.5** | 2393126 | **96.2** | 344081 | **99.2** |
| | IPACW | 2825843 | **99.7** | 2483940 | **99.8** | 344937 | **99.5** |
| | IPCW×IPACW | 2731968 | **96.4** | 2393242 | **96.2** | 340751 | **98.3** |
| | IPTW | 1901350 | **67.1** | 1912677 | **76.9** | 112135 | **32.3** |
| | Combined | 1594978 | **56.3** | 1614933 | **64.9** | 102749 | **29.6** |
| 3 | IPCW | 2716352 | **95.8** | 2390768 | **96.1** | 325887 | **94.0** |
| | IPACW | 2748327 | **96.9** | 2413952 | **97.0** | 338995 | **97.8** |
| | IPCW×IPACW | 2600174 | **91.7** | 2305439 | **92.6** | 306440 | **88.4** |
| | IPTW | 1901350 | **67.1** | 1912677 | **76.9** | 112135 | **32.3** |
| | Combined | 741071 | **26.1** | 654436 | **26.3** | 87376 | **25.2** |
| 5 | IPCW | 1770411 | **62.4** | 1527482 | **61.4** | 246409 | **71.1** |
| | IPACW | 1699997 | **60.0** | 1401274 | **56.3** | 329925 | **95.2** |
| | IPCW×IPACW | 1004635 | **35.4** | 829129 | **33.3** | 188460 | **54.4** |
| | IPTW | 1901350 | **67.1** | 1912677 | **76.9** | 112135 | **32.3** |
| | Combined | 12166 | **0.4** | 9338 | **0.4** | 62494 | **18.0** |
| 10 | IPCW | 749047 | **26.4** | 1406295 | **56.5** | 19422 | **5.6** |
| | IPACW | 137 | **0.0** | 106 | **0.0** | 312111 | **90.0** |
| | IPCW×IPACW | 44 | **0.0** | 34 | **0.0** | 7750 | **2.2** |
| | IPTW | 1901350 | **67.1** | 1912677 | **76.9** | 112135 | **32.3** |
| | Combined | 2 | **0.0** | 2 | **0.0** | 2173 | **0.6** |

**Table: Weight distributions**

| Follow-up (years) | Stage | Group | N | Median | P90 | P95 | P99 | P99.9 | Max |
|---|---|---|---|---|---|---|---|---|---|
| 1 | IPCW | Treated | 346705 | 1.089 | 1.180 | 1.226 | 1.360 | 1.767 | 30.1 |
| | IPACW | | | 1.128 | 1.223 | 1.265 | 1.410 | 1.748 | 10.8 |
| | IPCW x IPACW | | | 1.241 | 1.404 | 1.494 | 1.761 | 2.421 | 40.6 |
| | IPTW | | | 0.417 | 2.359 | 3.576 | 7.141 | 10.310 | 13.4 |
| | Combined | | | 0.518 | 3.008 | 4.647 | 9.713 | 15.572 | 51.9 |
| | IPCW | Untreated | 2488539 | 1.087 | 1.151 | 1.188 | 1.286 | 1.518 | 334.6 |
| | IPACW | | | 1.030 | 1.067 | 1.088 | 1.204 | 1.559 | 5.4 |
| | IPCW x IPACW | | | 1.126 | 1.202 | 1.249 | 1.436 | 1.835 | 338.0 |
| | IPTW | | | 0.918 | 1.098 | 1.281 | 2.676 | 7.823 | 166.7 |
| | Combined | | | 1.042 | 1.282 | 1.520 | 3.532 | 11.344 | 315.2 |
| 3 | IPCW | Treated | 346705 | 1.236 | 1.511 | 1.663 | 2.148 | 3.720 | 61.8 |
| | IPACW | | | 1.233 | 1.436 | 1.526 | 1.825 | 2.599 | 32.1 |
| | IPCW x IPACW | | | 1.572 | 2.053 | 2.338 | 3.273 | 5.956 | 81.7 |
| | IPTW | | | 0.417 | 2.359 | 3.576 | 7.141 | 10.310 | 13.4 |
| | Combined | | | 0.649 | 3.740 | 5.873 | 12.679 | 25.099 | 238.1 |
| | IPCW | Untreated | 2488539 | 1.240 | 1.435 | 1.546 | 1.877 | 2.692 | 175.5 |
| | IPACW | | | 1.079 | 1.186 | 1.246 | 1.503 | 2.932 | 84.3 |
| | IPCW x IPACW | | | 1.373 | 1.611 | 1.761 | 2.376 | 4.402 | 186.0 |
| | IPTW | | | 0.918 | 1.098 | 1.281 | 2.676 | 7.823 | 166.7 |
| | Combined | | | 1.273 | 1.683 | 2.025 | 4.891 | 20.852 | 2228.7 |
| 5 | IPCW | Treated | 346705 | 1.343 | 1.825 | 2.102 | 3.059 | 6.410 | 278.6 |
| | IPACW | | | 1.291 | 1.586 | 1.720 | 2.136 | 3.284 | 86.4 |
| | IPCW x IPACW | | | 1.827 | 2.707 | 3.237 | 5.131 | 11.142 | 597.9 |
| | IPTW | | | 0.417 | 2.359 | 3.576 | 7.141 | 10.310 | 13.4 |
| | Combined | | | 0.752 | 4.301 | 6.898 | 15.295 | 36.065 | 930.8 |
| | IPCW | Untreated | 2488539 | 1.369 | 1.715 | 1.905 | 2.545 | 4.176 | 1309.3 |
| | IPACW | | | 1.113 | 1.286 | 1.382 | 1.745 | 4.433 | 1266.8 |
| | IPCW x IPACW | | | 1.604 | 2.039 | 2.315 | 3.407 | 8.302 | 2670.7 |
| | IPTW | | | 0.918 | 1.098 | 1.281 | 2.676 | 7.823 | 166.7 |
| | Combined | | | 1.489 | 2.119 | 2.586 | 6.099 | 33.160 | 44023.4 |

| Follow-up (years) | Stage | Group | N | Median | P90 | P95 | P99 | P99.9 | Max |
|---|---|---|---|---|---|---|---|---|---|
| 10 | IPCW | Treated | 346705 | 1.423 | 2.488 | 3.027 | 5.203 | 13.442 | 3791.1 |
| | IPACW | | | 1.331 | 1.856 | 2.089 | 2.781 | 4.950 | 86.4 |
| | IPCW x IPACW | | | 1.948 | 4.324 | 5.467 | 10.548 | 30.998 | 9744.3 |
| | IPTW | | | 0.417 | 2.359 | 3.576 | 7.141 | 10.310 | 13.4 |
| | Combined | | | 0.929 | 5.347 | 8.974 | 22.357 | 69.464 | 17543.8 |
| | IPCW | Untreated | 2488539 | 1.544 | 2.361 | 2.737 | 4.244 | 8.826 | 885.2 |
| | IPACW | | | 1.155 | 1.474 | 1.648 | 2.279 | 7.982 | 340760.8 |
| | IPCW x IPACW | | | 1.896 | 3.170 | 3.775 | 6.353 | 22.728 | 1177150.9 |
| | IPTW | | | 0.918 | 1.098 | 1.281 | 2.676 | 7.823 | 166.7 |
| | Combined | | | 1.833 | 3.220 | 4.093 | 9.092 | 69.756 | 19403799.9 |

## 6.2.11 Overlap cohort 1 — Statins — Male — Age interactions: No

**Table: ESS diagnostics**

| Follow-up | Stage | ESS (overall) | ESS% (overall) | ESS (untreated) | ESS% (untreated) | ESS (treated) | ESS% (treated) |
|---|---|---|---|---|---|---|---|
| 1 | IPCW | 3044852 | **99.7** | 2700394 | **99.7** | 344467 | **99.6** |
| | IPACW | 3043922 | **99.7** | 2702561 | **99.8** | 344491 | **99.6** |
| | IPCW×IPACW | 3034186 | **99.4** | 2695872 | **99.6** | 342101 | **98.9** |
| | IPTW | 2036822 | **66.7** | 2014356 | **74.4** | 114330 | **33.1** |
| | Combined | 1716634 | **56.2** | 1689070 | **62.4** | 108272 | **31.3** |
| 3 | IPCW | 2980383 | **97.6** | 2646889 | **97.8** | 333568 | **96.4** |
| | IPACW | 2974956 | **97.4** | 2639790 | **97.5** | 340287 | **98.4** |
| | IPCW×IPACW | 2880133 | **94.3** | 2568057 | **94.9** | 320969 | **92.8** |
| | IPTW | 2036822 | **66.7** | 2014356 | **74.4** | 114330 | **33.1** |
| | Combined | 1010685 | **33.1** | 913858 | **33.8** | 99537 | **28.8** |
| 5 | IPCW | 2830112 | **92.7** | 2522756 | **93.2** | 307928 | **89.0** |
| | IPACW | 2726262 | **89.3** | 2391554 | **88.3** | 335468 | **97.0** |
| | IPCW×IPACW | 2418862 | **79.2** | 2144802 | **79.2** | 279505 | **80.8** |
| | IPTW | 2036822 | **66.7** | 2014356 | **74.4** | 114330 | **33.1** |
| | Combined | 307717 | **10.1** | 253988 | **9.4** | 86191 | **24.9** |
| 10 | IPCW | 1107983 | **36.3** | 2083843 | **77.0** | 24254 | **7.0** |
| | IPACW | 39286 | **1.3** | 29989 | **1.1** | 321735 | **93.0** |
| | IPCW×IPACW | 18971 | **0.6** | 14683 | **0.5** | 13300 | **3.8** |
| | IPTW | 2036822 | **66.7** | 2014356 | **74.4** | 114330 | **33.1** |
| | Combined | 1500 | **0.0** | 1216 | **0.0** | 5170 | **1.5** |

**Table: Weight distributions**

| Follow-up (years) | Stage | Group | N | Median | P90 | P95 | P99 | P99.9 | Max |
|---|---|---|---|---|---|---|---|---|---|
| 1 | IPCW | Treated | 345885 | 1.092 | 1.174 | 1.215 | 1.328 | 1.638 | 8.5 |
| | IPACW | | | 1.129 | 1.221 | 1.259 | 1.381 | 1.624 | 3.0 |
| | IPCW x IPACW | | | 1.244 | 1.392 | 1.467 | 1.693 | 2.166 | 10.7 |
| | IPTW | | | 0.423 | 2.253 | 3.375 | 6.687 | 10.448 | 13.1 |
| | Combined | | | 0.522 | 2.882 | 4.399 | 8.886 | 14.596 | 24.7 |
| | IPCW | Untreated | 2707273 | 1.089 | 1.155 | 1.192 | 1.284 | 1.485 | 22.2 |
| | IPACW | | | 1.027 | 1.064 | 1.085 | 1.198 | 1.531 | 4.9 |
| | IPCW x IPACW | | | 1.125 | 1.201 | 1.248 | 1.409 | 1.791 | 22.6 |
| | IPTW | | | 0.925 | 1.084 | 1.232 | 2.689 | 9.006 | 134.0 |
| | Combined | | | 1.049 | 1.262 | 1.450 | 3.587 | 12.894 | 263.9 |
| 3 | IPCW | Treated | 345885 | 1.242 | 1.501 | 1.638 | 2.040 | 3.100 | 29.3 |
| | IPACW | | | 1.233 | 1.433 | 1.518 | 1.775 | 2.359 | 5.9 |
| | IPCW x IPACW | | | 1.582 | 2.031 | 2.281 | 3.043 | 4.926 | 36.2 |
| | IPTW | | | 0.423 | 2.253 | 3.375 | 6.687 | 10.448 | 13.1 |
| | Combined | | | 0.648 | 3.607 | 5.567 | 11.527 | 21.636 | 69.8 |
| | IPCW | Untreated | 2707273 | 1.247 | 1.448 | 1.560 | 1.885 | 2.542 | 87.6 |
| | IPACW | | | 1.070 | 1.178 | 1.238 | 1.487 | 2.846 | 112.9 |
| | IPCW x IPACW | | | 1.369 | 1.607 | 1.758 | 2.283 | 4.177 | 153.7 |
| | IPTW | | | 0.925 | 1.084 | 1.232 | 2.689 | 9.006 | 134.0 |
| | Combined | | | 1.279 | 1.668 | 1.940 | 4.893 | 22.770 | 737.4 |
| 5 | IPCW | Treated | 345885 | 1.348 | 1.817 | 2.071 | 2.886 | 5.180 | 95.4 |
| | IPACW | | | 1.291 | 1.584 | 1.710 | 2.074 | 2.983 | 7.8 |
| | IPCW x IPACW | | | 1.835 | 2.689 | 3.163 | 4.733 | 9.169 | 173.3 |
| | IPTW | | | 0.423 | 2.253 | 3.375 | 6.687 | 10.448 | 13.1 |
| | Combined | | | 0.750 | 4.153 | 6.537 | 14.026 | 29.878 | 305.2 |
| | IPCW | Untreated | 2707273 | 1.379 | 1.739 | 1.930 | 2.566 | 3.853 | 215.9 |
| | IPACW | | | 1.099 | 1.272 | 1.370 | 1.727 | 4.311 | 183.2 |
| | IPCW x IPACW | | | 1.596 | 2.033 | 2.308 | 3.272 | 7.783 | 317.2 |
| | IPTW | | | 0.925 | 1.084 | 1.232 | 2.689 | 9.006 | 134.0 |
| | Combined | | | 1.493 | 2.101 | 2.517 | 5.923 | 36.686 | 2928.1 |

| Follow-up (years) | Stage | Group | N | Median | P90 | P95 | P99 | P99.9 | Max |
|---|---|---|---|---|---|---|---|---|---|
| 10 | IPCW | Treated | 345885 | 1.428 | 2.488 | 3.001 | 4.957 | 11.456 | 2678.2 |
| | IPACW | | | 1.331 | 1.855 | 2.079 | 2.700 | 4.395 | 24.4 |
| | IPCW x IPACW | | | 1.952 | 4.322 | 5.416 | 9.879 | 25.697 | 5895.1 |
| | IPTW | | | 0.423 | 2.253 | 3.375 | 6.687 | 10.448 | 13.1 |
| | Combined | | | 0.921 | 5.195 | 8.560 | 20.894 | 58.021 | 10379.3 |
| | IPCW | Untreated | 2707273 | 1.563 | 2.419 | 2.804 | 4.319 | 8.189 | 513.0 |
| | IPACW | | | 1.131 | 1.450 | 1.624 | 2.260 | 7.662 | 17231.0 |
| | IPCW x IPACW | | | 1.900 | 3.142 | 3.763 | 6.219 | 21.305 | 43996.6 |
| | IPTW | | | 0.925 | 1.084 | 1.232 | 2.689 | 9.006 | 134.0 |
| | Combined | | | 1.841 | 3.181 | 4.042 | 8.629 | 76.896 | 161677.7 |

### 6.2.12 Overlap cohort 1 — Statins — Male — Age interactions: Yes

**Table: ESS diagnostics**

| Follow-up | Stage | ESS (overall) | ESS% (overall) | ESS (untreated) | ESS% (untreated) | ESS (treated) | ESS% (treated) |
|---|---|---|---|---|---|---|---|
| 1 | IPCW | 2738710 | **96.6** | 2394807 | **96.2** | 344071 | **99.2** |
| | IPACW | 2825793 | **99.7** | 2483892 | **99.8** | 344936 | **99.5** |
| | IPCW×IPACW | 2733505 | **96.4** | 2394835 | **96.2** | 340734 | **98.3** |
| | IPTW | 1901244 | **67.1** | 1912579 | **76.9** | 112132 | **32.3** |
| | Combined | 1595397 | **56.3** | 1615639 | **64.9** | 102728 | **29.6** |
| 3 | IPCW | 2718422 | **95.9** | 2392803 | **96.2** | 325935 | **94.0** |
| | IPACW | 2748422 | **96.9** | 2414023 | **97.0** | 339020 | **97.8** |
| | IPCW×IPACW | 2602416 | **91.8** | 2307653 | **92.7** | 306552 | **88.4** |
| | IPTW | 1901244 | **67.1** | 1912579 | **76.9** | 112132 | **32.3** |
| | Combined | 745155 | **26.3** | 658605 | **26.5** | 87397 | **25.2** |
| 5 | IPCW | 1787709 | **63.1** | 1543184 | **62.0** | 247801 | **71.5** |
| | IPACW | 1701475 | **60.0** | 1402575 | **56.4** | 330019 | **95.2** |
| | IPCW×IPACW | 1019451 | **36.0** | 841698 | **33.8** | 190571 | **55.0** |
| | IPTW | 1901244 | **67.1** | 1912579 | **76.9** | 112132 | **32.3** |
| | Combined | 12443 | **0.4** | 9551 | **0.4** | 62474 | **18.0** |
| 10 | IPCW | 759725 | **26.8** | 1422875 | **57.2** | 19746 | **5.7** |
| | IPACW | 137 | **0.0** | 106 | **0.0** | 312253 | **90.1** |
| | IPCW×IPACW | 45 | **0.0** | 36 | **0.0** | 7891 | **2.3** |
| | IPTW | 1901244 | **67.1** | 1912579 | **76.9** | 112132 | **32.3** |
| | Combined | 2 | **0.0** | 2 | **0.0** | 2212 | **0.6** |

**Table: Weight distributions**

| Follow-up (years) | Stage | Group | N | Median | P90 | P95 | P99 | P99.9 | Max |
|---|---|---|---|---|---|---|---|---|---|
| 1 | IPCW | Treated | 346705 | 1.089 | 1.180 | 1.227 | 1.362 | 1.767 | 30.4 |
| | IPACW | | | 1.128 | 1.223 | 1.265 | 1.410 | 1.746 | 10.7 |
| | IPCW x IPACW | | | 1.241 | 1.404 | 1.494 | 1.762 | 2.422 | 40.9 |
| | IPTW | | | 0.417 | 2.360 | 3.576 | 7.144 | 10.313 | 13.4 |
| | Combined | | | 0.518 | 3.008 | 4.647 | 9.710 | 15.596 | 51.4 |
| | IPCW | Untreated | 2488491 | 1.087 | 1.152 | 1.188 | 1.287 | 1.519 | 331.2 |
| | IPACW | | | 1.030 | 1.067 | 1.088 | 1.204 | 1.559 | 5.4 |
| | IPCW x IPACW | | | 1.125 | 1.203 | 1.249 | 1.436 | 1.834 | 334.5 |
| | IPTW | | | 0.918 | 1.098 | 1.281 | 2.676 | 7.826 | 166.9 |
| | Combined | | | 1.042 | 1.282 | 1.520 | 3.532 | 11.362 | 311.9 |
| 3 | IPCW | Treated | 346705 | 1.236 | 1.511 | 1.664 | 2.150 | 3.707 | 61.9 |
| | IPACW | | | 1.233 | 1.436 | 1.526 | 1.826 | 2.592 | 31.8 |
| | IPCW x IPACW | | | 1.572 | 2.053 | 2.340 | 3.275 | 5.948 | 81.8 |
| | IPTW | | | 0.417 | 2.360 | 3.576 | 7.144 | 10.313 | 13.4 |
| | Combined | | | 0.648 | 3.741 | 5.876 | 12.695 | 25.167 | 236.5 |
| | IPCW | Untreated | 2488491 | 1.240 | 1.436 | 1.548 | 1.881 | 2.698 | 165.2 |
| | IPACW | | | 1.079 | 1.186 | 1.246 | 1.503 | 2.930 | 84.3 |
| | IPCW x IPACW | | | 1.373 | 1.612 | 1.762 | 2.379 | 4.409 | 175.1 |
| | IPTW | | | 0.918 | 1.098 | 1.281 | 2.676 | 7.826 | 166.9 |
| | Combined | | | 1.272 | 1.684 | 2.026 | 4.891 | 20.836 | 2212.9 |
| 5 | IPCW | Treated | 346705 | 1.342 | 1.826 | 2.104 | 3.063 | 6.470 | 268.2 |
| | IPACW | | | 1.291 | 1.586 | 1.720 | 2.137 | 3.288 | 85.6 |
| | IPCW x IPACW | | | 1.827 | 2.709 | 3.239 | 5.144 | 11.134 | 575.2 |
| | IPTW | | | 0.417 | 2.360 | 3.576 | 7.144 | 10.313 | 13.4 |
| | Combined | | | 0.752 | 4.301 | 6.897 | 15.297 | 36.151 | 923.4 |
| | IPCW | Untreated | 2488491 | 1.369 | 1.716 | 1.909 | 2.551 | 4.195 | 1320.0 |
| | IPACW | | | 1.113 | 1.286 | 1.382 | 1.746 | 4.434 | 1263.9 |
| | IPCW x IPACW | | | 1.604 | 2.041 | 2.319 | 3.412 | 8.299 | 2633.9 |
| | IPTW | | | 0.918 | 1.098 | 1.281 | 2.676 | 7.826 | 166.9 |
| | Combined | | | 1.488 | 2.121 | 2.588 | 6.104 | 33.186 | 43492.0 |

| Follow-up (years) | Stage | Group | N | Median | P90 | P95 | P99 | P99.9 | Max |
|---|---|---|---|---|---|---|---|---|---|
| 10 | IPCW | Treated | 346705 | 1.423 | 2.487 | 3.029 | 5.210 | 13.529 | 3758.5 |
| | IPACW | | | 1.331 | 1.856 | 2.089 | 2.779 | 4.956 | 85.6 |
| | IPCW x IPACW | | | 1.948 | 4.324 | 5.470 | 10.560 | 31.195 | 9655.2 |
| | IPTW | | | 0.417 | 2.360 | 3.576 | 7.144 | 10.313 | 13.4 |
| | Combined | | | 0.929 | 5.348 | 8.973 | 22.335 | 70.036 | 17381.2 |
| | IPCW | Untreated | 2488491 | 1.544 | 2.361 | 2.740 | 4.257 | 8.865 | 880.3 |
| | IPACW | | | 1.155 | 1.474 | 1.648 | 2.279 | 7.982 | 339493.5 |
| | IPCW x IPACW | | | 1.895 | 3.172 | 3.781 | 6.365 | 22.746 | 1152428.4 |
| | IPTW | | | 0.918 | 1.098 | 1.281 | 2.676 | 7.826 | 166.9 |
| | Combined | | | 1.832 | 3.222 | 4.098 | 9.103 | 69.673 | 19029558.8 |

## 6.2.13 Overlap cohort 1 — Antihypertensives — Female — Age interactions: No

**Table: ESS diagnostics**

| Follow-up | Stage | ESS (overall) | ESS% (overall) | ESS (untreated) | ESS% (untreated) | ESS (treated) | ESS% (treated) |
|---|---|---|---|---|---|---|---|
| 1 | IPCW | 7257921 | **99.4** | 6398298 | **99.4** | 863449 | **99.8** |
| | IPACW | 7210697 | **98.7** | 6429568 | **99.8** | 813692 | **94.0** |
| | IPCW×IPACW | 7157936 | **98.0** | 6397179 | **99.3** | 783239 | **90.5** |
| | IPTW | 2802901 | **38.4** | 2773637 | **43.1** | 164031 | **19.0** |
| | Combined | 1645771 | **22.5** | 1862635 | **28.9** | 96038 | **11.1** |
| 3 | IPCW | 6903781 | **94.5** | 6081543 | **94.4** | 849153 | **98.1** |
| | IPACW | 6742767 | **92.3** | 6307423 | **98.0** | 634367 | **73.3** |
| | IPCW×IPACW | 5488365 | **75.1** | 6011646 | **93.4** | 286001 | **33.0** |
| | IPTW | 2802901 | **38.4** | 2773637 | **43.1** | 164031 | **19.0** |
| | Combined | 284454 | **3.9** | 393008 | **6.1** | 14297 | **1.7** |
| 5 | IPCW | 6337362 | **86.8** | 5571026 | **86.5** | 821862 | **95.0** |
| | IPACW | 3811744 | **52.2** | 3857275 | **59.9** | 286938 | **33.2** |
| | IPCW×IPACW | 724451 | **9.9** | 3935692 | **61.1** | 13692 | **1.6** |
| | IPTW | 2802901 | **38.4** | 2773637 | **43.1** | 164031 | **19.0** |
| | Combined | 7744 | **0.1** | 8721 | **0.1** | 596 | **0.1** |
| 10 | IPCW | 2862546 | **39.2** | 2466857 | **38.3** | 577342 | **66.7** |
| | IPACW | 334 | **0.0** | 249 | **0.0** | 38423 | **4.4** |
| | IPCW×IPACW | 223 | **0.0** | 270 | **0.0** | 15 | **0.0** |
| | IPTW | 2802901 | **38.4** | 2773637 | **43.1** | 164031 | **19.0** |
| | Combined | 4 | **0.0** | 3 | **0.0** | 4 | **0.0** |

**Table: Weight distributions**

| Follow-up (years) | Stage | Group | N | Median | P90 | P95 | P99 | P99.9 | Max |
|---|---|---|---|---|---|---|---|---|---|
| 1 | IPCW | Treated | 865456 | 1.079 | 1.142 | 1.175 | 1.269 | 1.476 | 5.3 |
| | IPACW | | | 1.105 | 1.323 | 1.520 | 2.481 | 4.530 | 24.1 |
| | IPCW x IPACW | | | 1.199 | 1.479 | 1.727 | 3.015 | 6.091 | 40.3 |
| | IPTW | | | 0.152 | 2.891 | 5.385 | 9.390 | 10.745 | 13.5 |
| | Combined | | | 0.181 | 4.048 | 8.727 | 21.204 | 46.105 | 263.0 |
| | IPCW | Untreated | 6439280 | 1.148 | 1.293 | 1.335 | 1.415 | 1.586 | 7.3 |
| | IPACW | | | 1.021 | 1.054 | 1.076 | 1.184 | 1.478 | 6.9 |
| | IPCW x IPACW | | | 1.184 | 1.324 | 1.369 | 1.464 | 1.717 | 7.4 |
| | IPTW | | | 0.900 | 0.944 | 0.974 | 4.672 | 13.906 | 103.1 |
| | Combined | | | 1.079 | 1.210 | 1.274 | 5.893 | 21.798 | 392.1 |
| 3 | IPCW | Treated | 865456 | 1.212 | 1.397 | 1.488 | 1.767 | 2.477 | 14.5 |
| | IPACW | | | 1.189 | 1.549 | 1.811 | 3.429 | 9.791 | 219.0 |
| | IPCW x IPACW | | | 1.464 | 2.038 | 2.468 | 5.152 | 19.555 | 881.7 |
| | IPTW | | | 0.152 | 2.891 | 5.385 | 9.390 | 10.745 | 13.5 |
| | Combined | | | 0.225 | 4.919 | 9.954 | 29.872 | 138.782 | 7198.1 |
| | IPCW | Untreated | 6439280 | 1.357 | 1.912 | 2.128 | 2.584 | 3.478 | 45.7 |
| | IPACW | | | 1.050 | 1.138 | 1.194 | 1.452 | 2.446 | 90.8 |
| | IPCW x IPACW | | | 1.480 | 2.039 | 2.272 | 2.782 | 4.069 | 106.1 |
| | IPTW | | | 0.900 | 0.944 | 0.974 | 4.672 | 13.906 | 103.1 |
| | Combined | | | 1.360 | 1.915 | 2.245 | 7.626 | 36.628 | 6523.2 |
| 5 | IPCW | Treated | 865456 | 1.318 | 1.650 | 1.808 | 2.301 | 3.639 | 43.3 |
| | IPACW | | | 1.241 | 1.680 | 2.001 | 3.997 | 13.921 | 1291.0 |
| | IPCW x IPACW | | | 1.679 | 2.553 | 3.183 | 7.002 | 35.583 | 11562.7 |
| | IPTW | | | 0.152 | 2.891 | 5.385 | 9.390 | 10.745 | 13.5 |
| | Combined | | | 0.266 | 5.555 | 10.615 | 35.188 | 237.179 | 94394.0 |
| | IPCW | Untreated | 6439280 | 1.506 | 2.448 | 2.907 | 4.033 | 6.170 | 214.9 |
| | IPACW | | | 1.073 | 1.210 | 1.297 | 1.671 | 3.504 | 1417.1 |
| | IPCW x IPACW | | | 1.722 | 2.718 | 3.232 | 4.532 | 7.823 | 1807.9 |
| | IPTW | | | 0.900 | 0.944 | 0.974 | 4.672 | 13.906 | 103.1 |
| | Combined | | | 1.587 | 2.610 | 3.320 | 8.968 | 51.066 | 97161.3 |

| Follow-up (years) | Stage | Group | N | Median | P90 | P95 | P99 | P99.9 | Max |
|---|---|---|---|---|---|---|---|---|---|
| 10 | IPCW | Treated | 865456 | 1.443 | 2.232 | 2.570 | 3.635 | 6.779 | 672.0 |
| | IPACW | | | 1.307 | 1.828 | 2.341 | 4.937 | 22.742 | 5307.4 |
| | IPCW x IPACW | | | 1.979 | 3.780 | 4.969 | 11.734 | 87.232 | 960217.8 |
| | IPTW | | | 0.152 | 2.891 | 5.385 | 9.390 | 10.745 | 13.5 |
| | Combined | | | 0.355 | 6.540 | 11.898 | 44.972 | 506.002 | 7605931.7 |
| | IPCW | Untreated | 6439280 | 1.689 | 3.501 | 4.649 | 8.493 | 16.635 | 5132.1 |
| | IPACW | | | 1.097 | 1.354 | 1.515 | 2.134 | 6.085 | 458415.8 |
| | IPCW x IPACW | | | 1.959 | 4.300 | 5.697 | 10.604 | 23.686 | 909245.5 |
| | IPTW | | | 0.900 | 0.944 | 0.974 | 4.672 | 13.906 | 103.1 |
| | Combined | | | 1.841 | 4.215 | 5.901 | 13.714 | 90.992 | 42889033.1 |

## 6.2.14 Overlap cohort 1 — Antihypertensives — Female — Age interactions: Yes

**Table: ESS diagnostics**

| Follow-up | Stage | ESS (overall) | ESS% (overall) | ESS (untreated) | ESS% (untreated) | ESS (treated) | ESS% (treated) |
|---|---|---|---|---|---|---|---|
| 1 | IPCW | 3260236 | **51.0** | 2644088 | **47.8** | 855783 | **99.6** |
| | IPACW | 6313322 | **98.8** | 5525837 | **99.9** | 813530 | **94.7** |
| | IPCW×IPACW | 3093764 | **48.4** | 2468496 | **44.6** | 789908 | **92.0** |
| | IPTW | 2668687 | **41.7** | 2902870 | **52.5** | 144728 | **16.9** |
| | Combined | 1246785 | **19.5** | 1496634 | **27.0** | 85218 | **9.9** |
| 3 | IPCW | 6019259 | **94.2** | 5209969 | **94.2** | 823884 | **95.9** |
| | IPACW | 5997884 | **93.8** | 5461933 | **98.7** | 673817 | **78.5** |
| | IPCW×IPACW | 5379030 | **84.1** | 5152217 | **93.1** | 473273 | **55.1** |
| | IPTW | 2668687 | **41.7** | 2902870 | **52.5** | 144728 | **16.9** |
| | Combined | 453728 | **7.1** | 736338 | **13.3** | 26631 | **3.1** |
| 5 | IPCW | 5270068 | **82.4** | 4560055 | **82.4** | 733461 | **85.4** |
| | IPACW | 5255435 | **82.2** | 5226101 | **94.4** | 464746 | **54.1** |
| | IPCW×IPACW | 3240588 | **50.7** | 4255833 | **76.9** | 153228 | **17.8** |
| | IPTW | 2668687 | **41.7** | 2902870 | **52.5** | 144728 | **16.9** |
| | Combined | 74450 | **1.2** | 82653 | **1.5** | 7017 | **0.8** |
| 10 | IPCW | 3300262 | **51.6** | 3103507 | **56.1** | 254146 | **29.6** |
| | IPACW | 420399 | **6.6** | 328392 | **5.9** | 106837 | **12.4** |
| | IPCW×IPACW | 86870 | **1.4** | 98967 | **1.8** | 5528 | **0.6** |
| | IPTW | 2668687 | **41.7** | 2902870 | **52.5** | 144728 | **16.9** |
| | Combined | 81 | **0.0** | 53 | **0.0** | 281 | **0.0** |

**Table: Weight distributions**

| Follow-up (years) | Stage | Group | N | Median | P90 | P95 | P99 | P99.9 | Max |
|---|---|---|---|---|---|---|---|---|---|
| 1 | IPCW | Treated | 858861 | 1.074 | 1.152 | 1.193 | 1.304 | 1.584 | 11.3 |
| | IPACW | | | 1.104 | 1.286 | 1.484 | 2.431 | 4.355 | 13.7 |
| | IPCW x IPACW | | | 1.190 | 1.468 | 1.711 | 2.907 | 5.728 | 19.4 |
| | IPTW | | | 0.173 | 2.783 | 5.197 | 11.609 | 14.139 | 17.1 |
| | Combined | | | 0.205 | 3.861 | 8.648 | 23.693 | 54.728 | 251.4 |
| | IPCW | Untreated | 5533489 | 1.134 | 1.266 | 1.306 | 1.381 | 1.503 | 2822.0 |
| | IPACW | | | 1.023 | 1.062 | 1.088 | 1.197 | 1.417 | 5.1 |
| | IPCW x IPACW | | | 1.172 | 1.303 | 1.347 | 1.437 | 1.668 | 3107.0 |
| | IPTW | | | 0.886 | 0.941 | 0.978 | 4.768 | 11.538 | 105.9 |
| | Combined | | | 1.051 | 1.183 | 1.263 | 6.027 | 16.873 | 3222.6 |
| 3 | IPCW | Treated | 858861 | 1.198 | 1.415 | 1.538 | 1.924 | 3.042 | 92.4 |
| | IPACW | | | 1.188 | 1.486 | 1.751 | 3.422 | 9.485 | 114.4 |
| | IPCW x IPACW | | | 1.444 | 2.043 | 2.510 | 5.120 | 18.045 | 213.5 |
| | IPTW | | | 0.173 | 2.783 | 5.197 | 11.609 | 14.139 | 17.1 |
| | Combined | | | 0.251 | 4.654 | 10.293 | 33.241 | 146.092 | 3478.6 |
| | IPCW | Untreated | 5533489 | 1.331 | 1.798 | 1.985 | 2.408 | 2.966 | 319.7 |
| | IPACW | | | 1.055 | 1.158 | 1.223 | 1.496 | 2.266 | 85.9 |
| | IPCW x IPACW | | | 1.457 | 1.943 | 2.153 | 2.648 | 3.582 | 357.9 |
| | IPTW | | | 0.886 | 0.941 | 0.978 | 4.768 | 11.538 | 105.9 |
| | Combined | | | 1.315 | 1.812 | 2.168 | 8.023 | 27.976 | 3528.7 |
| 5 | IPCW | Treated | 858861 | 1.302 | 1.664 | 1.872 | 2.585 | 4.934 | 187.2 |
| | IPACW | | | 1.241 | 1.611 | 1.917 | 4.015 | 13.873 | 287.9 |
| | IPCW x IPACW | | | 1.655 | 2.564 | 3.273 | 7.268 | 33.495 | 997.5 |
| | IPTW | | | 0.173 | 2.783 | 5.197 | 11.609 | 14.139 | 17.1 |
| | Combined | | | 0.295 | 5.196 | 11.188 | 39.371 | 251.551 | 8742.4 |
| | IPCW | Untreated | 5533489 | 1.474 | 2.263 | 2.642 | 3.610 | 5.040 | 715.9 |
| | IPACW | | | 1.080 | 1.238 | 1.339 | 1.737 | 3.229 | 223.8 |
| | IPCW x IPACW | | | 1.693 | 2.565 | 3.007 | 4.197 | 6.548 | 851.0 |
| | IPTW | | | 0.886 | 0.941 | 0.978 | 4.768 | 11.538 | 105.9 |
| | Combined | | | 1.535 | 2.450 | 3.161 | 9.541 | 40.344 | 16838.2 |

| Follow-up (years) | Stage | Group | N | Median | P90 | P95 | P99 | P99.9 | Max |
|---|---|---|---|---|---|---|---|---|---|
| 10 | IPCW | Treated | 858861 | 1.441 | 2.228 | 2.624 | 4.135 | 9.790 | 1786.5 |
| | IPACW | | | 1.309 | 1.772 | 2.205 | 4.957 | 24.669 | 1440.1 |
| | IPCW x IPACW | | | 1.969 | 3.767 | 5.061 | 12.756 | 83.377 | 17014.5 |
| | IPTW | | | 0.173 | 2.783 | 5.197 | 11.609 | 14.139 | 17.1 |
| | Combined | | | 0.396 | 6.091 | 12.632 | 50.752 | 607.705 | 233297.7 |
| | IPCW | Untreated | 5533489 | 1.654 | 3.199 | 4.108 | 7.171 | 13.278 | 1548.6 |
| | IPACW | | | 1.114 | 1.396 | 1.580 | 2.257 | 5.948 | 7043.7 |
| | IPCW x IPACW | | | 1.956 | 4.045 | 5.244 | 9.448 | 20.061 | 23951.9 |
| | IPTW | | | 0.886 | 0.941 | 0.978 | 4.768 | 11.538 | 105.9 |
| | Combined | | | 1.821 | 3.964 | 5.521 | 13.440 | 75.927 | 1908821.5 |

# 6.2.15 Overlap cohort 1 — Antihypertensives — Male — Age interactions: No

**Table: ESS diagnostics**

| Follow-up | Stage | ESS (overall) | ESS% (overall) | ESS (untreated) | ESS% (untreated) | ESS (treated) | ESS% (treated) |
|---|---|---|---|---|---|---|---|
| 1 | IPCW | 7229486 | **99.4** | 6369821 | **99.4** | 863431 | **99.8** |
| | IPACW | 7182088 | **98.7** | 6400663 | **99.8** | 813842 | **94.0** |
| | IPCW×IPACW | 7130129 | **98.0** | 6368667 | **99.3** | 783823 | **90.6** |
| | IPTW | 2795605 | **38.4** | 2763783 | **43.1** | 164666 | **19.0** |
| | Combined | 1645285 | **22.6** | 1858612 | **29.0** | 96557 | **11.2** |
| 3 | IPCW | 6879700 | **94.6** | 6057051 | **94.5** | 849120 | **98.1** |
| | IPACW | 6721131 | **92.4** | 6280338 | **98.0** | 636655 | **73.6** |
| | IPCW×IPACW | 5507350 | **75.7** | 5988115 | **93.4** | 293707 | **33.9** |
| | IPTW | 2795605 | **38.4** | 2763783 | **43.1** | 164666 | **19.0** |
| | Combined | 294254 | **4.0** | 405206 | **6.3** | 14865 | **1.7** |
| 5 | IPCW | 6320495 | **86.9** | 5553211 | **86.6** | 821841 | **95.0** |
| | IPACW | 3996878 | **54.9** | 4092768 | **63.8** | 293064 | **33.9** |
| | IPCW×IPACW | 776343 | **10.7** | 4085817 | **63.7** | 14866 | **1.7** |
| | IPTW | 2795605 | **38.4** | 2763783 | **43.1** | 164666 | **19.0** |
| | Combined | 8676 | **0.1** | 9968 | **0.2** | 648 | **0.1** |
| 10 | IPCW | 2866559 | **39.4** | 2468684 | **38.5** | 577405 | **66.7** |
| | IPACW | 428 | **0.0** | 318 | **0.0** | 39749 | **4.6** |
| | IPCW×IPACW | 260 | **0.0** | 337 | **0.0** | 17 | **0.0** |
| | IPTW | 2795605 | **38.4** | 2763783 | **43.1** | 164666 | **19.0** |
| | Combined | 5 | **0.0** | 3 | **0.0** | 4 | **0.0** |

**Table: Weight distributions**

| Follow-up (years) | Stage | Group | N | Median | P90 | P95 | P99 | P99.9 | Max |
|---|---|---|---|---|---|---|---|---|---|
| 1 | IPCW | Treated | 865437 | 1.079 | 1.142 | 1.175 | 1.269 | 1.476 | 5.4 |
| | IPACW | | | 1.105 | 1.323 | 1.520 | 2.484 | 4.533 | 23.8 |
| | IPCW x IPACW | | | 1.199 | 1.479 | 1.726 | 3.014 | 6.073 | 39.7 |
| | IPTW | | | 0.153 | 2.891 | 5.362 | 9.361 | 10.723 | 13.3 |
| | Combined | | | 0.182 | 4.045 | 8.674 | 21.099 | 45.908 | 257.9 |
| | IPCW | Untreated | 6410348 | 1.147 | 1.292 | 1.334 | 1.414 | 1.586 | 7.4 |
| | IPACW | | | 1.021 | 1.054 | 1.076 | 1.185 | 1.477 | 6.7 |
| | IPCW x IPACW | | | 1.184 | 1.323 | 1.368 | 1.464 | 1.716 | 7.5 |
| | IPTW | | | 0.900 | 0.943 | 0.974 | 4.678 | 13.916 | 102.8 |
| | Combined | | | 1.079 | 1.209 | 1.273 | 5.902 | 21.835 | 386.9 |
| 3 | IPCW | Treated | 865437 | 1.213 | 1.397 | 1.489 | 1.768 | 2.477 | 14.6 |
| | IPACW | | | 1.189 | 1.549 | 1.811 | 3.429 | 9.815 | 213.2 |
| | IPCW x IPACW | | | 1.464 | 2.038 | 2.467 | 5.149 | 19.548 | 848.9 |
| | IPTW | | | 0.153 | 2.891 | 5.362 | 9.361 | 10.723 | 13.3 |
| | Combined | | | 0.226 | 4.916 | 9.910 | 29.803 | 137.833 | 6881.2 |
| | IPCW | Untreated | 6410348 | 1.356 | 1.907 | 2.121 | 2.578 | 3.471 | 45.0 |
| | IPACW | | | 1.050 | 1.139 | 1.195 | 1.454 | 2.444 | 83.1 |
| | IPCW x IPACW | | | 1.480 | 2.034 | 2.266 | 2.776 | 4.062 | 97.2 |
| | IPTW | | | 0.900 | 0.943 | 0.974 | 4.678 | 13.916 | 102.8 |
| | Combined | | | 1.358 | 1.911 | 2.240 | 7.649 | 36.591 | 6321.9 |
| 5 | IPCW | Treated | 865437 | 1.318 | 1.650 | 1.809 | 2.303 | 3.638 | 43.7 |
| | IPACW | | | 1.241 | 1.680 | 2.001 | 3.994 | 13.973 | 1246.1 |
| | IPCW x IPACW | | | 1.679 | 2.553 | 3.184 | 6.988 | 35.650 | 10955.2 |
| | IPTW | | | 0.153 | 2.891 | 5.362 | 9.361 | 10.723 | 13.3 |
| | Combined | | | 0.268 | 5.553 | 10.568 | 35.177 | 235.290 | 88802.1 |
| | IPCW | Untreated | 6410348 | 1.505 | 2.439 | 2.895 | 4.015 | 6.147 | 213.8 |
| | IPACW | | | 1.073 | 1.211 | 1.299 | 1.674 | 3.498 | 1227.6 |
| | IPCW x IPACW | | | 1.721 | 2.710 | 3.221 | 4.514 | 7.802 | 1568.7 |
| | IPTW | | | 0.900 | 0.943 | 0.974 | 4.678 | 13.916 | 102.8 |
| | Combined | | | 1.585 | 2.601 | 3.309 | 8.992 | 51.043 | 92560.5 |

| Follow-up (years) | Stage | Group | N | Median | P90 | P95 | P99 | P99.9 | Max |
|---|---|---|---|---|---|---|---|---|---|
| 10 | IPCW | Treated | 865437 | 1.443 | 2.233 | 2.571 | 3.638 | 6.775 | 676.8 |
| | IPACW | | | 1.307 | 1.828 | 2.341 | 4.931 | 22.736 | 5090.6 |
| | IPCW x IPACW | | | 1.980 | 3.782 | 4.970 | 11.707 | 86.782 | 910083.9 |
| | IPTW | | | 0.153 | 2.891 | 5.362 | 9.361 | 10.723 | 13.3 |
| | Combined | | | 0.356 | 6.528 | 11.868 | 44.817 | 500.573 | 7263318.5 |
| | IPCW | Untreated | 6410348 | 1.688 | 3.489 | 4.624 | 8.438 | 16.520 | 5096.5 |
| | IPACW | | | 1.097 | 1.356 | 1.518 | 2.140 | 6.089 | 382025.9 |
| | IPCW x IPACW | | | 1.959 | 4.292 | 5.677 | 10.550 | 23.594 | 759067.2 |
| | IPTW | | | 0.900 | 0.943 | 0.974 | 4.678 | 13.916 | 102.8 |
| | Combined | | | 1.841 | 4.206 | 5.883 | 13.678 | 91.221 | 35614047.4 |

## 6.2.16 Overlap cohort 1 — Antihypertensives — Male — Age interactions: Yes

**Table: ESS diagnostics**

| Follow-up | Stage | ESS (overall) | ESS% (overall) | ESS (untreated) | ESS% (untreated) | ESS (treated) | ESS% (treated) |
|---|---|---|---|---|---|---|---|
| 1 | IPCW | 3370751 | **53.0** | 2737615 | **49.7** | 855918 | **99.6** |
| | IPACW | 6278856 | **98.7** | 5497003 | **99.9** | 809471 | **94.2** |
| | IPCW×IPACW | 3201717 | **50.3** | 2561762 | **46.5** | 785245 | **91.4** |
| | IPTW | 2662678 | **41.8** | 2880443 | **52.3** | 146573 | **17.1** |
| | Combined | 1250917 | **19.7** | 1526865 | **27.7** | 83826 | **9.8** |
| 3 | IPCW | 5996662 | **94.2** | 5186645 | **94.2** | 824234 | **96.0** |
| | IPACW | 5411067 | **85.0** | 5436192 | **98.8** | 459312 | **53.5** |
| | IPCW×IPACW | 4595103 | **72.2** | 5132956 | **93.2** | 278471 | **32.4** |
| | IPTW | 2662678 | **41.8** | 2880443 | **52.3** | 146573 | **17.1** |
| | Combined | 301243 | **4.7** | 786862 | **14.3** | 14174 | **1.7** |
| 5 | IPCW | 5265492 | **82.7** | 4553788 | **82.7** | 734083 | **85.5** |
| | IPACW | 2383809 | **37.5** | 5213725 | **94.7** | 91878 | **10.7** |
| | IPCW×IPACW | 1233791 | **19.4** | 4268119 | **77.5** | 33462 | **3.9** |
| | IPTW | 2662678 | **41.8** | 2880443 | **52.3** | 146573 | **17.1** |
| | Combined | 21750 | **0.3** | 95249 | **1.7** | 1063 | **0.1** |
| 10 | IPCW | 3294975 | **51.8** | 3118269 | **56.6** | 246891 | **28.7** |
| | IPACW | 560792 | **8.8** | 565484 | **10.3** | 49914 | **5.8** |
| | IPCW×IPACW | 120242 | **1.9** | 181340 | **3.3** | 5518 | **0.6** |
| | IPTW | 2662678 | **41.8** | 2880443 | **52.3** | 146573 | **17.1** |
| | Combined | 149 | **0.0** | 95 | **0.0** | 311 | **0.0** |

**Table: Weight distributions**

| Follow-up (years) | Stage | Group | N | Median | P90 | P95 | P99 | P99.9 | Max |
|---|---|---|---|---|---|---|---|---|---|
| 1 | IPCW | Treated | 858996 | 1.074 | 1.152 | 1.193 | 1.304 | 1.584 | 11.5 |
| | IPACW | | | 1.104 | 1.286 | 1.486 | 2.438 | 4.413 | 44.0 |
| | IPCW x IPACW | | | 1.190 | 1.468 | 1.713 | 2.912 | 5.782 | 60.9 |
| | IPTW | | | 0.174 | 2.796 | 5.184 | 11.453 | 14.011 | 16.9 |
| | Combined | | | 0.206 | 3.881 | 8.624 | 23.408 | 53.974 | 732.3 |
| | IPCW | Untreated | 5504677 | 1.133 | 1.264 | 1.304 | 1.382 | 1.505 | 2704.6 |
| | IPACW | | | 1.023 | 1.063 | 1.088 | 1.198 | 1.419 | 4.7 |
| | IPCW x IPACW | | | 1.172 | 1.302 | 1.346 | 1.438 | 1.670 | 2978.7 |
| | IPTW | | | 0.886 | 0.940 | 0.977 | 4.778 | 11.544 | 100.6 |
| | Combined | | | 1.050 | 1.182 | 1.264 | 6.043 | 16.907 | 3085.5 |
| 3 | IPCW | Treated | 858996 | 1.198 | 1.415 | 1.539 | 1.926 | 3.053 | 89.2 |
| | IPACW | | | 1.188 | 1.487 | 1.754 | 3.426 | 9.617 | 731.4 |
| | IPCW x IPACW | | | 1.444 | 2.044 | 2.513 | 5.131 | 17.998 | 1389.4 |
| | IPTW | | | 0.174 | 2.796 | 5.184 | 11.453 | 14.011 | 16.9 |
| | Combined | | | 0.252 | 4.680 | 10.263 | 32.832 | 142.724 | 9792.7 |
| | IPCW | Untreated | 5504677 | 1.331 | 1.790 | 1.977 | 2.411 | 2.973 | 313.1 |
| | IPACW | | | 1.055 | 1.159 | 1.224 | 1.497 | 2.272 | 61.2 |
| | IPCW x IPACW | | | 1.456 | 1.936 | 2.146 | 2.651 | 3.593 | 350.6 |
| | IPTW | | | 0.886 | 0.940 | 0.977 | 4.778 | 11.544 | 100.6 |
| | Combined | | | 1.313 | 1.805 | 2.168 | 8.051 | 28.125 | 3531.3 |
| 5 | IPCW | Treated | 858996 | 1.302 | 1.664 | 1.873 | 2.590 | 4.969 | 170.2 |
| | IPACW | | | 1.241 | 1.611 | 1.918 | 4.030 | 13.965 | 3355.1 |
| | IPCW x IPACW | | | 1.655 | 2.566 | 3.278 | 7.286 | 33.247 | 7989.2 |
| | IPTW | | | 0.174 | 2.796 | 5.184 | 11.453 | 14.011 | 16.9 |
| | Combined | | | 0.297 | 5.219 | 11.129 | 38.976 | 244.428 | 77485.0 |
| | IPCW | Untreated | 5504677 | 1.473 | 2.250 | 2.625 | 3.613 | 5.059 | 734.8 |
| | IPACW | | | 1.080 | 1.239 | 1.341 | 1.739 | 3.244 | 195.2 |
| | IPCW x IPACW | | | 1.692 | 2.554 | 2.991 | 4.202 | 6.580 | 870.5 |
| | IPTW | | | 0.886 | 0.940 | 0.977 | 4.778 | 11.544 | 100.6 |
| | Combined | | | 1.533 | 2.438 | 3.155 | 9.575 | 40.565 | 13453.5 |

| Follow-up (years) | Stage | Group | N | Median | P90 | P95 | P99 | P99.9 | Max |
|---|---|---|---|---|---|---|---|---|---|
| 10 | IPCW | Treated | 858996 | 1.440 | 2.229 | 2.627 | 4.145 | 9.898 | 1797.4 |
| | IPACW | | | 1.308 | 1.773 | 2.207 | 4.965 | 24.474 | 3355.1 |
| | IPCW x IPACW | | | 1.968 | 3.769 | 5.067 | 12.769 | 83.797 | 15268.1 |
| | IPTW | | | 0.174 | 2.796 | 5.184 | 11.453 | 14.011 | 16.9 |
| | Combined | | | 0.398 | 6.115 | 12.561 | 50.497 | 589.132 | 208160.2 |
| | IPCW | Untreated | 5504677 | 1.653 | 3.183 | 4.074 | 7.131 | 13.390 | 1564.3 |
| | IPACW | | | 1.115 | 1.398 | 1.584 | 2.262 | 5.980 | 4554.2 |
| | IPCW x IPACW | | | 1.956 | 4.035 | 5.219 | 9.415 | 20.194 | 15730.1 |
| | IPTW | | | 0.886 | 0.940 | 0.977 | 4.778 | 11.544 | 100.6 |
| | Combined | | | 1.821 | 3.953 | 5.496 | 13.507 | 76.796 | 1268437.1 |

## 6.2.17 Overlap cohort 2 — Statins — Female — Age interactions: No

**Table: ESS diagnostics**

| Follow-up | Stage | ESS (overall) | ESS% (overall) | ESS (untreated) | ESS% (untreated) | ESS (treated) | ESS% (treated) |
|---|---|---|---|---|---|---|---|
| 1 | IPCW | 3044884 | **99.7** | 2700427 | **99.7** | 344466 | **99.6** |
| | IPACW | 3043905 | **99.7** | 2702543 | **99.8** | 344492 | **99.6** |
| | IPCW×IPACW | 3034216 | **99.4** | 2695900 | **99.6** | 342103 | **98.9** |
| | IPTW | 2036883 | **66.7** | 2014411 | **74.4** | 114332 | **33.1** |
| | Combined | 1716498 | **56.2** | 1688837 | **62.4** | 108284 | **31.3** |
| 3 | IPCW | 2980459 | **97.6** | 2646975 | **97.8** | 333559 | **96.4** |
| | IPACW | 2988531 | **97.9** | 2653848 | **98.0** | 340287 | **98.4** |
| | IPCW×IPACW | 2895220 | **94.8** | 2583893 | **95.4** | 320966 | **92.8** |
| | IPTW | 2037266 | **66.7** | 2015074 | **74.4** | 114293 | **33.0** |
| | Combined | 1055264 | **34.6** | 960587 | **35.5** | 99539 | **28.8** |
| 5 | IPCW | 2828593 | **92.7** | 2521358 | **93.1** | 307807 | **89.0** |
| | IPACW | 2871619 | **94.1** | 2540144 | **93.8** | 335461 | **97.0** |
| | IPCW×IPACW | 2571696 | **84.2** | 2305917 | **85.2** | 279323 | **80.8** |
| | IPTW | 2039613 | **66.8** | 2019273 | **74.6** | 114021 | **33.0** |
| | Combined | 404933 | **13.3** | 339122 | **12.5** | 85923 | **24.8** |
| 10 | IPCW | 1278723 | **41.9** | 2174320 | **80.4** | 30156 | **8.7** |
| | IPACW | 2421715 | **79.4** | 2100336 | **77.7** | 321620 | **93.0** |
| | IPCW×IPACW | 776372 | **25.5** | 1351842 | **50.0** | 24268 | **7.0** |
| | IPTW | 2019848 | **66.3** | 2005995 | **74.2** | 111841 | **32.3** |
| | Combined | 71809 | **2.4** | 58511 | **2.2** | 31748 | **9.2** |

**Table: Weight distributions**

| Follow-up (years) | Stage | Group | N | Median | P90 | P95 | P99 | P99.9 | Max |
|---|---|---|---|---|---|---|---|---|---|
| 1 | IPCW | Treated | 345885 | 1.092 | 1.174 | 1.215 | 1.328 | 1.635 | 8.5 |
| | IPACW | | | 1.129 | 1.221 | 1.259 | 1.381 | 1.626 | 3.0 |
| | IPCW x IPACW | | | 1.244 | 1.392 | 1.467 | 1.692 | 2.166 | 10.7 |
| | IPTW | | | 0.422 | 2.252 | 3.375 | 6.689 | 10.439 | 13.1 |
| | Combined | | | 0.522 | 2.881 | 4.399 | 8.876 | 14.602 | 24.8 |
| | IPCW | Untreated | 2707256 | 1.089 | 1.154 | 1.191 | 1.283 | 1.483 | 22.2 |
| | IPACW | | | 1.027 | 1.064 | 1.085 | 1.198 | 1.530 | 4.9 |
| | IPCW x IPACW | | | 1.125 | 1.201 | 1.247 | 1.408 | 1.791 | 22.5 |
| | IPTW | | | 0.925 | 1.084 | 1.232 | 2.689 | 9.003 | 134.0 |
| | Combined | | | 1.049 | 1.262 | 1.450 | 3.587 | 12.897 | 265.4 |
| 3 | IPCW | Treated | 345885 | 1.242 | 1.500 | 1.637 | 2.039 | 3.106 | 29.2 |
| | IPACW | | | 1.233 | 1.433 | 1.518 | 1.774 | 2.363 | 6.0 |
| | IPCW x IPACW | | | 1.582 | 2.031 | 2.280 | 3.039 | 4.947 | 36.1 |
| | IPTW | | | 0.422 | 2.252 | 3.375 | 6.691 | 10.444 | 13.1 |
| | Combined | | | 0.649 | 3.608 | 5.566 | 11.537 | 21.651 | 70.2 |
| | IPCW | Untreated | 2707209 | 1.247 | 1.448 | 1.558 | 1.883 | 2.522 | 88.5 |
| | IPACW | | | 1.070 | 1.178 | 1.238 | 1.488 | 2.841 | 26.8 |
| | IPCW x IPACW | | | 1.370 | 1.606 | 1.757 | 2.281 | 4.169 | 97.7 |
| | IPTW | | | 0.925 | 1.084 | 1.233 | 2.687 | 9.002 | 134.2 |
| | Combined | | | 1.280 | 1.667 | 1.940 | 4.891 | 22.771 | 614.5 |
| 5 | IPCW | Treated | 345881 | 1.348 | 1.816 | 2.069 | 2.882 | 5.181 | 96.9 |
| | IPACW | | | 1.291 | 1.584 | 1.710 | 2.073 | 2.986 | 7.8 |
| | IPCW x IPACW | | | 1.836 | 2.688 | 3.162 | 4.727 | 9.153 | 176.5 |
| | IPTW | | | 0.422 | 2.254 | 3.381 | 6.705 | 10.481 | 13.2 |
| | Combined | | | 0.749 | 4.156 | 6.548 | 14.075 | 29.817 | 312.1 |
| | IPCW | Untreated | 2706828 | 1.380 | 1.739 | 1.926 | 2.561 | 3.825 | 228.5 |
| | IPACW | | | 1.098 | 1.272 | 1.370 | 1.729 | 4.273 | 71.4 |
| | IPCW x IPACW | | | 1.597 | 2.031 | 2.306 | 3.269 | 7.699 | 256.1 |
| | IPTW | | | 0.925 | 1.084 | 1.233 | 2.682 | 9.019 | 138.4 |
| | Combined | | | 1.493 | 2.100 | 2.518 | 5.912 | 36.269 | 2909.8 |

| Follow-up (years) | Stage | Group | N | Median | P90 | P95 | P99 | P99.9 | Max |
|---|---|---|---|---|---|---|---|---|---|
| 10 | IPCW | Treated | 345786 | 1.428 | 2.486 | 2.998 | 4.932 | 11.329 | 3126.4 |
| | IPACW | | | 1.332 | 1.855 | 2.079 | 2.699 | 4.394 | 24.7 |
| | IPCW x IPACW | | | 1.952 | 4.320 | 5.407 | 9.835 | 25.516 | 5223.6 |
| | IPTW | | | 0.416 | 2.263 | 3.411 | 6.849 | 10.672 | 13.4 |
| | Combined | | | 0.910 | 5.218 | 8.643 | 21.074 | 58.520 | 2474.5 |
| | IPCW | Untreated | 2702758 | 1.564 | 2.421 | 2.803 | 4.304 | 8.149 | 271.5 |
| | IPACW | | | 1.130 | 1.450 | 1.626 | 2.263 | 7.130 | 129.2 |
| | IPCW x IPACW | | | 1.901 | 3.141 | 3.759 | 6.207 | 20.168 | 473.2 |
| | IPTW | | | 0.925 | 1.085 | 1.232 | 2.642 | 9.302 | 75.9 |
| | Combined | | | 1.840 | 3.175 | 4.039 | 8.578 | 74.029 | 10238.8 |

## 6.2.18 Overlap cohort 2 — Statins — Female — Age interactions: Yes

**Table: ESS diagnostics**

| Follow-up | Stage | ESS (overall) | ESS% (overall) | ESS (untreated) | ESS% (untreated) | ESS (treated) | ESS% (treated) |
|---|---|---|---|---|---|---|---|
| 1 | IPCW | 2825611 | **99.7** | 2481698 | **99.7** | 343959 | **99.2** |
| | IPACW | 2825842 | **99.7** | 2483939 | **99.8** | 344937 | **99.5** |
| | IPCW×IPACW | 2813141 | **99.2** | 2476861 | **99.5** | 340588 | **98.2** |
| | IPTW | 1901349 | **67.1** | 1912676 | **76.9** | 112135 | **32.3** |
| | Combined | 1617844 | **57.1** | 1645869 | **66.1** | 102736 | **29.6** |
| 3 | IPCW | 2719372 | **95.9** | 2393358 | **96.2** | 326312 | **94.1** |
| | IPACW | 2755159 | **97.2** | 2421044 | **97.3** | 338984 | **97.8** |
| | IPCW×IPACW | 2612801 | **92.2** | 2318109 | **93.2** | 306933 | **88.5** |
| | IPTW | 1901281 | **67.1** | 1912744 | **76.9** | 112109 | **32.3** |
| | Combined | 967817 | **34.1** | 899300 | **36.1** | 87321 | **25.2** |
| 5 | IPCW | 2601515 | **91.8** | 2316701 | **93.1** | 288977 | **83.4** |
| | IPACW | 2622951 | **92.5** | 2296672 | **92.3** | 329720 | **95.1** |
| | IPCW×IPACW | 2276885 | **80.3** | 2075054 | **83.4** | 237962 | **68.7** |
| | IPTW | 1913904 | **67.5** | 1930580 | **77.6** | 111856 | **32.3** |
| | Combined | 425684 | **15.0** | 365988 | **14.7** | 60411 | **17.4** |
| 10 | IPCW | 2197561 | **77.7** | 1962409 | **79.0** | 239300 | **69.2** |
| | IPACW | 2133035 | **75.4** | 1821191 | **73.3** | 313243 | **90.6** |
| | IPCW×IPACW | 1279807 | **45.2** | 1139085 | **45.9** | 149616 | **43.3** |
| | IPTW | 1898853 | **67.1** | 1932707 | **77.8** | 108327 | **31.3** |
| | Combined | 73118 | **2.6** | 59981 | **2.4** | 17011 | **4.9** |

**Table: Weight distributions**

| Follow-up (years) | Stage | Group | N | Median | P90 | P95 | P99 | P99.9 | Max |
|---|---|---|---|---|---|---|---|---|---|
| 1 | IPCW | Treated | 346705 | 1.089 | 1.180 | 1.226 | 1.360 | 1.770 | 32.3 |
| | IPACW | | | 1.128 | 1.223 | 1.265 | 1.410 | 1.748 | 10.8 |
| | IPCW x IPACW | | | 1.241 | 1.404 | 1.494 | 1.761 | 2.423 | 43.4 |
| | IPTW | | | 0.417 | 2.359 | 3.576 | 7.141 | 10.310 | 13.4 |
| | Combined | | | 0.518 | 3.008 | 4.647 | 9.713 | 15.572 | 51.9 |
| | IPCW | Untreated | 2488538 | 1.087 | 1.151 | 1.188 | 1.286 | 1.518 | 17.3 |
| | IPACW | | | 1.030 | 1.067 | 1.088 | 1.204 | 1.559 | 5.4 |
| | IPCW x IPACW | | | 1.126 | 1.202 | 1.249 | 1.436 | 1.835 | 17.7 |
| | IPTW | | | 0.918 | 1.098 | 1.281 | 2.676 | 7.823 | 166.7 |
| | Combined | | | 1.042 | 1.282 | 1.520 | 3.532 | 11.352 | 229.1 |
| 3 | IPCW | Treated | 346695 | 1.236 | 1.511 | 1.663 | 2.149 | 3.707 | 68.4 |
| | IPACW | | | 1.233 | 1.436 | 1.526 | 1.825 | 2.599 | 32.1 |
| | IPCW x IPACW | | | 1.572 | 2.053 | 2.339 | 3.275 | 5.965 | 90.7 |
| | IPTW | | | 0.417 | 2.359 | 3.576 | 7.143 | 10.318 | 13.4 |
| | Combined | | | 0.648 | 3.740 | 5.874 | 12.683 | 25.090 | 238.5 |
| | IPCW | Untreated | 2488502 | 1.240 | 1.435 | 1.546 | 1.878 | 2.694 | 230.2 |
| | IPACW | | | 1.079 | 1.186 | 1.246 | 1.503 | 2.932 | 86.9 |
| | IPCW x IPACW | | | 1.373 | 1.611 | 1.761 | 2.377 | 4.400 | 243.9 |
| | IPTW | | | 0.918 | 1.098 | 1.281 | 2.676 | 7.818 | 167.3 |
| | Combined | | | 1.272 | 1.683 | 2.025 | 4.890 | 20.819 | 588.6 |
| 5 | IPCW | Treated | 346612 | 1.342 | 1.826 | 2.103 | 3.062 | 6.315 | 176.7 |
| | IPACW | | | 1.291 | 1.586 | 1.720 | 2.136 | 3.292 | 91.0 |
| | IPCW x IPACW | | | 1.827 | 2.708 | 3.238 | 5.141 | 10.978 | 326.3 |
| | IPTW | | | 0.417 | 2.361 | 3.579 | 7.165 | 10.348 | 13.5 |
| | Combined | | | 0.751 | 4.303 | 6.906 | 15.324 | 36.025 | 995.1 |
| | IPCW | Untreated | 2488234 | 1.369 | 1.715 | 1.906 | 2.548 | 4.184 | 93.5 |
| | IPACW | | | 1.113 | 1.286 | 1.382 | 1.747 | 4.429 | 71.4 |
| | IPCW x IPACW | | | 1.604 | 2.040 | 2.317 | 3.411 | 8.281 | 134.6 |
| | IPTW | | | 0.918 | 1.098 | 1.281 | 2.672 | 7.796 | 171.2 |
| | Combined | | | 1.489 | 2.120 | 2.587 | 6.096 | 32.813 | 2013.4 |

| Follow-up (years) | Stage | Group | N | Median | P90 | P95 | P99 | P99.9 | Max |
|---|---|---|---|---|---|---|---|---|---|
| 10 | IPCW | Treated | 345715 | 1.423 | 2.484 | 3.021 | 5.165 | 12.824 | 162.0 |
| | IPACW | | | 1.332 | 1.857 | 2.089 | 2.785 | 4.973 | 79.6 |
| | IPCW x IPACW | | | 1.948 | 4.319 | 5.467 | 10.471 | 29.734 | 731.8 |
| | IPTW | | | 0.412 | 2.376 | 3.623 | 7.413 | 10.870 | 16.0 |
| | Combined | | | 0.919 | 5.379 | 9.087 | 22.748 | 72.002 | 4491.5 |
| | IPCW | Untreated | 2483950 | 1.544 | 2.362 | 2.740 | 4.251 | 8.782 | 294.3 |
| | IPACW | | | 1.154 | 1.475 | 1.651 | 2.286 | 7.442 | 261.9 |
| | IPCW x IPACW | | | 1.896 | 3.170 | 3.781 | 6.368 | 21.650 | 770.5 |
| | IPTW | | | 0.918 | 1.098 | 1.280 | 2.624 | 7.894 | 90.2 |
| | Combined | | | 1.831 | 3.219 | 4.096 | 9.017 | 65.786 | 9856.8 |

## 6.2.19 Overlap cohort 2 — Statins — Male — Age interactions: No

**Table: ESS diagnostics**

| Follow-up | Stage | ESS (overall) | ESS% (overall) | ESS (untreated) | ESS% (untreated) | ESS (treated) | ESS% (treated) |
|---|---|---|---|---|---|---|---|
| 1 | IPCW | 3044852 | **99.7** | 2700394 | **99.7** | 344467 | **99.6** |
| | IPACW | 3043922 | **99.7** | 2702561 | **99.8** | 344491 | **99.6** |
| | IPCW×IPACW | 3034186 | **99.4** | 2695872 | **99.6** | 342101 | **98.9** |
| | IPTW | 2036822 | **66.7** | 2014356 | **74.4** | 114330 | **33.1** |
| | Combined | 1716634 | **56.2** | 1689070 | **62.4** | 108272 | **31.3** |
| 3 | IPCW | 2980348 | **97.6** | 2646849 | **97.8** | 333573 | **96.4** |
| | IPACW | 2988535 | **97.9** | 2653852 | **98.0** | 340287 | **98.4** |
| | IPCW×IPACW | 2895153 | **94.8** | 2583804 | **95.4** | 320974 | **92.8** |
| | IPTW | 2037235 | **66.7** | 2015053 | **74.4** | 114291 | **33.0** |
| | Combined | 1054892 | **34.6** | 960249 | **35.5** | 99503 | **28.8** |
| 5 | IPCW | 2830124 | **92.7** | 2522882 | **93.2** | 307826 | **89.0** |
| | IPACW | 2871672 | **94.1** | 2540196 | **93.8** | 335464 | **97.0** |
| | IPCW×IPACW | 2572999 | **84.3** | 2307304 | **85.2** | 279318 | **80.8** |
| | IPTW | 2039555 | **66.8** | 2019218 | **74.6** | 114020 | **33.0** |
| | Combined | 404645 | **13.3** | 338869 | **12.5** | 85902 | **24.8** |
| 10 | IPCW | 1394635 | **45.7** | 2174379 | **80.4** | 35653 | **10.3** |
| | IPACW | 2415572 | **79.2** | 2094219 | **77.5** | 321646 | **93.0** |
| | IPCW×IPACW | 850487 | **27.9** | 1350053 | **50.0** | 28674 | **8.3** |
| | IPTW | 2019862 | **66.3** | 2006049 | **74.2** | 111838 | **32.3** |
| | Combined | 72152 | **2.4** | 58716 | **2.2** | 33437 | **9.7** |

**Table: Weight distributions**

| Follow-up (years) | Stage | Group | N | Median | P90 | P95 | P99 | P99.9 | Max |
|---|---|---|---|---|---|---|---|---|---|
| 1 | IPCW | Treated | 345885 | 1.092 | 1.174 | 1.215 | 1.328 | 1.638 | 8.5 |
| | IPACW | | | 1.129 | 1.221 | 1.259 | 1.381 | 1.624 | 3.0 |
| | IPCW x IPACW | | | 1.244 | 1.392 | 1.467 | 1.693 | 2.166 | 10.7 |
| | IPTW | | | 0.423 | 2.253 | 3.375 | 6.687 | 10.448 | 13.1 |
| | Combined | | | 0.522 | 2.882 | 4.399 | 8.886 | 14.596 | 24.7 |
| | IPCW | Untreated | 2707273 | 1.089 | 1.155 | 1.192 | 1.284 | 1.485 | 22.2 |
| | IPACW | | | 1.027 | 1.064 | 1.085 | 1.198 | 1.531 | 4.9 |
| | IPCW x IPACW | | | 1.125 | 1.201 | 1.248 | 1.409 | 1.791 | 22.6 |
| | IPTW | | | 0.925 | 1.084 | 1.232 | 2.689 | 9.006 | 134.0 |
| | Combined | | | 1.049 | 1.262 | 1.450 | 3.587 | 12.894 | 263.9 |
| 3 | IPCW | Treated | 345885 | 1.242 | 1.501 | 1.638 | 2.040 | 3.100 | 29.3 |
| | IPACW | | | 1.233 | 1.433 | 1.518 | 1.775 | 2.359 | 5.9 |
| | IPCW x IPACW | | | 1.582 | 2.031 | 2.281 | 3.043 | 4.926 | 36.1 |
| | IPTW | | | 0.422 | 2.253 | 3.376 | 6.690 | 10.442 | 13.1 |
| | Combined | | | 0.648 | 3.608 | 5.568 | 11.530 | 21.647 | 69.8 |
| | IPCW | Untreated | 2707227 | 1.247 | 1.448 | 1.560 | 1.885 | 2.542 | 87.7 |
| | IPACW | | | 1.070 | 1.178 | 1.238 | 1.488 | 2.842 | 26.8 |
| | IPCW x IPACW | | | 1.369 | 1.608 | 1.758 | 2.283 | 4.174 | 96.8 |
| | IPTW | | | 0.925 | 1.084 | 1.233 | 2.687 | 9.002 | 134.2 |
| | Combined | | | 1.279 | 1.668 | 1.940 | 4.891 | 22.745 | 615.8 |
| 5 | IPCW | Treated | 345881 | 1.348 | 1.817 | 2.071 | 2.885 | 5.176 | 96.6 |
| | IPACW | | | 1.291 | 1.584 | 1.710 | 2.074 | 2.982 | 7.8 |
| | IPCW x IPACW | | | 1.835 | 2.689 | 3.162 | 4.733 | 9.171 | 175.9 |
| | IPTW | | | 0.422 | 2.255 | 3.380 | 6.703 | 10.482 | 13.2 |
| | Combined | | | 0.749 | 4.156 | 6.546 | 14.072 | 29.915 | 310.8 |
| | IPCW | Untreated | 2706845 | 1.379 | 1.739 | 1.930 | 2.566 | 3.853 | 217.7 |
| | IPACW | | | 1.098 | 1.272 | 1.370 | 1.729 | 4.271 | 71.5 |
| | IPCW x IPACW | | | 1.596 | 2.033 | 2.309 | 3.273 | 7.720 | 244.1 |
| | IPTW | | | 0.925 | 1.084 | 1.233 | 2.683 | 9.022 | 138.4 |
| | Combined | | | 1.492 | 2.102 | 2.519 | 5.918 | 36.305 | 2920.7 |

| Follow-up (years) | Stage | Group | N | Median | P90 | P95 | P99 | P99.9 | Max |
|---|---|---|---|---|---|---|---|---|---|
| 10 | IPCW | Treated | 345787 | 1.428 | 2.486 | 2.998 | 4.939 | 11.389 | 2831.2 |
| | IPACW | | | 1.332 | 1.855 | 2.080 | 2.699 | 4.397 | 24.5 |
| | IPCW x IPACW | | | 1.952 | 4.319 | 5.411 | 9.855 | 25.589 | 4719.6 |
| | IPTW | | | 0.416 | 2.263 | 3.412 | 6.851 | 10.672 | 13.4 |
| | Combined | | | 0.910 | 5.219 | 8.641 | 21.102 | 58.523 | 2233.0 |
| | IPCW | Untreated | 2702773 | 1.563 | 2.420 | 2.805 | 4.320 | 8.186 | 258.6 |
| | IPACW | | | 1.130 | 1.450 | 1.626 | 2.262 | 7.125 | 128.8 |
| | IPCW x IPACW | | | 1.900 | 3.143 | 3.766 | 6.216 | 20.301 | 463.5 |
| | IPTW | | | 0.925 | 1.085 | 1.232 | 2.643 | 9.306 | 75.9 |
| | Combined | | | 1.839 | 3.179 | 4.044 | 8.590 | 74.182 | 10207.9 |

## 6.2.20 Overlap cohort 2 — Statins — Male — Age interactions: Yes

**Table: ESS diagnostics**

| Follow-up | Stage | ESS (overall) | ESS% (overall) | ESS (untreated) | ESS% (untreated) | ESS (treated) | ESS% (treated) |
|---|---|---|---|---|---|---|---|
| 1 | IPCW | 2825528 | **99.7** | 2481626 | **99.7** | 343948 | **99.2** |
| | IPACW | 2825792 | **99.7** | 2483891 | **99.8** | 344936 | **99.5** |
| | IPCW×IPACW | 2813049 | **99.2** | 2476792 | **99.5** | 340569 | **98.2** |
| | IPTW | 1901243 | **67.1** | 1912579 | **76.9** | 112132 | **32.3** |
| | Combined | 1617796 | **57.1** | 1645954 | **66.1** | 102714 | **29.6** |
| 3 | IPCW | 2722790 | **96.0** | 2396750 | **96.3** | 326363 | **94.1** |
| | IPACW | 2755182 | **97.2** | 2421038 | **97.3** | 339009 | **97.8** |
| | IPCW×IPACW | 2615791 | **92.3** | 2321113 | **93.3** | 307049 | **88.6** |
| | IPTW | 1901174 | **67.1** | 1912642 | **76.9** | 112106 | **32.3** |
| | Combined | 968171 | **34.1** | 899663 | **36.2** | 87344 | **25.2** |
| 5 | IPCW | 2601102 | **91.8** | 2315778 | **93.1** | 289363 | **83.5** |
| | IPACW | 2623228 | **92.5** | 2296841 | **92.3** | 329816 | **95.2** |
| | IPCW×IPACW | 2276444 | **80.3** | 2073508 | **83.3** | 238471 | **68.8** |
| | IPTW | 1913836 | **67.5** | 1930528 | **77.6** | 111853 | **32.3** |
| | Combined | 425068 | **15.0** | 365416 | **14.7** | 60377 | **17.4** |
| 10 | IPCW | 2183663 | **77.2** | 1949177 | **78.5** | 238415 | **69.0** |
| | IPACW | 2135201 | **75.5** | 1823235 | **73.4** | 313335 | **90.6** |
| | IPCW×IPACW | 1275373 | **45.1** | 1135567 | **45.7** | 148839 | **43.1** |
| | IPTW | 1898901 | **67.1** | 1932793 | **77.8** | 108324 | **31.3** |
| | Combined | 73413 | **2.6** | 60249 | **2.4** | 16969 | **4.9** |

**Table: Weight distributions**

| Follow-up (years) | Stage | Group | N | Median | P90 | P95 | P99 | P99.9 | Max |
|---|---|---|---|---|---|---|---|---|---|
| 1 | IPCW | Treated | 346705 | 1.089 | 1.180 | 1.227 | 1.362 | 1.770 | 32.5 |
| | IPACW | | | 1.128 | 1.223 | 1.265 | 1.410 | 1.746 | 10.7 |
| | IPCW x IPACW | | | 1.241 | 1.404 | 1.494 | 1.763 | 2.423 | 43.7 |
| | IPTW | | | 0.417 | 2.360 | 3.576 | 7.144 | 10.313 | 13.4 |
| | Combined | | | 0.518 | 3.008 | 4.647 | 9.710 | 15.597 | 51.4 |
| | IPCW | Untreated | 2488490 | 1.087 | 1.152 | 1.188 | 1.287 | 1.520 | 16.7 |
| | IPACW | | | 1.030 | 1.067 | 1.088 | 1.204 | 1.559 | 5.4 |
| | IPCW x IPACW | | | 1.125 | 1.203 | 1.249 | 1.436 | 1.834 | 17.1 |
| | IPTW | | | 0.918 | 1.098 | 1.281 | 2.676 | 7.826 | 166.9 |
| | Combined | | | 1.042 | 1.282 | 1.520 | 3.532 | 11.365 | 229.5 |
| 3 | IPCW | Treated | 346695 | 1.236 | 1.511 | 1.664 | 2.152 | 3.708 | 68.5 |
| | IPACW | | | 1.233 | 1.436 | 1.526 | 1.826 | 2.593 | 31.9 |
| | IPCW x IPACW | | | 1.572 | 2.054 | 2.340 | 3.278 | 5.956 | 90.8 |
| | IPTW | | | 0.417 | 2.360 | 3.577 | 7.148 | 10.319 | 13.4 |
| | Combined | | | 0.648 | 3.741 | 5.876 | 12.696 | 25.154 | 236.9 |
| | IPCW | Untreated | 2488453 | 1.240 | 1.436 | 1.549 | 1.881 | 2.701 | 215.8 |
| | IPACW | | | 1.079 | 1.186 | 1.246 | 1.504 | 2.929 | 86.9 |
| | IPCW x IPACW | | | 1.373 | 1.612 | 1.763 | 2.379 | 4.406 | 228.7 |
| | IPTW | | | 0.918 | 1.098 | 1.281 | 2.676 | 7.818 | 167.6 |
| | Combined | | | 1.272 | 1.684 | 2.026 | 4.889 | 20.791 | 589.6 |
| 5 | IPCW | Treated | 346611 | 1.342 | 1.826 | 2.104 | 3.065 | 6.328 | 177.6 |
| | IPACW | | | 1.291 | 1.586 | 1.720 | 2.137 | 3.293 | 90.1 |
| | IPCW x IPACW | | | 1.827 | 2.709 | 3.241 | 5.150 | 10.945 | 328.1 |
| | IPTW | | | 0.417 | 2.362 | 3.579 | 7.169 | 10.352 | 13.5 |
| | Combined | | | 0.751 | 4.303 | 6.906 | 15.333 | 36.080 | 986.1 |
| | IPCW | Untreated | 2488185 | 1.369 | 1.716 | 1.910 | 2.553 | 4.202 | 91.8 |
| | IPACW | | | 1.113 | 1.286 | 1.382 | 1.748 | 4.432 | 71.5 |
| | IPCW x IPACW | | | 1.603 | 2.042 | 2.320 | 3.418 | 8.286 | 134.5 |
| | IPTW | | | 0.918 | 1.098 | 1.281 | 2.672 | 7.796 | 171.5 |
| | Combined | | | 1.488 | 2.121 | 2.589 | 6.104 | 32.795 | 2020.1 |

| Follow-up (years) | Stage | Group | N | Median | P90 | P95 | P99 | P99.9 | Max |
|---|---|---|---|---|---|---|---|---|---|
| 10 | IPCW | Treated | 345708 | 1.423 | 2.484 | 3.023 | 5.179 | 12.891 | 161.7 |
| | IPACW | | | 1.332 | 1.857 | 2.089 | 2.784 | 4.988 | 79.5 |
| | IPCW x IPACW | | | 1.948 | 4.318 | 5.470 | 10.496 | 29.739 | 734.6 |
| | IPTW | | | 0.412 | 2.377 | 3.624 | 7.413 | 10.888 | 16.0 |
| | Combined | | | 0.919 | 5.380 | 9.087 | 22.755 | 71.759 | 4510.1 |
| | IPCW | Untreated | 2483899 | 1.544 | 2.362 | 2.743 | 4.263 | 8.821 | 279.1 |
| | IPACW | | | 1.154 | 1.475 | 1.651 | 2.286 | 7.440 | 262.6 |
| | IPCW x IPACW | | | 1.896 | 3.172 | 3.787 | 6.381 | 21.702 | 769.5 |
| | IPTW | | | 0.918 | 1.098 | 1.280 | 2.624 | 7.897 | 90.2 |
| | Combined | | | 1.830 | 3.221 | 4.102 | 9.030 | 65.757 | 9836.2 |

## 6.2.21 Overlap cohort 2 — Antihypertensives — Female — Age interactions: No

**Table: ESS diagnostics**

| Follow-up | Stage | ESS (overall) | ESS% (overall) | ESS (untreated) | ESS% (untreated) | ESS (treated) | ESS% (treated) |
|---|---|---|---|---|---|---|---|
| 1 | IPCW | 7257921 | **99.4** | 6398298 | **99.4** | 863449 | **99.8** |
| | IPACW | 7210697 | **98.7** | 6429568 | **99.8** | 813692 | **94.0** |
| | IPCW×IPACW | 7157936 | **98.0** | 6397179 | **99.3** | 783239 | **90.5** |
| | IPTW | 2802901 | **38.4** | 2773637 | **43.1** | 164031 | **19.0** |
| | Combined | 1645771 | **22.5** | 1862635 | **28.9** | 96038 | **11.1** |
| 3 | IPCW | 6903673 | **94.5** | 6081489 | **94.4** | 849111 | **98.1** |
| | IPACW | 6802250 | **93.1** | 6316696 | **98.1** | 657828 | **76.0** |
| | IPCW×IPACW | 5757201 | **78.8** | 6017136 | **93.4** | 337089 | **39.0** |
| | IPTW | 2803214 | **38.4** | 2774806 | **43.1** | 163845 | **18.9** |
| | Combined | 331185 | **4.5** | 413142 | **6.4** | 18671 | **2.2** |
| 5 | IPCW | 6337123 | **86.8** | 5570630 | **86.5** | 822658 | **95.1** |
| | IPACW | 6163076 | **84.4** | 6046531 | **93.9** | 504264 | **58.3** |
| | IPCW×IPACW | 3615743 | **49.5** | 5324520 | **82.7** | 118085 | **13.7** |
| | IPTW | 2801099 | **38.4** | 2778641 | **43.2** | 162311 | **18.8** |
| | Combined | 89273 | **1.2** | 108385 | **1.7** | 5759 | **0.7** |
| 10 | IPCW | 4510164 | **61.9** | 3932958 | **61.1** | 726778 | **85.1** |
| | IPACW | 4748812 | **65.2** | 4986738 | **77.5** | 325497 | **38.1** |
| | IPCW×IPACW | 1208450 | **16.6** | 3148913 | **48.9** | 25888 | **3.0** |
| | IPTW | 2761899 | **37.9** | 2860134 | **44.4** | 133570 | **15.6** |
| | Combined | 15816 | **0.2** | 24971 | **0.4** | 929 | **0.1** |

**Table: Weight distributions**

| Follow-up (years) | Stage | Group | N | Median | P90 | P95 | P99 | P99.9 | Max |
|---|---|---|---|---|---|---|---|---|---|
| 1 | IPCW | Treated | 865456 | 1.079 | 1.142 | 1.175 | 1.269 | 1.476 | 5.3 |
| | IPACW | | | 1.105 | 1.323 | 1.520 | 2.481 | 4.530 | 24.1 |
| | IPCW x IPACW | | | 1.199 | 1.479 | 1.727 | 3.015 | 6.091 | 40.3 |
| | IPTW | | | 0.152 | 2.891 | 5.385 | 9.390 | 10.745 | 13.5 |
| | Combined | | | 0.181 | 4.048 | 8.727 | 21.204 | 46.105 | 263.0 |
| | IPCW | Untreated | 6439280 | 1.148 | 1.293 | 1.335 | 1.415 | 1.586 | 7.3 |
| | IPACW | | | 1.021 | 1.054 | 1.076 | 1.184 | 1.478 | 6.9 |
| | IPCW x IPACW | | | 1.184 | 1.324 | 1.369 | 1.464 | 1.717 | 7.4 |
| | IPTW | | | 0.900 | 0.944 | 0.974 | 4.672 | 13.906 | 103.1 |
| | Combined | | | 1.079 | 1.210 | 1.274 | 5.893 | 21.798 | 392.1 |
| 3 | IPCW | Treated | 865390 | 1.212 | 1.397 | 1.488 | 1.767 | 2.477 | 14.5 |
| | IPACW | | | 1.189 | 1.549 | 1.811 | 3.428 | 9.789 | 108.2 |
| | IPCW x IPACW | | | 1.464 | 2.038 | 2.467 | 5.148 | 19.474 | 526.0 |
| | IPTW | | | 0.152 | 2.889 | 5.386 | 9.404 | 10.764 | 13.5 |
| | Combined | | | 0.225 | 4.915 | 9.953 | 29.865 | 139.409 | 4119.7 |
| | IPCW | Untreated | 6439273 | 1.357 | 1.912 | 2.128 | 2.584 | 3.478 | 45.7 |
| | IPACW | | | 1.050 | 1.138 | 1.194 | 1.452 | 2.445 | 70.8 |
| | IPCW x IPACW | | | 1.480 | 2.039 | 2.272 | 2.783 | 4.070 | 85.3 |
| | IPTW | | | 0.900 | 0.944 | 0.974 | 4.672 | 13.904 | 103.2 |
| | Combined | | | 1.360 | 1.915 | 2.245 | 7.625 | 36.605 | 6544.8 |
| 5 | IPCW | Treated | 864839 | 1.318 | 1.649 | 1.808 | 2.300 | 3.626 | 38.7 |
| | IPACW | | | 1.241 | 1.679 | 1.999 | 3.975 | 13.736 | 132.7 |
| | IPCW x IPACW | | | 1.679 | 2.551 | 3.179 | 6.966 | 35.115 | 1211.1 |
| | IPTW | | | 0.152 | 2.875 | 5.382 | 9.528 | 10.925 | 13.7 |
| | Combined | | | 0.266 | 5.521 | 10.624 | 35.282 | 240.410 | 9542.5 |
| | IPCW | Untreated | 6439050 | 1.506 | 2.448 | 2.907 | 4.034 | 6.176 | 216.4 |
| | IPACW | | | 1.073 | 1.210 | 1.297 | 1.673 | 3.505 | 115.4 |
| | IPCW x IPACW | | | 1.722 | 2.718 | 3.233 | 4.534 | 7.829 | 265.7 |
| | IPTW | | | 0.900 | 0.944 | 0.974 | 4.677 | 13.880 | 104.8 |
| | Combined | | | 1.587 | 2.609 | 3.319 | 8.962 | 50.964 | 25353.0 |

| Follow-up (years) | Stage | Group | N | Median | P90 | P95 | P99 | P99.9 | Max |
|---|---|---|---|---|---|---|---|---|---|
| 10 | IPCW | Treated | 854167 | 1.450 | 2.231 | 2.567 | 3.619 | 6.582 | 124.0 |
| | IPACW | | | 1.305 | 1.792 | 2.285 | 4.528 | 19.050 | 286.3 |
| | IPCW x IPACW | | | 1.978 | 3.748 | 4.903 | 11.037 | 68.309 | 4680.4 |
| | IPTW | | | 0.150 | 2.583 | 4.958 | 11.500 | 15.709 | 22.0 |
| | Combined | | | 0.345 | 5.787 | 11.647 | 42.845 | 448.608 | 66290.3 |
| | IPCW | Untreated | 6434678 | 1.689 | 3.502 | 4.651 | 8.500 | 16.647 | 239.6 |
| | IPACW | | | 1.097 | 1.353 | 1.514 | 2.147 | 6.031 | 235.5 |
| | IPCW x IPACW | | | 1.960 | 4.303 | 5.703 | 10.618 | 23.393 | 1169.8 |
| | IPTW | | | 0.900 | 0.944 | 0.976 | 4.730 | 13.770 | 125.9 |
| | Combined | | | 1.839 | 4.207 | 5.887 | 13.619 | 87.601 | 73706.4 |

## 6.2.22 Overlap cohort 2 — Antihypertensives — Female — Age interactions: Yes

**Table: ESS diagnostics**

| Follow-up | Stage | ESS (overall) | ESS% (overall) | ESS (untreated) | ESS% (untreated) | ESS (treated) | ESS% (treated) |
|---|---|---|---|---|---|---|---|
| 1 | IPCW | 6356872 | **99.4** | 5503442 | **99.5** | 855755 | **99.6** |
| | IPACW | 6313321 | **98.8** | 5525836 | **99.9** | 813530 | **94.7** |
| | IPCW×IPACW | 6272494 | **98.1** | 5501265 | **99.4** | 789876 | **92.0** |
| | IPTW | 2668669 | **41.7** | 2902840 | **52.5** | 144728 | **16.9** |
| | Combined | 1543629 | **24.1** | 2254962 | **40.8** | 85219 | **9.9** |
| 3 | IPCW | 6077375 | **95.1** | 5267097 | **95.2** | 822690 | **95.8** |
| | IPACW | 6002979 | **93.9** | 5467925 | **98.8** | 673799 | **78.5** |
| | IPCW×IPACW | 5430542 | **85.0** | 5220775 | **94.3** | 472164 | **55.0** |
| | IPTW | 2668992 | **41.8** | 2903480 | **52.5** | 144714 | **16.8** |
| | Combined | 467643 | **7.3** | 795638 | **14.4** | 26619 | **3.1** |
| 5 | IPCW | 5678729 | **88.8** | 4937278 | **89.2** | 756338 | **88.1** |
| | IPACW | 5486451 | **85.8** | 5318132 | **96.1** | 517172 | **60.2** |
| | IPCW×IPACW | 3772028 | **59.0** | 4731194 | **85.5** | 188734 | **22.0** |
| | IPTW | 2668466 | **41.7** | 2907207 | **52.5** | 144179 | **16.8** |
| | Combined | 89007 | **1.4** | 96986 | **1.8** | 8546 | **1.0** |
| 10 | IPCW | 4479910 | **70.2** | 3868362 | **69.9** | 659517 | **77.8** |
| | IPACW | 3470067 | **54.4** | 3364719 | **60.8** | 326131 | **38.5** |
| | IPCW×IPACW | 271499 | **4.3** | 217007 | **3.9** | 69469 | **8.2** |
| | IPTW | 2583438 | **40.5** | 2961974 | **53.6** | 121135 | **14.3** |
| | Combined | 31 | **0.0** | 20 | **0.0** | 2210 | **0.3** |

**Table: Weight distributions**

| Follow-up (years) | Stage | Group | N | Median | P90 | P95 | P99 | P99.9 | Max |
|---|---|---|---|---|---|---|---|---|---|
| 1 | IPCW | Treated | 858861 | 1.074 | 1.152 | 1.193 | 1.304 | 1.584 | 11.8 |
| | IPACW | | | 1.104 | 1.286 | 1.484 | 2.431 | 4.355 | 13.7 |
| | IPCW x IPACW | | | 1.190 | 1.468 | 1.711 | 2.907 | 5.728 | 19.4 |
| | IPTW | | | 0.173 | 2.783 | 5.197 | 11.609 | 14.139 | 17.1 |
| | Combined | | | 0.205 | 3.861 | 8.648 | 23.693 | 54.728 | 251.4 |
| | IPCW | Untreated | 5533488 | 1.134 | 1.266 | 1.306 | 1.381 | 1.503 | 14.7 |
| | IPACW | | | 1.023 | 1.062 | 1.088 | 1.197 | 1.417 | 5.1 |
| | IPCW x IPACW | | | 1.172 | 1.303 | 1.347 | 1.437 | 1.668 | 15.0 |
| | IPTW | | | 0.886 | 0.941 | 0.978 | 4.768 | 11.538 | 105.9 |
| | Combined | | | 1.051 | 1.183 | 1.263 | 6.027 | 16.873 | 361.3 |
| 3 | IPCW | Treated | 858849 | 1.198 | 1.415 | 1.538 | 1.925 | 3.047 | 103.5 |
| | IPACW | | | 1.188 | 1.486 | 1.751 | 3.422 | 9.484 | 114.4 |
| | IPCW x IPACW | | | 1.444 | 2.043 | 2.510 | 5.122 | 18.053 | 215.2 |
| | IPTW | | | 0.173 | 2.783 | 5.197 | 11.610 | 14.142 | 16.3 |
| | Combined | | | 0.251 | 4.654 | 10.293 | 33.245 | 146.160 | 3480.4 |
| | IPCW | Untreated | 5533468 | 1.331 | 1.798 | 1.985 | 2.408 | 2.966 | 139.8 |
| | IPACW | | | 1.055 | 1.158 | 1.223 | 1.496 | 2.266 | 38.8 |
| | IPCW x IPACW | | | 1.457 | 1.943 | 2.153 | 2.648 | 3.582 | 150.8 |
| | IPTW | | | 0.886 | 0.941 | 0.978 | 4.768 | 11.536 | 106.1 |
| | Combined | | | 1.315 | 1.812 | 2.168 | 8.022 | 27.933 | 3533.6 |
| 5 | IPCW | Treated | 858547 | 1.302 | 1.664 | 1.873 | 2.588 | 4.946 | 203.5 |
| | IPACW | | | 1.241 | 1.611 | 1.915 | 4.005 | 13.774 | 134.6 |
| | IPCW x IPACW | | | 1.655 | 2.564 | 3.274 | 7.258 | 33.084 | 622.9 |
| | IPTW | | | 0.173 | 2.779 | 5.203 | 11.650 | 14.266 | 16.5 |
| | Combined | | | 0.295 | 5.189 | 11.171 | 39.423 | 252.635 | 7858.7 |
| | IPCW | Untreated | 5533324 | 1.474 | 2.264 | 2.643 | 3.610 | 5.037 | 103.1 |
| | IPACW | | | 1.080 | 1.237 | 1.339 | 1.738 | 3.229 | 86.7 |
| | IPCW x IPACW | | | 1.693 | 2.566 | 3.007 | 4.198 | 6.546 | 179.6 |
| | IPTW | | | 0.886 | 0.941 | 0.978 | 4.766 | 11.531 | 107.5 |
| | Combined | | | 1.534 | 2.450 | 3.161 | 9.538 | 40.238 | 17167.9 |

| Follow-up (years) | Stage | Group | N | Median | P90 | P95 | P99 | P99.9 | Max |
|---|---|---|---|---|---|---|---|---|---|
| 10 | IPCW | Treated | 847601 | 1.447 | 2.230 | 2.629 | 4.142 | 9.603 | 130.8 |
| | IPACW | | | 1.307 | 1.744 | 2.153 | 4.543 | 21.821 | 304.7 |
| | IPCW x IPACW | | | 1.967 | 3.744 | 5.002 | 12.073 | 71.150 | 2378.2 |
| | IPTW | | | 0.170 | 2.535 | 4.964 | 13.085 | 19.328 | 30.2 |
| | Combined | | | 0.388 | 5.556 | 11.822 | 48.809 | 559.358 | 21677.2 |
| | IPCW | Untreated | 5530224 | 1.654 | 3.201 | 4.110 | 7.163 | 13.192 | 192.6 |
| | IPACW | | | 1.114 | 1.395 | 1.580 | 2.261 | 5.907 | 1894.7 |
| | IPCW x IPACW | | | 1.956 | 4.047 | 5.248 | 9.445 | 19.838 | 27878.0 |
| | IPTW | | | 0.886 | 0.942 | 0.982 | 4.741 | 11.578 | 151.0 |
| | Combined | | | 1.820 | 3.962 | 5.506 | 13.376 | 74.468 | 4208947.1 |

## 6.2.23 Overlap cohort 2 — Antihypertensives — Male — Age interactions: No

**Table: ESS diagnostics**

| Follow-up | Stage | ESS (overall) | ESS% (overall) | ESS (untreated) | ESS% (untreated) | ESS (treated) | ESS% (treated) |
|---|---|---|---|---|---|---|---|
| 1 | IPCW | 7229486 | **99.4** | 6369821 | **99.4** | 863431 | **99.8** |
| | IPACW | 7182088 | **98.7** | 6400663 | **99.8** | 813842 | **94.0** |
| | IPCW×IPACW | 7130129 | **98.0** | 6368667 | **99.3** | 783823 | **90.6** |
| | IPTW | 2795605 | **38.4** | 2763783 | **43.1** | 164666 | **19.0** |
| | Combined | 1645285 | **22.6** | 1858612 | **29.0** | 96557 | **11.2** |
| 3 | IPCW | 6879592 | **94.6** | 6056997 | **94.5** | 849077 | **98.1** |
| | IPACW | 6779035 | **93.2** | 6291023 | **98.1** | 658955 | **76.1** |
| | IPCW×IPACW | 5761453 | **79.2** | 5994358 | **93.5** | 343423 | **39.7** |
| | IPTW | 2795201 | **38.4** | 2764071 | **43.1** | 164477 | **19.0** |
| | Combined | 340102 | **4.7** | 424658 | **6.6** | 19186 | **2.2** |
| 5 | IPCW | 6320148 | **86.9** | 5552742 | **86.6** | 822590 | **95.1** |
| | IPACW | 6131698 | **84.3** | 6014324 | **93.8** | 503785 | **58.3** |
| | IPCW×IPACW | 3597922 | **49.5** | 5304839 | **82.8** | 118304 | **13.7** |
| | IPTW | 2792832 | **38.4** | 2767845 | **43.2** | 162859 | **18.8** |
| | Combined | 90408 | **1.2** | 109396 | **1.7** | 5876 | **0.7** |
| 10 | IPCW | 4512574 | **62.2** | 3933002 | **61.4** | 726635 | **85.1** |
| | IPACW | 4739684 | **65.3** | 4958612 | **77.4** | 328709 | **38.5** |
| | IPCW×IPACW | 1245555 | **17.2** | 3134762 | **48.9** | 27241 | **3.2** |
| | IPTW | 2749047 | **37.9** | 2845767 | **44.4** | 133488 | **15.6** |
| | Combined | 15432 | **0.2** | 22936 | **0.4** | 942 | **0.1** |

**Table: Weight distributions**

| Follow-up (years) | Stage | Group | N | Median | P90 | P95 | P99 | P99.9 | Max |
|---|---|---|---|---|---|---|---|---|---|
| 1 | IPCW | Treated | 865437 | 1.079 | 1.142 | 1.175 | 1.269 | 1.476 | 5.4 |
| | IPACW | | | 1.105 | 1.323 | 1.520 | 2.484 | 4.533 | 23.8 |
| | IPCW x IPACW | | | 1.199 | 1.479 | 1.726 | 3.014 | 6.073 | 39.7 |
| | IPTW | | | 0.153 | 2.891 | 5.362 | 9.361 | 10.723 | 13.3 |
| | Combined | | | 0.182 | 4.045 | 8.674 | 21.099 | 45.908 | 257.9 |
| | IPCW | Untreated | 6410348 | 1.147 | 1.292 | 1.334 | 1.414 | 1.586 | 7.4 |
| | IPACW | | | 1.021 | 1.054 | 1.076 | 1.185 | 1.477 | 6.7 |
| | IPCW x IPACW | | | 1.184 | 1.323 | 1.368 | 1.464 | 1.716 | 7.5 |
| | IPTW | | | 0.900 | 0.943 | 0.974 | 4.678 | 13.916 | 102.8 |
| | Combined | | | 1.079 | 1.209 | 1.273 | 5.902 | 21.835 | 386.9 |
| 3 | IPCW | Treated | 865370 | 1.213 | 1.397 | 1.489 | 1.768 | 2.476 | 14.6 |
| | IPACW | | | 1.189 | 1.549 | 1.811 | 3.427 | 9.828 | 105.9 |
| | IPCW x IPACW | | | 1.464 | 2.038 | 2.466 | 5.146 | 19.532 | 508.3 |
| | IPTW | | | 0.153 | 2.889 | 5.362 | 9.375 | 10.742 | 13.3 |
| | Combined | | | 0.226 | 4.912 | 9.907 | 29.798 | 138.478 | 3963.3 |
| | IPCW | Untreated | 6410341 | 1.356 | 1.907 | 2.121 | 2.578 | 3.471 | 45.0 |
| | IPACW | | | 1.050 | 1.139 | 1.195 | 1.454 | 2.444 | 66.8 |
| | IPCW x IPACW | | | 1.480 | 2.034 | 2.266 | 2.776 | 4.063 | 80.5 |
| | IPTW | | | 0.900 | 0.943 | 0.974 | 4.678 | 13.914 | 102.8 |
| | Combined | | | 1.358 | 1.911 | 2.240 | 7.648 | 36.572 | 6350.9 |
| 5 | IPCW | Treated | 864787 | 1.318 | 1.650 | 1.809 | 2.302 | 3.629 | 38.2 |
| | IPACW | | | 1.241 | 1.679 | 1.998 | 3.974 | 13.857 | 128.8 |
| | IPCW x IPACW | | | 1.679 | 2.552 | 3.179 | 6.950 | 35.189 | 1171.7 |
| | IPTW | | | 0.153 | 2.874 | 5.359 | 9.503 | 10.920 | 13.5 |
| | Combined | | | 0.267 | 5.517 | 10.571 | 35.223 | 239.147 | 9087.7 |
| | IPCW | Untreated | 6410114 | 1.505 | 2.440 | 2.895 | 4.016 | 6.154 | 215.4 |
| | IPACW | | | 1.073 | 1.211 | 1.299 | 1.675 | 3.506 | 113.3 |
| | IPCW x IPACW | | | 1.721 | 2.710 | 3.221 | 4.516 | 7.805 | 260.4 |
| | IPTW | | | 0.900 | 0.943 | 0.974 | 4.683 | 13.890 | 104.5 |
| | Combined | | | 1.585 | 2.601 | 3.308 | 8.987 | 50.998 | 24772.0 |

| Follow-up (years) | Stage | Group | N | Median | P90 | P95 | P99 | P99.9 | Max |
|---|---|---|---|---|---|---|---|---|---|
| 10 | IPCW | Treated | 853927 | 1.450 | 2.233 | 2.569 | 3.621 | 6.574 | 122.9 |
| | IPACW | | | 1.305 | 1.792 | 2.284 | 4.520 | 19.006 | 269.6 |
| | IPCW x IPACW | | | 1.978 | 3.749 | 4.903 | 11.014 | 67.357 | 4439.6 |
| | IPTW | | | 0.150 | 2.581 | 4.934 | 11.448 | 15.866 | 22.2 |
| | Combined | | | 0.347 | 5.778 | 11.610 | 42.674 | 445.531 | 62345.9 |
| | IPCW | Untreated | 6405767 | 1.688 | 3.490 | 4.626 | 8.446 | 16.539 | 229.0 |
| | IPACW | | | 1.097 | 1.355 | 1.516 | 2.153 | 6.039 | 221.7 |
| | IPCW x IPACW | | | 1.960 | 4.295 | 5.683 | 10.564 | 23.302 | 1197.1 |
| | IPTW | | | 0.900 | 0.944 | 0.976 | 4.739 | 13.787 | 125.3 |
| | Combined | | | 1.838 | 4.197 | 5.868 | 13.587 | 87.734 | 75529.2 |

## 6.2.24 Overlap cohort 2 — Antihypertensives — Male — Age interactions: Yes

**Table: ESS diagnostics**

| Follow-up | Stage | ESS (overall) | ESS% (overall) | ESS (untreated) | ESS% (untreated) | ESS (treated) | ESS% (treated) |
|---|---|---|---|---|---|---|---|
| 1 | IPCW | 6328656 | **99.4** | 5475035 | **99.5** | 855888 | **99.6** |
| | IPACW | 6278821 | **98.7** | 5497002 | **99.9** | 809445 | **94.2** |
| | IPCW×IPACW | 6237551 | **98.0** | 5472752 | **99.4** | 785170 | **91.4** |
| | IPTW | 2662643 | **41.8** | 2880434 | **52.3** | 146569 | **17.1** |
| | Combined | 1522589 | **23.9** | 2238342 | **40.7** | 83788 | **9.8** |
| 3 | IPCW | 6053866 | **95.1** | 5242967 | **95.2** | 822985 | **95.8** |
| | IPACW | 5964770 | **93.7** | 5436226 | **98.8** | 670146 | **78.0** |
| | IPCW×IPACW | 5407436 | **85.0** | 5191696 | **94.3** | 474525 | **55.2** |
| | IPTW | 2660666 | **41.8** | 2882302 | **52.4** | 146117 | **17.0** |
| | Combined | 474685 | **7.5** | 786772 | **14.3** | 27475 | **3.2** |
| 5 | IPCW | 5658528 | **88.9** | 4915265 | **89.3** | 758136 | **88.3** |
| | IPACW | 5459804 | **85.8** | 5297912 | **96.2** | 514935 | **60.0** |
| | IPCW×IPACW | 3778846 | **59.4** | 4719259 | **85.7** | 191589 | **22.3** |
| | IPTW | 2653839 | **41.7** | 2886412 | **52.4** | 144533 | **16.8** |
| | Combined | 100499 | **1.6** | 121960 | **2.2** | 8427 | **1.0** |
| 10 | IPCW | 4456635 | **70.2** | 3849973 | **70.0** | 649442 | **76.7** |
| | IPACW | 3987175 | **62.8** | 4029404 | **73.2** | 338716 | **40.0** |
| | IPCW×IPACW | 538274 | **8.5** | 464482 | **8.4** | 73905 | **8.7** |
| | IPTW | 2561417 | **40.3** | 2936065 | **53.4** | 120730 | **14.3** |
| | Combined | 62 | **0.0** | 40 | **0.0** | 2259 | **0.3** |

**Table: Weight distributions**

| Follow-up (years) | Stage | Group | N | Median | P90 | P95 | P99 | P99.9 | Max |
|---|---|---|---|---|---|---|---|---|---|
| 1 | IPCW | Treated | 858995 | 1.074 | 1.152 | 1.193 | 1.305 | 1.586 | 11.9 |
| | IPACW | | | 1.104 | 1.286 | 1.486 | 2.438 | 4.411 | 44.3 |
| | IPCW x IPACW | | | 1.190 | 1.468 | 1.713 | 2.912 | 5.777 | 61.3 |
| | IPTW | | | 0.174 | 2.796 | 5.184 | 11.453 | 14.012 | 16.9 |
| | Combined | | | 0.206 | 3.881 | 8.624 | 23.410 | 53.980 | 738.6 |
| | IPCW | Untreated | 5504676 | 1.133 | 1.264 | 1.304 | 1.382 | 1.506 | 14.6 |
| | IPACW | | | 1.023 | 1.063 | 1.088 | 1.198 | 1.419 | 4.7 |
| | IPCW x IPACW | | | 1.172 | 1.302 | 1.346 | 1.438 | 1.670 | 15.0 |
| | IPTW | | | 0.886 | 0.940 | 0.977 | 4.778 | 11.544 | 100.6 |
| | Combined | | | 1.050 | 1.182 | 1.264 | 6.043 | 16.906 | 330.8 |
| 3 | IPCW | Treated | 858875 | 1.198 | 1.415 | 1.539 | 1.927 | 3.060 | 99.8 |
| | IPACW | | | 1.188 | 1.486 | 1.753 | 3.419 | 9.588 | 108.7 |
| | IPCW x IPACW | | | 1.444 | 2.044 | 2.513 | 5.124 | 17.983 | 207.6 |
| | IPTW | | | 0.174 | 2.792 | 5.186 | 11.490 | 14.077 | 16.3 |
| | Combined | | | 0.252 | 4.675 | 10.267 | 32.885 | 143.398 | 3298.7 |
| | IPCW | Untreated | 5504655 | 1.331 | 1.790 | 1.977 | 2.411 | 2.973 | 129.0 |
| | IPACW | | | 1.055 | 1.159 | 1.224 | 1.497 | 2.272 | 61.1 |
| | IPCW x IPACW | | | 1.456 | 1.936 | 2.146 | 2.651 | 3.593 | 139.2 |
| | IPTW | | | 0.886 | 0.940 | 0.978 | 4.776 | 11.541 | 101.7 |
| | Combined | | | 1.313 | 1.805 | 2.168 | 8.048 | 28.088 | 3532.5 |
| 5 | IPCW | Treated | 858277 | 1.301 | 1.665 | 1.874 | 2.595 | 4.972 | 184.2 |
| | IPACW | | | 1.241 | 1.610 | 1.915 | 4.004 | 13.733 | 128.5 |
| | IPCW x IPACW | | | 1.654 | 2.565 | 3.276 | 7.255 | 32.542 | 596.3 |
| | IPTW | | | 0.173 | 2.781 | 5.190 | 11.623 | 14.351 | 17.6 |
| | Combined | | | 0.296 | 5.192 | 11.119 | 39.249 | 245.538 | 7605.9 |
| | IPCW | Untreated | 5504505 | 1.473 | 2.250 | 2.626 | 3.613 | 5.058 | 91.7 |
| | IPACW | | | 1.080 | 1.239 | 1.341 | 1.740 | 3.244 | 81.7 |
| | IPCW x IPACW | | | 1.692 | 2.555 | 2.992 | 4.203 | 6.574 | 178.6 |
| | IPTW | | | 0.886 | 0.940 | 0.978 | 4.771 | 11.533 | 105.4 |
| | Combined | | | 1.533 | 2.438 | 3.155 | 9.563 | 40.506 | 14263.0 |

| Follow-up (years) | Stage | Group | N | Median | P90 | P95 | P99 | P99.9 | Max |
|---|---|---|---|---|---|---|---|---|---|
| 10 | IPCW | Treated | 847088 | 1.447 | 2.232 | 2.631 | 4.154 | 9.676 | 154.8 |
| | IPACW | | | 1.306 | 1.742 | 2.152 | 4.514 | 21.280 | 272.1 |
| | IPCW x IPACW | | | 1.967 | 3.745 | 5.008 | 12.051 | 69.986 | 2234.2 |
| | IPTW | | | 0.171 | 2.532 | 4.955 | 13.053 | 19.641 | 29.5 |
| | Combined | | | 0.389 | 5.545 | 11.787 | 48.678 | 548.224 | 19003.1 |
| | IPCW | Untreated | 5501399 | 1.653 | 3.184 | 4.075 | 7.122 | 13.304 | 177.3 |
| | IPACW | | | 1.115 | 1.398 | 1.583 | 2.266 | 5.934 | 1160.2 |
| | IPCW x IPACW | | | 1.956 | 4.037 | 5.222 | 9.413 | 19.977 | 18094.2 |
| | IPTW | | | 0.885 | 0.941 | 0.982 | 4.745 | 11.596 | 152.2 |
| | Combined | | | 1.819 | 3.949 | 5.480 | 13.430 | 75.055 | 2753911.1 |

# 6.3 Interpretation of Diagnostic Results

## 6.3.1 Statins

We start by discussing the results for statins.

- We don't have enough support to estimate the relative risk ratio of sustaining treatment for 10 years. This is highlighted by the effective sample sizes in the entire cohort for 10 years follow-up. Extreme violation of the positivity assumption.

- We try to deal with violation of the positivity assumption through creating overlap cohorts.
- This changes the estimand to target a cohort/population where we should have better support for to estimate. The first overlap cohort filters to individuals who have >1% chance of receiving either treatment, i.e., improve overlap between treatment groups. The second overlap cohort also filters to those with > 1% chance of not being artificially censored (sustaining treatment for follow-up period), and those with > 1% chance of being censored by the normal censoring mechanism (probability of being observed for the follow-up period and therefore able to obtain an estimate).
- There is very little difference between the first and second overlap cohort in terms of number of individuals, as few (in the hundreds/low thousands) have probabilities < 1%. However, this does make a notable impact to effective sample sizes induced by the IPACW weights in the longer follow-up times.
- Even in overlap cohort 2, ESS is a very small % of N (see Relative Risk Estimates). This indicates that the relative risk estimates are still being dominated by a relatively small proportion of observations. Note, while any individual weight (IPCW, IPACW, IPTW) in overlap 2 cohort will be smaller than 100, the combined weights still have the potential to be very large. This can be observed in the table in Diagnostics.
- The tables in Diagnostics indicate that the violations of positivity arise predominately through the artificial censoring mechanism (reflected in drop in ESS caused by IPACW weights alone), followed by a lack of overlap in treatment at baseline (reflected by drop in ESS caused by IPTW weights alone).

- At five-year follow-up, the positivity issues largely remain, in the entire cohort and overlap cohorts.

- At three year follow-up, there is evidence of residual positivity problems in the entire cohort, but there is improvement in the overlap cohorts.
- The ESS is approximately ~25-35% of the original N (overlap cohort 2).
- The diagnostics suggest that a lack of overlap in treatment at baseline (reflected by diagnostics for each weight alone) contributes the most to remaining positivity problems (see Diagnostics).
  - This is reasonable. As follow-up gets shorter, the probability of observing individuals sustaining treatment for this time period increases and will results in fewer positivity issues driven by the artificial censoring.
- At one year follow-up, there is evidence of residual positivity problems in the entire cohort, but there is improvement in the overlap cohorts.
- The ESS is approximately ~30-60% of the original N (overlap cohort 2).
- Again, the diagnostics suggest that a lack of overlap in treatment at baseline contributions the most to remaining positivity problems (see Diagnostics).

The other key assumption which may be failed is exchangeability, in particular with respect to treatment at baseline. Our treated group is a prevalent user cohort, rather than a new user design (Ray 2003; Lund, Richardson, and Sturmer 2015; Webster-Clark, Ross, and Lund 2021). This means the decision to initiate treatment may have been made long before the index date. Consequently, baseline covariates may not adequately capture the clinical information that originally determined treatment initiation, in particular because clinical information may have itself been modified by the intervention. New-user designs align baseline covariate assessment with treatment initiation specifically to avoid this problem. Consequently, in a prevalent-user cohort, the conditional exchangeability assumption required for the estimation of the IPTWs is less plausible.

## 6.3.2 Antihypertensives

The pattern in terms of IPCW and IPACW is broadly the same for antihypertensives as it was for statins. The reason this has been split out, is that there are greater positivity issues surrounding being on/off antihypertensives at baseline, highlighted by the IPTWs, and the large drop in ESS driven by the IPTW's alone.

There is evidence of positivity violations driven by the nature of antihypertensive prescribing: treatment is determined largely by blood pressure exceeding a clinical threshold, meaning the propensity score distribution in the treated group is highly concentrated. Many treated individuals have high probabilities of being treated or untreated, producing large IPTW weights.

The overlap restriction removes the most extreme propensity scores but cannot resolve the underlying heterogeneity within the overlap region. Even after restriction, the weighting procedure remains highly unstable, suggesting limitation overlap even after restriction, reflected in the persistent ESS collapse.

# Sensitivity analysis: direct effect of statins

26 August 2026

**Associated manuscript: A flexible approach to sequential prediction under intervention**

# 1 Introduction

This document contains the results from a sensitivity analysis where we assume a pleiotropic effect for statins exists, meaning statins has a direct (protective) effect on cardiovascular risk.

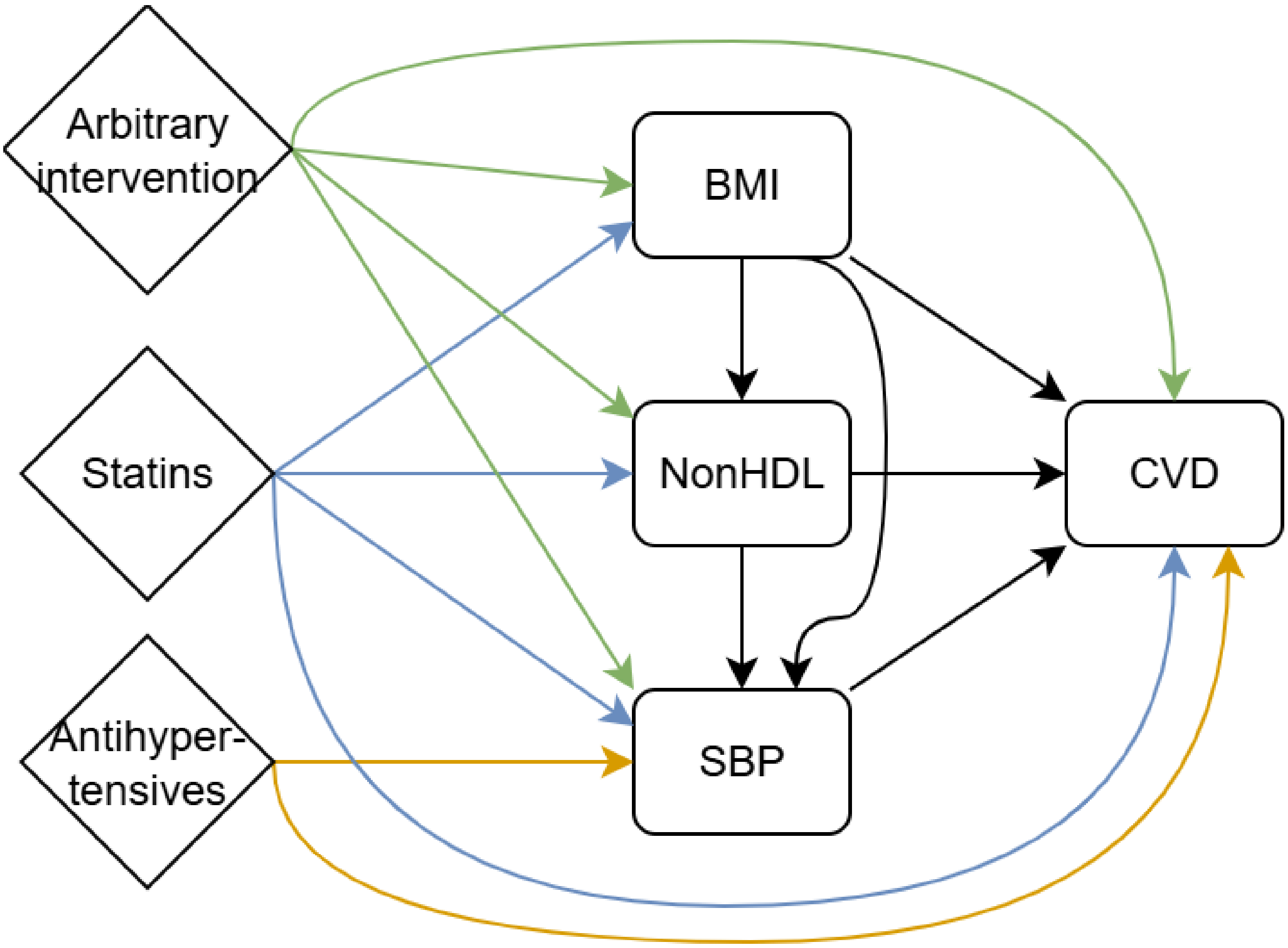

During the estimation of the DAG (see supplementary material file 3), we found that the sum of the indirect effects through body mass index (BMI), non-HDL cholesterol and systolic blood pressure (SBP) was equal to the total effect of statins reported from a meta-analysis of systematic reviews. We are not claiming this provides strong evidence that there is no pleiotropic effect of statins, as the DAG was formed using data from a number of sources, which are not consistent with each other in terms of outcome/estimand. However, this did inform the decision to not include a pleiotropic effect of statins when building the two-component model (i.e. we assumed the direct effect of statins represented in the DAG, had no effect). In the main manuscript, we claim it is possible to include direct effects of interventions within the two-component model framework. Given there is evidence supporting the existence of a pleiotropic effect of statins, we implement a sensitivity analysis to showcase the inclusion of a direct effect.

In other situations, if the sum of the indirect effects does not match a reported total effect, it may be appropriate to estimate the size of the direct effect based on this information. However, in our case, the introduction of a direct effect will cause a disagreement between the total effect of statins as calculated from the DAG (sum of the direct and indirect effects), and the total effect as reported from an overview of systematic reviews.(Karmali et al. 2016) There is no one size fits all approach for what to do when there is conflicting information supporting the derivation of the DAG. If some pathways are deemed to be based on information from lower quality sources, it may be sensible to derive the effect of these pathways based off information from stronger sources in the rest of the DAG. In our example, the effect of statins on cardiovascular risk mediated by non-HDL cholesterol and SBP comes from high quality sources. The effect of statins on cardiovascular disease mediated through BMI comes from lower quality sources, but the effect is assumed to be very small. It would therefore be unreasonable to modify the effect of statins on cardiovascular risk mediated through BMI to obtain agreement between the total effect estimated from the DAG (including hypothesised direct effect), and the total effect reported from the meta-analysis. However such an approach may be reasonable in other scenarios.

Instead, we include the direct effect of statins without modifying the indirect pathways, and take note that this results in a total effect of statins that is too big compared to Karmali et al. (2016). There is evidence indicating that statins has a pleiotropic effect on cardiovascular risk,(Ayyalu et al. 2025; Chen et al. 2024; Schooling et al. 2020) but there variation in the size of this effect. On the contrat, there are a number of studies indicating either no pleiotropic effect, or that have inconclusive findings.(B. A. Ference et al. 2019; Labos et al. 2018; Robinson et al. 2005; Byrne et al. 2022) Given the inconsistent evidence, we choose the size of the direct effect purely as a sensitivity analysis. The size is chosen so that the total effect of statins on cardiovascular risk is 20% bigger than previously (on the hazard ratio scale), with all other effects in the DAG remaining the same. This aligns with supplementary data file 4, where model performance is assessed under misspecification of the intervention effects by 20%.

We repeat the analyses from the main paper for the two component model, including model validation, and the worked example.

# 2 Model evaluation

It can be seen the model is well calibrated and there is no drop in performance compared to the model fitted in the main manuscript with no pleiotropic effect. This is to be expected, given the pleiotropic effect is relatively small.

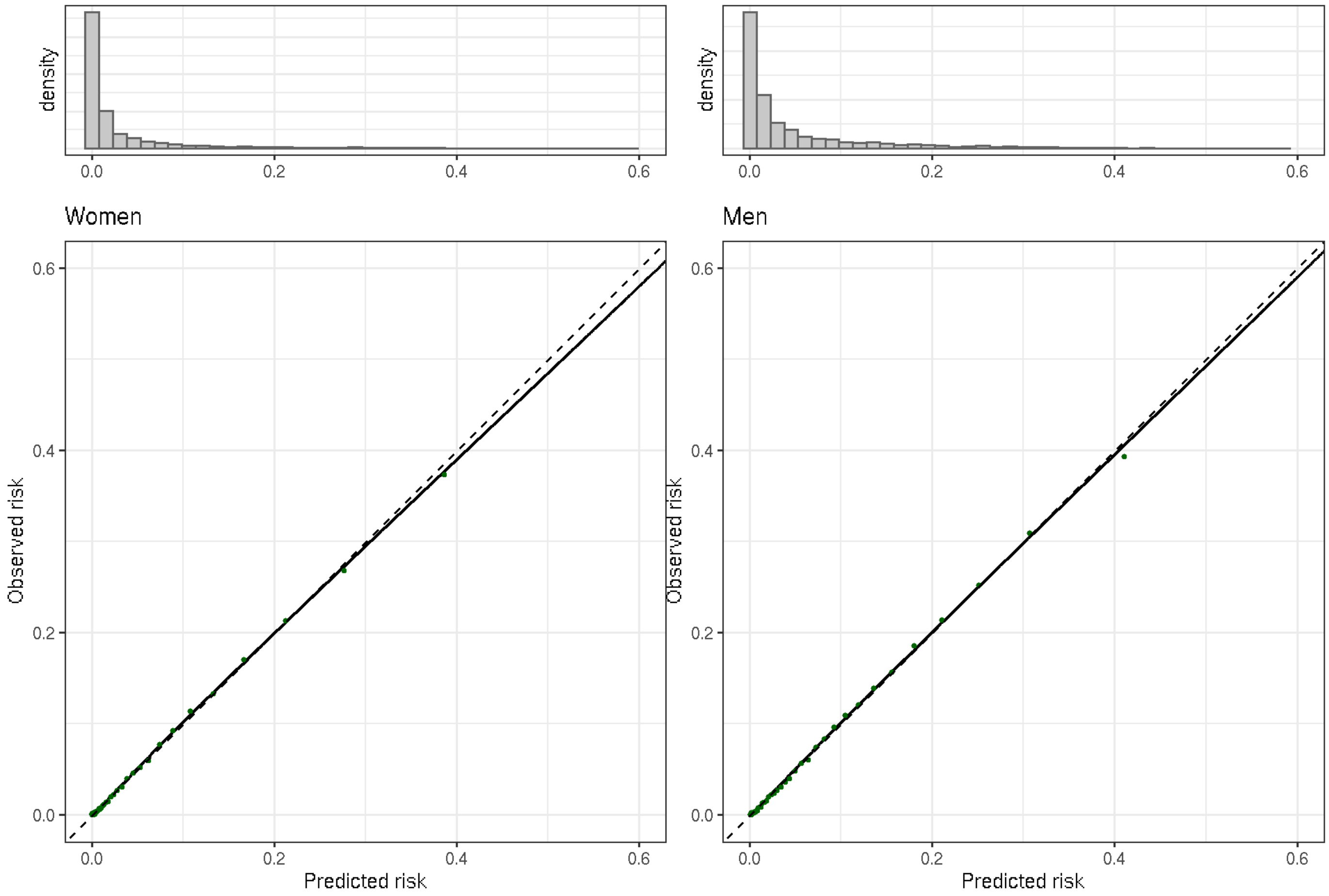

| gender | metric | value | lower CI | upper CI |
|---|---|---|---|---|
| Female | ICI | 0.001 | 0.001 | 0.001 |
| Female | E50 | 0.001 | 0.000 | 0.001 |
| Female | E90 | 0.002 | 0.001 | 0.003 |
| Female | C_index | 0.872 | 0.871 | 0.872 |
| Male | ICI | 0.001 | 0.001 | 0.001 |
| Male | E50 | 0.001 | 0.000 | 0.001 |
| Male | E90 | 0.002 | 0.001 | 0.004 |
| Male | C_index | 0.854 | 0.854 | 0.854 |

# 3 A worked example

We consider the same individuals from the worked example in the main manuscript. We compare predicted risks under the statin intervention from the model with the pleiotropic effect, with the model used in the main manuscript (no pleiotropic effect).

At visit 0, we estimate risk if we don't intervene, and risk under statin treatment. We then consider visit 1 (1-year later). At follow-up visit 1, we estimate the updated risk if we had not intervened. We then compare two further scenarios. The first is where statins has achieved an expected reduction in each of the modifiable risk factors (referred to as 'statins effective' scenario). The second, is where statins has had no effect on the modifiable risk factors (referred to as 'statins ineffective' scenario). This highlights the key difference between the two models, as the pleiotropic model still shows a benefit of the intervention at follow-up visit 1, even when statins have had no effect on the modifiable risk factors, whereas the original model does not. We can also see when statins does have the expected impact on the modifiable risk factors ('statins effective' scenario), the predicted risks under the pleiotropic model are lower, due to the added direct effect of statin use on cardiovascular risk.

| Scenario | Risk original model (%) | Risk pleiotropic model (%) |
|---|---|---|
| Visit 0: Do not intervene | 16.88 | 16.93 |
| Visit 0: Risk under statin intervention | 12.64 | 11.81 |
| Follow-up visit 1: Risk if we hadn't intervened | 17.72 | 17.77 |
| Follow-up visit 1: Risk based on achieved modifiable risk factors (statins ineffective) | 17.72 | 16.59 |
| Follow-up visit 1: Risk based on achieved modifiable risk factors (statins effective) | 13.28 | 12.42 |

# Sensitivity Analysis: Temporal Validation

26 August 2026

**Associated manuscript: A flexible approach to sequential prediction under intervention**

# 1 Overview and methodology

This report summarises performance of the two-component model under temporal validation, where the index dates are defined 1, 2, 3, 4 or 5 years after the start of follow-up.

Start of follow-up is defined as the maximum of: date turned 18; 1 year up to standard follow-up in CPRD; 1st January 2005. This is what was used as the index date in the main manuscript when developing and validating the models. For $t_{fup}$ in $\{1, 2, 3, 4, 5\}$, validation cohorts were restricted to individuals who remained event free and under follow-up within $t_{fup}$ years of their start of follow-up. Time until incident cardiovascular event was then defined relative to these new index dates. We refer to these as temporal validation cohorts.

Model validation was then implemented in the temporal validation cohorts. Predicted risks are under the "continue on current treatment strategy" counterfactual. We evaluate performance of 5-year rather than 10-year predicted risks. This is due to when evaluating performance at 4/5 years post index dates, there would be very few individuals with a further 10 years follow-up. Counterfactual survival times are estimated in the same way as was done in the main manuscript. First statin and antihypertensive use at baseline was reassessed. Counterfactual survival times under the treatment strategy of "do nothing" were then estimated and used to estimate calibration and discrimination.

Section 2 contains a summary table of calibration (ICI, E50, E90) and discrimination (C-statistic) metrics, with 95% confidence intervals. Section 3 contains calibration plots (plus grouped Kaplan-Meier estimates vs mean predicted risk in 25 groups defined by predicted risk).

# 2 Summary Table of Performance Metrics

We find that as the index date gets further away from start of follow-up, the calibration gets worse (symptomatic of calibration drift), however not by a large amount. There is a more considerable drop in discrimination (0.863 to 0.838 for females, 0.845 to 0.814 for males). This is likely due to higher risk individuals having events in the first 5 years and being removed from the validation cohort. This will result in a healthier cohort, within which it is harder to discriminate between high and low risk individuals.

Calibration and discrimination metrics (95% CI)

| Follow-up (years) | N | ICI | E50 | E90 | C-statistic |
|---|---|---|---|---|---|
| Male | | | | | |

| Calibration and discrimination metrics (95% CI) | | | | | |
|---|---|---|---|---|---|
| Follow-up (years) | N | ICI | E50 | E90 | C-statistic |
| 1 | 850304 | 0.001 (0.001, 0.001) | 0.000 (0.000, 0.001) | 0.001 (0.001, 0.002) | 0.845 (0.845, 0.845) |
| 2 | 727831 | 0.001 (0.001, 0.002) | 0.001 (0.000, 0.001) | 0.003 (0.002, 0.004) | 0.837 (0.837, 0.837) |
| 3 | 628389 | 0.002 (0.002, 0.003) | 0.001 (0.001, 0.002) | 0.005 (0.003, 0.006) | 0.829 (0.829, 0.829) |
| 4 | 547071 | 0.003 (0.002, 0.003) | 0.002 (0.001, 0.002) | 0.007 (0.004, 0.008) | 0.821 (0.821, 0.821) |
| 5 | 481009 | 0.004 (0.003, 0.004) | 0.002 (0.002, 0.003) | 0.009 (0.007, 0.010) | 0.814 (0.814, 0.814) |
| Female | | | | | |
| 1 | 826655 | 0.001 (0.001, 0.001) | 0.000 (0.000, 0.001) | 0.002 (0.001, 0.003) | 0.863 (0.863, 0.863) |
| 2 | 692698 | 0.002 (0.001, 0.002) | 0.001 (0.000, 0.001) | 0.004 (0.002, 0.005) | 0.855 (0.855, 0.855) |
| 3 | 590933 | 0.002 (0.002, 0.003) | 0.001 (0.001, 0.001) | 0.006 (0.004, 0.007) | 0.849 (0.849, 0.849) |
| 4 | 515915 | 0.003 (0.002, 0.003) | 0.001 (0.001, 0.002) | 0.008 (0.006, 0.009) | 0.843 (0.843, 0.843) |
| 5 | 455388 | 0.004 (0.003, 0.004) | 0.001 (0.001, 0.002) | 0.011 (0.008, 0.012) | 0.838 (0.838, 0.838) |

# 3 Calibration Plots

The plots below show the smoothed calibration curve (red line) together with grouped Kaplan-Meier estimates (points, with 95% confidence intervals) across 25 risk groups, for each gender and follow-up time.

We observe a calibration drift, and find that as the index date gets further away from start of follow-up, the model begins to over predict. This aligns with previous research which highlights a downwards secular trend in the incidence of cardiovascular disease observed in UK primary care (Pate, Staa, and Emsley 2020). The magnitude of the misscalibration is not large

# 3.1 Female

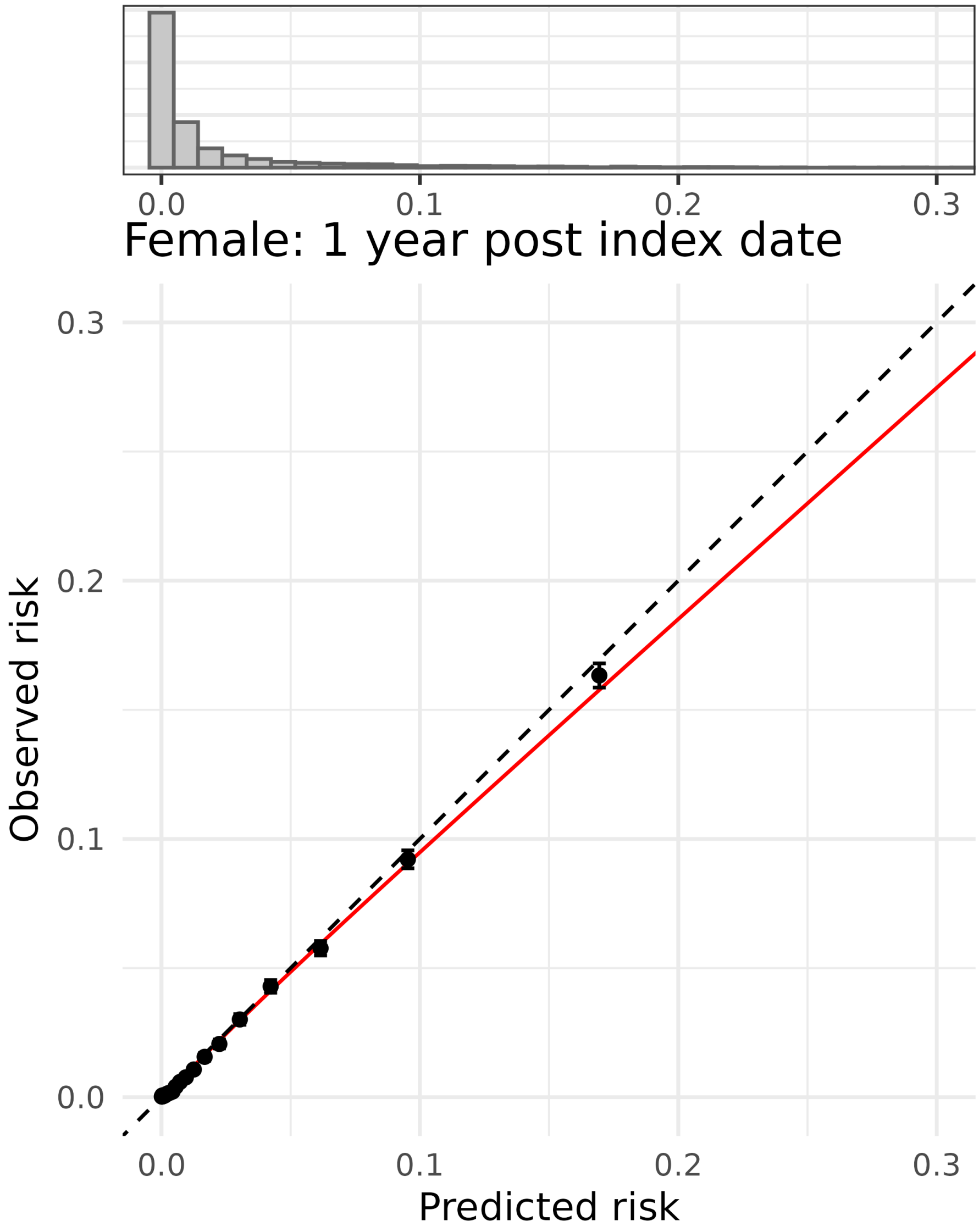
0.0
0.1
0.2
0.3
Female: 1 year post index date
Observed risk
Predicted risk

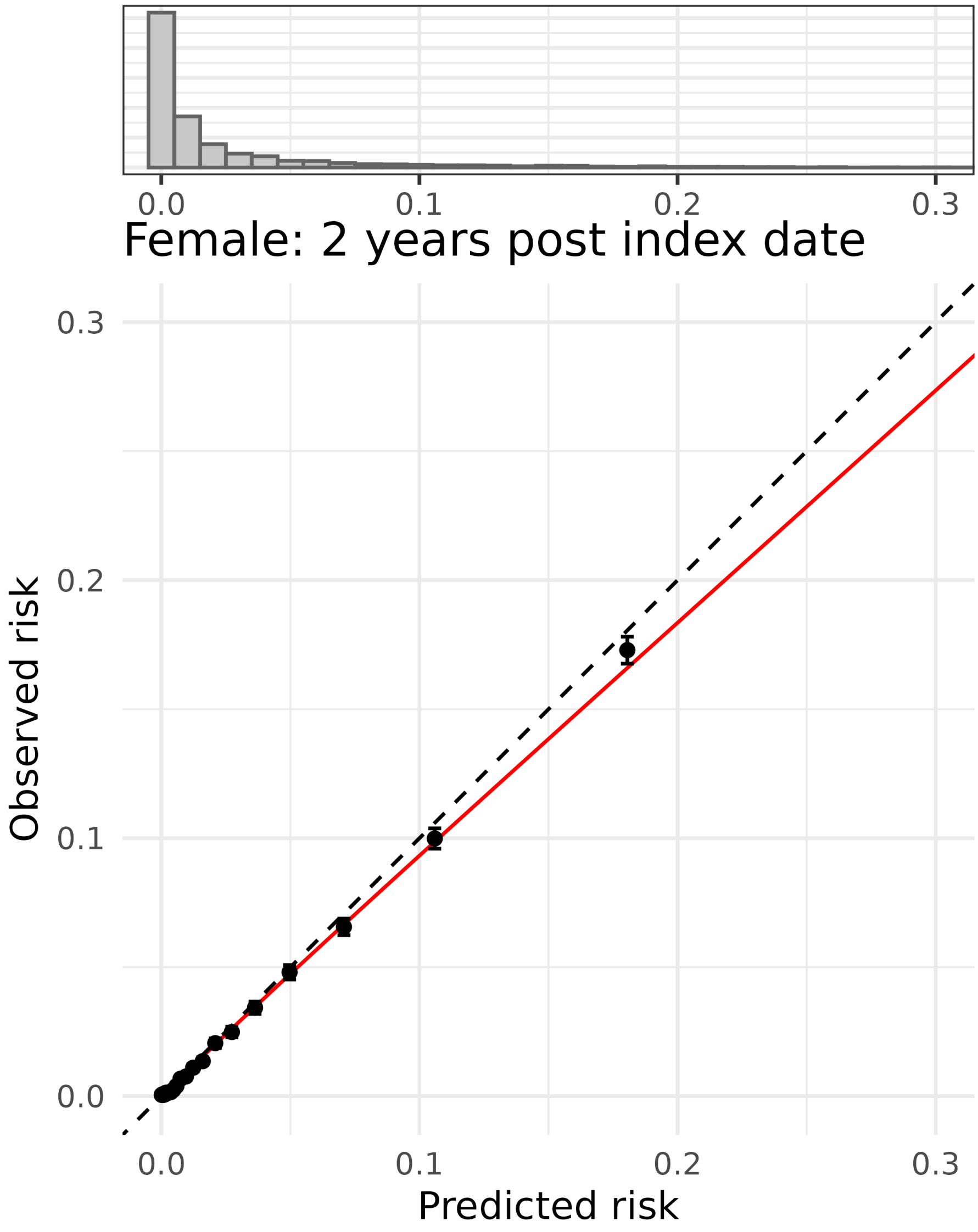
Female: 2 years post index date
0.0
0.1
0.2
0.3
Observed risk
Predicted risk

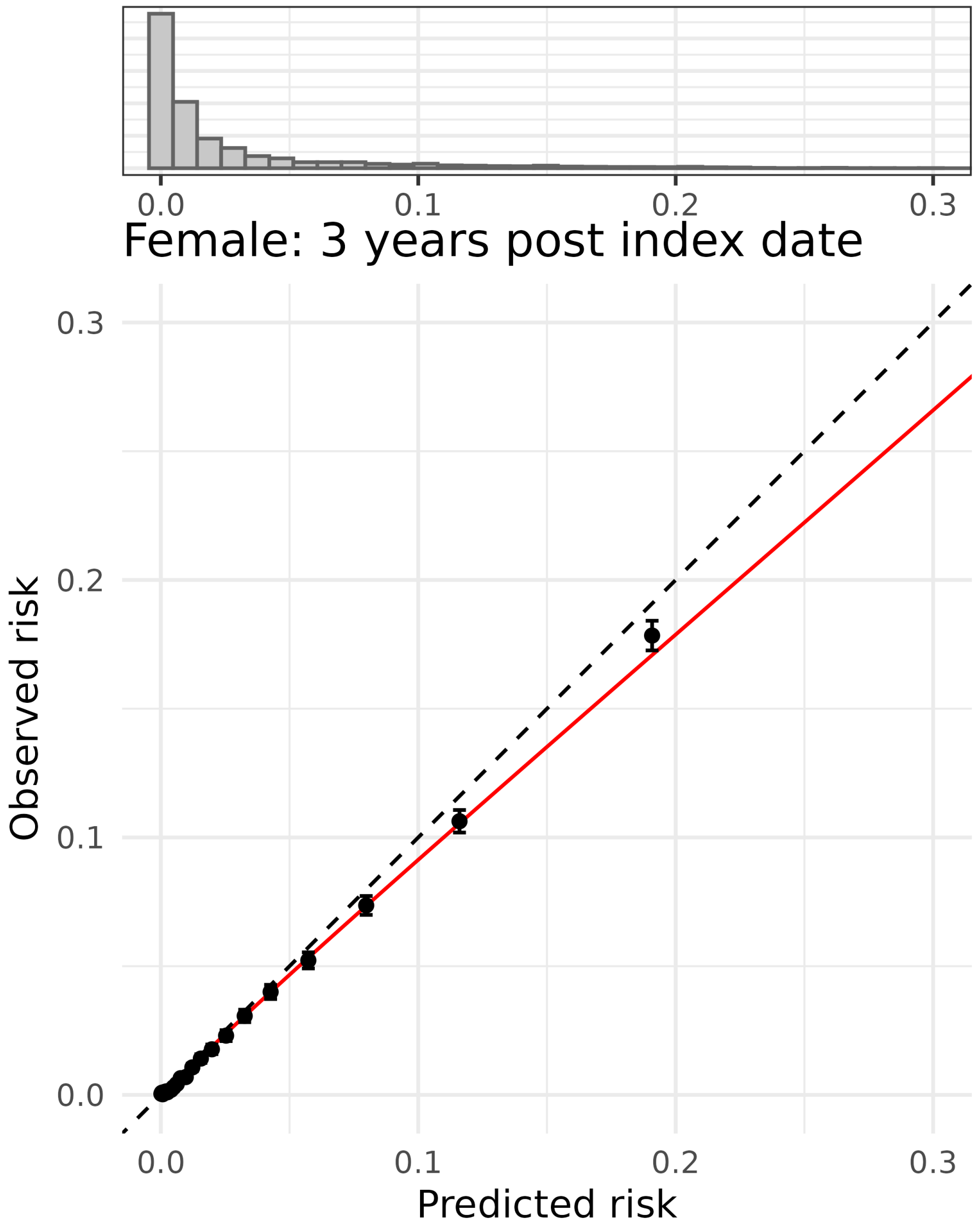
0.0
0.1
0.2
0.3
Female: 3 years post index date
Observed risk
Predicted risk

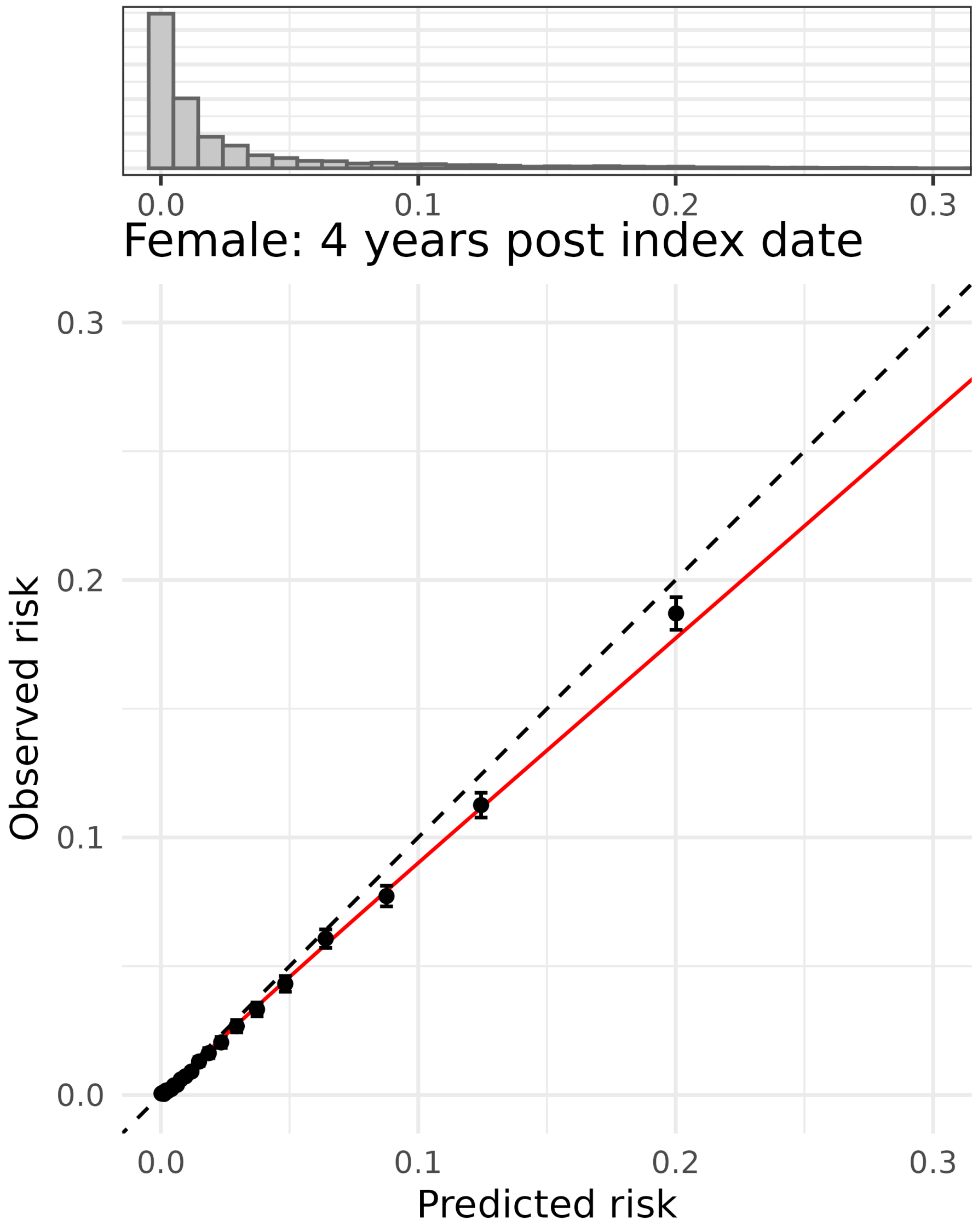
0.0
0.1
0.2
0.3
Female: 4 years post index date
Observed risk
Predicted risk

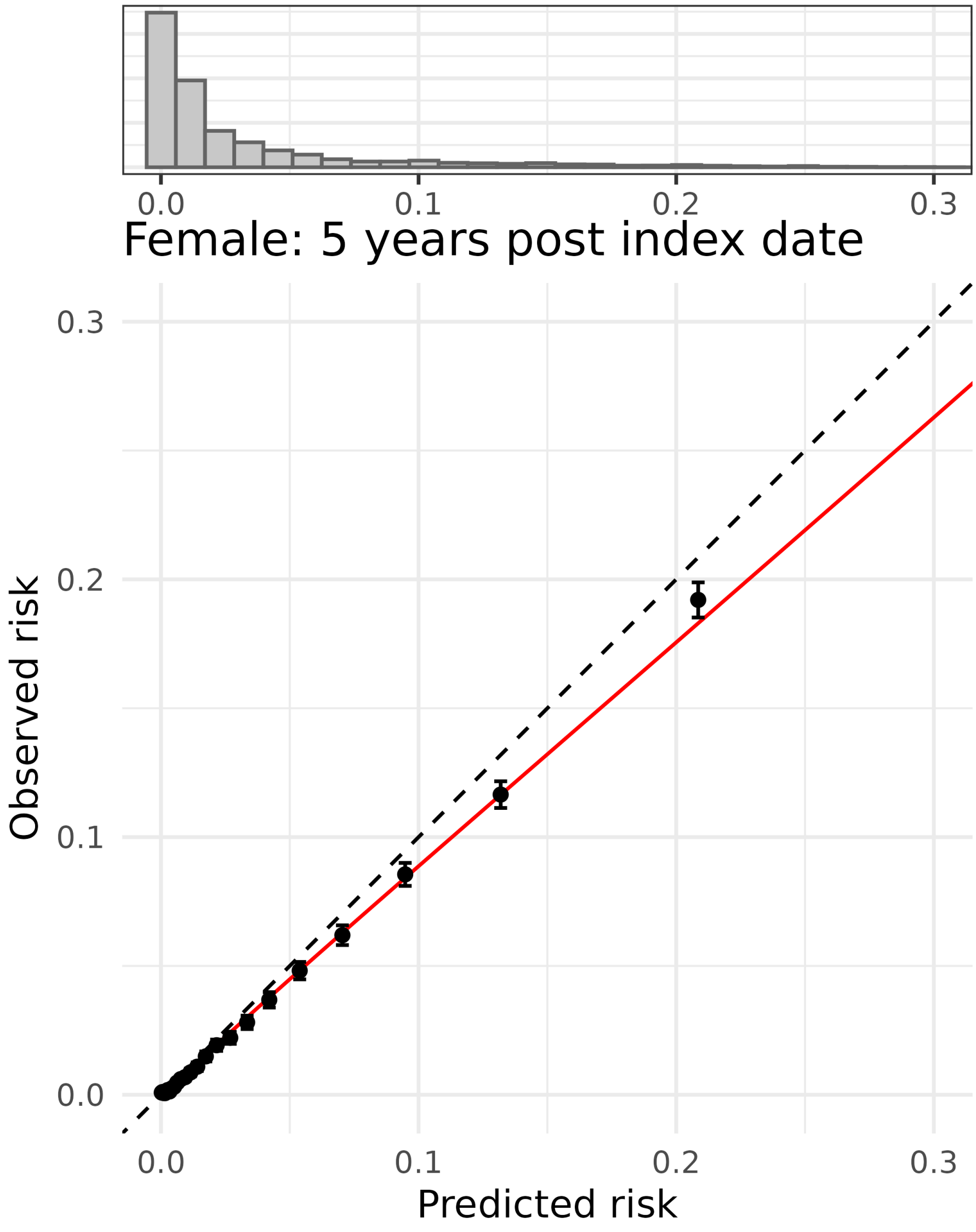
Female: 5 years post index date
Observed risk
Predicted risk
0.0
0.1
0.2
0.3

# 3.2 Male

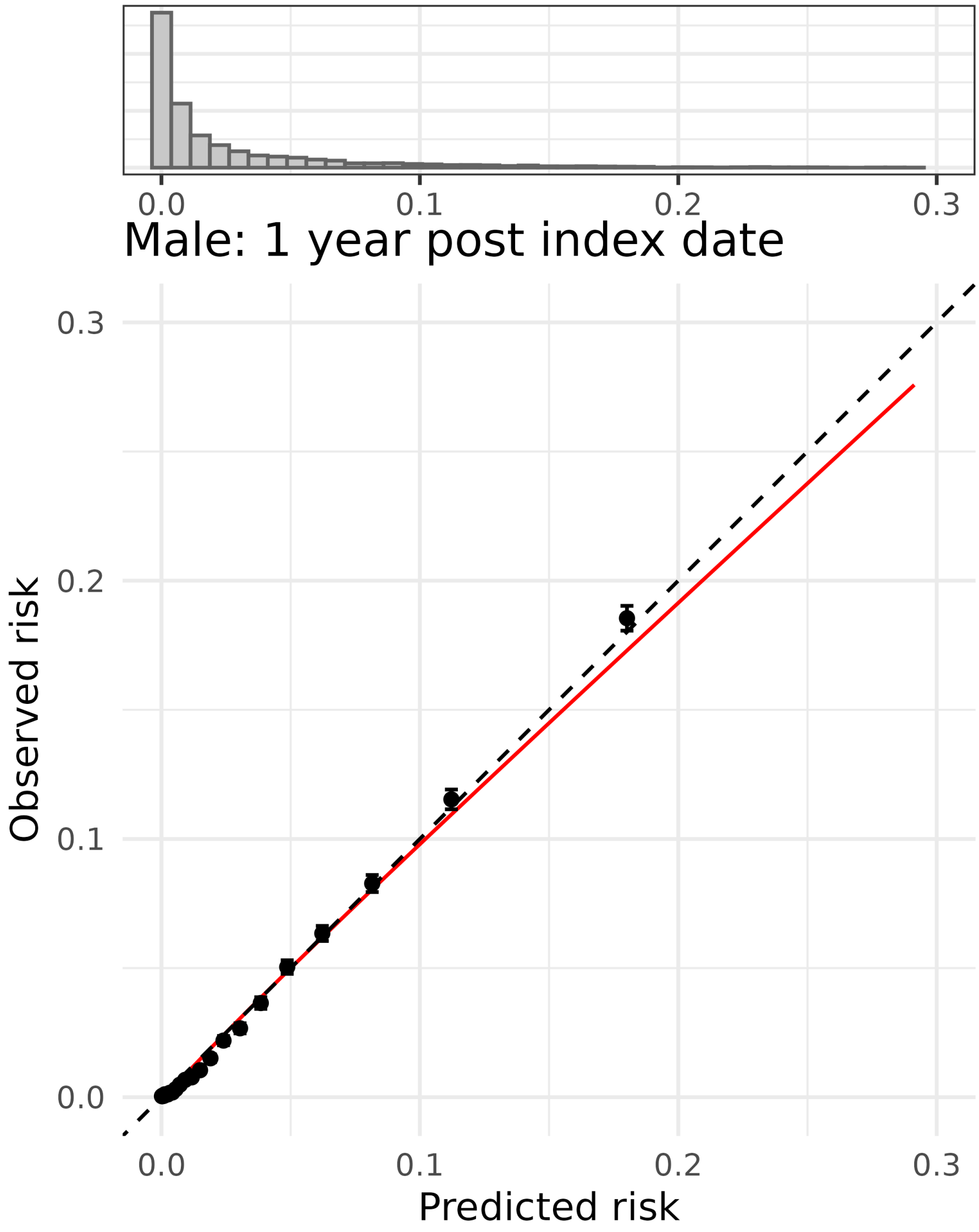
0.0
0.1
0.2
0.3
Male: 1 year post index date
Observed risk
Predicted risk

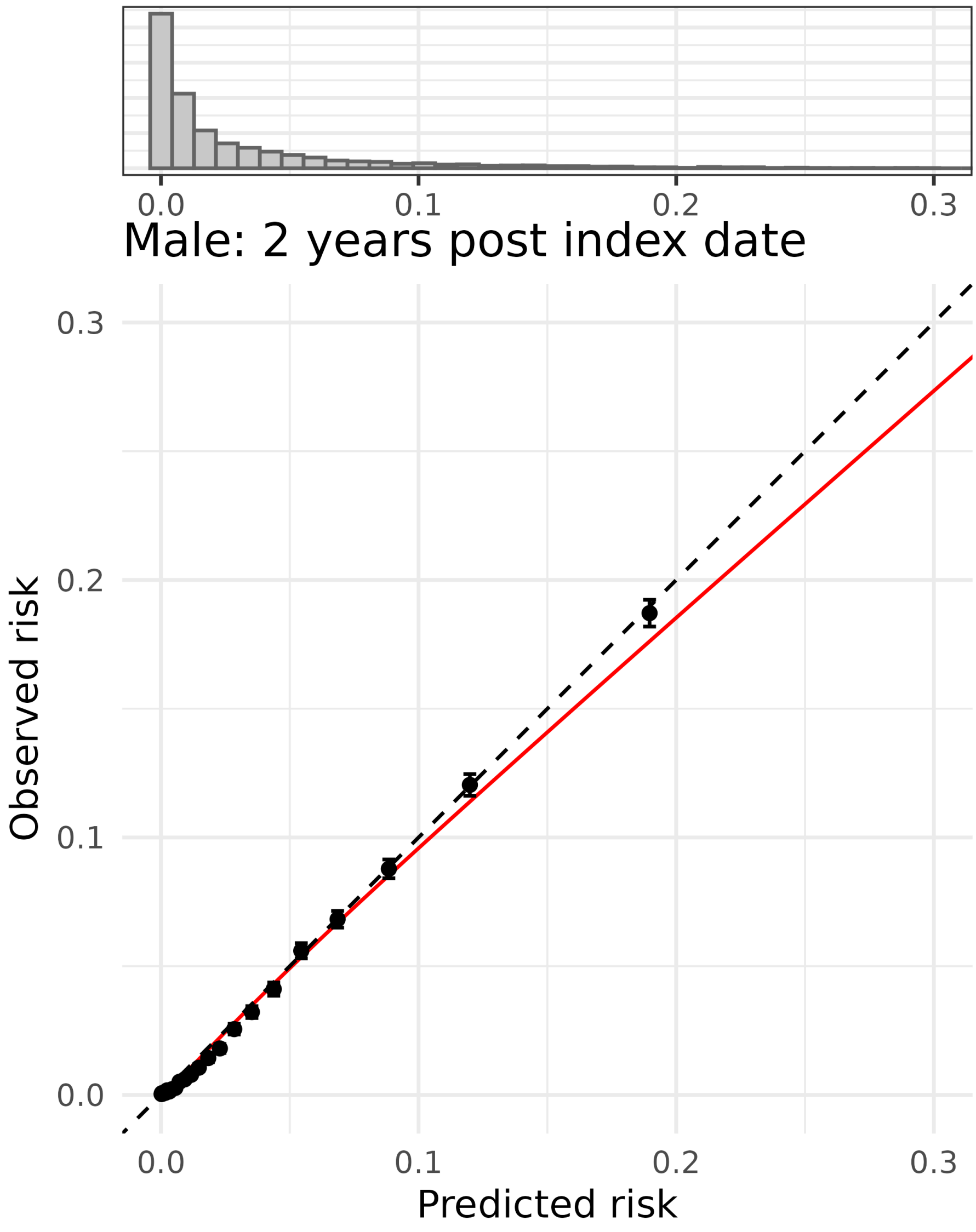
Male: 2 years post index date
0.0
0.1
0.2
0.3
Observed risk
Predicted risk

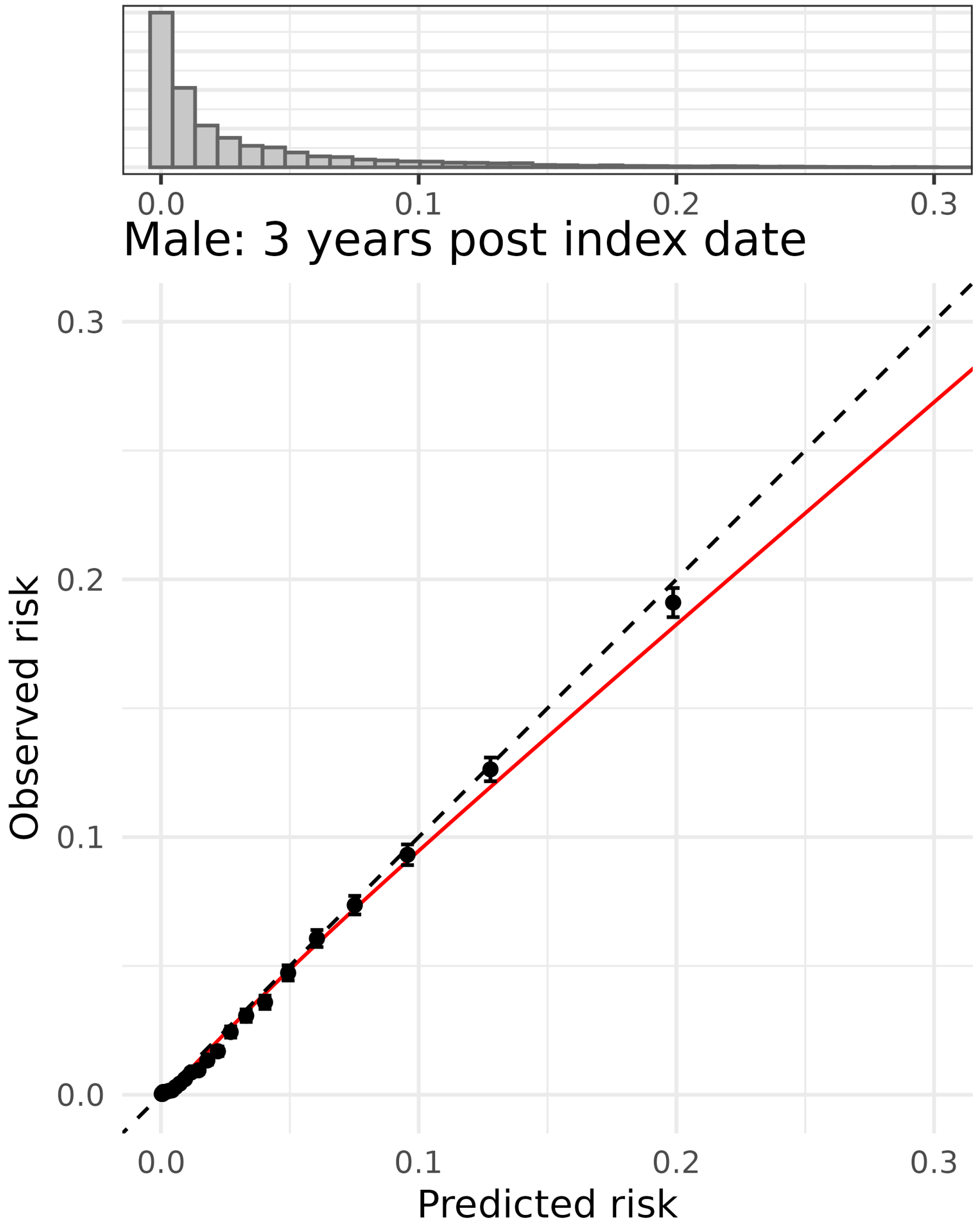
0.0
0.1
0.2
0.3
Male: 3 years post index date
0.3
0.2
0.1
0.0
Observed risk
0.0
0.1
0.2
0.3
Predicted risk

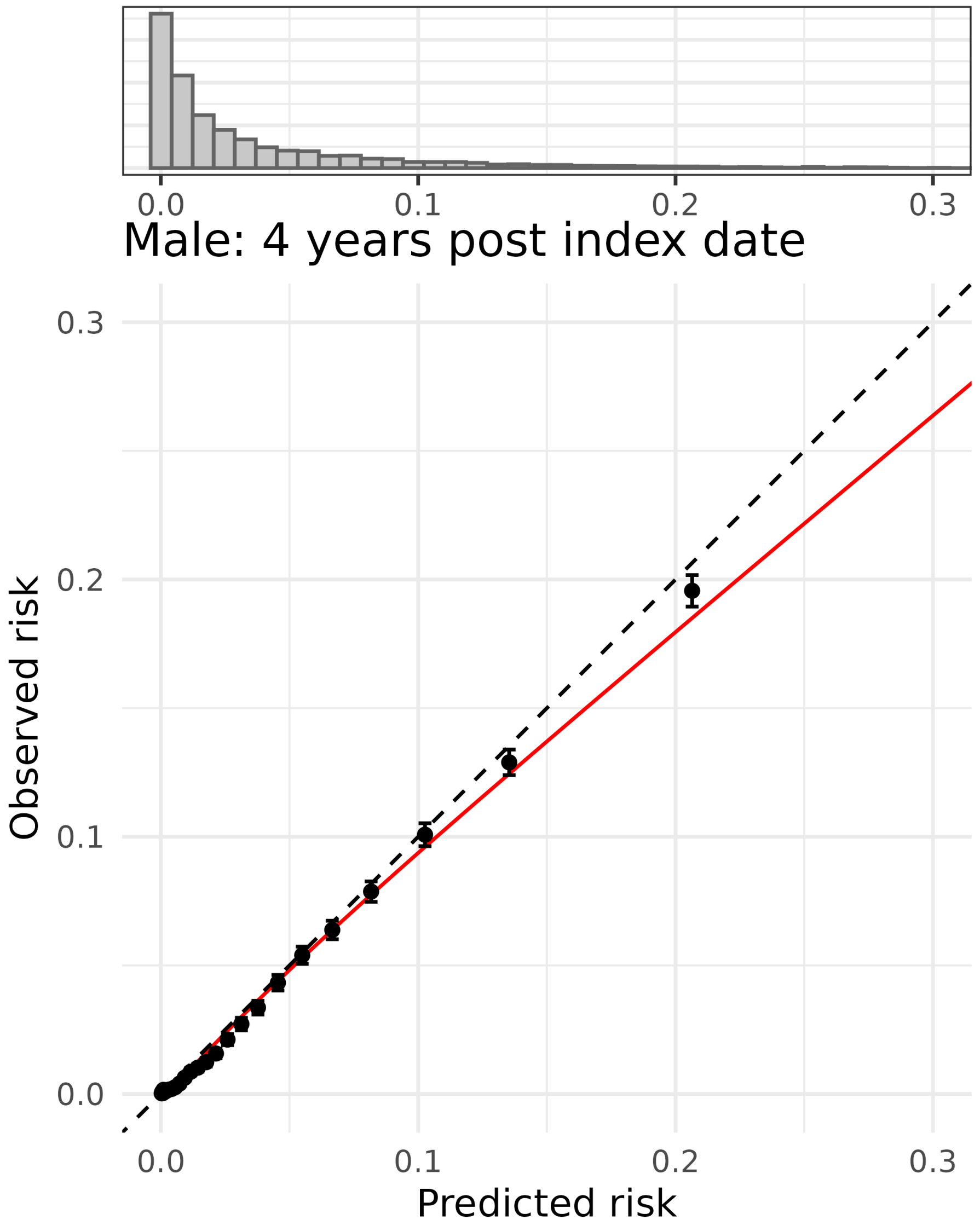
0.0
0.1
0.2
0.3
Male: 4 years post index date
0.3
0.2
0.1
0.0
Observed risk
0.0
0.1
0.2
0.3
Predicted risk

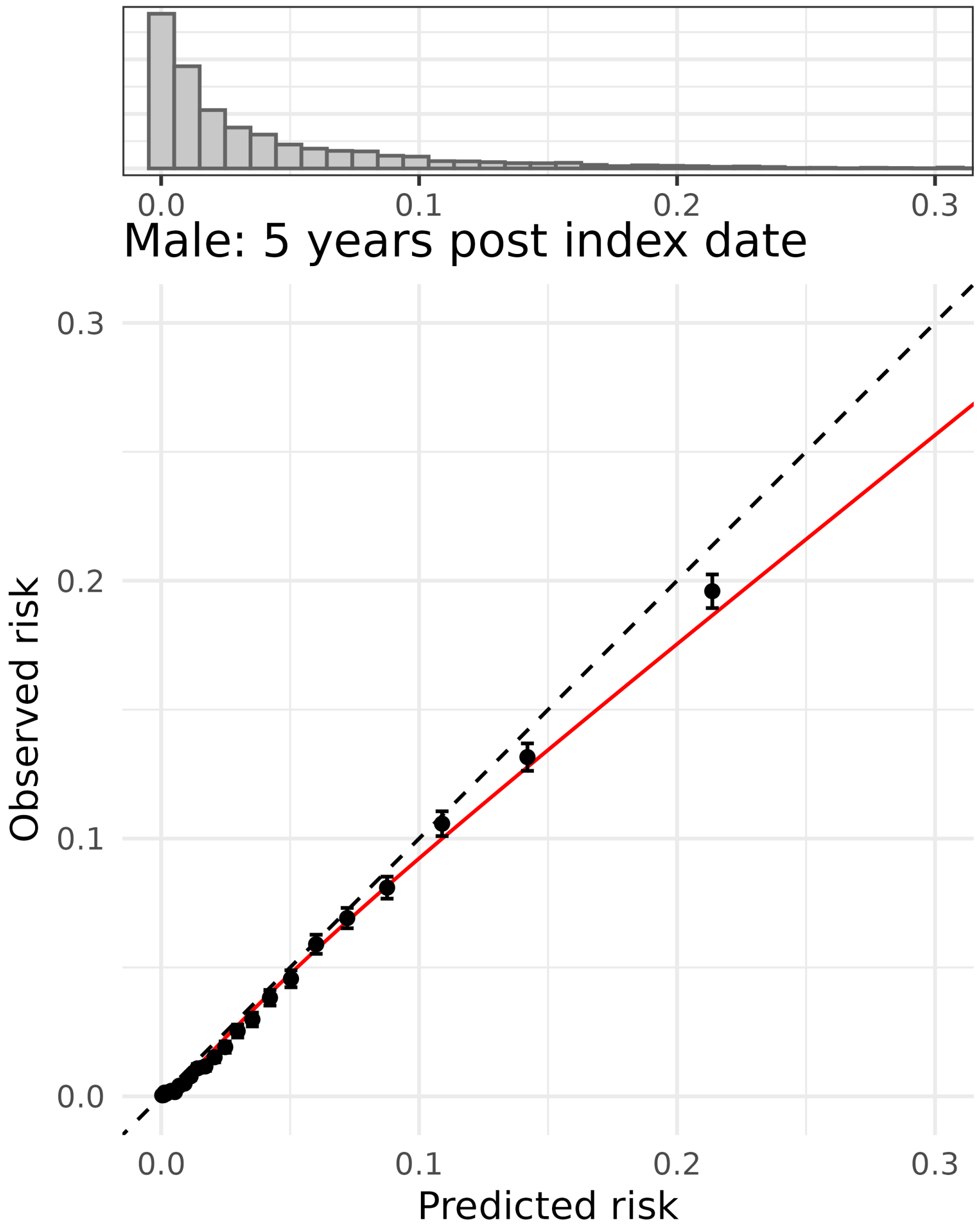